\documentclass[12pt,oneside]{book}
\usepackage{latexsym,graphicx,amssymb,natbib,afterpage,aasmod,aas_stuff}

\bibpunct{(}{)}{;}{a}{}{,}

\begin{document}

\frontmatter

\thispagestyle{empty}
\hbox{ }
\vspace{-2.0cm}
\begin{center}
\Large{\textbf{On the Development and Applications of}}\\
\Large{\textbf{Automated Searches for Eclipsing Binary Stars}}\\
\end{center}
\vspace{0.8cm}
\normalsize
\centerline{A thesis presented}
\vspace{0.8cm}
\centerline{by}
\vspace{0.8cm}
\centerline{\Large{Jonathan Devor}}
\vspace{0.8cm}
\normalsize
\centerline{to}
\vspace{0.8cm}
\centerline{The Department of Astronomy}
\vspace{0.8cm}
\centerline{in partial fulfillment of the requirements}
\setlength{\baselineskip}{24pt}
\centerline{for the degree of}
\centerline{Doctor of Philosophy}
\centerline{in the subject of}
\centerline{Astronomy}
\vspace{0.8cm}
\centerline{Harvard University}
\centerline{Cambridge, Massachusetts}
\vspace{0.8cm}
\centerline{13 February 2008}
\newpage

\thispagestyle{empty} \vspace*{3.8in}
\centerline{\ooalign{\mathhexbox20D\crcr\hfil\raise.07ex\hbox{c}\hfil}~2008, by Jonathan Devor}
\centerline{All rights reserved.}
\newpage

\addcontentsline{toc}{section}{Abstract}

\noindent Dissertation Advisor: Prof.\ David B.\ Charbonneau \hfill Jonathan~Devor

\begin{center}
\large
\bf
On the Development and Applications of\\ Automated Searches for Eclipsing Binary Stars
\end{center}

\vskip 0.4cm
{\centerline {\bf Abstract}}

{ \parskip 0pt
Eclipsing binary star systems provide the most
accurate method of measuring both the masses and radii of stars.
Moreover, they enable testing tidal synchronization and
circularization theories, as well as constraining models of
stellar structure and dynamics. With the recent availability of
large-scale multi-epoch photometric datasets we were able to study
eclipsing binary stars en masse. In this thesis, we analyzed
185,445 light curves from ten TrES fields, and 218,699 light
curves from the OGLE~II bulge fields. In order to manage such
large quantities of data, we developed a pipeline with which we
systematically identified eclipsing binaries, solved for their
geometric orientations, and then found their components' absolute
properties. Following this analysis we assembled catalogs of
eclipsing binaries with their models, computed statistical
distributions of their properties, and located rare cases for
further follow-up, including T-Cyg1-03378, which has unusual
eclipse timing variations. Of particular importance are low-mass 
eclipsing binaries, which are rare yet critical for resolving the 
ongoing mass-radius discrepancy between theoretical models and
observations. To this end, we have discovered over a dozen new
low-mass eclipsing binary candidates and spectroscopically
confirmed the masses of five of them. One of these confirmed
candidates, T-Lyr1-17236, is especially interesting because of its
long orbital period. We examined T-Lyr1-17236 in detail and found 
that it is consistent with the magnetic disruption hypothesis. 
Both the source code of our pipeline and the complete list of 
our candidates are freely available.}
\clearpage

\singlespace
\tableofcontents
\newpage

\doublespace    

\thispagestyle{empty}
\vspace*{0.5in}
\begin{flushright}
\singlespace
\large

\emph{``The only way to realize the full scientific\\}
\emph{benefit of our observations is to share the\\}
\emph{data with our competition.''\ \ \ \ \ \ \ \ \ \ \ \ \ \ \ \ \ \ \ \\}
\vspace*{0.25in}
--Bohdan Paczy\'{n}ski {\small \ (1940 - 2007)}\\

\vspace*{2.5in}
\bf{For my Family.}
\doublespace
\normalsize
\end{flushright}
\newpage

\newpage

\thispagestyle{plain}

\addcontentsline{toc}{section}{Acknowledgments}

\vskip 0.5cm
{\centerline {\Large \bf Acknowledgments}}
\vskip 0.5cm

\normalsize

Alfr$\rm\acute{e}$d R$\rm\acute{e}$nyi once said, and many others
subsequently paraphrased him, that ``a mathematician is a device
for turning coffee into theorems.'' Indeed, in the hard-nosed
professional world we live in, be it academic or otherwise, it is
easy to forget all the human elements that must fall into place
for one to be able to produce original work.

In the years leading up to graduate school, I was extraordinarily
fortunate to be surrounded by friends, family, and teachers who
provided me with continuous support and guidance. My parents, who
would patiently answer an unending stream of questions from a very
persistent child, would later encourage me to follow my heart's
desire, wherever it might lead me. My decision to spend over a
half-decade in school to study Astronomy, was never once
questioned. However, as so often the case, my path to Astronomy
was by no means a straightforward one.

Carl Sagan's \textit{Cosmos} provided me with the first major push
in this direction, and numerous visits to the San Francisco
Exploratorium and the Boston Museum of Science cemented my love
for the physical sciences at a young age. However, if it was not
for an inspirational high school physics teacher named Reuben
Tel-Dan, I would probably never have considered further studying
physics in college. After serving in the Israeli military, I
enrolled at the Hebrew University of Jerusalem, where I would meet
a second inspirational teacher-- Avishai Dekel. I first met
Professor Dekel at a talk he gave during a freshman survey course.
In that lecture I was first introduced to the concepts of dark
matter, dark energy, and the concordance model of cosmology. He
concluded with the current attempts to simulate the Universe at
the grandest scales. I was immediately hooked, and proceeded to
work in his group until my graduation. It is due to this
experience that I chose to go to graduate school, while most of my
friends were pursuing jobs at high tech companies.

My first encounter with Harvard University resulted in a bit of a
culture shock due to its immense range of possibilities, both
academically and otherwise. In the following years I would make a
point of trying as many activities as I could, sometimes to the
dismay of my professors, from rowing in crew and Dragon Boat
teams, to building robots that play soccer. When it came time to
choose a research topic I had a very difficult time settling on a
single project. I was naturally drawn to the search for
Exoplanets, which was at that time and probably still is, ranked
as the sexiest topic in Astronomy. However, following the sage
advice of Bob Noyse and Dimitar Sasselov, I decided to start with
something easier and more of a ``sure thing''-- finding eclipsing
binary stars. This topic was intended to be a stepping stone, to
get my feet wet, however the more I learned about eclipsing
binaries, the more interested I got. The two most attractive
aspects of this field were the fact that the necessary
observational data were already available from the OGLE team and
others, and the fact that even a quick perusal through this data
reveals strange and curious variable stars that no one had ever
studied before.

My first attempts at automatic clustering and classification the
OGLE data were seen as ``butterfly collecting,'' since I had no
well-defined scientific goal in mind. However, the idea of
automated searches showed great promise, and I was advised to
identify specific types of targets with well established
scientific motivations. This revised approach led me to the
starting point of this thesis, the development of DEBiL. With the
completion of the DEBiL pipeline in 2004, I was intently thinking
of how to leverage DEBiL's capabilities in order to gain access to
the absolute physical properties of binaries. The solution to this
came to me during the $205^{th}$ meeting of the American
Astronomical Society in San Diego. This idea was implemented to
form MECI-express, and later incrementally refined to form MECI.
The implementation of MECI was very exciting, since I saw it as my
most significant, entirely original, scientific contribution; and
indeed much of this thesis centers around it. I would soon
realize, however, that to confirm the accuracy of the DEBiL/MECI
pipeline I would need to make direct observations, but because the
OGLE targets I was studying at the time were so dim, they would be
very difficult to follow-up. That predicament would be solved with
the arrival of Dave Charbonneau, who brought with him the TrES
dataset of comparably bright stars.

Even before Dave's arrival to the CfA, I heard many impressive
stories from senior graduate students who remembered him as a
Harvard grad. Indeed, he is one of the most charismatic professors
I have ever met, and has a real skill for explaining intricate
ideas in a clear and simple way. Needless to say, I was thrilled
when he accepted me as an advisee. The following years would be
quite transformative. With Dave's guidance, I learned to switch
from a mostly engineering-oriented mindset to one that is more
centered on addressing the scientific questions at hand. Thus even
though I continue to enjoy building things that work, I learned to
construct the machinery around the research needs, and not the
other way around. One can see this incremental progression
throughout this thesis, and as a consequence, this dissertation
will hopefully better benefit the astronomical community.

\newpage

\mainmatter

\chapter{Introduction
\label{chapter1}}

Astronomical research has traditionally been divided into two
interrelated categories: Observational Astronomy, which confirms
or refutes hypotheses, constrains models, and finds new empirical
relations, and Theoretical Astronomy (or Astrophysics), which
constructs and refines physical models, and in so doing provides
direction and motivation for further observations. In recent
decades, a third category has been emerging: Computational
Astronomy. This category is not simply an elaboration of
theoretical tools, nor is it just a larger scale approach to
observational data reduction, but rather it is a methodology that
promises a fundamentally new way of solving astronomical problems.
In this thesis, we introduce the reader to the advances that a
computational approach can achieve in the field of eclipsing
binary analysis, which until recently has been considered by many
to be a mature and largely solved area of research.

A binary star system consists of two stars that orbit one another.
When the orbital plane of such a binary is nearly parallel to an
observer's line of sight (i.e. with an inclination of $i \simeq
90^\circ$), the binary components will be seen periodically
eclipsing one another. If such a system is fortuitously so aligned
with respect to Earth, it is termed an eclipsing binary (EB). The
probability that an observed binary will be seen eclipsing is on
the order of 1 in a 100, however this fraction varies by as much
as an order of magnitude in different regions of the sky
\citep{Devor05, Devor08}. The observed fraction, however, is
biased downwards by the fact that many EBs have very long orbital
periods or shallow eclipses, making them difficult to identify.
The most direct way to identify EBs is by compiling series of
multi-epoch photometric observations of stars, thus monitoring how
their brightness changes as a function of time. Such monitoring is
often done for large fields on the sky, thereby photometrically
capturing many thousands of stars simultaneously. These time
series are called light curves (LCs), and since for EBs they
typically repeat themselves with a fixed period, they are usually
shown folded modulo their respective periods. Such folded LCs are
called phased LCs, and by convention, their time-axes are labels
in units of their period. Binaries can also be identified through
their motion in the sky (astrometric binaries) and through the
doppler shift of their spectra (spectroscopic binaries). Both
these techniques can be used to identify binaries that are not
eclipsing, however, applying them to observe large numbers of
stars is technically very challenging. Furthermore, despite recent
advances in measuring additional properties of stars that do not
eclipse [e.g., \citet{Perrin03, Zucker07a, Zucker07b}], EBs remain
unique in that they enable the accurate measurement of both the
masses and radii of a large number of stars.

EBs are generally divided into three classes according to the
shape of their LC: Algol-type, $\beta$~Lyrae-type, and
W~Ursae~Majoris-type variables. Algol-type variables (EA) are EBs
whose brightness remains almost constant between eclipses. In this
type of EB, the eclipse duration is a small fraction of the
binary's orbital period, indicating that the binary components are
detached from one another. In contrast, $\beta$~Lyrae-type
variables\footnote{The accepted symbol for this class is EB,
however we will be using this symbol throughout this thesis as an
acronym for ``eclipsing binary''.} vary their brightness
continuously between eclipses, indicating that their components
are sufficiently close to bring about significant tidal
distortions. In extreme cases, an evolving component will fill its
Roche Lobe and then spill over. Such binaries that undergo one-way
mass transfer are called semi-detached. However, like Algol-type
variables, the eclipses of these variables do not typically have
the same depth and shape. Finally, W~Ursae~Majoris-type variables
(EW) are similar to $\beta$~Lyrae variables in that their
brightness varies continuously, however the eclipses of these
variables are nearly identical, indicating that the components
have similar temperatures. This is often due to the fact that the
components are in contact and have a common envelope. In this
thesis, we limit ourselves to Algol-type variables, as the other
two classes involve far more complex models, and therefore require
significantly more computational resources for their study.

Because we are able to accurately measure the absolute properties
of the EB components (i.e. their mass, radii, luminosity, etc.),
they have become critical tools for the study of star formation,
stellar structure, stellar evolution, and stellar dynamics.
Furthermore, EBs have been used as standard candles
\citep{Stebbing10, Paczynski97} to both determine the size and
structure of the Galaxy, as well as to constrain the cosmological
distance ladder \citep{Bonanos06}. Ongoing work will likely help
resolve current uncertainties regarding binary formation and
parameter distribution \citep{Duquennoy91}, and may also improve
our understanding of Type~Ia supernovae \citep{Iben84}. However,
despite their great importance, only a small fraction of the
observable EBs have been identified, and only a comparably small
number of these systems have been analyzed.

In this thesis, we demonstrate how one can greatly increase the
speed and efficiency of EB analysis by constructing automated
pipelines. Such pipelines will ultimately become necessary to take
advantage of the exponentially growing number of photometric LCs
being made available \citep{Szalay01}. We hope to show here that
not only can this be achieved without excessive human effort and
computational resources, but also that the derived data contain a
wealth of scientifically interesting information, which can and
should be used to solve open questions in astronomy and to
facilitate novel insights.

\section {A Brief Historical Overview of Eclipsing Binary Analysis}
\label{subsecHistory}

The first person to record an eclipse of a binary was Geminiano
Montanari, who noted a minimum in the brightness of Algol ($\beta$
Persei) on November 8, 1670, however, the variability of Algol is
likely to have been noticed far earlier than that\footnote{For a
more comprehensive survey of the history of binary star
observations, see \citet{Aitken64} and \citet{Kopal90}.}.
Nevertheless, it was not until 1783 when the British amateur
astronomer John Goodricke first noticed that Algol's minima were
periodic and suggested that Algol was being eclipsed by a large
opaque body. The Royal Society of London found this report to be
of such importance that they awarded Goodricke their highest
award, the Copley Medal. However, it would take over a century
before the first systematic efforts to quantify the photometric
brightness of variable stars were made, first by eye and later
using photosensitive devices, which allowed significant progress
in modeling such variable stars. One of the most notable successes
during this period was that of \citet{Stebbing10}, who used a
Selenium detector to discover Algol's secondary eclipse, thereby
confirming that it was an EB whose secondary component was simply
a dimmer star. Soon after, Henry Norris Russell developed more
generalized analytical tools for modeling such EBs
\citep{Russell12a, Russell12b}. Russell, together with his student
Harlow Shapley, then built upon this work to include the effects
of stellar limb darkening \citep{Russell12c, Russell12d}. During
the decades that followed, an increasing number of theoreticians
continued this work and constructed models that include star spots
and surface inhomogeneities, stellar non-sphericity due to tidal
distortions, gravity darkening, mutual reflections, gravitational
perturbation by additional stellar components, relativistic
effects, and others. Accounting for these effects required
increasingly sophisticated mathematical techniques, as well as
tedious numerical calculation and the construction of many lookup
tables. These efforts culminated in the definitive work done by
the Czech-born astronomer Zden\v{e}k Kopal, beginning in the 1930s
and continuing throughout the following 60 years [see e.g.,
\citet{Kopal59}].

Soon after the arrival of the first programmable computers,
researchers began writing codes that generate EB LCs, as well as
routines that fit the model parameters so that the generated LC
will match the observational data (see
\S\ref{subsecParameterExtraction}). With the growing availability
and speed of computers, research efforts slowly shifted from
idealized models that can be calculated using only
paper-and-pencil arithmetic, to more complex models that require
iterative numerical computations. The first widely accepted code
if this kind was WD, originally developed by \citet{Wilson71}, and
named for its authors. This code has been incrementally improved
and built upon numerous times [e.g., \citet{Wilson94, Prsa05}] and
is generally considered the most comprehensive EB modeling code
available today. Shortly thereafter, based on the \citet{Nelson72}
model, the EBOP (Eclipsing Binary Orbit Program) code was released
\citep{Etzel81, Popper81b}. EBOP has also been continuously
improved, and now contains elaborate methods for parameter
uncertainty estimation \citep{Southworth04a, Southworth04b,
Southworth05}.

Today, much of the computational effort has been redirected from
creating ever more comprehensive binary models that attempt to
include every possible phenomenon to more simplified, faster codes
that are more robust and can reliably analyze large datasets of
LCs with minimal human intervention. \citet{Wyithe01, Wyithe02}
produced the first such code based on WD. Soon afterwards,
\citet{Tamuz06} repeated this effort, building upon the EBOP code
and designated the end product EBAS (Eclipsing Binary Automated
Solver). Included among these is the DEBiL (Detached Eclipsing
Binary Light curve) fitter code \citep{Devor04, Devor05}. DEBiL
was the starting point of this thesis, and was developed from the
ground up to be simpler and more robust than any of the previous
codes, thus enabling it to systematically analyze the largest LC
datasets available, with essentially no need for human
intervention at all. In the upcoming chapters of this thesis we
will describe DEBiL, as well as how it can be used to extract
scientifically interesting information out of the vast photometric
datasets available today.

\section {A Primer on Parameter Extraction}
\label{subsecParameterExtraction}

It is important to underline the difference between the analytic
parameter extraction and the numerical/iterative methodology that
superseded it. The analytic approach attempts to measure specific
geometric features of the LC, such as primary and secondary
eclipse depths and durations, ingress and egress slopes, and times
of the center of the eclipses, and uses these results to directly
estimate the physical attributes of the binary system. In
contrast, numerical methods use analytic extractions as an initial
parameter guess, but then perturb these parameters through many
iterations, generating and comparing each LC with the observed
data. Though the LC generators are the most conspicuous part of
these codes and are often used as benchmarks, the success of the
parameter fitting code may in fact be more dependent on having a
reliable optimization algorithm. These optimization algorithms
direct the parameter perturbations based on previous results, and
strive to converge the model LC to match the observations.

In order to run an optimization algorithm, one must first define
an optimization statistic that describes, in a single real value,
how well the model LC fits the observations. Thus, given an
observed LC, this statistic simply becomes a function of the model
parameters. It is convenient to visualize this statistic as a
contour plot over the parameter space, where by convention, a
lower value indicates a better model fit. By far the most accepted
statistic for this purpose is the $\chi^2$ statistic, which is
defined as:

\begin{equation}
\chi^2 = \sum_i{\left(x_{model,i} - x_{obs,i}\right)^2 /
\sigma^2_{obs,i}}\ ,
\end{equation}

where $x_{obs}$ indicates the value of observations made,
$\sigma_{obs}$ indicates their respective uncertainties, and
$x_{model}$ indicates the model's predicted values during the time
of the observations. However, many other statistics have been
suggested, including the `scatter score' \citep{Devor05}, which is
based upon the correlation between neighboring residuals, and the
`alarm statistic' described by \citet{Tamuz06}, which is based
upon the distribution of positive and negative residual runs.

The choice of optimization algorithms is, however, far more
complex than the choice of statistic. This is because optimization
algorithms have subtle underlying behaviors that might make them
successful at solving one problem, but perform poorly with
another. Optimization algorithms generally fall into one of three
categories: (1) Steepest Descent algorithms [e.g.,
\citet{Nelder65, Press92}], also known as ``greedy'' optimization
algorithms, which perturb the model in the way that will bring
about the maximum improvement at each iteration. These algorithms
generally converge to a solution very quickly, however they will
often settle in a local minimum instead of the global minimum.
This is especially troublesome since highly non-linear, high
dimension parameter spaces such as these, have a large, if not
infinite, number of local minima. Therefore, Steepest Descent
algorithms are highly sensitive to the initial guess and require
very close supervision by an expert user to make sure that they
converge to a physically realistic result. We note that, in
principle, due to observational noise, instrumental systematics,
and imperfections in the model, the global minimum may not exactly
point to the true physical solution. However, with the absence of
any additional information, the global minimum is the point that
is most likely to be the true solution. (2) Genetic algorithms
\citep{Holland75} are inspired by biological evolution through
natural selection. They attempt to reduce the sensitivity of the
initial parameter choice by initiating many searches in parallel
from randomized starting points. Searches that are in an
unpromising region of the parameter space (i.e. represent poor
model fits) are rapidly terminated, whereas those in more
promising regions are allowed to multiply. The progeny are usually
created by averaging a number of successful parameter space
points, and adding small random variances (mutations) that place
them in new locations. Genetic algorithms have the advantage of
being easily parallelizable and therefore well suited for
multi-CPU computers, however despite a number of attempts to
introduce them [e.g., \citet{Charbonneau95, Metcalfe99}] they have
not yet been widely adopted by the astronomical community. (3)
Steepest descent with Simulated Annealing \citep{Kirkpatrick83,
Press92} includes a critical improvement over the earlier
described Steepest Descent algorithm, in that with varying
probabilities, it makes ``leaps of faith'' in directions that may
initially worsen the model fit, but enable searching regions of
the parameter space that would otherwise be cordoned off. Such
``leaps of faith'' are usually performed liberally in early
iterations, and become increasingly conservative in later
iterations, thus becoming more like the simple Steepest Descent
algorithms. These ``leaps of faith'' have the effect of smoothing
small scale bumps in the contours of the parameter space. At early
iterations the search is guided into large-scale depressions, and
only later is influenced by the smaller-scale divots at the bottom
of the depression. In this thesis, we adopted this third approach,
and implemented it with an extremely fast and simple LC generator
(see \S\ref{subsecLCgenerator}) that is built directly into the
optimization code, so as to achieve maximum efficiency. The
optimization algorithm can thus afford to scour many regions of
the parameter space in a large number of iterations (up to
$10^4$), while requiring only modest computational resources.

Finally, once the algorithm converges and produces a model that
fits the observed LC, one must estimate the resulting parameter
uncertainties. Traditional analytic methods, such as analysis of
variance (ANOVA), estimate the uncertainty from the local
curvature or width of the minimum. However, such formal
uncertainties will almost always underestimate the parameter
errors, sometimes by several orders of magnitude, since as
mentioned earlier, the rough small-scale terrain around the
minimum will usually have far steeper slopes than the large-scale
structure of the parameter space. To remedy this problem, the
formal uncertainty is often multiplied by an empirical ``fudge
factor'', however, a far more rigorous solution is to repeat the
fit many times using Bootstrapping or Monte Carlo simulations
\citep{Press92, Southworth05} and then assess the distribution of
the resulting solutions. Bootstrapping repeats the fitting
procedure with randomized subsets (i.e. resampling) of the
observational data, whereas Monte Carlo randomizes the initial
guesses and parameter perturbations. These methods have been shown
to return far more robust estimations of the uncertainties and can
indicate correlations and degeneracies between the parameters.
Furthermore, unlike most analytic methods, these randomized
estimators do not assume that the errors have normal distributions
(i.e. white noise). While Bootstrapping methods are generally
simpler to implement, they have a tendency to generate overly
optimistic uncertainty estimates, and thus have fallen out of
favor, being replaced by Monte Carlo methods. In this thesis, we
use an empirical multiplier to determine the parameter
uncertainties at the initial pipeline phase, when modeling very
large numbers of EBs. Then, once a smaller and more manageable
group of systems has been selected for follow-up, we refine the
uncertainty estimates using Monte Carlo simulations.

\section {A Simple Binary Light Curve Generator}
\label{subsecLCgenerator}

Over the past century EB models and LC generation codes have
become increasingly sophisticated and complex (see
\S\ref{subsecHistory}). In this thesis, however, we chose to adopt
an extremely simple model, which can be rapidly calculated, and
optimized over many iterations without requiring large
computational resources (see \S\ref{subsecParameterExtraction}).
This approach is the basis of the Detached Eclipsing Binary Light
curve (DEBiL) fitter code, whose implementation is described in
detail in the first chapter of this thesis. The model we use
consists of two limb darkened spherical stars orbiting in a
Newtonian orbit and thus describes a perfectly detached binary.
Obviously, a perfectly detached system cannot exist, since all
binary components will produce some small amount of tidal
distortion on their sibling component. However, in photometric
surveys with durations of more than a few months, such as the ones
used in this thesis, we find that the majority of the EB LCs can
be successfully fit with the DEBiL model. The non-detached
systems, which either have very short-periods (typically
$\lesssim$ 1 day) or contain evolved components, are identified,
and then either removed from the pipeline or are tagged as systems
with unreliable parameters.

Since it is a central component of this thesis, we provide here a
first-principle derivation of the underlying DEBiL equations. We
begin with the equation of motion of a Newtonian 2-body system
(Kepler's equation) on a Cartesian plane:

\begin{eqnarray}
x &=& \cos E - e\\
y &=& \left(1-e^2\right)^{1/2} \sin E \ ,
\end{eqnarray}

where $E$ is the eccentric anomaly, and $e$ is the binary's
eccentricity. The origin marks the system's center of mass (i.e.
the focus of the eclipse), and the unit distance is defined as the
sum of the binary components' semi-major axes ($a$). We then
calculate the system's eccentric anomaly at any given time ($t$)
using:

\begin{equation}
E - e \sin E = 2 \pi \left(t - t_0\right) / P \ ,
\end{equation}

where $t_0$ is the epoch of periastron, and $P$ is the system's
orbital period. Though the value of the eccentric anomaly cannot
be calculated directly, it can be estimated numerically to a
sufficient degree of precision in only a few iterations. Next, we
rotate the coordinate system by the argument of perihelion ($\omega$):

\begin{eqnarray}
x' &=& x \cos \omega - y \sin \omega\\
y' &=& x \sin \omega + y \cos \omega \ ,
\end{eqnarray}

and finally tilt the plane by a given inclination angle ($i$):

\begin{eqnarray}
x'' &=& x'\\
y'' &=& y' \cos i\ .
\end{eqnarray}

After modest algebraic manipulations, we then arrive at a formula
describing the projected distance ($D$) between the stellar
components' centers:

\begin{eqnarray}
D^2 &=& \left(x''\right)^2 +  \left(y''\right)^2 = \\
&=& \left( 1- e \cos E \right)^2 - \left[ \left(\cos E - e \right) \sin \omega + \left( 1 - e^2 \right)^{1/2} \sin E \cos \omega \right]^2 \sin^2 i\ .
\nonumber
\end{eqnarray}

\begin{figure}
\includegraphics[width=5in]{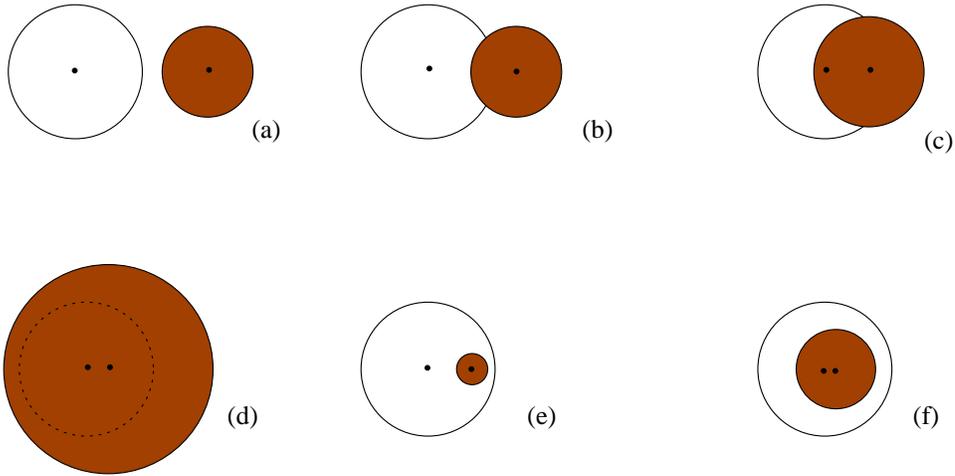}
\caption{\singlespace The six classes of geometric alinement of
the binary components' disks: (a) no intersection, (b) partial
intersection, where the front component is not over the back
component's center, (c) partial intersection, where the front
component is over the back component's center, (d) the back
component is completely eclipsed by front component, (e) the front
component is entirely within the disk of the back component, but
it is not over the back component's center, and (f) the front
component is entirely within the disk of the back component, and
it covers the back component's center.}
\label{figLC_circ}
\end{figure}

\begin{figure}
\includegraphics[width=3in]{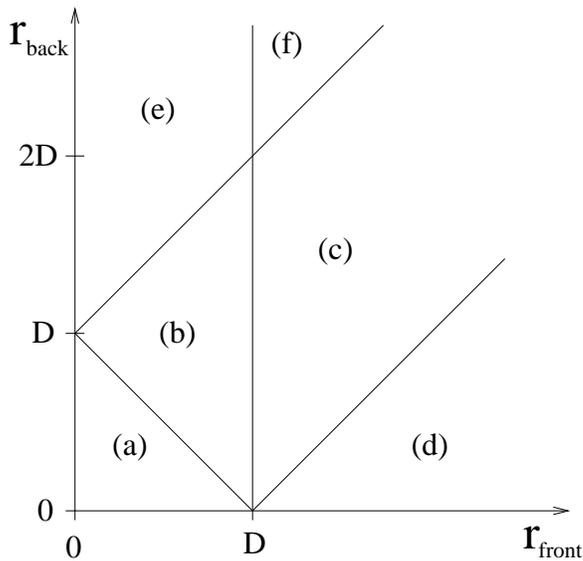}
\caption{\singlespace This chart classifies which alignment class
(see Figure \ref{figLC_circ}) a binary would belong to at some
moment, given the radii of its components ($r_{fr,bk}$) and the
distance between its components' centers ($D$). This
classification proves that there are no additional alignment
classes.}
\label{figLC_lines}
\end{figure}

We use this projected distance to calculate the unit area flux
received from the eclipsed (back) component. For this to be done
accurately, we must know the limb darkening function of the binary
stellar components. We adopted the quadratic law approximation
\citep{Claret03} as it allows for faster computation:

\begin{equation}
I(\cos \theta) = I_0 \left[1 - \tilde{a}(1-\cos\theta) - \tilde{b}(1-\cos\theta)^2 \right]\ ,
\end{equation}

where $\theta$ is the angle between the line of sight and the
emergent flux, $I_0$ is the flux at the center of the stellar
disk, and $\tilde{a}$, $\tilde{b}$ are the quadratic law
coefficients of the given star. We next observe the geometry of
the stellar disks during eclipse. All the points at a radius of
$r$ will emit a uniform flux per unit area of:

\begin{equation}
I_{bk}(r) = I\left(\sqrt{1 - (r/r_{bk})^2}\right)\ ,
\end{equation}

where $r_{bk}$ is the radius of the eclipsed (back) star.
Furthermore, the exposed (i.e. non-eclipsed) points at a radius of
$r$ form an angle equal to:

\begin{equation}
\alpha(r) = 2 \arccos \left[\frac{r^2_{fr} - D^2 - r^2}{2Dr}\right]\ ,
\end{equation}

where $r_{fr}$ is the radius of the eclipsing (front) star. We
remind the readers that our unit distance was defined earlier as
the binary's semi-major axis ($a$), therefore all the radii
parameters indicated here are in fact fractional radii. Finally,
we are able to integrate the flux emitted from the exposed portion
of the eclipsed stellar disk. However, the binary may be in one of
six orientations (see Figures \ref{figLC_circ} and
\ref{figLC_lines}), where each orientation will result in a
slightly different integration expression:

\begin{eqnarray}
(a)\ \ F_{bk} &=& 2\pi \!\int_0^{r_{bk}}\!\!\!r\:I_{bk}(r)\:dr\\
(b)\ \ F_{bk} &=& 2\pi \!\int_0^{D-r_{fr}}\!\!\!r\:I_{bk}(r)\:dr + \int_{D-r_{fr}}^{r_{bk}}\!\!\!r\:I_{bk}(r)\alpha(r)\:dr\\
(c)\ \ F_{bk} &=& \int_{r_{fr}-D}^{r_{bk}}\!\!\!r\:I_{bk}(r)\alpha(r)\:dr\\
(d)\ \ F_{bk} &=& 0\\
(e)\ \ F_{bk} &=& 2\pi \!\int_0^{D-r_{fr}}\!\!\!\!\!\!\!\!r\:I_{bk}(r)\:dr + \int_{D-r_{fr}}^{D+r_{fr}}\!\!\!\!\!\!\!\!r\:I_{bk}(r)\alpha(r)\:dr + 2\pi \!\int_{D+r_{fr}}^{r_{bk}}\!\!\!\!\!\!\!r\:I_{bk}(r)\:dr\\
(f)\ \ F_{bk} &=& \int_{r_{fr}-D}^{r_{fr}+D}\!\!\!r\:I_{bk}(r)\alpha(r)\:dr + 2\pi \!\int_{D+r_{fr}}^{r_{bk}}\!\!\!r\:I_{bk}(r)\:dr\ .
\end{eqnarray}

Because we chose to use a quadratic limb darkening law, the
integrals not involving $\alpha(r)$ can be solved analytically.
The remaining integrals are be solved numerically, however this
typically requires only a few iterations, since their associated
functions are comparably smooth. We note that \citet{Mandel02}
solved these integrals using complete elliptical integrals of the
third kind, however the added complexity of this approach is only
justified when one must generate LCs with very high precision.
Finally, we integrate the flux of the eclipsing star ($F_{fr}$)
and sum the results to form the total expected flux from the
entire binary systems ($F_{binary}$):

\begin{eqnarray}
I_{fr}(r) &=& I\left(\sqrt{1 - (r/r_{fr})^2}\right)\\
F_{fr} &=& 2\pi \!\int_0^{r_{fr}}\!\!\!r\:I_{fr}(r)\:dr\\
F_{binary} &=& F_{fr} + F_{bk}\ .
\end{eqnarray}

Note that the values of both $F_{fr}$ and case (a) of $F_{bk}$
need only be evaluated once. These parameters will then swap, as
each component assumes the role of the eclipsing star and then the
eclipsed star, in the course of each orbital rotation.

\section {Chapter Summaries}

This thesis contains seven chapters. These chapters are
chronological, closely following the actual development of the
pipeline, so that each chapter builds upon the results of the
previous ones. The first chapter is this introduction, which
attempts to provide the reader with both the historical and
computational context that lead to the research done in this
thesis. \textbf{Chapter~2} describes the Detached Eclipsing Binary
Light-curve (DEBiL) fitter and how it can be used. DEBiL was
developed from the ground up with the intention of robustly
processing large datasets of noisy LCs, which contain both EB and
non-EB systems. We demonstrate how DEBiL successfully reproduces
the model parameters of both previously published and simulated
LCs. We then proceed to use DEBil to process over $10^4$ OGLE II
LCs, thus creating the largest database of EB solutions to date.
In \textbf{Chapter~3} we sketch our first efforts at building upon
DEBiL in a fundamentally novel way, by incorporating our
theoretical understanding of stellar structure and evolution. We
thus constrain the models to physically realistic stars, and by
doing so greatly reduce the parameter space that needs to be
searched. Furthermore, abnormal systems, such as binaries that
underwent mass transfer, can be identified by their poor model
fits. We describe two competing approaches to the implementation
of our mass determination algorithm. One is based on isochrone
tables [we used \citet{Baraffe98, Yi01}] and is designated the
Method for Eclipsing Component Identification (MECI). The other
approach is based on the empirical properties of stellar spectral
classes [we used \citet{Cox00}] and is far simpler and faster but
more crude than MECI, thus we call it MECI-express\footnote{This
approach was developed first but was later given this name to make
clear its standing in relation to MECI.}. \textbf{Chapter~4}
describes our further development and testing of MECI. We verify
that MECI can indeed be used to reliably determine the masses of
both binary components using only photometric data, and thus can
be a powerful tool in analyzing large LC datasets. We then use it
to identify rare and interesting systems for follow-up. In
\textbf{Chapter~5} we develop a pipeline for rapid EB LC analysis,
which incorporates both DEBiL and MECI. We then use this pipeline
to analyze all the LCs within 10 fields of the TrES survey, thus
identifying 773 EBs and determining the absolute properties of
most of them. In this single effort, we were able to significantly
increase the number of EBs with known absolute properties, perhaps
doubling their number. Furthermore, the fact that these systems
were discovered and analyzed systematically, enables us to make
meaningful statistical inferences regarding the distribution of
stellar properties and their evolution. We then categorized these
systems, and point out groups of interesting candidates: eccentric
systems, low-mass systems, and abnormal systems. In
\textbf{Chapter~6} we chose one particularly promising long-period
low-mass candidate, T-Lyr1-17236 and further analyzed the system
using additional photometric and spectroscopic observations. These
observations confirm the MECI results, making this systems the
longest period confirmed low-mass EB currently known, by a factor
of three. As such, this system is a valuable test case for the
magnetic disruption hypothesis \citep{Ribas06, Torres06}, which is
used to explain the disparity between theoretical and
observational mass-radius relations for stars at the bottom of the
main-sequence. Finally, in \textbf{Chapter~7} we present
preliminary work on nine additional low-mass EB candidates.
Unfortunately, we do not have sufficient observations to
accurately determine all their absolute properties, however, in
many cases we were able to provide strong constraints on their
component masses. Furthermore, we present two EBs with interesting
additional properties. One has large O-C eclipse timing
variations, and the other exhibits pulsations with a period of 1.9
hours. We would like these targets to be available to the
community, so that with additional observations, their properties
can become better determined, and thus their physics will be
better understood.

\chapter{Solutions for 10,000 Eclipsing Binaries in the Bulge Fields of OGLE~II Using DEBiL
\label{chapter2}}

\title{Solutions for 10,000 Eclipsing Binaries in the Bulge Fields of OGLE~II Using DEBiL}

J.~Devor 2005, \emph{The Astrophysical Journal}, {\bf 628}, 411$-$425

\section*{Abstract}

We have developed a fully-automated pipeline for systematically
identifying and analyzing eclipsing binaries within large datasets
of light curves. The pipeline is made up of multiple tiers that
subject the light curves to increasing levels of scrutiny. After
each tier, light curves that did not conform to a given criteria
were filtered out of the pipeline, reducing the load on the
following, more computationally intensive tiers. As a central
component of the pipeline, we created the fully automated Detached
Eclipsing Binary Light curve fitter (DEBiL), which rapidly fits
large numbers of light curves to a simple model. Using the results
of DEBiL, light curves of interest can be flagged for follow-up
analysis. As a test case, we analyzed the 218,699 light curves
within the bulge fields of the OGLE~II survey and produced 10,862
model fits\footnote{The list of OGLE~II bulge fields model
solutions, as well as the latest version of the DEBiL source code
are available online at: http://cfa-www.harvard.edu/$\sim
$jdevor/DEBiL.html}. We point out a small number of extreme
examples as well as unexpected structure found in several of the
population distributions. We expect this approach to become
increasingly important as light curve datasets continue growing in
both size and number.

\section{Introduction}

Light curves of eclipsing binary star systems provide the only
known direct method for measuring the radii of stars without
having to resolve their stellar disk. These measurements are
needed for better constraining stellar models. This is especially
important for such cases as low-mass dwarfs, giants,
pre-main-sequence stars, and stars with non-solar compositions,
for which we currently have a remarkably small number of
well-studied examples. Other important benefits of locating binary
systems include more accurate calibration of the local distance
ladder \citep{Paczynski97, Kaluzny98}, constraining the low-mass
IMF, and discovering new extrasolar planets\footnote{Although
there are additional complications, extrasolar planets can be seen
as the limiting case where one of the binary components has zero
brightness.}. In order to obtain measurements of the stars' radii
and masses, one needs to incorporate both the light curve
(photometric observations) and radial velocities (spectroscopic
observations) of the system. Since making large-scale
spectroscopic surveys is significantly more difficult than making
photometric surveys, it is far more efficient to begin with a
photometric survey and later follow-up with spectroscopic
observation only on systems of interest.

During the past decade, there have been numerous light curve
surveys [e.g., OGLE: \citet{Udalski94}; EROS: \citet{Beaulieu95};
DUO: \citet{Alard97}; MACHO: \citet{Alcock98}] that take advantage
of advances in photometric analysis, such as difference image
analysis \citep{Crotts92,Phillips95,Alard98}. The original goal of
many of these surveys was not to search for eclipsing binaries,
but rather to search for gravitational microlensing events
\citep{Paczynski86}. Fortunately, the data derived from these
surveys are ideal for eclipsing binary searches as well. More
recently, there have also been mounting efforts to create
automated light curve surveys [e.g., ROTSE: \citet{Akerlof00};
HAT: \citet{Bakos04} ; TrES: \citet{Alonso04}] using small robotic
telescopes for extrasolar planet searches. The upcoming large
synoptic surveys [e.g., Pan-STARRS: \citet{Kaiser02}; LSST:
\citet{Tyson02}], spurred by the decadal survey of the National
Academy of Sciences, are expected to dwarf all the surveys that
precede them. Put together, these surveys provide an exponentially
growing quantity of photometric data \citep{Szalay01}, with a
growing fraction becoming publicly available.

\section{Motivation}
\label{secMotivation}

For over 30 years many codes have been developed with the aim of
fitting increasingly complex models to eclipsing binary light
curves [e.g., \citet{ Wilson71,Nelson72,Wilson79,Etzel81}]. These
codes have had great success at accurately modeling the observed
data, but also require a substantial learning curve to fully
master their operation. The result is that up till now there is no
comprehensive catalog of reliable elements for eclipsing binaries
\citep{Cox00}.

\begin{figure}
\includegraphics[width=5in]{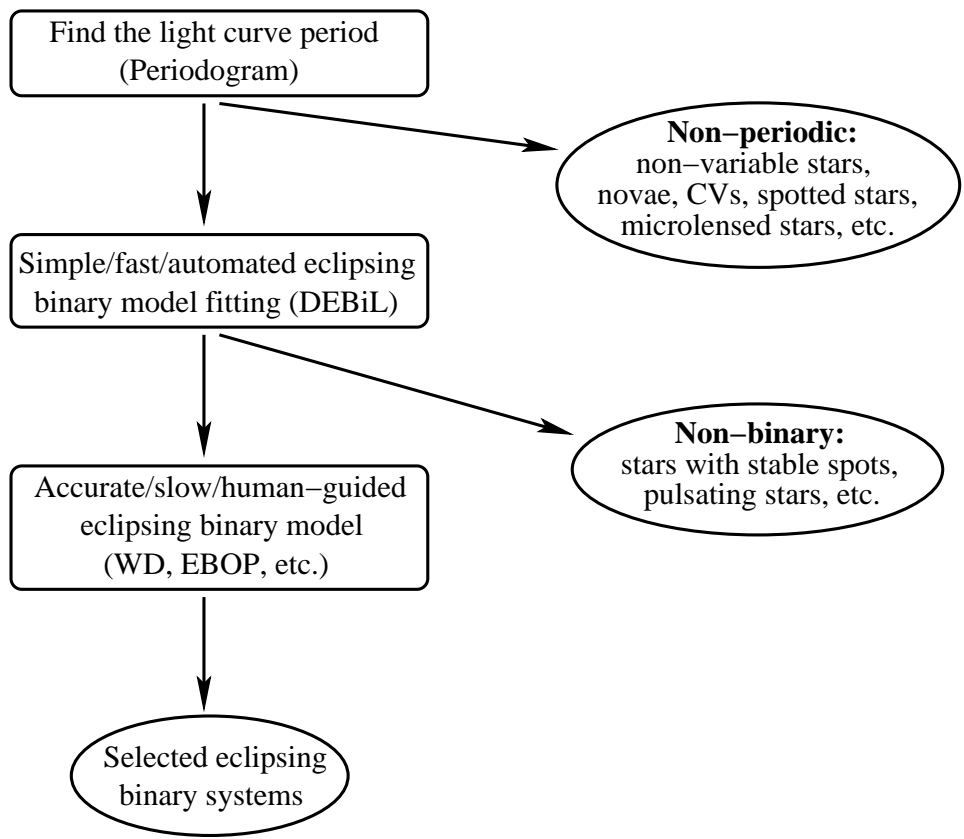}
\caption{Diagram of a multi-tiered model-fitting pipeline.}
\label{figBoxChart}
\end{figure}

In order to take full advantage of the large-scale survey
datasets, one must change the traditional approach of manual light
curve analysis. The traditional method of painstakingly fitting
models one by one is inherently limited by the requirement of
human guidance. Ideally, fitting programs should be both
physically accurate and fully automated. Many have cautioned
against full automation \citep{Popper81a,Etzel91,Wilson94} since
it is surprisingly difficult, without the aid of a human eye, to
recognize when a fit is ``good.'' In addition, it is essentially
impossible to resolve certain parameter degeneracies without
a~priori knowledge and extensive user experience. Despite these
challenges, there have been a small number of pioneering attempts
at such automated programs \citep{Wyithe01, Wyithe02}. However,
the large numerical requirements of these programs make it
computationally expensive to perform full fits (i.e., without
having some parameters set to a constant) of large-scale datasets.
Moore's law, which stated in effect that CPU speed doubles every
18 months, cannot in itself solve this problem, since the quantity
of data to be analyzed is also growing at a similar exponential
rate \citep{Szalay01}. Instead, we advocate replacing the approach
of using monolithic automated fitting programs with a multi-tiered
pipeline (see Figure~\ref{figBoxChart}). In such a pipeline, a
given light curve is piped through a set of programs that analyze
it with increasing scrutiny at each tier. Light curves that are
poorly fit or do not comply with set criteria are filtered out of
the pipeline, thus passing a far smaller number of candidates on
to the following, more computationally demanding tier. Such a
pipeline, coupled with efficient analysis programs, can increase
the effective speed of fitting models by a few orders of magnitude
compared to monolithic fitting programs. This approach makes it
practical to perform full fits of the largest light curve
datasets, with only moderate computational resources. Using a
single-CPU SUN UltraSPARC 5 workstation (333 MHz), the average
processing time of our pipeline was $\sim$1 minute per light
curve, where each light curve typically contains a few hundred
photometric observations\footnote{The processing time scales
linearly with the average number of observations in the light
curves.}. We must emphasize that even at these speeds, we still
need a few CPU-months to fully process an OGLE-like survey
containing $10^5$ light curves. In about a decade, the large
synoptic surveys are slated to create datasets that are more than
4 orders of magnitude larger than that \citep{Tyson02}.

The Detached Eclipsing Binary Light curve fitter (DEBiL) is a
program we created to serve as an intermediate tier for such a
multi-tiered pipeline. DEBiL is a fully automated program for
fitting eclipsing binary light curves, designed to rapidly fit a
large dataset of light curves in an effort to locate a small
subset that match given criteria. The matched light curves can
then be more carefully analyzed using traditional fitters.
Conversely, one can use DEBiL to filter out the eclipsing binary
systems in order to study other periodic systems (e.g., spotted or
pulsating stars). In order to achieve speed and reliable
automation, DEBiL employs a simple model of a perfectly detached
binary system: limb darkened spherical stars with no reflections
or third light, in a classical 2-body orbit. Given the system's
period and quadratic limb darkening coefficients (the default is
solar limb darkening), the DEBiL fitter will fit the following
eight parameters:

\begin{quote}
\begin{itemize}
\item Radius of primary star
\item Radius of secondary star
\item Brightness of primary star
\item Brightness of secondary star
\item Orbital eccentricity
\item Orbital inclination
\item Epoch of periastron
\item Argument of periastron
\end{itemize}
\end{quote}

Since we do not have an absolute length-scale, we measure both
stars' radii as a fraction of the sum of their orbital semimajor
axes ($a$).

This simple model allows the use of a nimble convergence
algorithm that is comprised of many thousands of small steps that
scour a large portion of the problem's phase space. Admittedly,
this model can only give a crude approximation for semidetached
and contact binaries, but it can easily identify such cases and
can flag them for an external fitting procedure. In addition to
pipeline filtration, DEBiL can also provide an initial starting
guess for these external fitters, a task that usually has to be
performed manually.

\section{Method}

This section describes our implementation of a multi-tiered light
curve-fitting pipeline. We fine-tuned and tested our design using
light curves from the bulge fields of the second phase Optical
Gravitational Lensing Experiment [OGLE~II ;
\citet{Udalski97,Wozniak02}]. Our resulting pipeline consists of
the following six steps:

\begin{quote}
First tier: Periodogram
\begin{enumerate}
\item [(1)] Find the light curve period
\item [(2)] Filter out non-periodic light curves
\end{enumerate}

Second tier: DEBiL
\begin{enumerate}
\item [(3)] Find an ``initial guess'' for the eclipsing binary
model parameters
\item [(4)] Filter out non-eclipsing systems (i.e. pulsating stars)
\item [(5)] Numerically fit the parameters of a detached
eclipsing binary model
\item [(6)] Filter out unsuccessful fits
\end{enumerate}
\end{quote}

For OGLE II data, we found that about half of the total CPU-time
was spent on step (1) and about half on step (5). The remaining
steps required an insignificant amount of CPU-time. Note that step
(5), which typically takes more than 10 times longer to run than
step (1), was only run on less than $10\%$ of the light curves. By
filtering out more than $90\%$ of the light curves at earlier
steps, the pipeline was able to run $\sim$10 times faster than it
would have been able to otherwise. Both steps (1) and (5) can
themselves be speeded up, but at a price of lowering their
reliability and accuracy. The third tier, in which the light
curves of interest are fitted using physically accurate models, is
dependent on the research question being pursued and will not be
further discussed here.

\subsection{The First Tier -- Finding the Period}
\label{subsecCh2firstTier}

Step (1) is performed using an ``off-the-shelf'' period search
technique. All the periodogram algorithms that we have tested give
comparable results, and we adopted an analysis of variance
\citep{SchwarzenbergCzerny89, SchwarzenbergCzerny96} as it appears
to do a good job of handling the aliasing in OGLE light curves. In
our implementation, we scanned periods from 0.1 days up until the
full duration of the light curve ($\sim$1000 days for OGLE~II). We
then selected the period that minimizes the variance around a
second order polynomial fit within eight phase bins. Aliases pose
a serious problem for period searches since they can prevent the
detection of weak periodicities, or periods that are close to an
alias. The ``raw'' period distribution of the OGLE~II light curves
showed aliases with a typical widths ranging from 0.001 days for
the shortest periods, and up to 0.04 days for longest periods. We
suppressed the 12 strongest aliases, over which the results were
dominated by false positives, and had the period finder return the
next best period. Fewer than $1\%$ of the true light curve periods
are expected to have been affected by this alias suppression.
Finally, once a period was located, rational multiples of it, with
numerators and denominators of 1 through 19 were also tested to
see whether they provided better periods.

Step (2) filters out all the light curves that are not periodic.
The analysis of variance from step (1) provides us with a measure
of ``scatter'' in a light curve, after being folded into a given
period. Ideally, when the period is correct, the folded data are
neatly arranged, with minimal scatter due to noise. In contrast,
when the light curve is folded into an incorrect period, the data
are randomized and the scatter is increased. In order to quantify
this, we measure the amount of scatter in each of the tested
periods, and calculate the number of sigmas the minimum scatter is
from the mean scatter. We call this quantity the ``periodic
strength score''. In an attempt to minimize the number of
non-eclipsing binaries that continue to the next step, while
maximizing the number of eclipsing binaries that pass through, we
chose a minimum periodic strength score cutoff of 6.5. In addition
to this, we set a requirement that the variables' period be no
longer than 200 days, which guarantees at least four foldings.
These two criteria filtered out approximately $90\%$ of all the
light curves in the OGLE~II dataset.

\begin{figure}
\includegraphics[width=5in]{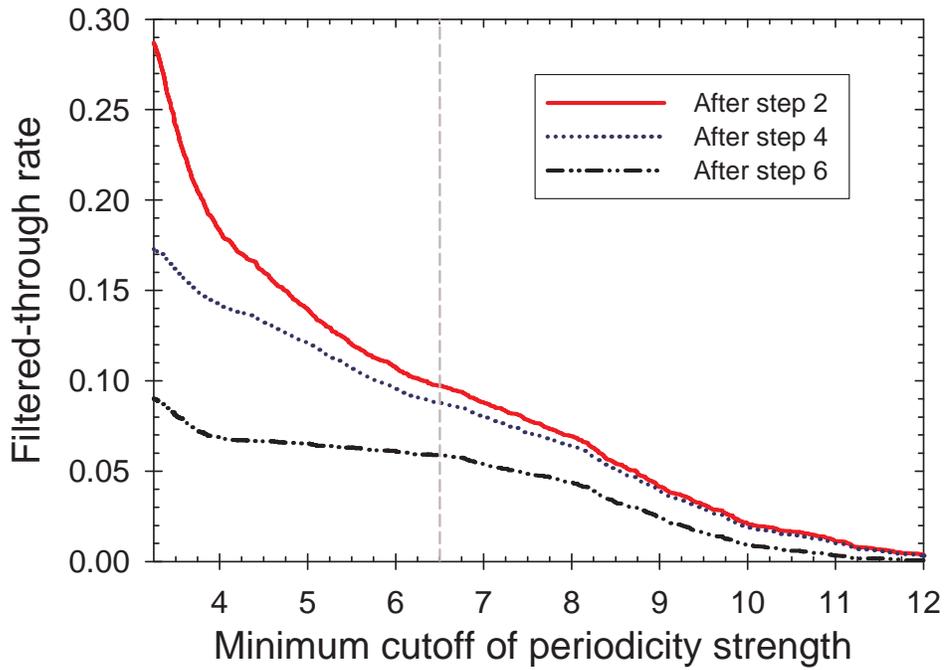}
\caption{The pipeline filtration fractions of the variables within
bulge field 33 of OGLE II (N=4526), with varying periodicity
strength cutoffs. The vertical dashed line indicates the chosen
cutoff for our pipeline ($>6.5$).}
\label{figFilters}
\end{figure}

In order to test the effectiveness of these filtration criteria,
we measured the filtration rates for field 33, a typical OGLE~II
bulge field (see Figure~\ref{figFilters}). We then repeated this
measurement for a range of periodicity strength cutoffs (the
200-day criterion remained unchanged). For periodicity strength
cutoffs up to 4, there is a sharp reduction in the number of
systems, as non-variable light curves are filtered out. Users
should be aware that constant light curves can be well-fit by
degenerate DEBiL models. By filtering out systems with such low
periodicity strength we correctly remove these systems, and in so
doing noticeably lower the total number of well-fitted systems.
Raising the filtration cutoff, up until about 6.5, will continue
reducing the filter-through rate, but with only a small impact on
the number of well-fit systems. Further raising the cutoff will
again reduce the number of fitted systems, this time removing good
systems. We can conclude from this test that the optimal
periodicity strength cutoff is between 4 and 6.5, so that
non-variable systems are mostly filtered out while eclipsing
binaries are mostly filtered through. Since the filter-through
rate monotonically decreases as the cutoff is raised, the pipeline
becomes significantly more computationally efficient at the high
end of this range, thus bringing us to our cutoff choice of 6.5.
Users with more computing power at their disposal may consider
lowering the cutoff to the lower end of this range. In so doing
they slightly reduce the risk of filtering out eclipsing binaries,
at the price of significantly lowering the pipeline computational
efficiency.

\subsection{The Second Tier -- DEBiL Fitter}
\label{subsecDEBiLfitter}

 Steps (3) through (5) are performed within the
DEBiL program. Step (3) provides an ``initial guess'' for the
model parameters. It identifies and measures the phase, depth and
width of the two flux dips that occur in each orbit. Using a set
of equations that are based on simplified analytic solutions for
detached binary systems \citep{Danby64,MallenOrnelas03,Seager03},
DEBiL produces a starting point for the fitting optimization
procedure. Step (4) filters out light curves with out-of-bound
parameters, so as to protect the following step. In practice this
step is remarkably lax. It typically filters out only a small
fraction of the light curves that pass through it, but those that
are filtered out are almost certain not to be eclipsing binaries.
Step (5) fits each light curve to the 8-parameter DEBiL model (see
\S\ref{secMotivation}), fine-tunes it, and estimates its parameter
uncertainties. This step starts with the light curve that would
have been seen with the ``initial guess'' parameters. It then
systematically varies the parameters according to an optimization
algorithm, so as to minimize the square of the residuals in an
attempt to converge to the best fit. The optimization algorithm
chosen for the DEBiL fitter is the downhill simplex method
combined with simulated annealing \citep{Nelder65,Kirkpatrick83,
Vanderbilt83,Otten89,Press92}. This algorithm was selected for its
simplicity, speed, and relatively long history of reliably solving
similar problems, which involve locating a global minimum in a
high-dimensional parameter space. Other methods that were
considered are gradient-based \citep{Press92} and genetic
algorithms \citep{Holland75,Charbonneau95}. Gradient-based
(steepest descent) algorithms can converge very quickly to a local
minimum, but are not designed for finding the global minimum. In
addition, the difficulty in calculating the gradient of
non-analytic function slows these algorithms considerably and
causes them to be less robust. Genetic algorithms provide a
promising new approach for locating global minima, with the unique
advantage of being parallelizable. Unfortunately, their
implementations are more complicated, while not having a
significant advantage in speed or reliability over the downhill
simplex method.

Determining a convergence threshold at which to stop optimization
algorithms is known to be a difficult problem
\citep{Charbonneau95}. This is because the convergence process of
these algorithms will go through fits and stops. The length of the
``stops,'' whereby the convergence does not significantly improve,
becomes longer with time, ultimately approaching infinity. Since
there is no known way to generally predict these fits and stops,
we chose simply to have the algorithm always run for 10,000
iterations (this choice can be adjusted at the command line). This
number was found to be adequate for OGLE light curves (see
\S\ref{secTests}), with larger numbers not showing a significant
improvement in the convergence. Using a constant number of
iterations has the significant benefit of enabling the user to
make accurate predictions of the total computing time that will be
required. At the end of the 10,000 iterations, the best solution
encountered so far is further fine-tuned, so as to guarantee that
it is very close to the bottom of the current minimum. In our
implementation we make sure every parameter is within $0.1\%$ of
the minimum ($p_{min}$).

Finally, DEBiL attempts to estimate the uncertainties of the
fitted parameters. This is done by perturbing each parameter by a
small amount ($\Delta p$) and measuring how sensitive the model's
reduced chi square ($\chi_\nu^2$) is to that parameter. In our
implementation we set the perturbation to be $0.5\%$ of $p_{min}$.
At each parameter perturbation measurement, the remaining
parameters are re-fine-tuned\footnote{We limited the number of
iterations for this task, so that it will not become a
computational bottleneck. But since the perturbations are small,
we could use a greedy fitting algorithm, for which this is rarely
a problem.}, so as to take into account the parameters'
covariances. We then use a second order Taylor expansion to derive
the second derivative of $\chi_\nu^2$ at the minimum, which is
used to extrapolate the local shape of the $\chi_\nu^2$-surface:

\begin{equation}
\chi_\nu^2 \left( p_{min} + \Delta p \right) \simeq \chi_\nu^2
\left( p_{min}  \right) + \frac{1}{2} \cdot \frac{\partial^2
\chi_\nu^2}{\partial p_{min}^2 } \cdot \left( {\Delta p } \right)^2
\end{equation}

\begin{figure}
\includegraphics[width=5in]{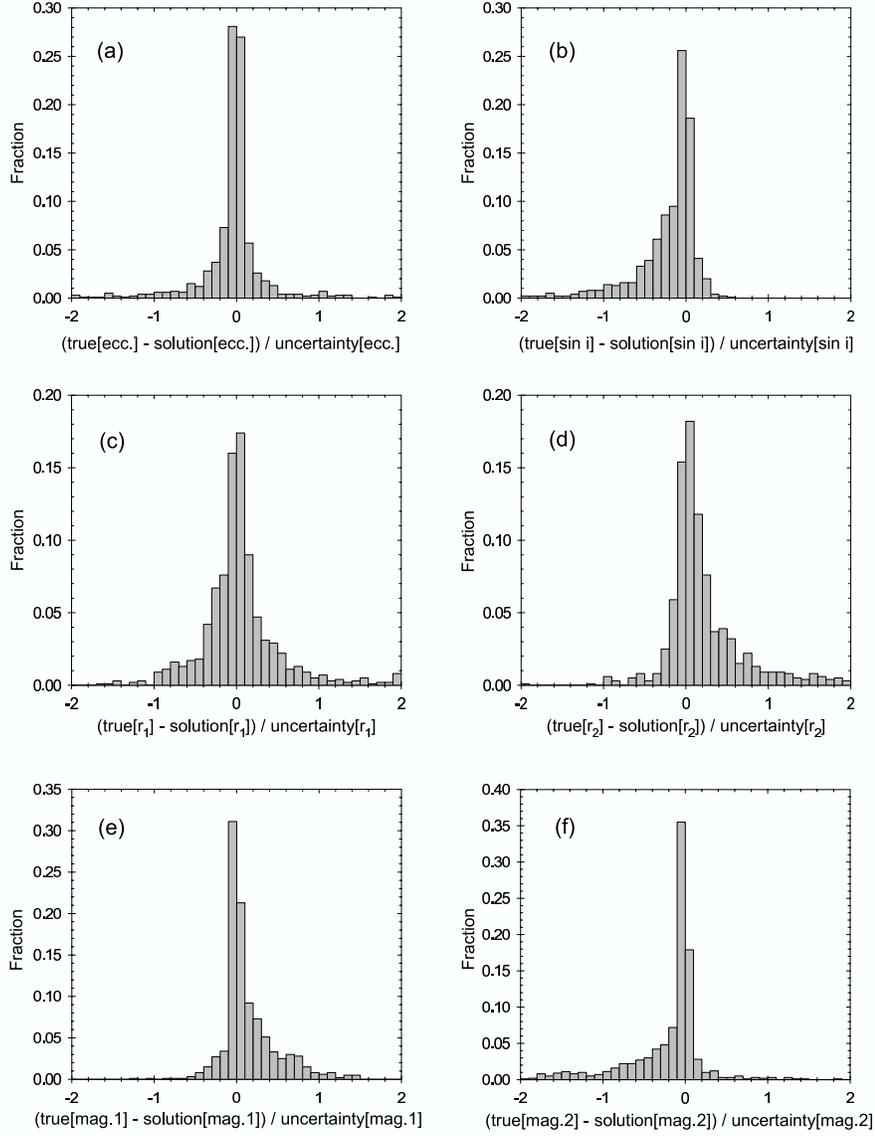}
\caption{Histograms of the error
distribution in DEBiL fitted parameters. This plot was created by
fitting 1000 simulated eclipsing light curves with 5\% Gaussian
photometric noise. The fitting error of each parameter was
normalized by its estimated uncertainty (as defined in
\S\ref{subsecDEBiLfitter}). The distributions seen here are not
Gaussian, but rather have a slender peak, and long tails (i.e.
large kurtosis). The distributions also have varying degrees of
skewness, which is discussed in \S\ref{subsecLimitations}.}
\label{figAll05Hist}
\end{figure}

We chose to employ a non-standard definition for the DEBiL
parameter uncertainties, which seems to describe the errors of all
the fitted parameters (see Figure~\ref{figAll05Hist}) far better
than the standard definition \citep{Press92}. In the standard
definition, the uncertainty of a parameter is equal to the size of
the perturbation from the parameter's best-fit value, which will
raise $\chi_\nu^2$ by $1/\nu$, while fitting the remaining
parameters. The reasoning behind this definition implicitly
assumes that $\chi_\nu^2$ is a smooth function. But in fact the
$\chi_\nu^2$-surface of this problem is jagged with numerous local
minima. These minima will fool the best attempts at converging to
the global minimum and cause the parameter errors to be far larger
than the standard uncertainty estimate would have us believe. For
this reason we adopted a non-standard empirical definition for the
parameter uncertainties ($\varepsilon_p$). In our variant, the
uncertainty of a parameter is equal to the size of the
perturbation from the parameter's best-fit value, which will
\textit{double} $\chi_\nu^2$, while fitting the remaining
parameters:

\begin{equation}
\chi_\nu^2 \left( p_{min} + \varepsilon_p \right) = 2 \chi_\nu^2
\left( p_{min} \right)
\end{equation}

This definition assigns larger uncertainties to more poorly-fit
models. In addition, it is insensitive to systematic over- or
under-estimated photometric uncertainties, which are all too
common in many light curve surveys. Using the previous two
equations, we can estimate $\varepsilon_p$ as:

\begin{equation}
\varepsilon_p  \simeq {\Delta p \cdot \sqrt {\frac{\chi_\nu ^2
\left( p_{min} \right)} {\chi_\nu ^2 \left( {p_{min} + \Delta p}
\right) - \chi_\nu^2 \left( p_{min} \right)}} }
\end{equation}

Note that the standard uncertainty is approximately:
$\varepsilon_p / \sqrt {\chi ^2}$ , so users can easily convert to
it, if so desired.

Step (6) is the final gatekeeper of our pipeline. It evaluates the
model solutions and filters out all but the ``good'' models
according to some predefined criteria. DEBiL users are expected to
configure this step so as to fit the particular needs of their
research. To this end, DEBiL provides a number of auxiliary tests
designed to quantify how well the model fits the data. The reduced
chi-squared and fitness score values measure the overall quality
of the fit, while the scatter score and waviness values measure
local systematic departures of the model from the data (see
appendix A).

Additional filtering criteria are usually needed in order to
remove non-eclipsing-binary light curves that have either (a) have
overestimated uncertainties that produce low reduced chi-squared
results, or (b) look deceptively similar to eclipsing binary light
curves. Filtration criteria should be placed with great care in
order to minimize filtering out ``good'' light curves. In order to
handle overestimated uncertainties (a), one can filter out models
with low fitness scores. Handling non-eclipsing-binary light
curves that look like binary light curves (b) is considerably more
difficult. Many of these problematic light curves are created by
pulsating stars (e.g., RR-Lyrae type~C), which have sinusoidal
light curves that resemble those of contact binaries. To this end,
DEBiL also provides the reduced chi-squared of a best-fit
sinusoidal function of each light curve. If this value is similar
or lower than the model's reduced chi-squared, then is it likely
that the light curve indeed belongs to a pulsating star.

\subsection{Limitations}
\label{subsecLimitations}

In the previous subsection we discussed the considerable
difficulties in finding the global minimum in the jagged structure
of the $\chi_\nu^2$-surface. At this point we must further add,
that since the data are noisy and the model is imperfect, the true
solution might not be at the global minimum. For the lack of
better information, we can only use the global minimum as the
point in parameter space that is the most \textit{likely} to be
the true solution.

Another source of errors are systematic fitting biases, which must
especially be taken into account when making detailed population
studies. Two main sources of these biases are imperfect models and
asymmetric $\chi_\nu^2$-minima. Almost all models are imperfect,
but when effects not included in the model become significant, the
optimization algorithm will often try to compensate for this by
erroneously skewing some of the parameters within the model. An
example of this is seen in semidetached binaries. The tidal
distortions of these stars are not modeled by DEBiL, so as a
result DEBiL will compensate for this by overestimating their
radii. Most model imperfections are flagged by a large reduced
chi-squared. Surprisingly, for tidal distortions, the reduced
chi-squared is not significantly increased. For this reason we
provide a ``detached system'' criterion that will be described in
\S\ref{secResults}. The effects of asymmetric $\chi_\nu^2$-minima
are more subtle. In such cases a perturbation to one side of a
$\chi_\nu^2$-minimum raises $\chi_\nu^2$ less than a perturbation
to the other side. Thus random noise in the data will cause the
parameters to be systematically shifted more often in the former
direction than in the latter. We believe that these biases, which
are universal to all fitting programs, can be corrected after the
fact. We chose not to do this in order to avoid having to insert
any fudge factors into DEBiL. But we acknowledge that this may be
necessary in order to extend the regimes in which DEBiL is
reliable without significantly reducing its speed.

Possibly the most problematic fitting errors are those that are
caused by mistaken light curve periods. For eclipsing binaries
with strong periodicity scores, the main cause of this is the
confusion between two types of light curves: (a) light curves with
very similar and equally-spaced eclipses, and (b) light curves
with an undetected eclipse. Both these types of light curves will
appear to have a single eclipse in their phased light curve. Light
curves with similar eclipses (a) will cause the period finder to
return a period that is half the correct value\footnote{Strictly
speaking, this is not an error on the part of the period finder,
since it is in fact returning the best period from its
standpoint.}, folding the primary and secondary eclipse over one
another. In contrast to this, light curves with an undetected
eclipse (b), either because it's hidden within the noise or
because it's in a phase coverage gap, will have the correct
period. Because the undetected eclipse is necessary for
determining a number of the fitting parameter, we are not able to
model this type of light curve. Fortunately, light curves with
similar eclipses (a) can be easily modeled by simply doubling
their period. Since we can not generally distinguish between these
two types of light curves, the DEBiL fitter treats all the
single-eclipse light curves as type (a), doubling their period and
fitting them as best it can. Whenever such a period-doubling
occurs, DEBiL inserts a warning message into the log file. These
light curves should be used with increased scrutiny. This problem
is further compounded in surveys such as the OGLE II bulge fields,
which consist of many non-eclipsing-binary variables. As mentioned
in \S\ref{subsecDEBiLfitter}, some of these systems look
deceptively similar to eclipsing binaries. Since their brightness
oscillation typically consists of a single minima, they too will
have their period doubled. In conclusion, unless the systems with
doubled periods are filtered out, there will be an erroneous
excess of light curves with similar primary and secondary
eclipses. In turn, this will manifests itself in an excess of
model solutions with stars of approximately equal surface
brightness.

Even though our discussion of possible causes and remedies for the
limitations stems from our experience with one particular
pipeline, many of these points are also likely to apply to other
similar pipelines and fitting procedures.

\section{Tests}
\label{secTests}

\begin{figure}
\includegraphics[width=5in]{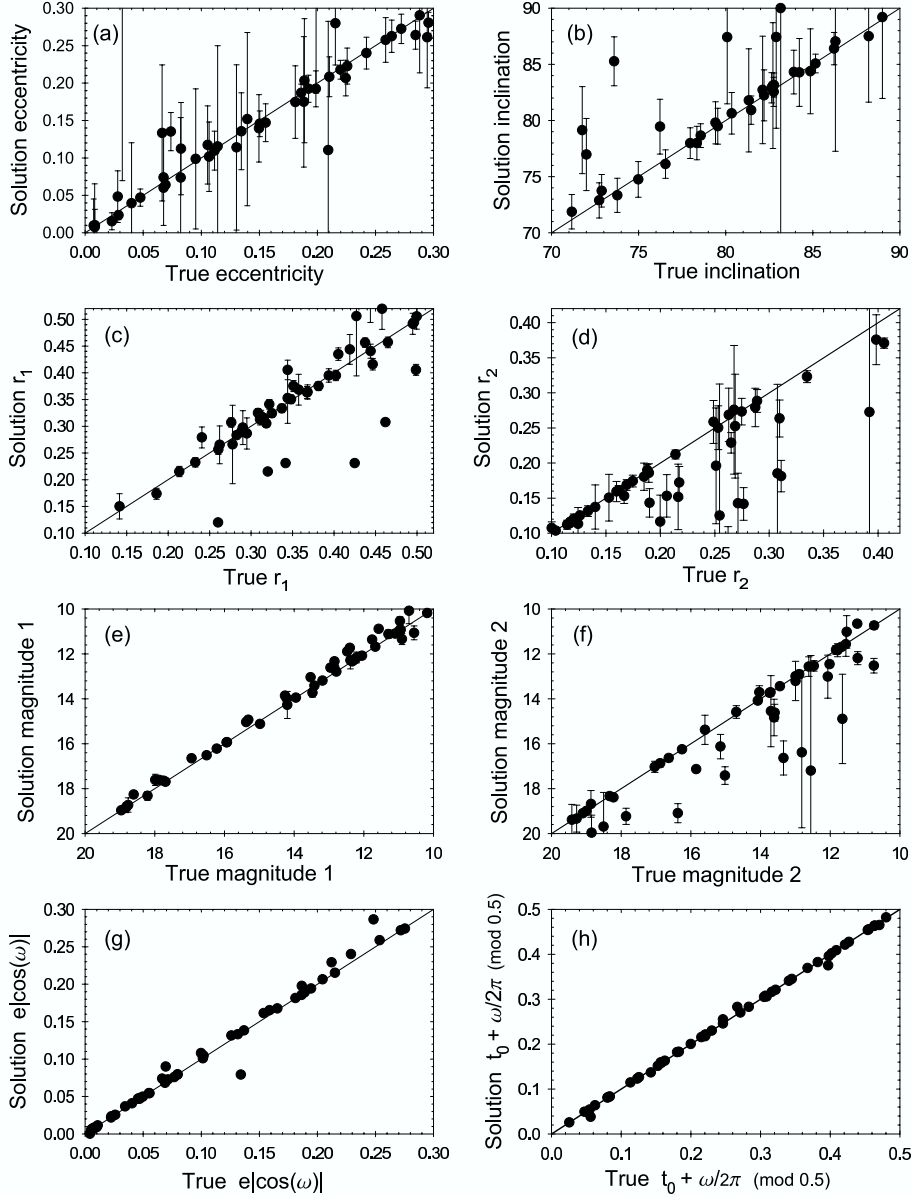}
\caption{The fitted solutions
versus the true solution in 50 simulated eclipsing binary light
curves with Gaussian photometric noise of 1\%. We simulated a
uniform distribution of parameters, with the only requirement
being that both eclipse dips were detectable through the noise.
Panels (a) through (f) show the fits of DEBiL model parameters
with their uncertainties, as defined in \S\ref{subsecDEBiLfitter}.
Panels (g) and (h) combine parameters so that they describe
prominent features of the light curve (respectively, the
separation and offset of the eclipse centers). In these
combination, the parameter errors largely cancel out, so that the
formal uncertainties should not be used.}
\label{figAll01}
\end{figure}

In order to test the DEBiL fitter, we ran it both on simulated
light curves and on published, fully analyzed, observed light
curves \citep{Lacy00,Lacy02,Lacy03}. Figure~\ref{figAll05Hist}
shows the results of fitting 1000 simulated light curves, with
$5\%$ Gaussian photometric noise. Figure~\ref{figAll01} provides a
more detailed look at the fits to 50 simulated light curves, with
$1\%$ Gaussian photometric noise, giving results comparable to
those of \citet{Wyithe01}. Not surprisingly, when less noise was
inserted into the light curve, the fitter estimated considerably
smaller uncertainties.

\begin{figure}
\includegraphics[width=5in]{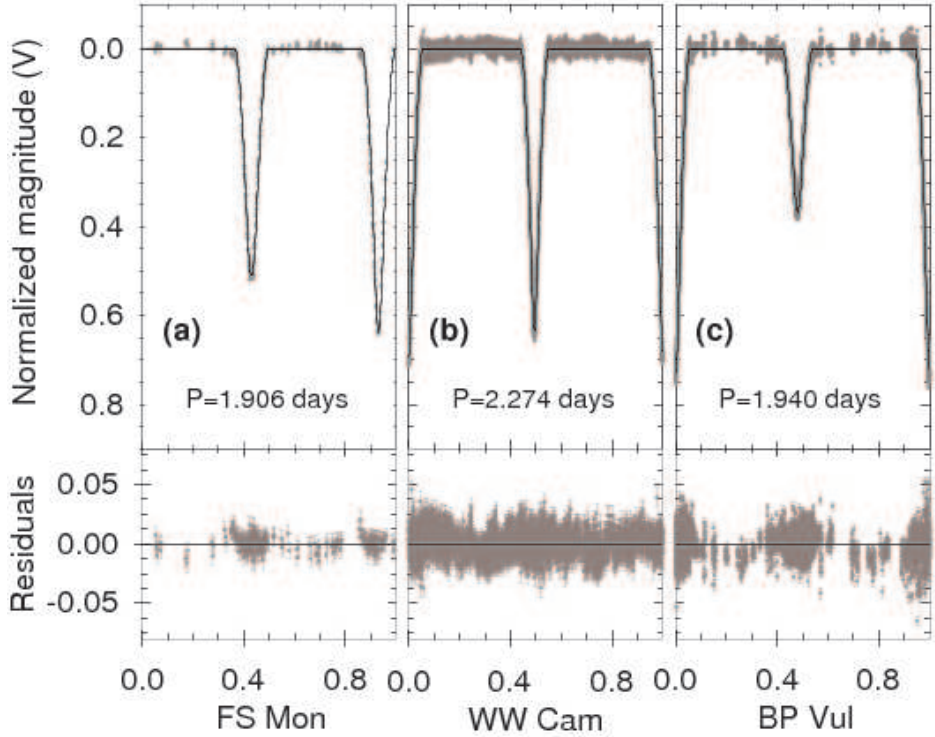}
\caption{The phased light curves with the DEBiL model best fit (solid line) and its residuals,
for the eclipsing binary systems: (a) FS Monocerotis; (b) WW Camelopardalis; (c) BP Vulpeculae.}
\label{figLacy}
\end{figure}

While the simulated light curves are easy to produce and have
known parameter values, observed light curves are the only ones
that can provide a true reality check. To this end, we also
reanalyze three published light curves (see Figure~\ref{figLacy}):

\begin{itemize}
\item Table~\ref{tableLacyA}: FS Monocerotis \citep{Lacy00}
\item Table~\ref{tableLacyB}: WW Camelopardalis \citep{Lacy02}, and
\item Table~\ref{tableLacyC}: BP Vulpeculae \citep{Lacy03}.
\end{itemize}

We present here a comparison between the aforementioned published
photometric fits, using the Nelson-Davis-Etzel model implemented
by EBOP \citep{Etzel81,Popper81b} and the DEBiL fits. For all
three cases, we used the $V$-band observational data and set DEBiL's
limb darkening quadratic coefficients to the solar $V$-band values
\citep{Claret03}. We found that when applying the physically
correct limb darkening coefficients, the improvements in the
best-fit model were negligible compared to the uncertainties. For
this reason, we chose to use solar limb darkening coefficients
throughout this project.

\begin{deluxetable}{lcccc}
\tabletypesize{\scriptsize}
\tablecaption{FS Monocerotis (N = 249)}
\tablewidth{0pt}
\tablehead{\colhead{Parameters} & \colhead{Symbol} & \colhead{\citep{Lacy00}}
& \colhead{DEBiL best fit} & \colhead{Relative error}}
\startdata
Radius of primary (larger) star& $R_1/a$ & 0.2188 $\pm $ 0.0005& 0.222 $\pm $ 0.003& 1.3 {\%} \\
Radius of secondary (smaller) star& $R_2/a$ & 0.173 $\pm $ 0.003& 0.179 $\pm $ 0.006& 3.5 {\%} \\
Surface brightness ratio& $J_s$ & 0.903 $\pm $0.003& 0.916 $\pm $ 0.05& 1.5 {\%} \\
Orbital inclination& $i$& 87.48 $\pm $ 0.08& 87.86 $\pm $ 0.015& 0.4 {\%} \\
Eccentricity& $e$& 0.0 (fixed)& 0.001 $\pm $ 0.01& \\
\enddata
\label{tableLacyA}
\end{deluxetable}

\begin{deluxetable}{lcccc}
\tabletypesize{\scriptsize}
\tablecaption{WW Camelopardalis (N = 5759)}
\tablewidth{0pt}
\tablehead{\colhead{Parameters} & \colhead{Symbol} & \colhead{\citep{Lacy02}} &
\colhead{DEBiL best fit} & \colhead{Relative error}}
\startdata
Radius of primary (larger) star & $R_1/a$ & 0.168 $\pm $ 0.0013& 0.169 $\pm $ 0.018& 0.5 {\%} \\
Radius of secondary (smaller) star & $R_2/a$ & 0.159 $\pm $ 0.016& 0.165 $\pm $ 0.014& 3.4 {\%} \\
Surface brightness ratio & $J_s$ & 0.950 $\pm $ 0.003& 0.949 $\pm $ 0.08& 0.1 {\%} \\
Orbital inclination & $i$& 88.29 $\pm $ 0.06& 88.35 $\pm $ 0.03& 0.1 {\%} \\
Eccentricity& $e$& 0.0099 $\pm $ 0.0007& 0.01 $\pm $ 0.05& \\
\enddata
\label{tableLacyB}
\end{deluxetable}

\begin{deluxetable}{lcccc}
\tabletypesize{\scriptsize}
\tablecaption{BP Vulpeculae (N = 5236)}
\tablewidth{0pt}
\tablehead{\colhead{Parameters} & \colhead{Symbol} &
\colhead{\citep{Lacy03}} & \colhead{DEBiL best fit} &
\colhead{Relative error}}
\startdata
Radius of primary (larger) star& $R_1/a$ & 0.1899 $\pm $ 0.0008& 0.190 $\pm $ 0.006& 0.1 {\%} \\
Radius of secondary (smaller) star& $R_2/a$ & 0.161 $\pm $ 0.009& 0.166 $\pm $ 0.009& 3.1 {\%} \\
Surface brightness ratio& $J_s$ & 0.624 $\pm $ 0.0013& 0.614 $\pm $ 0.08& 1.5 {\%} \\
Orbital inclination& $i$& 86.71 $\pm $ 0.09& 86.50 $\pm $ 0.012& 0.2 {\%} \\
Eccentricity& $e$& 0.0355 $\pm $ 0.0005& 0.04 $\pm $ 0.03&\\
\enddata
\label{tableLacyC}
\end{deluxetable}

\section{Results}
\label{secResults}

 We used the aforementioned pipeline to identify
and analyze the eclipsing binary systems within the bulge fields
of OGLE~II \citep{Udalski97,Wozniak02}. The final result of our
pipeline contained only about 5{\%} of the total number of light
curves we started with. The filtration process progressed as
follows:

\begin{itemize}
\item Total number of OGLE~II (bulge fields) variables: 218,699
\item After step (2), with strong periodicity and periods of 0.1-200 days: 19,264
\item After step (4), the output of the DEBiL program: 17,767
\item After step (6), with acceptable fits to binary models ($\chi_\nu^2 <4$): 10,862
\end{itemize}

\begin{figure}
\centerline{
\begin{tabular}{c@{\hspace{2pc}}c}
\includegraphics[width=2.4in]{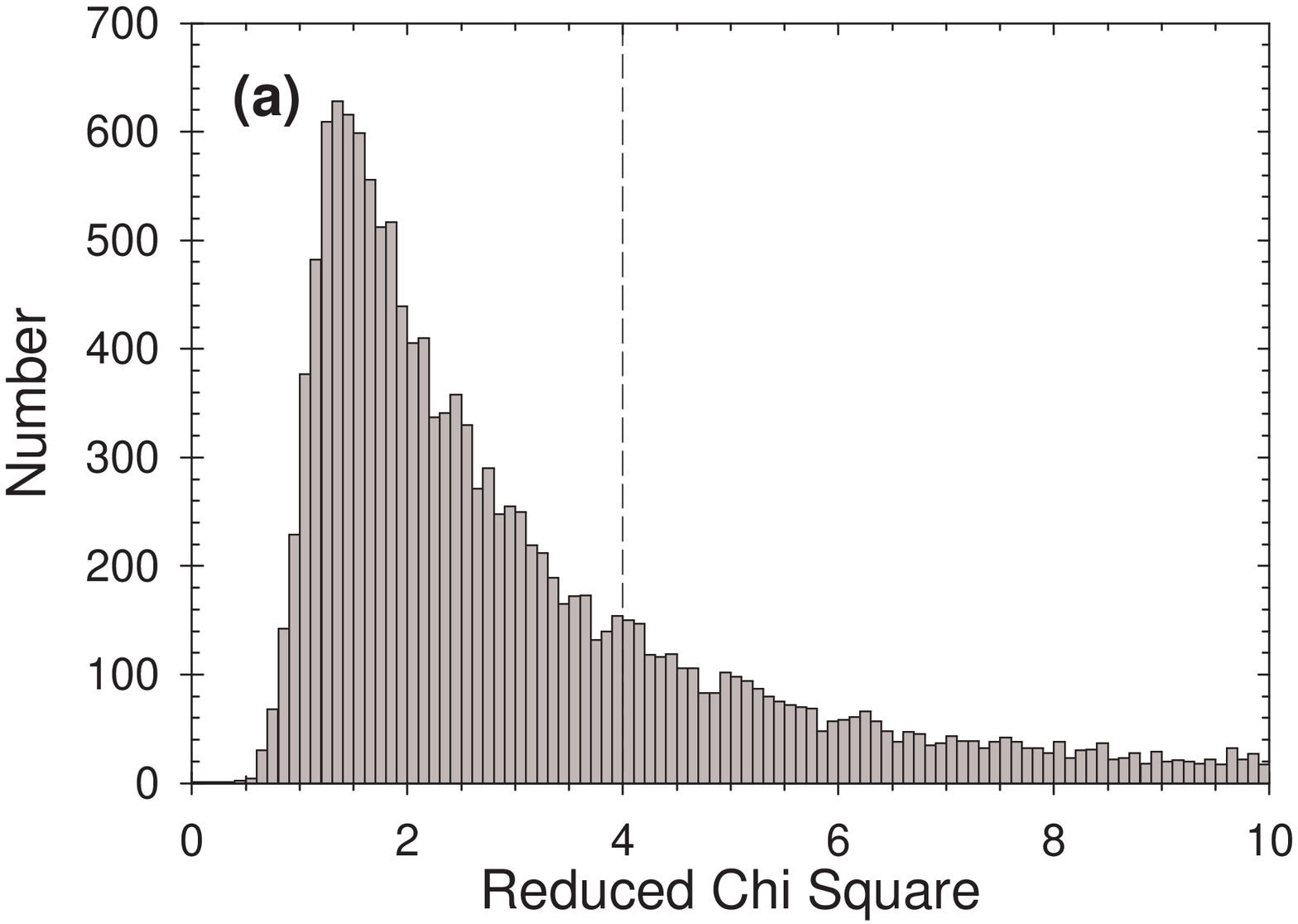} &
\includegraphics[width=2.4in]{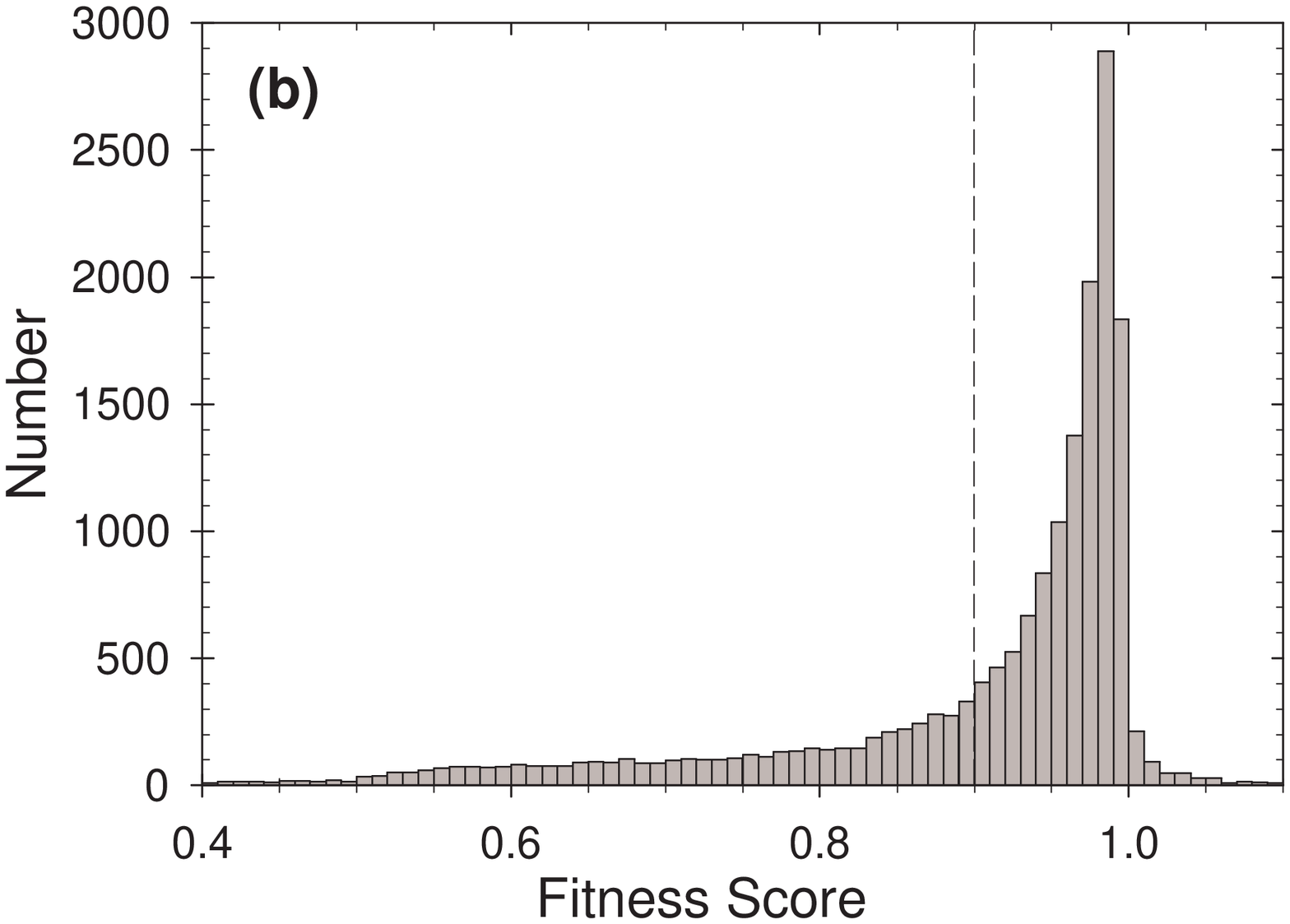} \\
a. Reduced chi-squared ($\chi_\nu^2$) distribution & b. Fitness
score distribution
\end{tabular}}
\caption{Results
of the DEBiL model fits for OGLE~II bulge (see appendix A). The
vertical dashed lines mark the filtration thresholds used in our
pipeline ($<4$ and $>0.9$ respectively). Both tests show a
definite peak near 1, indicating that it is more likely that
DEBiL will produce a ``good'' fit than a ``bad'' fit.}
\label{figHistChi}
\end{figure}

\begin{figure}
\includegraphics[width=5in]{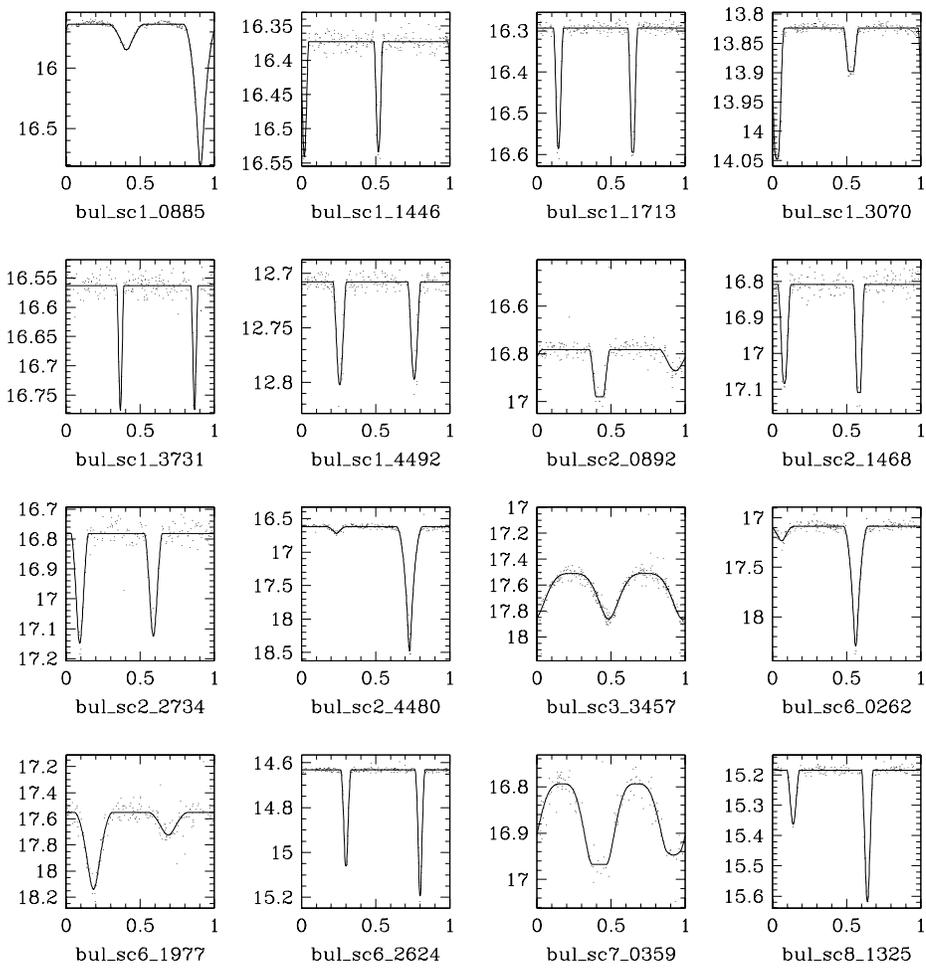}
\caption{Selected examples of OGLE~II bulge field light curves, with their DEBiL fits.}
\label{figCatalog}
\end{figure}

Most of the fits that reach step (6) can be considered successful
(see Figure~\ref{figHistChi}). It is then up to the user to choose
criteria for light curves that are of interest, and to define a
threshold for the quality of the fits. In our pipeline we chose a
very liberal quality threshold ($\chi_\nu^2 <4$), so as to allow
through light curves with photometric uncertainties that are too
small (a common occurrence), and to leave users with a large
amount of flexibility in their further filtrations. We list the
first 15 DEBiL fits that passed through step (6) in
table~\ref{tableDEBiL}. Figure~\ref{figCatalog} shows another
sampling of models, with their corresponding phased light curves.
The complete dataset of OGLE II bulge models, both plotted and in
machine readable form, is available online.

Since the filter at step (6) may not be stringent enough for many
application, we also provide our results after each of two further
levels of filtration:

\begin{itemize}
\item Non-pulsating (fitness score $> 0.9$; non-sinusoidal light curves): 8,471
\item Detached systems (both stars are within their Roche limit): 3,170
\end{itemize}

For non-sinusoidal light curves, we require that the DEBiL fit
have a smaller reduced chi square than the best-fit sinusoidal
model. For the Roche limit calculation, we assumed an early
main-sequence mass-radius power law relation: $R \propto
M^{0.652}\;$ \citep{Gorda98} and set it in a third order
approximation of the Roche radius \citep{deLoore92}. The resulting
approximations are:

\begin{eqnarray}
R_{Roche,1}/a & \simeq & 0.37771+0.31054x+0.04324x^2+0.08208x^3\label{eq_roche1}\\
R_{Roche,2}/a & \simeq & \label{eq_roche2}
\cases{0.37710-0.32684x-0.01882x^2-0.023812x^3, & if x $\ge 0.65$
\cr 0.37771-0.31054x+0.04324x^2-0.08208x^3, &otherwise\cr}
\end{eqnarray}

Where $x\equiv \log \left( {R_1 /R_2 } \right)\;$, assuming $R_1 \ge R_2$.

\begin{deluxetable}{cccccccccccccc}
\tabletypesize{\scriptsize}
\rotate
\tablecaption{Selected parameters from the DEBiL dataset of eclipsing binary systems in the galactic bulge.}
\tablewidth{0pt}
\tablehead{\colhead{Field} & \colhead{Object} & \colhead{Period} &
\colhead{e} & \colhead{$R_1/a$} & \colhead{$R_2/a$} &
\colhead{$I_1$ [mag.]} & \colhead{$I_2$ [mag.]} &
\colhead{$\sin(i)$} & \colhead{$t_0$ \tablenotemark{a}} &
\colhead{$\omega$ [deg.]} & \colhead{$\chi_\nu ^2$} &
\colhead{\begin{tabular}{c} Corrected \\ I [mag.] \tablenotemark{b} \end{tabular}} &
\colhead{\begin{tabular}{c} Corrected \\ V-I [mag.] \tablenotemark{b,c} \end{tabular}}}
\startdata
1& 39& 129.656& 0.074& 0.773& 0.118& 13.56& 17.21& 0.9208
& 0.665& 256.0& 1.13& 12.81& 1.93 \\
1& 45& 0.55677& 0.025& 0.589& 0.386& 17.36& 18.20& 0.9979
& 0.724& 271.4& 1.06& 16.21& -1000 \tablenotemark{d} \\
1& 53& 2.52158& 0.099& 0.412& 0.103& 12.20& 16.15& 1.0000
& 0.418& 91.4& 3.07& 11.38& 0.18 \\
1& 108& 1.53232& 0.005& 0.516& 0.301& 16.80& 19.12& 0.9143
& 0.497& 254.1& 0.93& 15.84& 0.48 \\
1& 112& 0.35658& 0.000& 0.514& 0.486& 17.90& 17.87& 0.9763
& 0.429& 40.6& 1.56& 16.36& 0.65 \\
1& 155& 0.96092& 0.014& 0.683& 0.303& 16.15& 17.84& 0.9203
& 0.199& 294.5& 1.12& 15.10& 0.45 \\
1& 183& 0.57793& 0.009& 0.555& 0.338& 17.99& 20.03& 0.9828
& 0.055& 56.4& 1.29& 17.14& 0.41 \\
1& 201& 0.67241& 0.101& 0.509& 0.246& 17.26& 20.08& 0.9971
& 0.473& 264.3& 0.87& 16.44& 0.32 \\
1& 202& 4.51345& 0.174& 0.309& 0.206& 17.58& 17.08& 0.9968
& 0.035& 270.9& 2.68& 15.82& 0.64 \\
1& 215& 0.48925& 0.000& 0.770& 0.230& 15.93& 18.62& 0.9212
& 0.870& 150.5& 1.54& 15.06& 0.20 \\
1& 221& 0.45013& 0.004& 0.558& 0.434& 18.23& 18.87& 0.9332
& 0.107& 100.7& 0.92& 16.92& 0.58 \\
\enddata
\tablecomments{This table is published in its entirety in the
electronic edition of the {\it Astrophysical Journal}.}
\tablenotetext{a}{Phased epoch of periastron: heliocentric Julian
date, minus 2450000.0, folded by the period.}
\tablenotetext{b}{Extinction corrected using the \citet{Sumi04}
extinction map of the galactic bulge.}
\tablenotetext{c}{The combined binary color was taken from \citep{Udalski02}.}
\tablenotetext{d}{The ``-1000'' values indicate missing magnitude or color data.}
\label{tableDEBiL}
\end{deluxetable}

\begin{figure}
\includegraphics[width=5in]{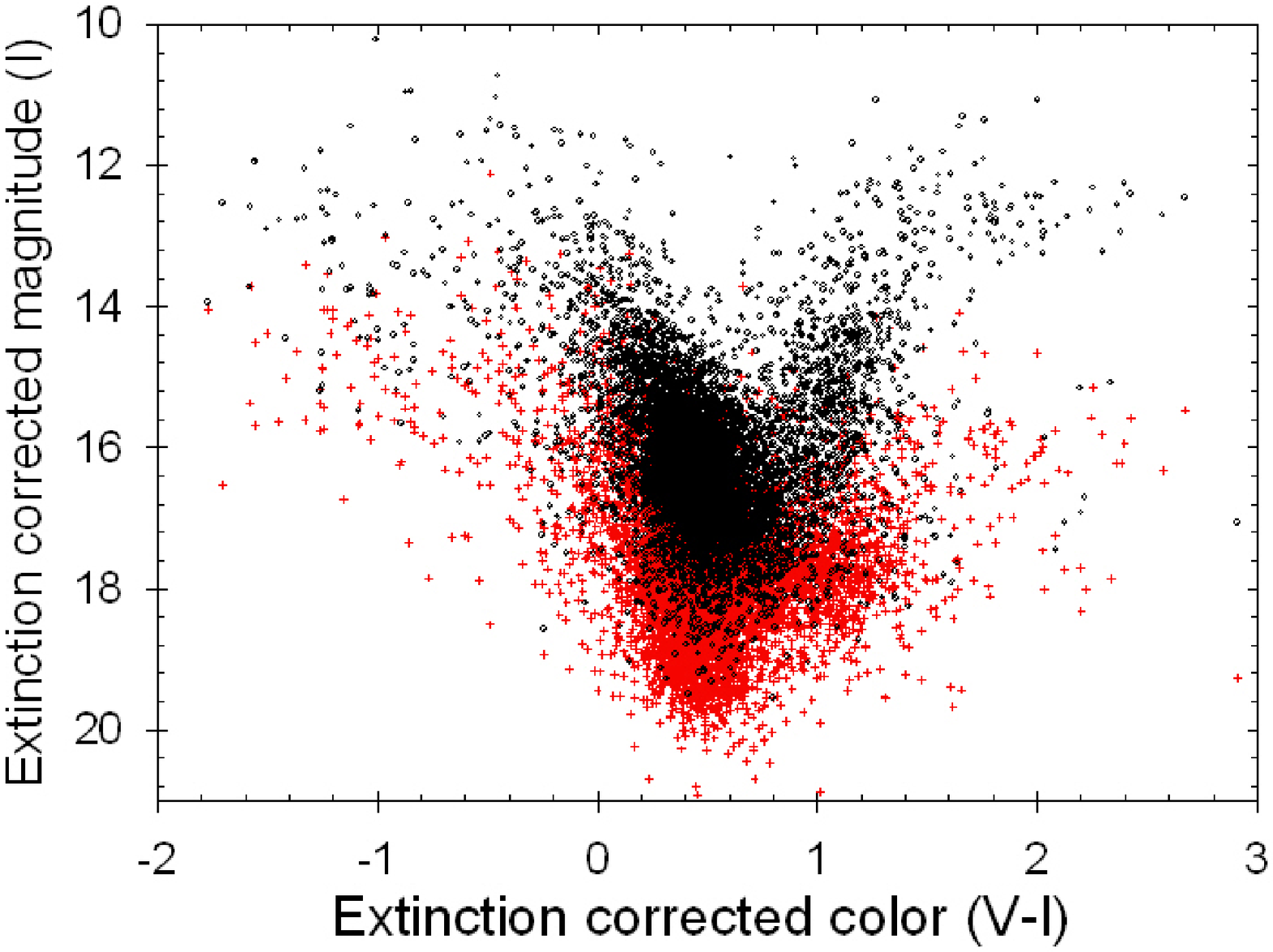}
\caption{The color-magnitude diagram of both the primary (circles) and secondary (crosses) stars of the DEBiL models.
The color of each star is the combined color of the binary, from \citep{Udalski02}.}
\label{figColorMag}
\end{figure}

\begin{figure}
\includegraphics[width=5in]{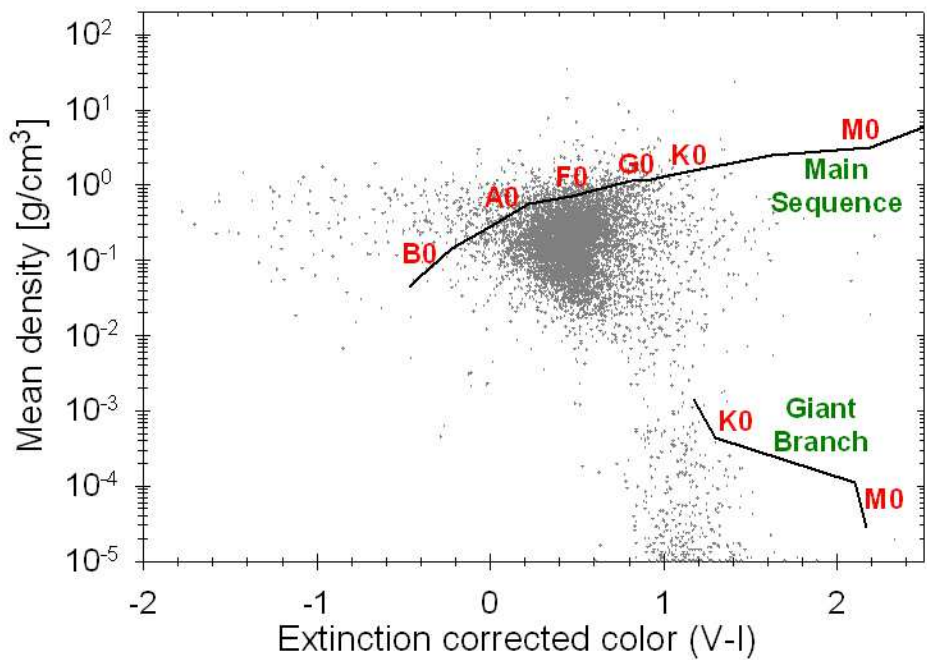}
\caption{Color-density diagram of DEBiL models. The solid lines
trace the main-sequence stars and giants \citep{Cox00}. Notice the
strong observational selection bias for main-sequence F-type and
G-type stars in the OGLE~II bulge fields.} \label{figColorDens}
\end{figure}

In order to better interpret the data derived by the DEBiL
pipeline, we combined it with color ($V-I$) and magnitude
information \citep{Udalski02}, as well as an extinction map of the
galactic bulge \citep{Sumi04}. Thus for each eclipsing system, we
also have its extinction corrected combined $I$-band magnitude and
V-I color. Since many of the stars in the bulge fields are not in
the galactic bulge but rather in the foreground, it is likely they
will be overcorrected, making them too blue and too bright. These
stars can be seen in both the color-magnitude diagram (see
Figure~\ref{figColorMag}) and in the color-density diagram (see
Figure~\ref{figColorDens}). With this qualification in mind, the
color-density diagram provides a distance independent tool for
identifying star types. Because the measured values result from a
combination of the two stars in the binary, the values are not
expected to precisely match either one of the stars in a binary.
Remarkably, using the maximum density measure instead of the mean
density (appendix B) this problem seems to be considerably
lessened.

\subsection{Population Distributions}

Due to the limitations of the OGLE observations, analysis and
subsequent filtrations, there are a myriad of complex selection
effects that need to be accounted for. For this reason, we
hesitate to make any definite population statements in this paper,
although there are a number of suggestive clusterings and trends
that merit further scrutiny.

\begin{figure}
\includegraphics[width=5in]{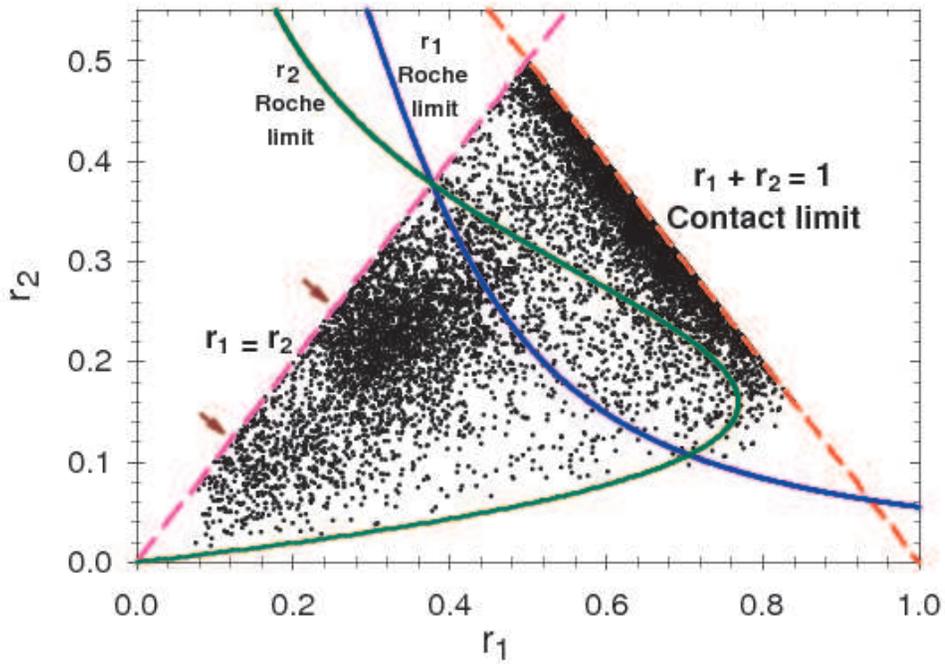}
\caption{Radius-radius plot. The dashed lines
mark the outer limit of the data. The left side is bounded by the
fact that by definition: $r_1 \geq r_2$ , while the right side is
bounded by the physical contact limit of the stars. The two arrows
mark anomalous clusterings. The two solid curves approximate the
location where the primary and secondary stars reach their
respective Roche limit (see Equations \ref{eq_roche1} \& \ref{eq_roche2}).
Systems between the two solid
curves are semidetached, systems to their left are detached, and
systems to their right are contact or overcontact systems.}
\label{figR1R2}
\end{figure}

When considering the distribution of $r_{1,2} \equiv R_{1,2} / a$
for eclipsing binary systems, one would expect a comparably smooth
distribution as determined by the binary star IMF, binary orbit
dynamics and observational/detection selection effects. This
distribution can be best seen in a radius-radius plot (see
Figure~\ref{figR1R2}). This plot has a few features that merit
discussion. One feature is that the number of systems rapidly
dwindles as their radii become smaller. This should not be
surprising, as selection effects dominate the expected number of
observed systems for small radii (this point will be elaborated at
the end of this subsection). A second, far more surprising
feature, is the appearance of three clusterings: along the contact
limit, around $\left(r_1 = 0.33, r_2 = 0.23 \right)$, and around
$\left(r_1 = 0.15, r_2 = 0.09 \right)$. These three clusterings
are most likely artificial, caused by two technical limitations
that are discussed in \S\ref{subsecLimitations}. The first two
clusterings were likely formed when many of the semidetached
systems that populated the region between the clusterings, were
swept into the contact limit, because their radii were
overestimated. The third clustering, although less pronounced,
seems to echo the structure of the second clustering, only with
about half its radii. This hints at the possibility that the
periods of some of these systems were doubled when they should not
have been, probably because of an undetected eclipse. It is worth
mentioning that both the second and third clusterings are centered
at a significant distance from $r_1 = r_2$, which is where the
clusters would have been located, if $r_1$ and $r_2$ had been
independent variables.

\begin{figure}
\includegraphics[width=5in]{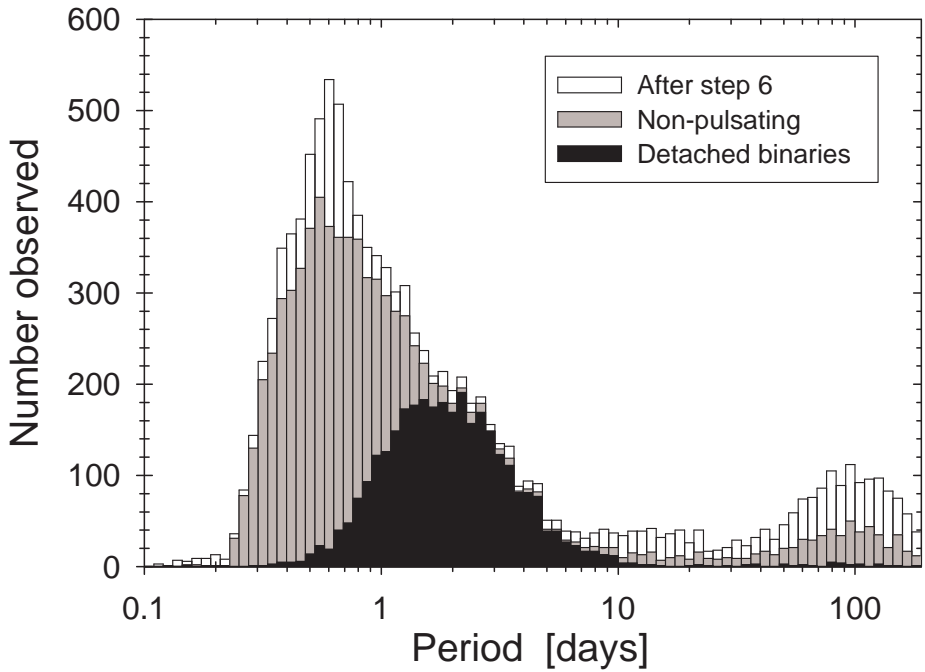}
\caption{The period distribution of OGLE~II bulge eclipsing binaries, following various stages of filtration.}
\label{figPeriod1}
\end{figure}

\begin{figure}[H]
\centerline{
\begin{tabular}{c@{\hspace{2pc}}c}
\includegraphics[width=2.4in]{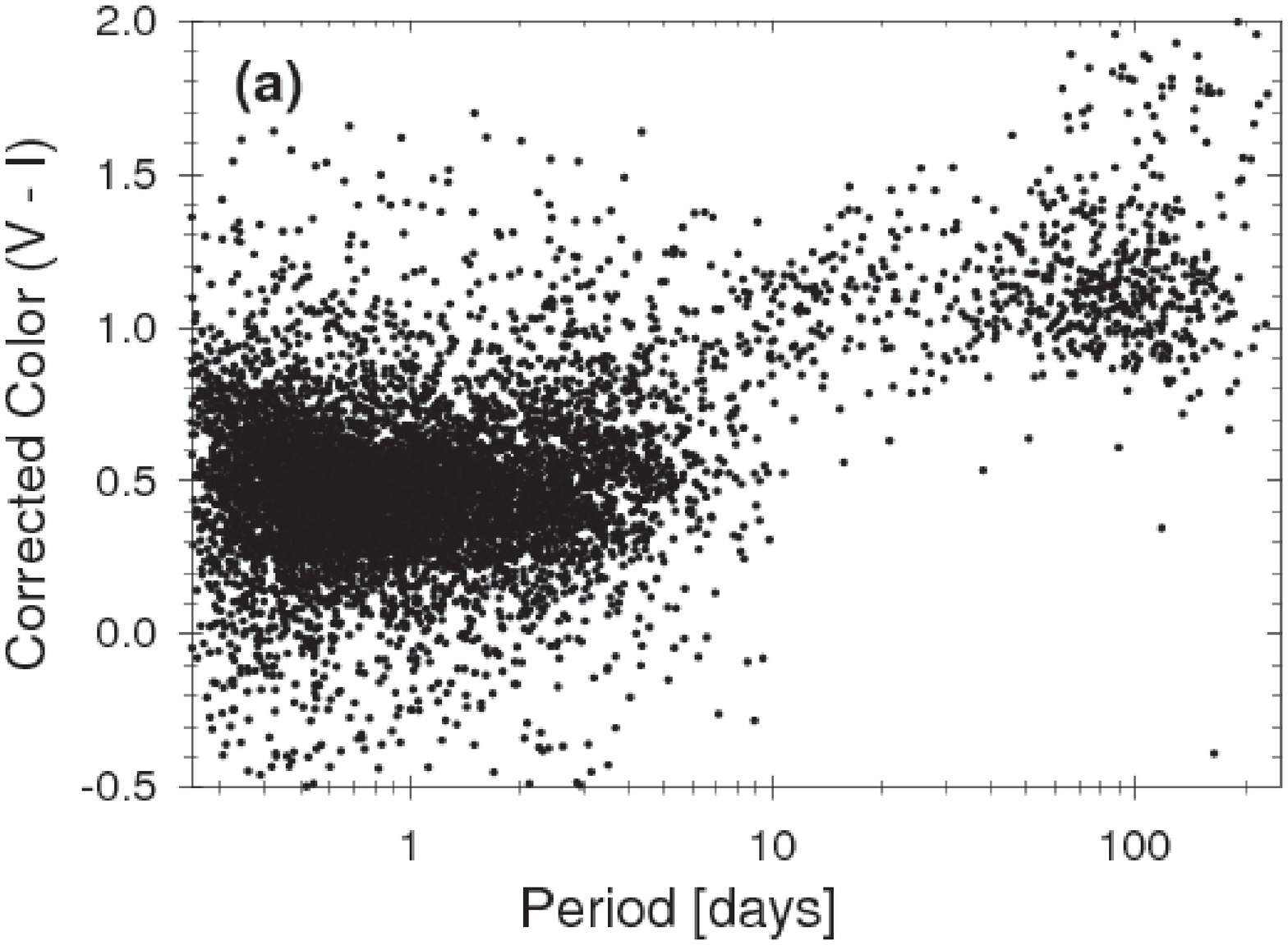} &
\includegraphics[width=2.4in]{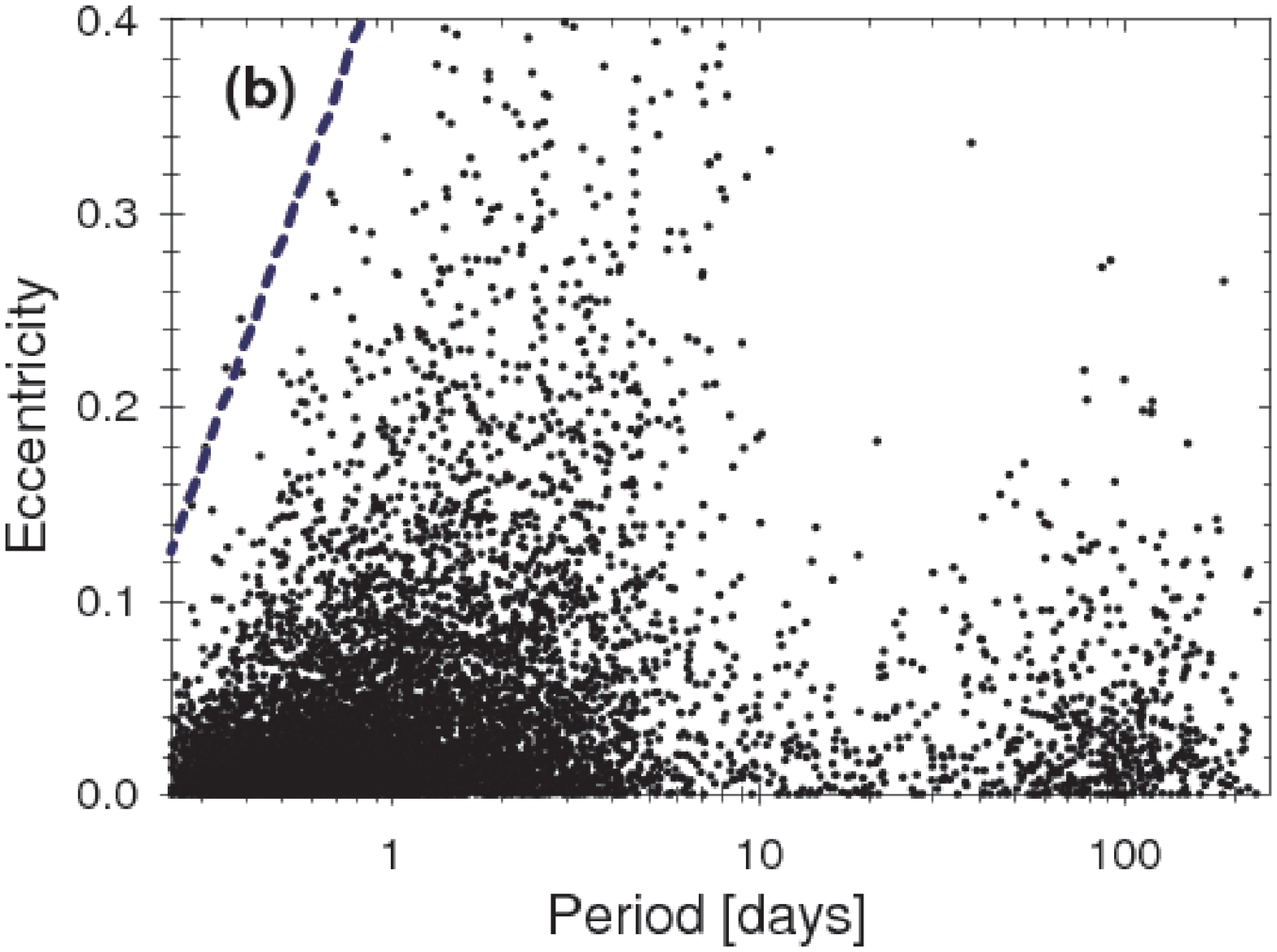} \\
a. Color-period plot & b. Eccentricity-period plot
\end{tabular}}
\caption{The relation of the binary V-I color and eccentricity to
its period. The upper limit on the eccentricities of short-period
binaries (dashed line) is probably due to tidal circularization.}
\label{figEccColorPeriod}
\end{figure}

Perhaps even more surprising is the period distribution (see
Figure~\ref{figPeriod1}). Using only the default filters of our
pipeline, this distribution is bimodal, peaking at approximately 1
and 100 days, and with a desert around a 20-day period. To
understand this phenomenon one should look at the scatter plots of
figure \ref{figEccColorPeriod}. In this figure we can see that
long-period binaries (period $> 20$ days) are significantly redder
than average and have low eccentricities. This is possibly due to
red giants, which cannot be in short-period systems. This
possibility is problematic since it contradicts the unimodal
results of previous period distribution studies
\citep{Farinella79,Antonello80,Duquennoy91}. In addition,
figure~\ref{figPeriod1} shows that with additional filtrations,
the grouping of systems around the 100-day period is greatly
reduced. All this leads us to conclude that the $\sim$100 day
period peak is probably erroneous, created by a population
contamination of pulsating stars, probably mostly semi-regulars,
which can easily be confused with contact binaries. A similar
phenomenon is seen in the spike in the number of systems around a
period of 0.6 days. This increase is probably also due to
pulsating stars, only this time they are likely RR-Lyrae type
variables. They too were largely filtered out using the techniques
described in the previous section.

Finally, we unraveled the geometric selection effects by weighting
each eclipsing light curve by the inverse of the probability of
observing it as eclipsing. For example, if with a random
orientation there is a $1\%$ chance of a given binary system being
seen as eclipsing, we will give it a weight of 100. In this way we
tabulated each observed occurrence as representing 100 such binary
systems, the remaining of which exist but are not seen eclipsing
and so are not included in our sample of variable stars. A binary
system with a circular orbit ($e = 0$) will eclipse when its
inclination angle ($i$) obeys: $\cos (i) < (R_1 + R_2)/a $. If the
orbital orientations are randomly distributed, then the
inclination angles are distributed as: $p(i)\propto \left| \sin
(i) \right|$. Therefore, the probability that such a binary system
will eclipse becomes:

\begin{equation}
p_{eclipse}(R_{1,2}/a,e=0) = \frac{R_1 + R_2}{a}
\end{equation}

\begin{figure}
\includegraphics[width=5in]{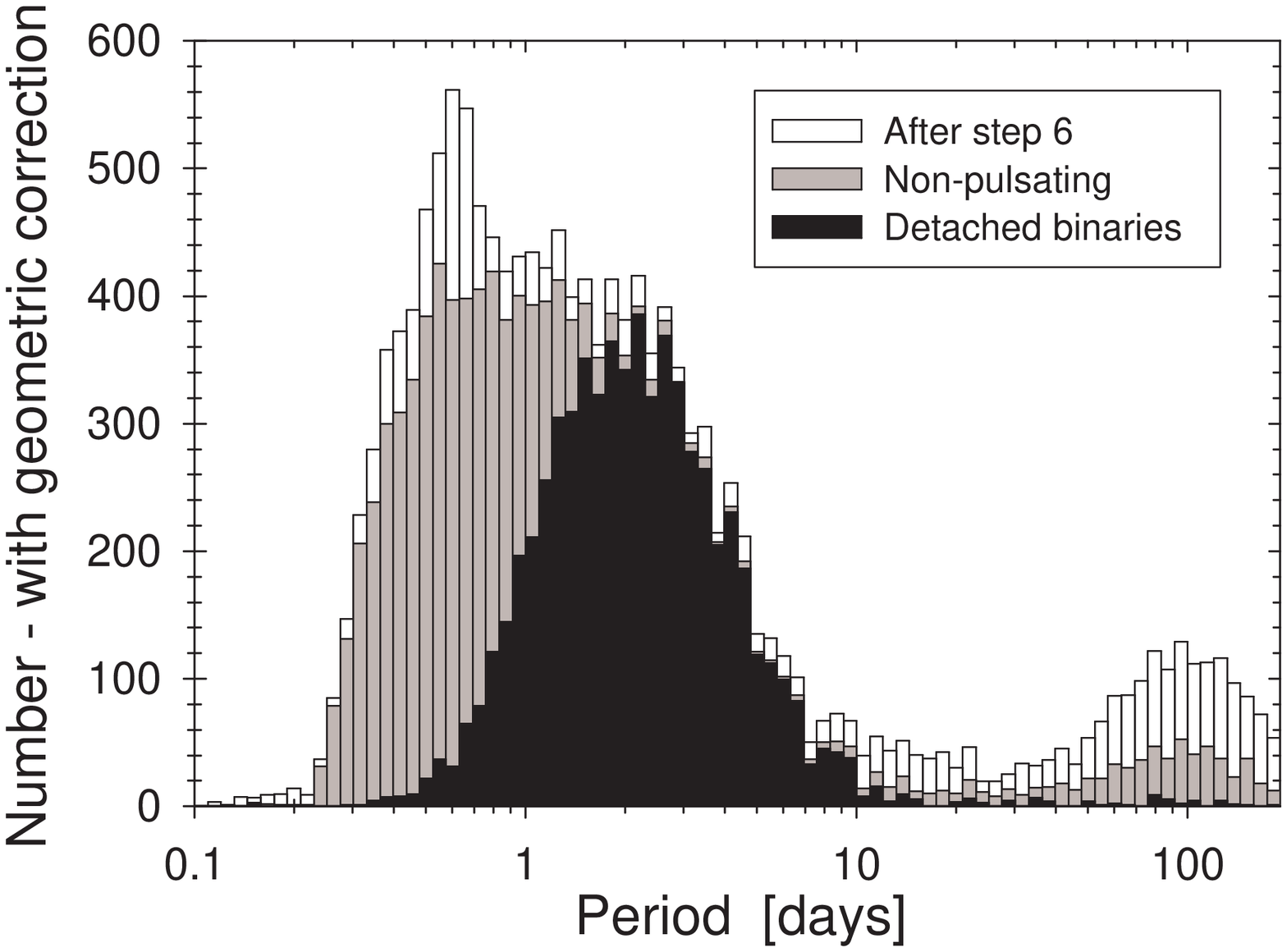}
\caption{The period distribution of OGLE~II bulge eclipsing binaries,
following various stages of filtration, after correcting for their geometric selection effect.}
\label{figPeriod2}
\end{figure}

The probability of eccentric systems (e $> 0$) eclipsing is more
difficult to calculate. We used a Monte-Carlo approach to
calculate a 1000$\times$1000 $[(R_1 + R_2)/a , e]$ probability
table, which was then used to interpolate the probability of each
fitted light curve. When applying this correction for the
geometric selection effect (see Figure~\ref{figPeriod2}) we see
that, as expected, it primarily effects the detached binaries. The
period ($P$) distribution of detached systems, both before
($p_{eclipse}$) and after ($p_{all}$) the geometric correction are
remarkably similar, and can both be well fit to a log-normal
distribution:

\begin{equation}
p_{eclipse,all}(P) = \frac{1}{\sqrt{2\pi} \sigma P} \exp
\left({-\frac{\ln^2(P/P_0)}{2 \sigma^2}} \right)
\end{equation}

Before the geometric correction ($p_{eclipse}$), we get a fit of
$P_0 = 1.83 \pm 0.01$ days and $\sigma = 0.593 \pm 0.005$ ($r^2
\simeq 0.9926$). After the correction ($p_{all}$), we get a fit of
$P_0 = 2.06 \pm 0.01$ days and $\sigma = 0.621 \pm 0.005$ ($r^2
\simeq 0.9928$). Such log-normal distributions are generally
indicative of many independent multiplicative processes taking
place, possibly both in the formation of binary systems as well as
in their observational/detection selection effects.

One should be very careful when using this method to calculate the
total number of binaries. Simply comparing the number of binaries
before and after the geometric correction will result in the
conclusion that we are observing as eclipsing binaries about half
of the total number of binaries. However, this fraction should in
fact be smaller than that, since we cannot detect eclipsing
binaries with periods longer than a few hundred days with the
OGLE~II survey.

\subsection{Extreme Systems}

In the previous subsection we considered the way large sets of
eclipsing systems are distributed. Now we will consider individual
systems. As examples we chose to locate extreme binary systems,
systems with a parameter well outside the normal range. For this
to be done properly, we need to take great care in avoiding the
pitfalls that arise from parameter estimation errors, both
systematic and non-systematic. One pitfall is that some of the
extreme systems may have large systematic errors since they will
contain additional phenomena that the fitted model neglects.
Another, perhaps more problematic pitfall, is that we will
retrieve non-extreme systems with large errors that happen to
shift the parameter in question beyond our filtering criterion
threshold. Some examples of possible sources of large errors are
inaccuracies in the correction for dust reddening (see the
beginning of this section) and difficulties in the estimation of
the argument of periastron, which directly affects the
determination of binary's orbital eccentricity \citep{Etzel91}.
Because of these pitfalls, we expect that the human eye will be
required as the final decision maker for this task in the
foreseeable future.

We present candidates of the most extreme eclipsing systems within
the OGLE~II bulge field dataset in five categories
(table~\ref{tableExtreme}). We chose examples for which we were
comparably confident, though we would need to follow them up with
spectroscopic measurements to be certain of their designations.

\begin{deluxetable}{lccccccccccccc}
\tabletypesize{\scriptsize}
\rotate
\tablecaption{Extreme eclipsing binary system candidates}
\tablewidth{0pt}
\tablehead{\colhead{Category} & \colhead{Field} & \colhead{Object} &
\colhead{Period} & \colhead{e} & \colhead{$R_1/a$} & \colhead{$R_2/a$} &
\colhead{$I_1$ [mag.]} & \colhead{$I_2$ [mag.]} & \colhead{$\chi_\nu ^2$} &
\colhead{$\bar{\rho }$} & \colhead{$\rho_{\max }$} &
\colhead{\begin{tabular}{c} Corrected \\ I [mag.] \end{tabular}} &
\colhead{\begin{tabular}{c} Corrected \\ V-I [mag.] \end{tabular}}}
\startdata
High density&      21& 5952&   1.468& 0.001& 0.137& 0.039& 15.21& 17.78& 1.30&    3.372&   153.6& 14.26&  0.48 \\
High density&      23& 1774&   0.505& 0.018& 0.245& 0.162& 16.91& 17.41& 3.17&    3.925&   17.45& 14.87& -1000 \\
Low density&        3& 8264& 151.026& 0.023& 0.487& 0.228& 17.39& 18.95& 2.53& 0.000007& 0.00007& 15.46&  1.33 \\
Low density&       21& 2568& 186.496& 0.265& 0.472& 0.127& 15.28& 18.18& 1.82& 0.000005& 0.00026& 14.47&  1.04 \\
High eccentricity&  2& 547&    2.419& 0.231& 0.211& 0.077& 12.12& 14.66& 1.89&    0.330&   7.049& 11.47& -0.37 \\
High eccentricity& 38& 4059&   2.449& 0.454& 0.288& 0.179& 17.81& 20.65& 1.61&    0.107&   0.553& 16.94&  0.68 \\
Blue&              21& 3797&   2.653& 0.003& 0.193& 0.104& 12.33& 14.35& 3.03&    0.322&   2.373& 11.48& -0.37 \\
Blue&              30& 1778&   6.442& 0.084& 0.089& 0.028& 12.81& 15.53& 2.39&    0.627&   21.37& 11.81& -0.52 \\
Short period&      18& 3424&   0.179& 0.119& 0.670& 0.211& 16.73& 19.28& 2.10&    1.895&  62.905& 15.53&  0.13 \\
Short period\tablenotemark{e}&
                   46&  797&   0.198& 0.072& 0.444& 0.426& 16.67& 16.60& 2.80&    2.913&   6.208& 14.87&  1.36 \\
Short period&      49&  538&   0.228& 0.008& 0.728& 0.264& 16.09& 18.35& 1.07&    0.904&  19.929& 15.03&  0.31 \\
Short period&      42& 2087&   0.233& 0.021& 0.568& 0.336& 17.94& 19.40& 1.27&    1.580&   9.242& 16.68&  0.85 \\
\enddata
\tablenotetext{e}{After finding this candidate using our pipeline,
we identified it as the eclipsing binary BW3 V38, which was
discovered and extensively studied by \citet{Maceroni97}. For
consistency, we listed the DEBiL fitted parameters, though their
accuracy is considerably worse than what is currently available in
the literature \citep{Maceroni04}. Specifically, the DEBiL fits
for the binary components' radii are overestimated by $\sim$25\%
due to their tidal distortions (see \S\ref{subsecLimitations}).}

\label{tableExtreme}
\end{deluxetable}

In the case of the high-density, low-density, and blue systems the
measured value of the characteristic is a weighted mean of the two
stars in the binary system. The weighting of the high- and
low-density systems is described in appendix B. The weighting of
the blue systems is determined, approximately, by the bolometric
luminosity of each of the stars in the binary system.

\section{Conclusions}

We present a new multi-tiered method for systematically analyzing
eclipsing binary systems within large-scale light curve surveys.
In order to implement this method, we have developed the DEBiL
fitter, a program designed to rapidly fit a large number of light
curves to a simplified detached eclipsing binary model. Using the
results of DEBiL one can select small subsets of light curves for
further follow-up. Applying this approach, we have analyzed
218,699 light curves from the bulge fields of OGLE~II, resulting
in 10,862 model fits. From these fits we identified unexpected
patterns in their parameter distribution. These patterns are
likely caused by selection effects and/or biases in the fitting
program. The DEBiL model was designed to fit only fully detached
systems, so users should use fits of semidetached and contact
binary systems with caution. One can probably find corrections for
the parameters of these systems, although it is best to refit them
using more complex models, which also take into account mutual
reflection and tidal effects. Even so, the DEBiL fitted parameters
will likely prove useful for quickly previewing large datasets,
classification, fitting detached systems, and providing an initial
guess for more complex model fitters.

\section*{Acknowledgments}

We would like to thank Robert Noyes, Krzysztof Stanek, Dimitar
Sasselov and Guillermo Torres for many useful discussions and
critiques. In addition we would like to thank Takahiro Sumi and
the OGLE collaboration for providing us with the data used in this
paper. Finally, we would like to thank Lisa Bergman for both her
editorial help and utmost support throughout this project. This
work was supported in part by NASA grants NAG5-10854 and
NNG04GN74G.

\section{Appendix - Statistical Tests}

\subsection{Fitness Score}

One of the problems with using the reduced chi-squared test is
that the light curve uncertainties may be systematically
overestimated (underestimated), causing the reduced chi-squared to
be too small (large). An easy way to get around this problem is by
comparing the reduced chi-squared of the DEBiL model being
considered, with the reduced chi-squared of an alternative model.
We used two simple alternative models:

- A constant, set to the average amplitude of the data.

- A smoothed spline, derived from a second order polynomial fit
within a sliding kernel\footnote{In our implementation, we used a
rectangular kernel whose width varies so as to cover a constant
number of data points. This is needed to robustly handle sparsely
sampled regions of the phased light curve.} over the phased light
curve.

The constant model should have a larger reduced chi-squared than
the best-fit model, while the spline model should usually have a
smaller reduced chi-squared. In a way similar to an F-test, we
define the fitness score as:

\begin{equation}
\mbox{Fitness\ Score} \equiv \frac{\chi_\nu^2 (\mbox{const}) - \chi_\nu^2
(\mbox{DEBiL})}{\chi_\nu^2 (\mbox{const}) - \chi_\nu^2 (\mbox{spline})}
\end{equation}

This definition is useful since gross over- or underestimates of
the uncertainties will largely cancel out. Light curves that reach
step 6 will mostly have fitness scores between 0 and 1. If the
reduced chi-squared of the DEBiL model equals the constant model's
reduced chi-squared, the fitness score will be 0, and if it equals
the spline model's reduced chi-squared, the fitness score will be
1. The fact that most of the DEBiL models have fitness scores
close to 1, and sometimes even surpassing it (see
Figure~\ref{figHistChi}), provides a validation for the fitting
algorithm used.

\subsection{Scatter Score}

This test quantifies the systematic scatter of data above or below
the model, using the correlation between neighboring residuals.
The purpose of this test is to quantify the quality of the model
fit independently of the reduced chi-squared test. While the
reduced chi-squared test considers the amplitude of the residuals,
the scatter score considers their distribution. The scatter score
is defined after folding the $n$ data points into a phase curve:

\begin{equation}
\mbox{Scatter\ Score} \equiv \frac{\Delta X_n \cdot \Delta X_1
+\sum\limits_{i=2}^n {\Delta X_{i-1} \cdot \Delta X_i }
}{\sum\limits_{i=1}^n {\Delta X_i^2 } }=\frac{\left\langle {\Delta
X_{i-1} \cdot \Delta X_i } \right\rangle }{\left\langle {\Delta
X_i^2 } \right\rangle }
\end{equation}

Where: $\Delta X_i \equiv X_i (\mbox{data})-X_i (\mbox{model})$

The scatter score will always be between 1 and -1. A score close
to 1 occurs when all $\Delta X_i$ are approximately equal. In
practice, this represent a severe systematic error, where the
model is entirely above or entirely below the data. When there is
no systematic error, the data are distributed randomly around the
model, generating a scatter score approaching 0. Scores close to
-1, although theoretically possible when $\Delta X_i \simeq
-\Delta X_{i-1}$, can be considered unphysical in that they are
unlikely to occur through systematic or non-systematic errors.

\subsection{Waviness}

This is a special case of the scatter score (see previous subsection).
Here, we consider only data points in the light curve's plateau
(i.e. the region in the phased curve between the eclipsing dips,
where both stars are fully exposed). The Waviness score is the
scatter score of these data points around their median. The purpose
of this test is to get a model-independent measure of irregularities
in the binary brightness, out of eclipse. A large value may indicate
such effects as stellar elongation, spots, or flares.

\section{Appendix - Density Estimation}

One of the most important criteria for selecting binaries for
follow-up is its stellar density. Unfortunately, the parameters
that can be extracted from the light curve fitting do not provide
us with enough information for deducing the density of any one of
the stars in the binary, but only a combined value. We define the
mean density as the sum of the stars' masses divided by the sum of
their volumes:

\begin{equation}
\bar {\rho }\equiv \frac{M_1+M_2}{\left( {4\pi /3} \right)\left(
{R_1^3+R_2^3 } \right)}=\frac{3\pi }{GP^2\left( {r_1^3 +r_2^3 }
\right)}\simeq \frac{0.01893\,g\,{cm}^{-3}}{P_{day}^2 \left(
{r_1^3+r_2^3 } \right)}\simeq \frac{0.01344\,
\rho_{\sun}}{P_{day}^2 \left( {r_1^3 + r_2^3} \right)}
\end{equation}

Where: $r_{1,2} \equiv R_{1,2} /a$ \ , and from Kepler's law:
$a^3=G\left( {M_1+M_2} \right)\left( P/{2\pi} \right)^2$

It should be noted here that if the stars' have very different
sizes, their mean density will be dominated by the larger one,
according to the weighted average:

\begin{equation}
\bar{\rho } = \frac{\left( {r_1 / r_2} \right)^3\rho_1 + \rho_2
}{\left({r_1 / r_2} \right)^3 + 1}
\end{equation}

Similarly, assuming $R_1 \ge R_2 $, the maximum possible density
is:

\begin{equation}
\rho_{\max } \equiv \frac{M_1+M_2}{\left( {4\pi /3} \right)R_2^3 }
= \bar {\rho } \left( {1+\left( {r_1 /r_2} \right)^3}
\right)\simeq \frac{0.01893\,g\,{cm}^{-3}}{P_{day}^2 r_2^3 }\simeq
\frac{0.01344\,\rho_{\sun}}{P_{day}^2 r_2^3}
\end{equation}

Adding the assumption that the more dense star of the binary is
the less massive component, we can reduce the upper limit of its
density to $\rho_{\max } / 2$.

\chapter{A Novel Approach to Analyzing Eclipsing Binaries in Large Photometric Datasets
\label{chapter3}}

\title{A Novel Approach to Analyzing Eclipsing Binaries in Large Photometric Datasets}

J.~Devor \& D.~Charbonneau 2006, \emph{Astrophysics and Space Science}, {\bf 304}, 351$-$354\\
\\
The original title of this paper was:\\
``A Method For Eclipsing Component Identification In Large Photometric Datasets.''

\section*{Abstract}

We describe an automated method for assigning the most likely
physical parameters to the components of an eclipsing binary (EB),
using only its photometric light curve and combined colors. In
traditional methods (e.g., WD and EBOP) one attempts to optimize a
multi-parameter model over many iterations, so as to minimize the
chi-squared value. We suggest an alternative method, where one
selects pairs of coeval stars from a set of theoretical stellar
models, and compares their simulated light curves and combined
colors with the observations. This approach greatly reduces the EB
parameter space over which one needs to search, and allows one to
determine the components' masses, radii and absolute magnitudes,
without spectroscopic data. We have implemented this method in an
automated program using published theoretical isochrones and limb
darkening coefficients. Since it is easy to automate, this method
lends itself to systematic analyses of datasets consisting of
photometric time series of large numbers of stars, such as those
produced by OGLE, MACHO, TrES, HAT, and many other surveys.

\section{Introduction}

Eclipsing double-lined spectroscopic binaries provide the only
method by which both the masses and radii of stars can be
estimated without having to resolve spatially the binary or rely
on astrophysical assumptions. Despite the large variety of models
and parameter-fitting implementations (e.g., WD and EBOP), their
underlying methodology is essentially the same. Photometric data
provides the light curve of the EB, and spectroscopic data provide
the radial velocities of its components. The depth and shape of
the light curve eclipses constrain the components' brightness and
fractional radii, while the radial velocity sets the length scale
of the system. In order to characterize fully the components of
the binary, one needs to combine all of this information.
Unfortunately, only a small fraction of all binaries eclipse, and
spectroscopy with sufficient resolution can be performed only for
bright stars. The intersection of these two groups leaves a
pitifully small number of stars.

In the past decade, there has been a dramatic growth in the number
of stars with high-quality, multi-epoch, photometric data. This
has been due to major advances in both CCD detectors and the
implementation of image-difference analysis techniques
\citep{Crotts92, Alard98, Alard00}, which enables simultaneous
photometric measurements of tens of thousands of stars in a single
exposure. Today, there are many millions of light curves available
from a variety of surveys, such as OGLE, \citet{Udalski94}; MACHO,
\citet{Alcock98}; TrES, \citet{Alonso04}; and HAT,
\citet{Bakos04}. But there has not been a corresponding growth in
the quantity of spectroscopic data, nor is this likely to occur in
the near future. Thus, the number of fully-characterized EBs has
not grown significantly. In recent years there has been a growing
effort to mine the wealth of available photometric data, by
employing simplified EB models in the absence of spectroscopic
observations \citep{Wyithe01,Wyithe02, Devor04, Devor05}.

In this paper we present a novel approach, which utilizes
theoretical models of stellar properties to estimate the orbital
parameters as well as the masses, radii, and absolute magnitudes
of the stars, while requiring \emph{only} a photometric light
curve and an estimate of the binary's combined color. This
approach can be used to characterize quickly large numbers of
eclipsing binaries, however it is not sufficient to improve
stellar models since underlying isochrones must be assumed. We
have created two implementations of this idea. The first, which we
have named MECI-express, and is described in section
\ref{secMECIexp}, is a ``quick and dirty'' program that is
designed as a simple extension to the Detached Eclipsing Binary
Light curve (DEBiL) fitter \citep{Devor04, Devor05}. The second,
which we have named MECI, and is described in section
\ref{secMECI}, is considerably more accurate, but also more
computationally demanding. The source code for both MECI-express
and MECI will be provided upon request.

\section{Express Method for Eclipsing Component Identification (MECI-express)}
\label{secMECIexp}

The primary application of MECI-express is to identify the stellar
components of a given EB. It operates after a conventional EB
model-fitting program has already analyzed the given EB's light
curve. In our implementation, we chose to employ DEBiL
\citep{Devor04, Devor05} since it is simple, fast, and fully
automated. The fitted parameters are the orbital period (P), the
apparent magnitudes ($mag_{1,2}$), and the fractional radii
($r_{1,2}$) of the binary components. A fractional radius is
defined as the radius ($R_{1,2}$) divided by the sum of the
components' semimajor axes ($a$). In MECI-express we iterate
through a large group of MK spectral type pairings, to each of
which we associate typical stellar parameters \citep{Cox00}. These
stellar parameters are the masses ($M_{1,2}$), the radii
($R_{1,2}$), and the absolute magnitudes ($Mag_{1,2}$) of the
binary components. If the assumed values of the stellar parameters
match the true values, then the stellar and fitted parameters
should obey to the following equations:

\begin{eqnarray}
\frac{4\pi^2R_1^3}{G(M_1 + M_2)} &=& P^2r_1^3 \label{eq_r1}\\
\frac{4\pi^2R_2^3}{G(M_1 + M_2)} &=& P^2r_2^3 \label{eq_r2}\\
Mag_1 - Mag_2 &=& mag_1 - mag_2 \label{eqMag}
\end{eqnarray}

We also may have additional constraints from the observed
out-of-eclipse combined colors of the system. For example, in the
case of OGLE~II targets, we have the estimated V-I color:

\begin{equation}
Mag_V - Mag_I = mag_V - mag_I \label{eqColor}
\end{equation}

\begin{figure}[H]
\centerline{
\begin{tabular}{ccc}
\includegraphics[width=1.75in]{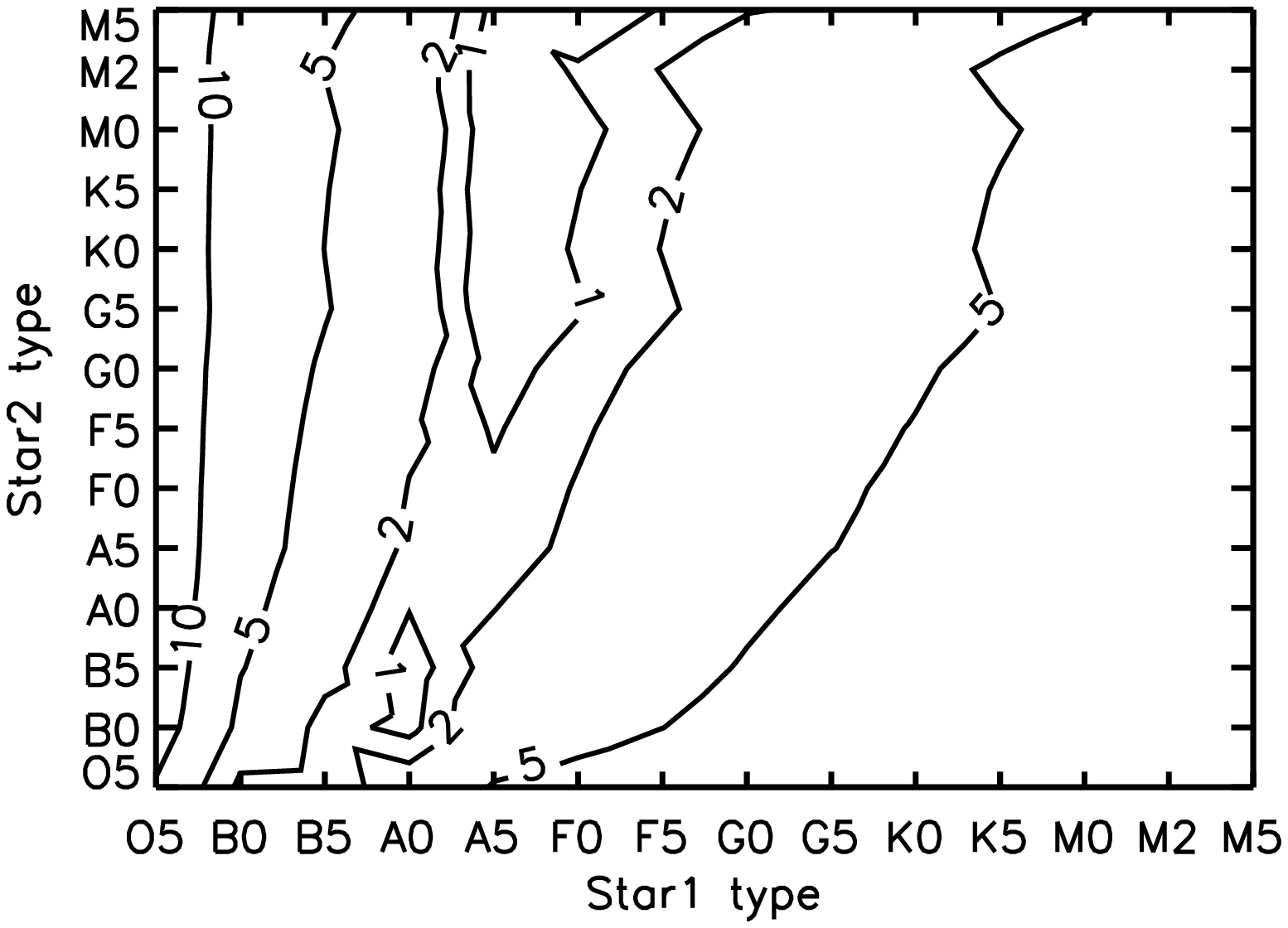} &
\includegraphics[width=1.75in]{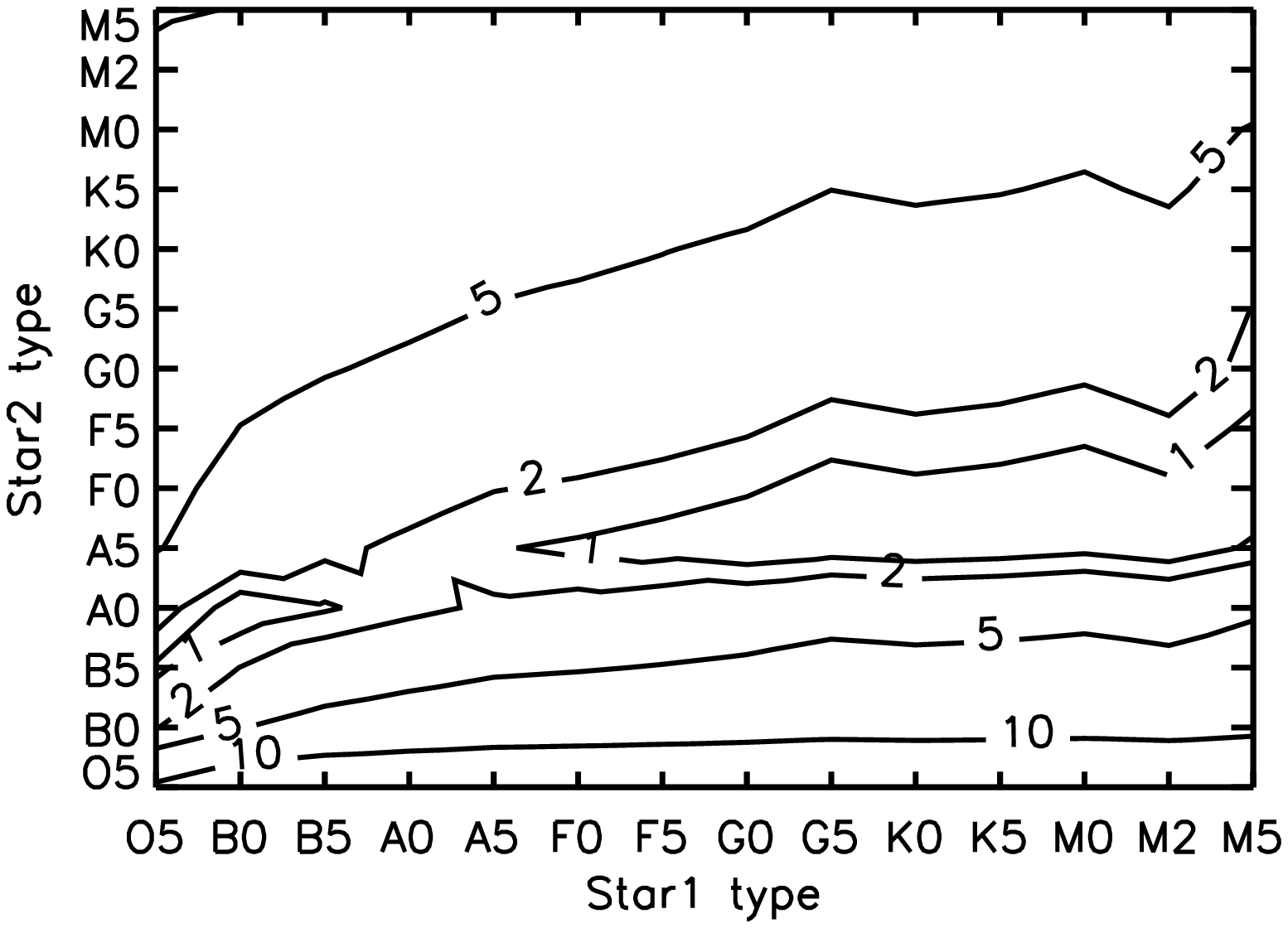} &
\includegraphics[width=1.75in]{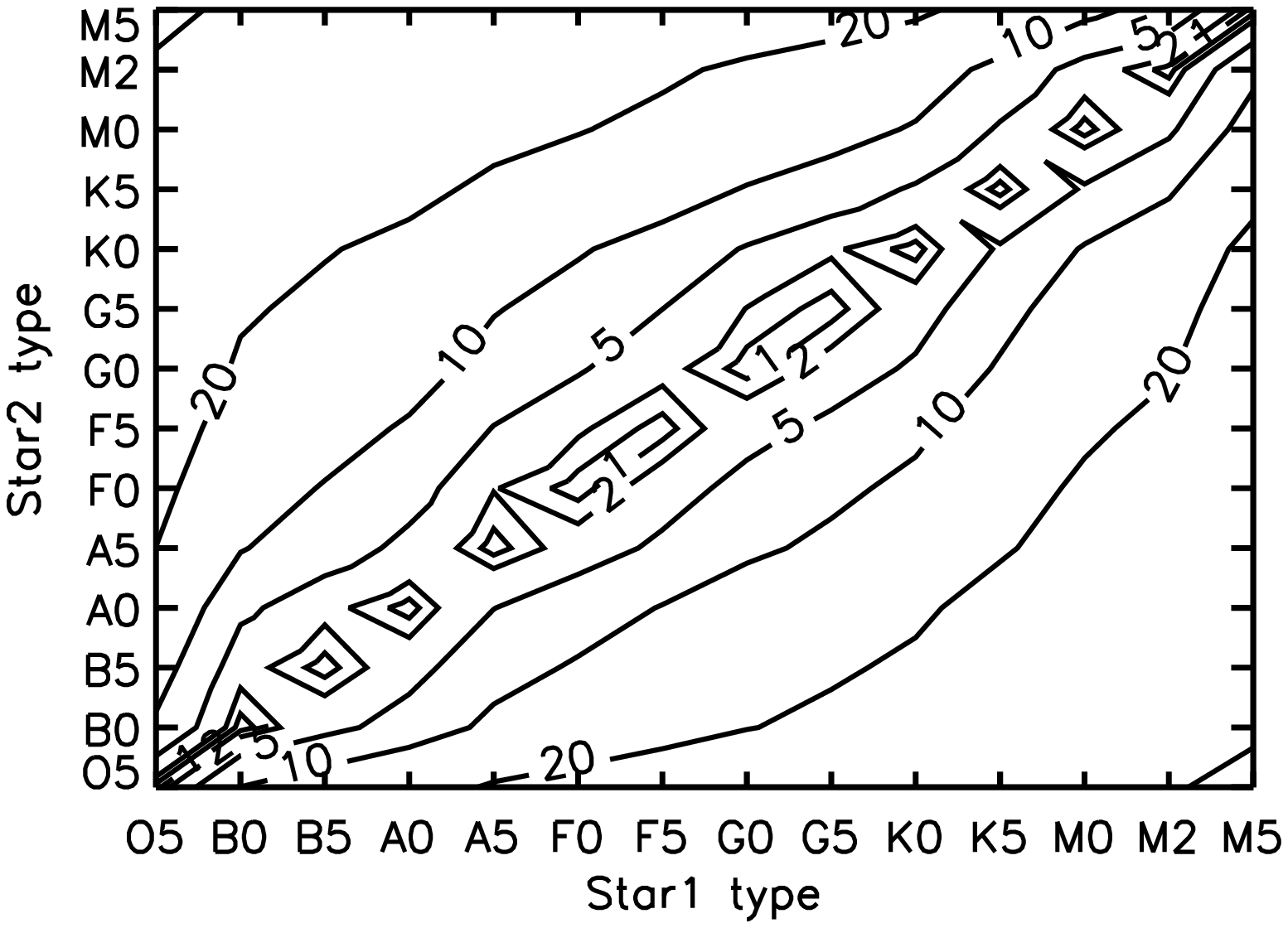}\\
a. Constraints from Eq. \ref{eq_r1} & b. Constraints from Eq. \ref{eq_r2} &
c. Constraints from Eq. \ref{eqMag}
\end{tabular}}
\caption{Contour plots of the absolute difference between the
left-hand-side and the right-hand-side of each equation, divided
by its uncertainty, as applied to the WW Camelopardalis light
curve \citep{Lacy02}. Adding these results in quadrature, produces
the likelihood plot shown in Figure~\ref{fig2}.a.}
\label{fig1}
\end{figure}

\begin{figure}[H]
\centerline{
\begin{tabular}{c@{\hspace{2pc}}c}
\includegraphics[width=2.4in]{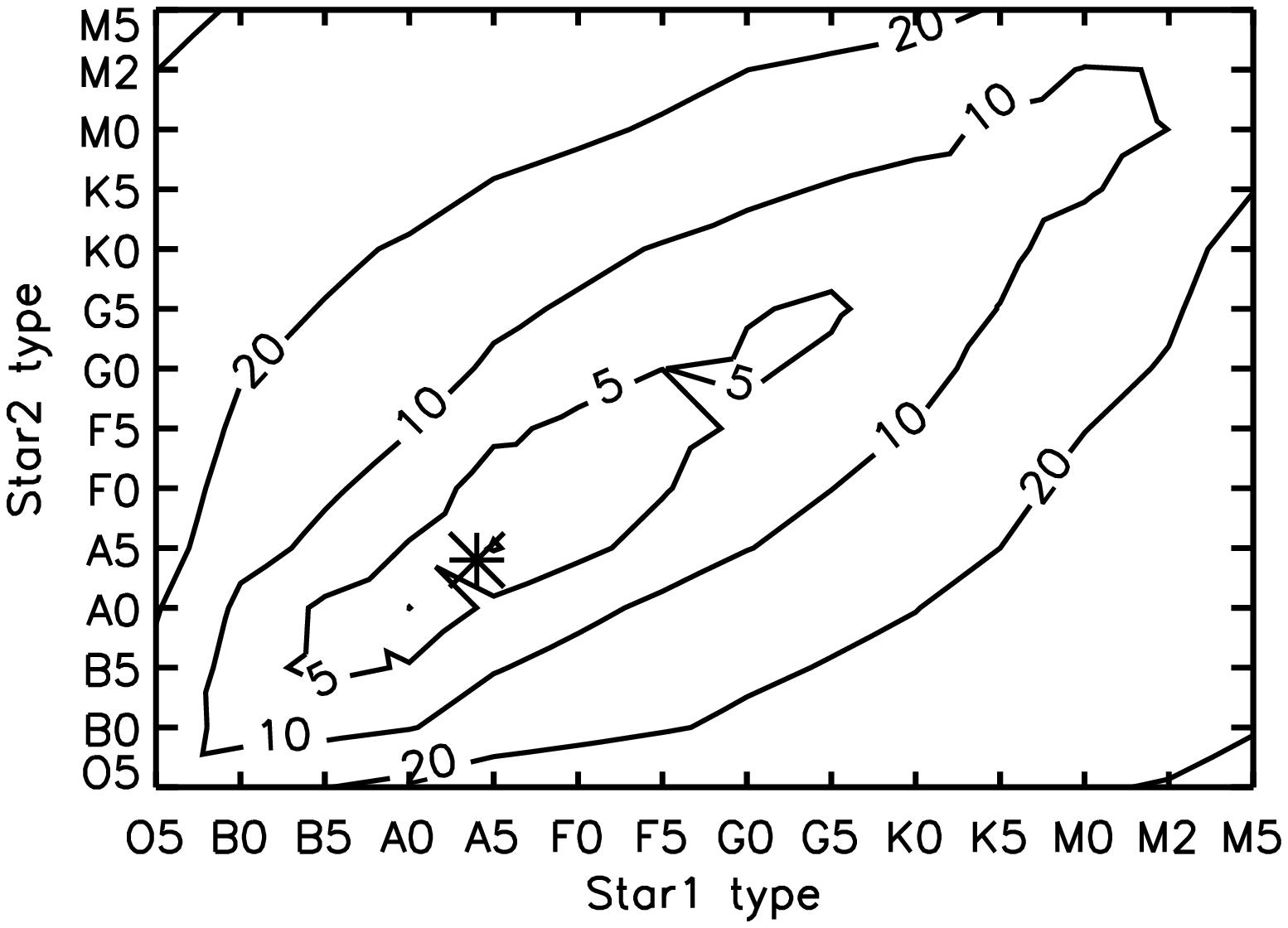} &
\includegraphics[width=2.4in]{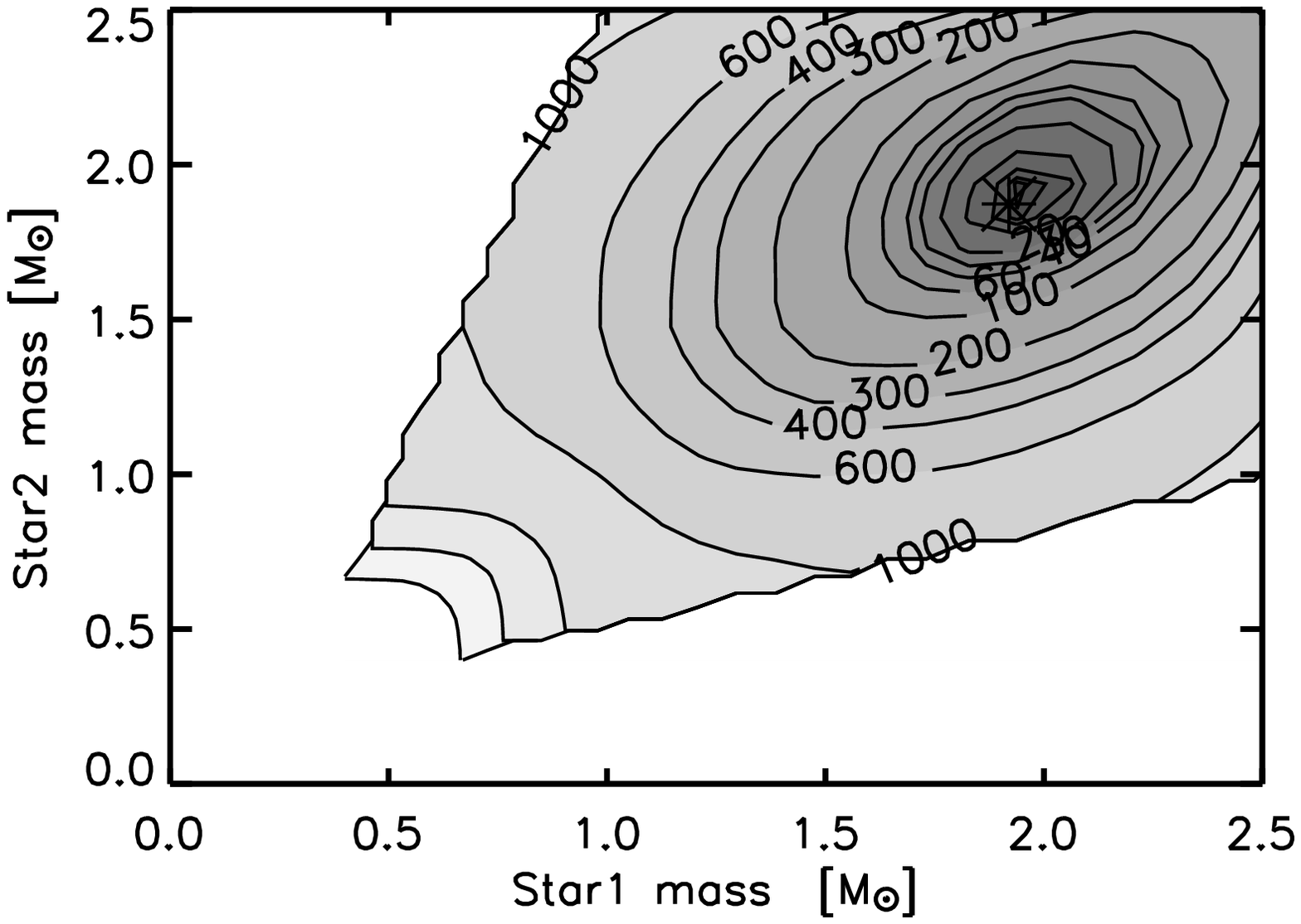} \\
a. MECI-express likelihood plot & b. MECI likelihood plot (age = 0.6 Gyr)
\end{tabular}}
\caption{A comparison of the MECI-express (left) and MECI (right)
likelihood contour plots for WW Camelopardalis. The value of the
contours are described in the body of the text. The asterisk marks
the solution of \citet{Lacy02}.}
\label{fig2}
\end{figure}

We assume that the color has been corrected for reddening and that
no systematic errors are present, so any inequalities would be due
to an incorrect choice for the component pairing. The likelihood
of each pairing is assessed by calculating the difference between
the left-hand-side (stellar parameters) and right-hand-side
(fitted parameters) of each equation. These differences are
divided by their uncertainties, and added in quadrature. The
pairing with the smallest sum is deemed the most likely pairing.
For each given EB light curve, MECI-express returns the list of
the top ranked (most likely) binary pairings, with their
corresponding sums. MECI-express can also be used to create a
contour plot of the probability distribution for all pairings. We
illustrate an example of individual MECI-express components in
Figures~\ref{fig1}.a-c, which are then combined to create the
result shown in Figure~\ref{fig2}.a.

\section{Method for Eclipsing Component Identification (MECI)}
\label{secMECI}

MECI was developed to improve significantly upon the accuracy of
MECI-express (see Table 1). This was done as follows: We replaced
the use of spectral types with the more fundamental (and
continuous) quantities of mass and age. Furthermore, in MECI we
assume that the two binary components are coeval, thus replacing
the 2-dimensional spectral type - spectral type grid, with a
3-dimensional mass-mass-age grid. Finally, we no longer rely on
parameter fits of the components' apparent magnitudes and
fractional radii directly from the light curve, which are often
very uncertain, nor do we assume constant limb darkening
coefficients. Instead, we interpolate these values for the given
mass-mass-age pairing, from precalculated tables [Yonsei-Yale
isochrones \citep{Kim02} ; ATLAS limb darkening coefficients
\citep{Kurucz92}, used when $T_{eff} \geq 10000K$ or $\log g \leq
3.5$  ; PHOENIX limb darkening coefficients \citep{Claret98,
Claret00}, used when $T_{eff} < 10000K$ and $\log g > 3.5$]. Thus,
by assuming the masses ($M_{1,2}$) of the EB components and the
system's age, we can look-up the radii ($R_{1,2}$), the absolute
magnitudes ($Mag_{1,2}$), and the limb darkening coefficients for
the binary components. We then use these values, as well as the
observationally-determined period (P) and combined magnitude out
of eclipse ($mag_{comb}$), to calculate the apparent magnitudes
($mag_{1,2}$) and factional radii ($r_{1,2}$) of the EB
components, as follows:

\begin{eqnarray}
mag_1 &=& mag_{comb} + 2.5\log \left[1 + 10^{-0.4(Mag_2 - Mag_1)}\right]\\
mag_2 &=& mag_1 + (Mag_2 - Mag_1)\\
a &=& [G(M_1 + M_2)(P/2\pi)^2]^{1/3} \simeq \nonumber\\
  & & 4.206 R_{\odot}(M_1/M_{\odot} + M_2/M_{\odot})^{1/3} P^{2/3}_{day}\\
r_{1,2} &=& R_{1,2} / a
\end{eqnarray}

Besides the epochs of eclipses, which can be determined directly
from the EB light curve, there are only two additional parameters
required for us to simulate the light curves of the given pairing:
the orbital eccentricity (e) and inclination (i). For binaries
with short periods ($\lesssim$~2~days) and a secondary eclipse
precisely half an orbit after the primary eclipse, it is
reasonable to assume a circular orbit ($e=0$). Otherwise, one
should use the eccentricity derived by an EB model-fitting program
(we use DEBiL). Finding the inclination robustly is more
difficult. We employ a bracket search \citep{Press92}, which
returns the inclination that produces the best resulting fit.

To summarize, for every combination of component masses and system
age of an EB, we can look-up, calculate, or fit all the parameters
needed to simulate its light curve (P, limb darkening
coefficients, mag$_{1,2}$, r$_{1,2}$, epochs of eclipses, e, i),
as well as its apparent combined color. We systematically iterate
through many such combinations. For each one we compare the
expected light curve with the observations, and calculate the
reduced chi-squared value ($\chi^2_\nu$). We also compare each
observed color ($O_c \pm \epsilon_c$) with its calculated value
($C_c$), and combine them by defining: $score\equiv(w\chi^2_\nu +
\sum_{c=1}^N [(O_c - C_c)/\epsilon_c]^2)/(w+N)$. Where $w$ is the
$\chi^2_\nu$ information weighting. We use $w=1$, and assume that
the smaller the score, the more likely it is that we have chosen
the correct binary pairing. One can visualized this result using a
series of $score(M_1, M_2)$ contour plots, each with a constant
age (e.g., Figure~\ref{fig2}.b).

\section{Conclusions}

We have described a novel method for identifying an EB's
components using only its photometric light curve and combined
color. By utilizing theoretical isochrones and limb darkening
coefficients, this method greatly reduces the EB parameter space
over which one needs to search. This approach seeks to estimate
the masses, radii and absolute magnitudes of the components,
without spectroscopic data. We described two implementations of
this method, MECI-express and MECI, which enable systematic
analyses of datasets consisting of photometric time series of
large numbers of stars, such as those produced by OGLE, MACHO,
TrES, HAT, and many others. Such techniques are expected to grow
in importance with the next generation surveys, such as Pan-STARRS
\citep{Kaiser02} and LSST \citep{Tyson02}.


\begin{deluxetable}{lcc|ccc|ccc}
\tabletypesize{\scriptsize}
\tablecaption{A comparison of the results produced by MECI-express,
MECI, and conventional analyses with their uncertainties \citep{Lacy00, Lacy02, Lacy03}.
The square brackets with numerical values
indicate the deviation of our results from those of the conventional approach.}
\tablewidth{0pt}
\tablehead{ & \multicolumn{2}{c}{MECI-express} & \multicolumn{3}{c}{MECI} & \multicolumn{3}{c}{\citet{Lacy00, Lacy02, Lacy03}}}
\startdata
Parameter & Mass 1       & Mass 2        & Mass 1        & Mass 2        & Age   & Mass 1        & Mass 2        & Age \\
      &[$M_{\odot}$] & [$M_{\odot}$] & [$M_{\odot}$] & [$M_{\odot}$] & [Gyr] & [$M_{\odot}$] & [$M_{\odot}$] & [Gyr] \\
\hline
FS Mon & 2.9 (A0) & 2.0 (A5) & 1.62     & 1.52     &  1.4  & 1.632      & 1.462      & 1.6      \\
       & [77.7\%] & [36.8\%] & [0.6\%] & [4.1\%] & [0.2] & $\pm$0.012 & $\pm$0.010 & $\pm$0.3 \\
WW Cam & 2.0 (A5) & 2.0 (A5) & 1.97     & 1.89     &  0.5  & 1.920      & 1.873      & 0.5      \\
       & [4.2\%]  & [6.8\%]  & [2.8\%] & [1.0\%] & [0.0] & $\pm$0.013 & $\pm$0.018 & $\pm$0.1   \\
BP Vul & 2.0 (A5) & 1.6 (F0) & 1.77     & 1.48     &  0.7  & 1.737      & 1.408      & 1.0      \\
       & [15.1\%] & [13.6\%] & [2.1\%] & [5.4\%] & [0.3] & $\pm$0.015 & $\pm$0.009 & $\pm$0.2
\enddata
\end{deluxetable}

We are grateful to Guillermo Torres for many helpful
conversations.

\chapter{MECI: A Method for Eclipsing Component Identification
\label{chapter4}}

\title{MECI: A Method for Eclipsing Component Identification}

J.~Devor \& D.~Charbonneau 2006, \emph{The Astrophysical Journal}, {\bf 653}, 647$-$656

\section*{Abstract}

We describe an automated method for assigning the most probable
physical parameters to the components of an eclipsing binary,
using only its photometric light curve and combined colors. With
traditional methods, one attempts to optimize a multi-parameter
model over many iterations, so as to minimize the chi-squared
value. We suggest an alternative method, where one selects pairs
of coeval stars from a set of theoretical stellar models, and
compares their simulated light curves and combined colors with the
observations. This approach greatly reduces the parameter space
over which one needs to search, and allows one to estimate the
components' masses, radii and absolute magnitudes, without
spectroscopic data. We have implemented this method in an
automated program using published theoretical isochrones and limb
darkening coefficients. Since it is easy to automate, this method
lends itself to systematic analyses of datasets consisting of
photometric time series of large numbers of stars, such as those
produced by OGLE, MACHO, TrES, HAT, and many others surveys.

\section{Introduction}

Eclipsing double-lined spectroscopic binaries provide the only
method by which both the masses and radii of stars can be
estimated without having to resolve the binary spatially or rely
on astrophysical assumptions. Despite the large variety of models
and parameter-fitting implementations [e.g., WD \citep{Wilson71}
and EBOP \citep{Etzel81, Popper81b}], their underlying methodology
is essentially the same. Photometric data provide the light curve
of the eclipsing binary (EB), and spectroscopic data provide the
radial velocities of its components. The depth and shape of the
light curve eclipses constrain the components' brightness and
fractional radii, while the radial velocity sets the length scale
of the system. In order to fully characterize the components of
the binary, one needs to combine all of this information. Only a
small fraction of all binaries eclipse, and spectra with
sufficient resolution and signal-to-noise ratios can be gathered
only for bright stars. The intersection of these two groups leaves
a small number of stars.

Over the past decade, the number of stars with high-quality,
multi-epoch, photometric data has grown dramatically due to the
growing interest in finding gravitational lensing events
\citep{Wambsganss06} and eclipsing extrasolar planets
\citep{Charbonneau07}. In addition, major technical improvements
in both CCD detectors and implementations of image-difference
analysis techniques \citep{Crotts92, Alard98, Alard00} enable
simultaneous photometric measurements of tens of thousands of
stars in a single exposure. Today, there are many millions of
light curves available from a variety of surveys, such as OGLE
\citep{Udalski94}, MACHO \citep{Alcock98}, TrES \citep{Alonso04},
HAT \citep{Bakos04}, and XO \citep{McCullough06}. Despite the
increase in photometric data, there has not been a corresponding
growth in the quantity of spectroscopic data, nor is this growth
likely to occur in the near future. Thus, the number of fully
characterized EBs has not grown at a rate commensurate with the
available photometric datasets.

In recent years, there has been a growing effort to mine the
wealth of available photometric data, by employing automated
pipelines that use simplified EB models in the absence of
spectroscopic observations and hence without a fixed physical
length scale or absolute luminosity \citep{Wyithe01, Wyithe02,
Devor04, Devor05}. In this paper, we present a method that
utilizes theoretical isochrones and multi-epoch photometric
observations of the binary system to estimate the physical
parameters of the component stars, while still not requiring
spectroscopic observations.

Our Method for Eclipsing Component Identification\footnote{The
source code can be downloaded from:
http://cfa-www.harvard.edu/$\sim$jdevor/MECI.html} (MECI), finds
the most probable masses, radii, and absolute magnitudes of the
stars. The input for MECI is an EB's photometric light curve and
out-of-eclipse colors (we note that in the absence of color
information, the accuracy in the estimation of the stellar
parameters is significantly reduced; \S\ref{subsecSimulate}). This
approach can be used to quickly characterize large numbers of
eclipsing binaries; however it is not sufficient to improve
stellar models, since underlying isochrones must be assumed.

In a previous paper \citep{Devor06a}, we outlined the ideas behind
both MECI and a closely related, ``quick and dirty'' alternative,
which we called ``MECI-express.'' Although MECI-express is much
faster and easier to implement, it is also far less accurate. For
this reason we will not discuss it further, and instead
concentrate exclusively on MECI. We discuss its applications
(\S\ref{secCh4Motivation}), aspects of its implementation
(\S\ref{secCh4Method}), tests of its accuracy
(\S\ref{secCh4Tests}), and finally summarize our findings
(\S\ref{secConclusions}).

\section{Motivation}
\label{secCh4Motivation}

\subsection{Characterizing the Binary Stellar Population}
\label{subsecStellarPop}

First and foremost, MECI is designed as a high throughput means to
systematically estimate the masses of large numbers of stars.
Although the result in each system is uncertain, by statistically
analyzing large catalogs, one can reduce the non-systematic
errors. Much work has already been invested into characterizing
binary systems through spectroscopic binary surveys
\citep[e.g.,][]{Duquennoy91, Pourbaix04}, yet the limited data and
their large uncertainties have led to inconsistent results
\citep{Mazeh05}. The driving questions that have spurred debate in
the community include: What are the initial mass functions of the
primary and secondary components? How do they relate to the
initial mass function of single stars? What is the distribution of
the components' mass ratio, $q$, and in particular, does it peak
at unity? This lack of understanding is further highlighted by the
fact that most of the stars in our galaxy are members of binary
systems, and that these questions have lingered for over a
century. MECI may help sort this out by systematically
characterizing the component stars of many EB systems.

By requiring only photometric data, a survey using MECI can study
considerably fainter binary systems than spectroscopic surveys,
and thus remain complete to a far larger volume. As an
illustrative example, the difference image analysis of the bulge
fields of OGLE II, using the Las Campanas 1.3m Warsaw telescope
in a drift-scan mode (an effective exposure time of 87 seconds),
attained a median noise level of 0.1 mag, for $I = 18$ binaries,
even in moderately crowded fields \citep{Wozniak00}. In contrast
to this, the CfA Digital Speedometer on the 1.5m F.~L.~Whipple
Observatory telescope has a spectral resolution of \textsf{R} $\simeq
35,000$ (at $5177$\AA) and typically yields a radial velocity
precision of 0.5 ${\rm km \, s^{-1}}$, with a faint magnitude
limit of $V = 13$ \citep{Latham92}. Although the limiting magnitudes
are very much dependent on the throughput of the relevant
instruments and the precision one wishes to achieve, this $5$
magnitude difference for telescopes of similar aperture
corresponds to a factor of 10 in distance or 1000 in volume,
and illustrates the significant expansion that can be achieved by
purely photometric surveys. Conversely, one can achieve the same
magnitude limit with an aperture $10$ times smaller. The success
of this approach has been demonstrated by several automated
observatories, such as TrES \citep{Alonso04} and HAT
\citep{Bakos04}, which each use networks of observatories with
10-cm camera lenses to monitor stars to $V \simeq 13$.

\subsection{Identifying Low-Mass Main-Sequence EBs}
\label{subsecIdLowMass}

One of the most compelling applications of MECI will be to quickly
sort thousands of EBs present in large photometric surveys, and to
subsequently select a small subset of objects from the resulting
catalog for further study. In particular, lower main-sequence
stars that are partially or fully convective have not been studied
with a level of detail remotely approaching that of solar-type
(and more massive) stars. This is particularly troubling since
late-type stars are the most common in the Galaxy, and dominate
its stellar mass. It has been shown that models underestimate the
radii of low-mass stars by as much as 15-20\% \citep{Lacy77b,
Torres02}; a significant discrepancy considering that for
solar-type stars the agreement with the observations is typically
within 1-2\% \citep{Andersen98}. Similar problems exist for the
effective temperatures predicted theoretically for low-mass stars.
Progress in this area has been hampered by the lack of suitable
M-dwarf binary systems with accurately determined stellar
properties, such as mass, radius, luminosity, surface temperature,
and metallicity. Detached eclipsing systems are ideal for this
purpose, but only five are known among M-type stars: CM Dra
\citep{Lacy77a, Metcalfe96}, YY Gem \citep{Kron52, Torres02}, CU
Cnc \citep{Delfosse99, Ribas03}, OGLE BW3 V38 \citep{Maceroni97,
Maceroni04}, and TrES-Her0-07621 \citep{Creevey05}. They range in
mass from about 0.25 $M_{\sun}$ (CM Dra) to 0.6 $M_{\sun}$ (YY
Gem).  The number of such objects could be greatly increased by
using tools such as MECI to mine the extant photometric datasets
and locate these elusive low-mass systems.

\subsection{EBs as Standard Candles}
\label{secStandardCandles}

\begin{figure}
\includegraphics[width=5in]{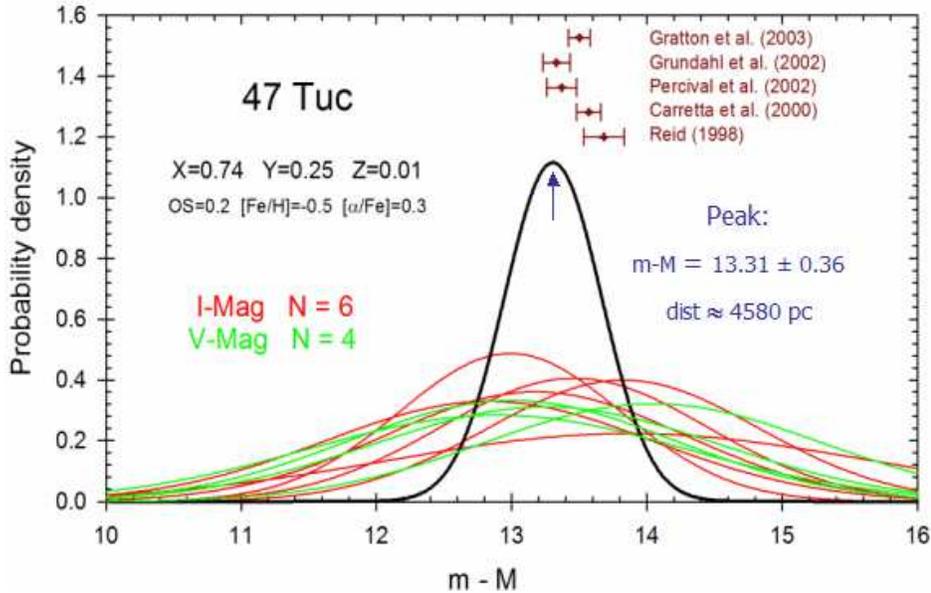}
\caption{This figure was not originally part of this paper. We
include it here since it nicely illustrates the point discussed in
\S\ref{secStandardCandles}. In this figure we estimate the
distance modulus of 47 Tucanae, by combining the results of ten
MECI analyses of EB LCs ($I$-band or $V$-band) observed in this
globular cluster \citep{Albrow01}. We show the results of each of
these MECI analyses as a Gaussian distribution. The product of
these distributions produced the far narrower Gaussian
distribution indicated by the arrow. The Peak of the combined
result is consistent with a variety of other methods that were
used to derive the distance modulus \citep{Reid98, Carretta00,
Percival02, Grundahl02, Gratton03}.}
\label{figDistMod}
\end{figure}

Using MECI, we are able to estimate the absolute magnitude of the
binary system. This, together with its extinction-corrected
out-of-eclipse apparent magnitude, allows us to then calculate the
distance modulus to any given EB. The estimation of distances to
EBs dates back to \citet{Stebbing10}, and their use as distance
candles in the modern astrophysical context was recently
elucidated by \citet{Paczynski97}. However, unlike these studies,
MECI does not require spectroscopy and therefore is able to
analyze binaries that are significantly less luminous (see
\S\ref{subsecStellarPop}). Although the distance estimation from
MECI will be uncertain, in many cases this will still be an
improvement over existing methods. For example, if there are many
EBs in a stellar cluster, the distance estimate can be greatly
improved by combining their results, reducing the non-systematic
errors by a factor of the square root of the number of systems.
Following \citet{Guinan96}, one might be able to use such
clustered EB standard candles to better constrain the distance to
the LMC and SMC, and thus be able to further constrain the bottom
of the cosmological distance ladder. In the case of MECI, the
uncertainties of each distance measurement will be considerably
larger, but as suggested by T. Mazeh (2005, private
communication), this will be compensated for by the far larger
number of measurements that can be made (see, e.g.,
Figure~\ref{figDistMod}). Another intriguing application of such
EB standard candles is to map large scale structures in the
Galaxy, such as the location and orientation of the Galactic bar,
arms, and merger remnants [see, e.g., \citet{Vallee05} and
references therein].

\section{Method}
\label{secCh4Method}

The EB component identification is performed in two stages. First
the orbital parameters of the EB are estimated
(\S\ref{subsecOrbitalParams}), then the most likely stellar
parameters are identified (\S\ref{subsecStellarParams}). Our
implementation of MECI has the option to fix the estimates of the
orbital parameters, or to fine-tune them for each
stellar pairing considered in the second stage. The average
running time for MECI to analyze a $1000$-point light curve on a
single 3.4GHz Intel Xeon CPU is 0.4 minutes. If we permit fine
tuning of the orbital parameters for each pairing, the running
time grows to 6 minutes per light curve.

\subsection{Stage 1: Finding the Orbital Parameters}
\label{subsecOrbitalParams}

In the first stage, we estimate the EB's orbital parameters from
its light curve.  Many EBs have orbital periods of a few
days or less, owing to the greater probability for such systems
to present mutual eclipses, and to the limited baselines in the
datasets from which they are identified. Most of these
systems will have orbits that have been circularized due to
tidal effects. For such circular orbits, the only
parameters we seek are the orbital period, $P$, and epoch of
periastron, $t_0$. For non-circular orbits we
also fit the orbital eccentricity, $e$, and the argument of periastron,
$\omega$. The period is determined using a periodogram, and the
remaining parameters are obtained through fitting the offset,
duration and time interval between the light curve's eclipses (see
below). Holding these parameters fixed at these initial estimates
significantly reduces the computational requirements of MECI.

We postpone fitting the orbital inclination, $i$, until the second
stage, since it is difficult to determine this parameter robustly
without first assuming values for the stellar radii and masses.
This difficulty arises because it is often difficult to
distinguish a small secondary component from a large secondary
component in a grazing orbit. In stage 2, additional information,
such as the theoretical stellar mass-radius relation and colors
are used to help resolve this degeneracy.

The procedure for fitting the aforementioned parameters from the
EB light curve is a well-studied problem \citep{Kopal59, Wilson71,
Etzel91}. We chose to estimate the period with a variant of the
analysis of variances (AOV) periodogram\footnote{The source code
and running examples of both the AOV periodogram and the DEBiL
fitter can be downloaded from:
http://cfa-www.harvard.edu/$\sim$jdevor/DEBiL.html} by
\citet{SchwarzenbergCzerny89, SchwarzenbergCzerny96}. We then use
the Detached Eclipsing Binary Light curve (DEBiL)
fitter\footnotemark[3] by \citet{Devor04, Devor05} for fitting the
remaining orbital parameters. For non-circular systems, following
\citet{Kopal59} and \citet{Kallrath99}, we estimate the orbital
eccentricity and argument of periastron from the orbital period,
the duration of the eclipses, $\Theta_{1,2}$, and the time
interval between the eclipse centers, $\Delta t$, as follows:
\begin{eqnarray}
\omega & \simeq & \arctan \left[\frac{2}{\pi} \left(\frac{\Theta_1 - \Theta_2}{\Theta_1 + \Theta_2} \right)\left(\frac{\Delta t}{P} - \frac{1}{2}\right)^{-1}\right], {\rm \ and}\\
e & \simeq & \frac{\pi}{2 \cos \omega} \left| \frac{\Delta t}{P} - \frac{1}{2} \right|.
\end{eqnarray}

In practice, it is difficult to accurately determine the eclipse
duration. We estimate this duration by first calculating
the median flux outside the eclipses, then estimating the midpoints and
depths of the eclipses using a spline. We then assign
the duration of each eclipse to be the time elapsed
from the moment at which the light curve during ingress crosses the midpoint between the
out-of-eclipse and bottom-of-eclipse fluxes, until the moment
at which the light curve crosses the corresponding point during egress.

\subsection{Stage 2: Finding the Absolute Stellar Parameters}
\label{subsecStellarParams}

\begin{figure}
\includegraphics[width=5in]{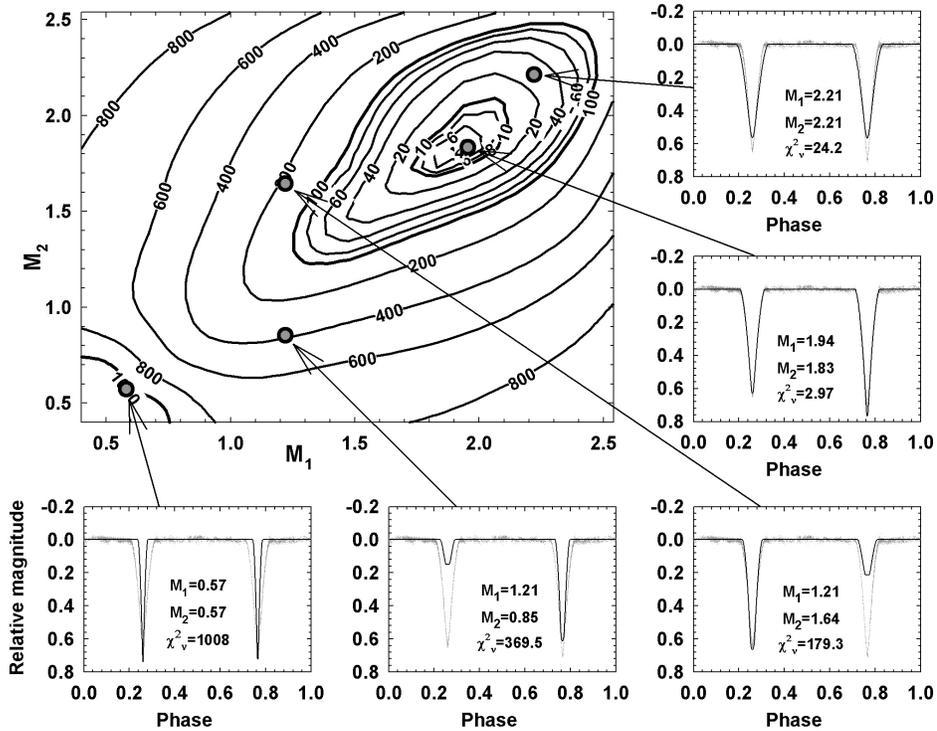}
\caption{The large upper-left panel shows the MECI $\chi^2_\nu$ surface as
a function of the assumed masses (in solar units)
of the component stars in the WW Camelopardalis system.  The model light curve
at five locations in the grid is shown in the smaller panels, overplotted on
the observed light curve from \citet{Lacy02}.}
\label{figMECIpanels}
\end{figure}

\begin{figure}
\includegraphics[width=5in]{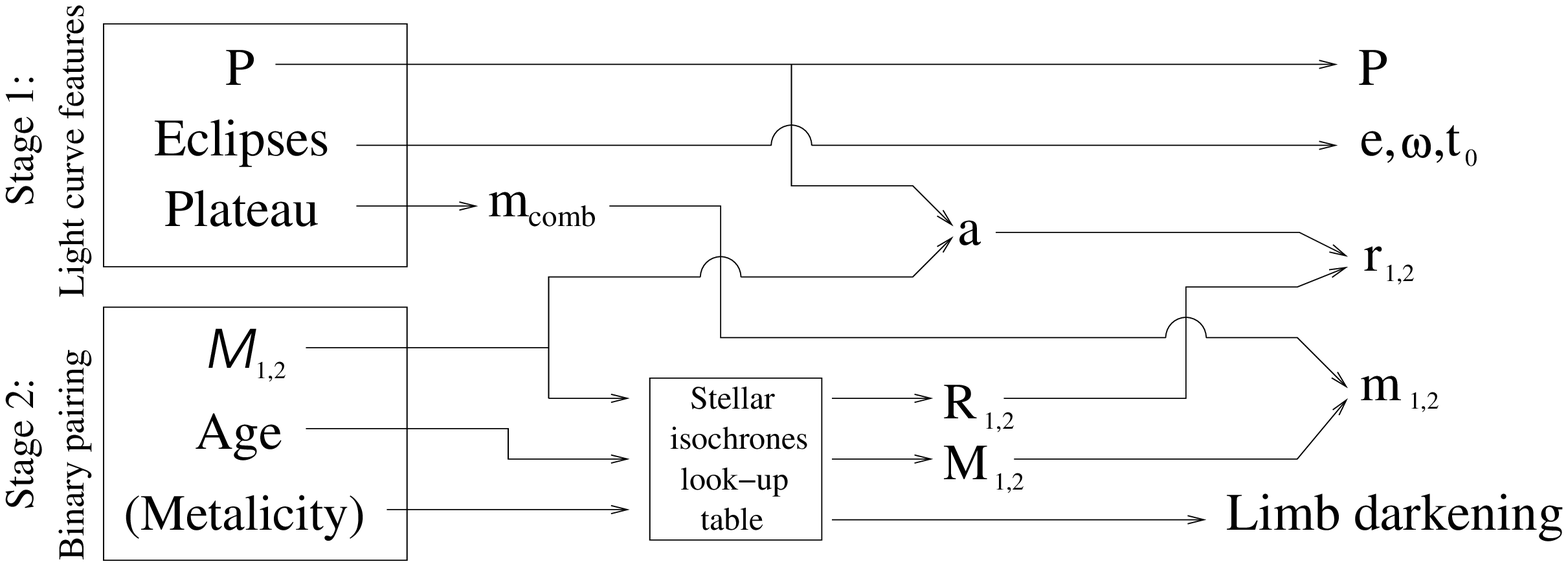}
\caption{A flow diagram demonstrating the process by which
MECI assigns the parameters to an EB based on its observed
light curve.  The details of stages~1 \& 2 are described
in \S\ref{subsecOrbitalParams} and \S\ref{subsecStellarParams},
respectively.}
\label{figMethod}
\end{figure}

In the second stage, we estimate the EB's absolute stellar
parameters by iterating through many possible stellar pairings,
simulating their expected light curves (see
Figure~\ref{figMECIpanels}), and finding the pairing that minimizes
the $\chi^2_\nu$ function (see \S\ref{subsecLikelihoodScore}). The
parameters we fit are the masses of the two EB components,
$M_{1,2}$, their age (the components are assumed to be coeval),
and their orbital inclination, $i$. Optionally, we can also
fine-tune the orbital parameters obtained from the first stage.
This option is necessary only for binaries with eccentric orbits,
since varying their inclination will affect the fit of their
previously estimated orbital parameters.  The flow diagram for the
entire procedure is shown in Figure~\ref{figMethod}.

If an estimate of the out-of-eclipse combined apparent magnitude,
$mag_{comb}$, of the EB (i.e. the light curve plateau) is
available, we may also estimate the distance modulus. If
$mag_{comb}$ is not available (e.g., if the light curve has been
normalized), the distance modulus cannot be evaluated unless an
independent measurement of the out-of-eclipse brightness is
available.  In either case, this procedure does not affect our
estimates of the stellar parameters.

Once we assume the masses and age of the binary components, we use
pre-calculated theoretical tables to
look up their absolute stellar parameters, namely their radii,
$R_{1,2}$, and absolute magnitudes, $Mag_{1,2}$.  We use the Yonsei-Yale
isochrones of solar metallicity \citep{Kim02} to specify the
binary components' radii and absolute magnitudes, in a range of
filters $(U,B,V,R,I)_{Cousins}$ and $(J,H,K)_{ESO}$.
We note that the Yonsei-Yale isochrones do not extend below
$0.4\, M_\odot$. To consider stars with masses below this value we
constructed tables from the isochrones of \citet{Baraffe98}, which
are generally more reliable for masses below $0.75\, M_\odot$.

Together with the orbital parameters
(\S\ref{subsecOrbitalParams}), we have all the information
required to simulate the EB light curve. The fractional radii,
$r_{1,2}$, and apparent magnitudes, $mag_{1,2}$, of the binary
components, which are needed for this calculation, are calculated
as follows:
\begin{eqnarray}
a &=& [G(M_1 + M_2)(P/2\pi)^2]^{1/3} \simeq\\
  & & 4.206 R_{\odot}(M_1/M_{\odot} + M_2/M_{\odot})^{1/3} (P/day)^{2/3}\nonumber,\\
r_{1,2} &=& R_{1,2} / a,\\
mag_1 &=& mag_{comb} + 2.5\log \left[1 + 10^{-0.4(Mag_2 - Mag_1)}\right], {\rm \ and}\\
mag_2 &=& mag_1 + (Mag_2 - Mag_1).
\end{eqnarray}

We create model light curves using DEBiL, which has a fast light
curve generator. DEBiL assumes that the EB is detached, with
limb-darkened spherical components (i.e. no tidal distortions or
reflections). To describe the stellar limb darkening, it employs
the quadratic law \citep{Claret95}:
\begin{equation}
I(\theta ) = I_0 \left[ {1-\tilde{a}(1-\cos \theta)-\tilde{b}(1-\cos \theta )^2} \right],
\end{equation}
where $\theta$ is the angle between the line of sight and the
emergent flux, $I_0$ is the flux at the center of the stellar
disk, and $\tilde{a}$, $\tilde{b}$ are coefficients that define
the amplitude of the center-to-limb variations.  We use the ATLAS
\citep{Kurucz92} and PHOENIX \citep{Claret98, Claret00} tables to
look up the quadratic limb darkening coefficients, for high-mass
($T_{eff} \geq 10000$K or $\log g \leq 3.5$) and low-mass
($T_{eff} < 10000$K and $\log g > 3.5$) main-sequence stars
respectively.

Finally, the orbital inclination is fit at each iteration so as to
make the simulated light curve most similar to the observations.
For this we employed the robust ``golden section'' bracket search
algorithm \citep{Press92}. This inner loop dominates the
computational time required. In the case of non-circular orbits,
it is often necessary to iterate the estimates of the orbital
parameters ($e$, $t_0$, $\omega$, $i$). When this option is
enabled, MECI employs the rolling simplex algorithm
\citep{Nelder65, Press92}, which fits all four orbital parameters
simultaneously.

\subsection{Assessing the Likelihood of a Binary Pairing}
\label{subsecLikelihoodScore}

The observational data for each EB consists of $N_{lc}$ observed
magnitudes $O_i$, each with an associated uncertainty $\epsilon_i$,
as well as $N_{colors}$ out-of-eclipse colors $\tilde{O}_c$,
each with an uncertainty $\tilde{\epsilon}_c$.  Our model
yields the corresponding predicted light curve magnitudes
$C_i$ and out-of-eclipse colors $\tilde{C}_c$.
We define the goodness-of-fit function to be:
\begin{equation}
\chi^2_\nu = \frac{1}{w+N_{colors}}\left[\frac{w}{N_{lc}}\sum_{i=1}^{N_{lc}} \left(\frac{O_i - C_i}{\epsilon_i}\right)^2 +
\sum_{c=1}^{N_{colors}} \left(\frac{\tilde{O}_c - \tilde{C}_c}{\tilde{\epsilon}_c}\right)^2\right],
\end{equation}
where $w$ is a factor that describes the relative weights assigned
to the light curve and color data (see below). The value of
$\chi^2_\nu$ should achieve unity if the assumed model accurately
describes the data, and the errors are Gaussian-distributed and
are estimated correctly.

In practice, typical light curves may have $N_{lc} > 1000$ points,
whereas only $1 \leq N_{colors} \leq 5$ might be available. We
have found it necessary to select a value for $w$ that increases
the relative weight of the color information to obtain reliable
results ($w < N_{lc}$). In general, the optimal value for $w$ will
depend on the accuracy of the observed colors $\tilde{O}_c$ and
the degree to which the EB light curve deviates from the
assumption of two well-detached, limb-darkened spherical
components. Based on the tests described in \S\ref{secCh4Tests}, we
find that a wide range of values for $w$ produces similar results,
and that values in the range $10 \leq w \leq 100$ most accurately
recover the correct values for the stellar parameters.

We identify the global minimum of $\chi^2_\nu$ in three steps:
First, we calculate the value of $\chi^2_\nu$ at all points in a
coarse $N \times N$ grid at each age slice. The $N$ mass values
are selected to be spaced from the lowest mass value present in
the models to the greatest values at which the star has not yet
evolved off the main-sequence. Next, we identify any local minima,
and refine their values by evaluating all available intermediate
mass pairings. Finally, we identify the global minimum from the
previous step, and fit an elliptic paraboloid to the local
$\chi^2_\nu$ surface around the lowest minimum.  We assign the
most likely values for the stellar masses and age to be the
location of the minimum of the paraboloid. The curvature of the
paraboloid in each axis provides the estimates of the
uncertainties in these parameters. In practice, these formal
uncertainties underestimate the true uncertainties since they do
not consider the systematic errors due to (1) the over-simplified
EB model, (2) errors in the theoretical stellar isochrones and
limb darkening coefficients, and (3) sources of non-Gaussian noise
in the data.

When choosing the value of $N$ above, we must balance
computational speed considerations with the risk of missing the
global minimum by under-sampling the $\chi^2_\nu$ surface. For
most main-sequence EBs, the $\chi^2_\nu$ surface contains only
one, or at most a few local minima, and our experience is that
$N=10$ usually suffices (see \S\ref{subsecSimulate}). For systems
that are either very young or in which a component has begun to
evolve off the main-sequence, the $\chi^2_\nu$ surface requires a
much denser sampling. Evolved components, which may be present in
as many as a third of the EBs of a magnitude-limited photometric
survey \citep{Alcock97}, introduce an additional challenge if
their isochrones intersect other isochrones on the color-magnitude
diagram. At such intersection points, stars of different masses
will have approximately equal sizes and effective temperatures,
creating degenerate regions on the $\chi^2_\nu$ surface. This
degeneracy can, in principle, be broken with sufficient color
information, which will probe differences in the stars' limb
darkening and absorption features, both of which vary with surface
gravity.

We also note that multiple local minimum may result for light curves
with very small formal uncertainties.  In this case, numerical
errors in the simulated light curve dominate.  This problem can
be mitigated by increasing the number of
iterations used in fitting the orbital parameters (see
\S\ref{subsecStellarParams}).

\subsection {Optimization}
\label{secOptimization}

We implemented a number of optimizations to increase the speed of
MECI. First, since each light curve is independent, we parsed the
data set and ran MECI in parallel on multiple CPUs. Second, we
reduced the number of operations by identifying and skipping
unphysical stellar pairings. Specifically, we required ($r_1 + r_2
< 0.8$) to preclude binaries that were not well detached.  In
addition, for EBs with clear primary and secondary eclipses, we
skipped high-contrast-ratio pairings for which the maximum depth
of the primary eclipse, $\Delta mag_{1}$, or the maximum depth of
the secondary eclipse, $\Delta mag_{2}$, fell below a specified
threshold, $\Delta mag_{\rm cutoff}$. In particular, we skipped
over pairings for which $\min \left( {\Delta mag_{1}}, {\Delta
mag_{2}} \right) \le {\Delta mag_{\rm cutoff}}$, where
\begin{eqnarray}
\Delta mag_{1} &\simeq& 2.5 \log \left[1 + \frac{(R_2/R_1)^2}{1 - (R_2/R_1)^2 + 10^{0.4(Mag1 - Mag2)}}\right], {\rm \ and}\\
\Delta mag_{2} &=& 2.5 \log \left[1 + 10^{0.4(Mag1 - Mag2)}\right].
\end{eqnarray}

These estimates assume equatorial eclipses, since we seek to
evaluate the maximum possible eclipse depths. The first expression
is approximate because it neglects the effects of limb darkening
on the eclipse depth. In practice, the chosen value for $\Delta
mag_{\rm cutoff}$ will depend on the typical precision and cadence
of the data set in question.

We note here a special case that we revisit in
\S\ref{secPitfalls}. For EB light curves with equally spaced
eclipses of equal depth, we must also consider the possibility
that our assumed period is double the true value, and hence the
secondary eclipse is undetected. When we identified such cases, we
analyzed the light curve as usual but removed the above
requirement.  In such cases, we can place only an upper limit on
the mass of the secondary component.

\section {Testing MECI}
\label{secCh4Tests}

In order to establish the accuracy and reliability of MECI under a
variety of scenarios, we conducted two distinct tests.

\subsection {Observed Systems}
\label{ObservedSystems}

The first test was to run MECI on several observed
light curves of eclipsing binary systems whose stellar parameters
had been precisely determined from detailed
photometric and spectroscopic studies.

\begin{figure}
\includegraphics[width=5in]{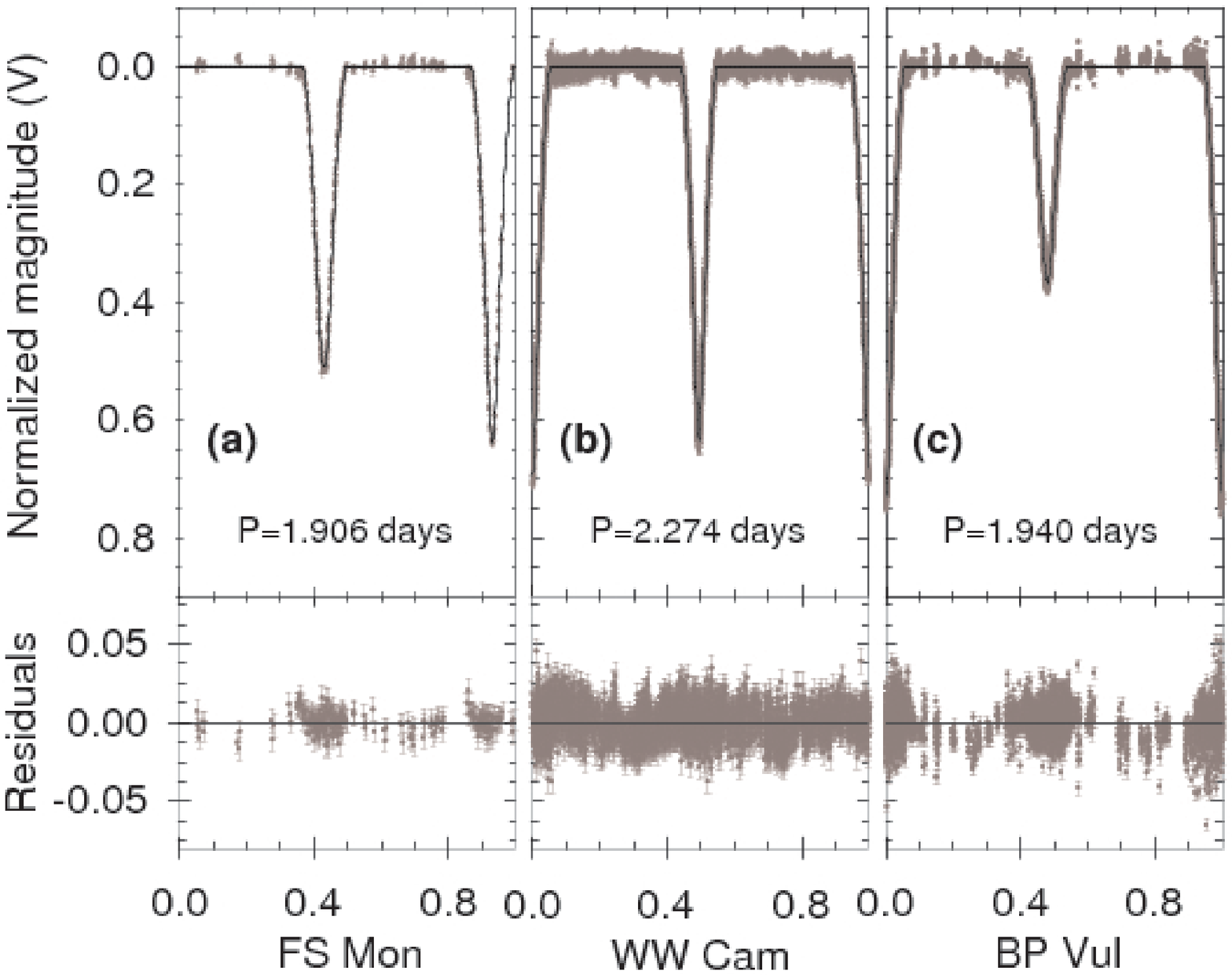}
\caption{The observed light curves of FS~Monocerotis \citep{Lacy00}, WW~Camelopardalis
\citep{Lacy02}, and BP~Vulpeculae \citep{Lacy03}, each overplotted with the
best-fit model DEBiL solution used in our MECI algorithm.  The masses
and ages corresponding to these solutions are listed in Table~\ref{tableLacy}. The residuals
to each fit are shown in the lower panels.}
\label{figLacyLC}
\end{figure}

\begin{figure}
\includegraphics[width=5in]{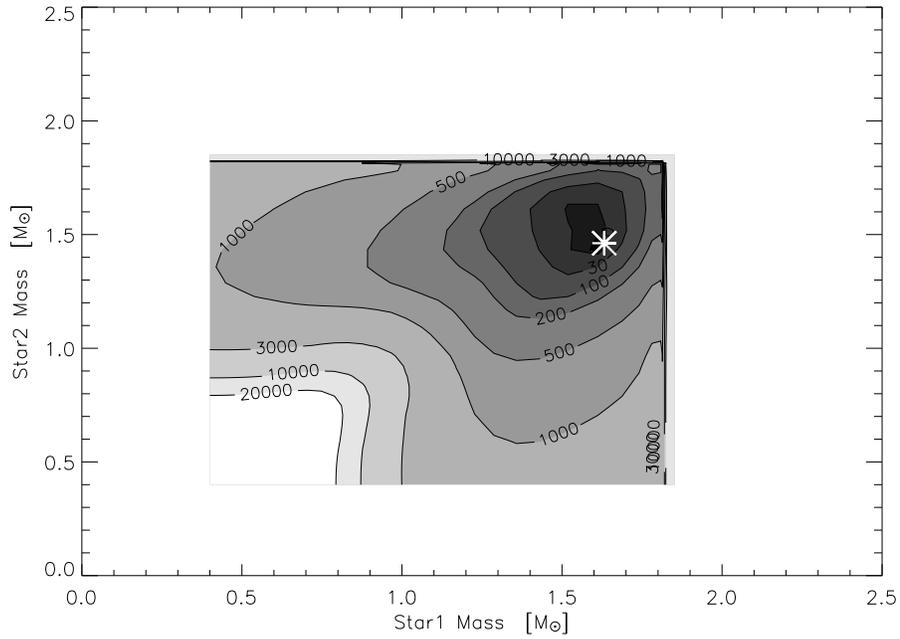}
\caption{The MECI $\chi^2_\nu$ surface to the FS~Monocerotis light curve and colors \citep{Lacy00},
assuming an age of $1.6$~Gyr and fixing $w=10$. The estimate of the stellar masses \citep{Lacy00}
from a combined analysis of the light curve and spectroscopic observations is indicated by
a white asterisk, and is near to the minimum identified by MECI.
Note the erratic behavior of the contours at the upper end of the mass range, which
results from the rapid evolution of stars of those masses at this age.}
\label{figLacy00}
\end{figure}

\begin{figure}
\includegraphics[width=5in]{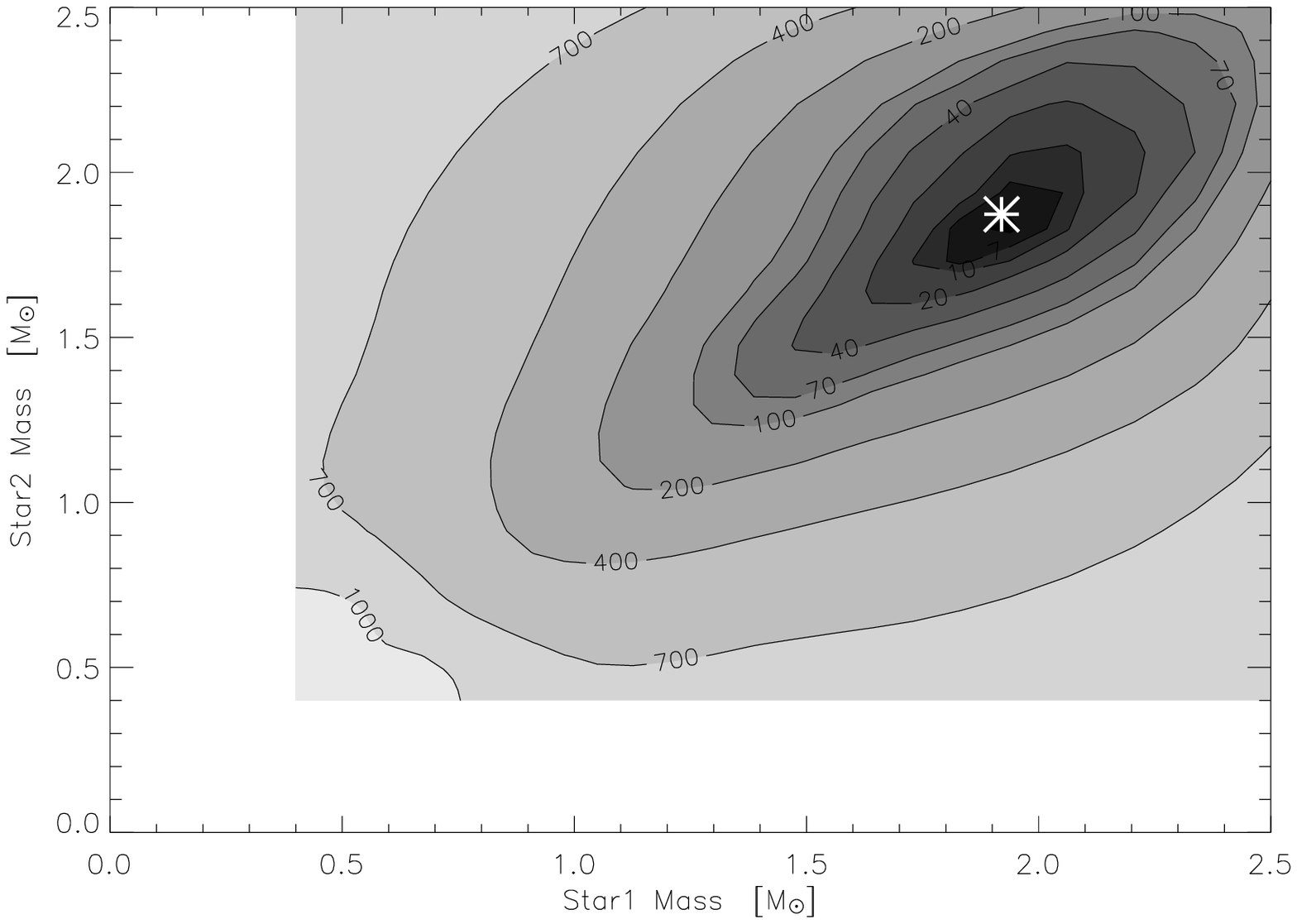}
\caption{The MECI $\chi^2_\nu$ surface to the WW~Camelopardalis light curve and colors \citep{Lacy02},
assuming an age of $0.6$~Gyr and fixing $w=10$. The estimate of the stellar masses \citep{Lacy02}
from a combined analysis of the light curve and spectroscopic observations is indicated by
a white asterisk, and is extremely close to the solution identified by MECI.}
\label{figLacy02}
\end{figure}

\begin{figure}
\includegraphics[width=5in]{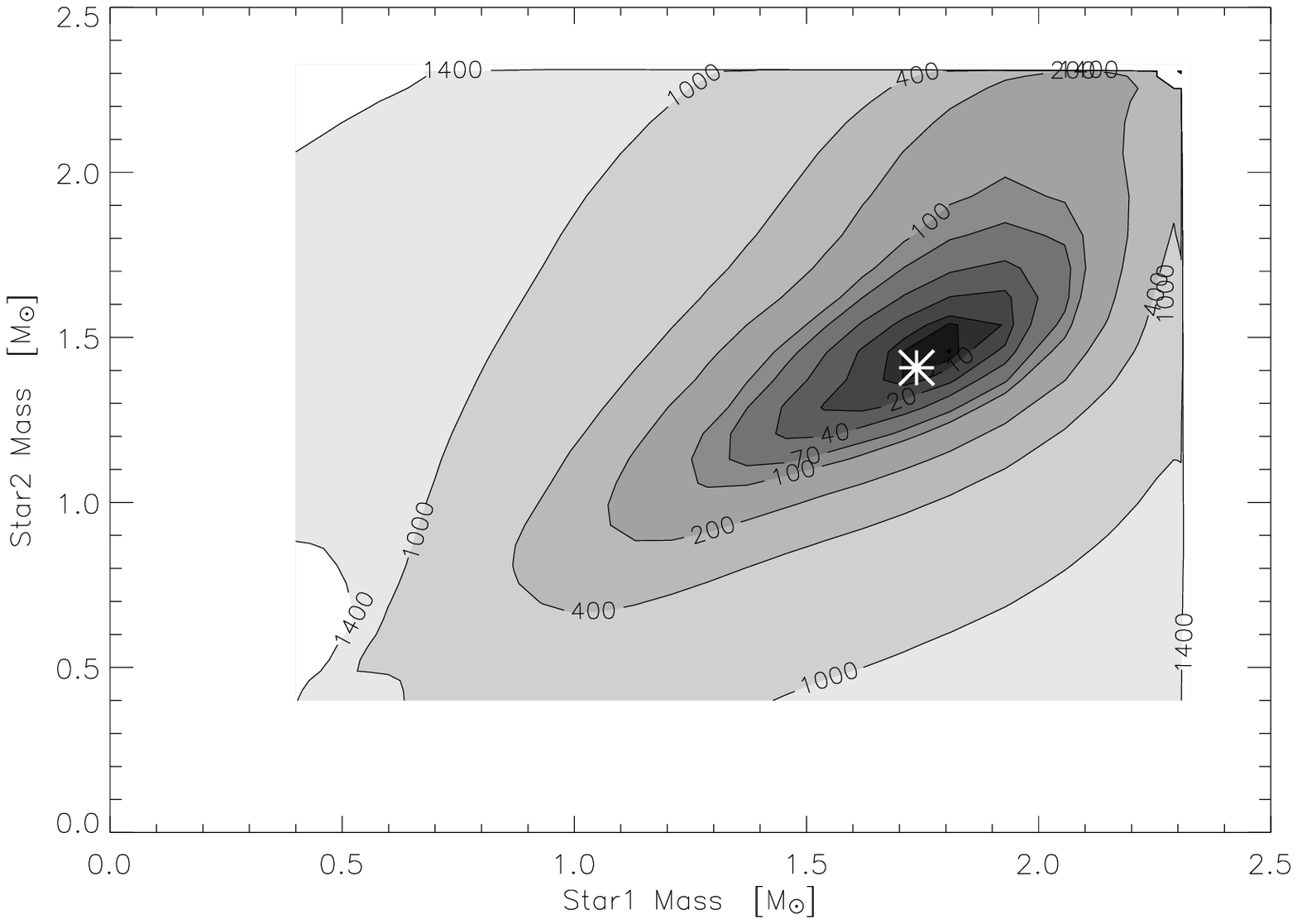}
\caption{The MECI $\chi^2_\nu$ surface to the BP~Vulpeculae light curve and colors \citep{Lacy03},
assuming an age of $0.8$~Gyr and fixing $w=10$. The estimate of the stellar masses \citep{Lacy03}
from a combined analysis of the light curve and spectroscopic observations is indicated by
a white asterisk, and is extremely close to the solution identified by MECI.}
\label{figLacy03}
\end{figure}

We examined three well-studied EBs. The first was FS Monocerotis
\citep{Lacy00}, for which we modeled the published light curve,
which had $N_{lc}=249$ data points, as well as the published $U-B$
and $B-V$ colors. The second was WW Camelopardalis \citep{Lacy02},
for which we modeled the published light curve, which had
$N_{lc}=5759$ observations, as well as the $B-V$ color.  Finally,
we studied BP Vulpeculae \citep{Lacy03}, for which we modeled the
published light curve, which had $N_{lc}=5236$ observations, as
well as the $B-V$ color. All three published light curves were
observed in $V$-band and are plotted in Figure~\ref{figLacyLC}.
The colors had been corrected for reddening. The contour plots of
the ${\chi}^2_\nu$ surfaces resulting from our MECI analysis
(setting the weighting $w=10$) are shown in
Figures~\ref{figLacy00}, \ref{figLacy02}, and \ref{figLacy03}.
Note that FS~Mon is more tightly constrained due to its greater
color information. Furthermore, the asymmetry in the BP Vul
contour is due to its unequal eclipse depths. In all cases, the
${\chi}^2_\nu$ surface has a single minimum, which is close to the
published values. In Table~\ref{tableLacy}, we tabulate the
results of our analysis and compare these to the published values.

We then changed the weighting factor to $w=100$ and repeated this
procedure. The MECI results for FS~Mon and BP~Vul were essentially
identical to our earlier findings for $w=10$. In the case of
WW~Cam, the results for $w=10$ were significantly closer to the
published values. This is likely due to the fact that it is a
young system ($age=500$~Myr), for which the brightness and radii
at constant mass vary significantly. Thus, the lower light curve
information weighting brought about smoother $\chi^2_\nu$ contours
(see \S\ref{subsecLikelihoodScore}).

\begin{deluxetable}{lccc|ccc|ccc}
\tabletypesize{\scriptsize}
\tablecaption{Accuracy of MECI parameter estimates for 3 well-studied binaries.}
\tablewidth{0pt}
\tablehead{ & \multicolumn{3}{c}{MECI ($w=10$)} & \multicolumn{3}{c}{MECI ($w=100$)} & \multicolumn{3}{c}{\citet{Lacy00, Lacy02, Lacy03}}}
\startdata
System  & Mass 1        & Mass 2        & Age     & Mass 1        & Mass 2        & Age     & Mass 1        & Mass 2        & Age     \\
        & $[M_{\odot}]$ & $[M_{\odot}]$ & $[Gyr]$ & $[M_{\odot}]$ & $[M_{\odot}]$ & $[Gyr]$ & $[M_{\odot}]$ & $[M_{\odot}]$ & $[Gyr]$ \\
\hline
FS Monocerotis    & 1.58     & 1.47     &  1.6    & 1.57     & 1.47     &  1.6    & 1.632      & 1.462      & 1.6      \\
($N_{lc}=249$)   & [3.3\%]  & [0.5\%]  & [0.3\%] & [3.6\%]  & [0.5\%]  & [0.1\%] & $\pm0.012$ & $\pm0.010$ & $\pm0.3$ \\
WW Camelopardalis & 1.92     & 1.86     &  0.5    & 2.10     & 2.02     &  0.4    & 1.920      & 1.873      & 0.5      \\
($N_{lc}=5759$)  & [0.2\%]  & [0.9\%]  & [3\%]   & [9.6\%]  & [8.0\%]  & [17\%]  & $\pm0.013$ & $\pm0.018$ & $\pm0.1$ \\
BP Vulpeculae     & 1.78     & 1.48     &  0.7    & 1.77     & 1.48     &  0.8    & 1.737      & 1.408      & 1.0      \\
($N_{lc}=5236$)  & [2.2\%]  & [5.3\%]  & [26\%]  & [1.9\%]  &
[5.2\%]  & [22\%]  & $\pm0.015$ & $\pm0.009$ & $\pm0.2$
\enddata
\tablecomments{The rightmost columns list the masses, ages, and errors
of the component stars as determined by a combined analysis
of their light curves and spectroscopic orbits \citep{Lacy00, Lacy02, Lacy03}.
The leftmost columns list the estimates of these
quantities produced by MECI assuming $w=10$, and the central
columns list the estimates from MECI assuming $w=100$.
The square brackets indicate the fractional errors of the MECI results
with respect to the numbers in the rightmost columns.}
\label{tableLacy}
\end{deluxetable}

\subsection {Simulated Systems}
\label{subsecSimulate}

In our second test, we produced large numbers of simulated EB
light curves with various levels of injected noise, and
subsequently analyzed these photometric datasets with MECI. We
then compared the input and derived estimates of the stellar
masses and ages in order to quantify the accuracy of the MECI
analysis.

We selected the orbital and stellar parameters of each simulated
EB as follows. First, we drew an age at random from a uniform
probability distribution between $200$~Myr and $10$~Gyr. We then
selected the masses of the two EB components independently from a
flat distribution from $0.4M_{\odot}$ and the maximum mass at
which stars of this age would still be located on the
main-sequence. We then assigned the orbital period by drawing a
number from a uniform probability distribution spanning $0 < P
\le 10$~days. Similarly, we assigned the epoch of perihelion by
drawing from a uniform probability distribution spanning $0 \le
t_0 < P$, and the orbital inclination from a uniform
distribution within the range that produces eclipses,
$\arccos(r_1+r_2) \le i \le {\pi}/2$. For the tests of eccentric
systems, we also randomly selected an eccentricity, uniformly from
$0 \le e \le 0.1$, and randomly selected the angle of perihelion,
uniformly from $0 \le \omega < 2{\pi}$. Finally, we rejected any
EB system if its components were overlapping or in near contact,
$r_1 + r_2 \geq 0.8$. We also filtered out EBs with undersampled
eclipses, or for which one of the eclipse depths was smaller than
the assumed $1 \, \sigma$ noise level.

Each simulated light curve contained 1000 $R$-band data points, to
which we injected Gaussian-distributed noise. When color
information was required, we computed the out-of-eclipse
photometric colors for each EB, and injected a $0.02$~mag
Gaussian-distributed error to this value. The colors we considered
were $(V-I)_{Cousins}$, which is similar to the color provided by
the OGLE II catalog \citep{Wozniak02}, as well as $(J-H)_{ESO}$
and $(H-K)_{ESO}$, which are similar to the colors provided by the
2MASS catalog\footnote{The 2MASS catalog uses custom J, H, and
$K_s$ filters, which can be approximately converted to the ESO
standard using linear transformations \citep{Carpenter01}.}
\citep{Kleinmann94}.

We simulated 8 sets of 2500 systems each,
with the sets differing in the following respects
(see Table~\ref{tableSimulations}):  (1) circular or
eccentric orbits, (2) the number of points in the search
grid, (3) the value of $w$, which describes the relative
weight between the color and photometric data, and (4) the
availability of color information.

\begin{figure}
\includegraphics[width=5in,height=5.5in]{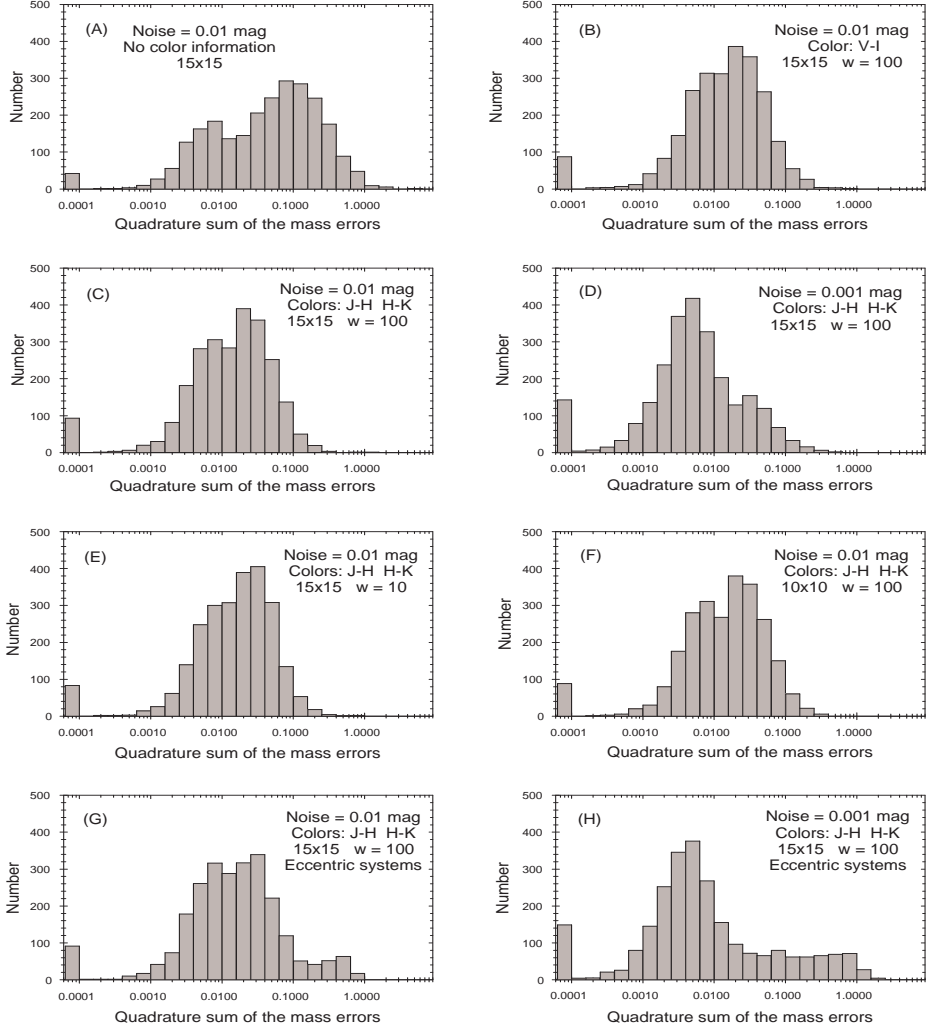}
\caption{Histograms of the quadrature sums of the relative
differences in the assumed and calculated masses for the stellar
components, for each of the sets of simulated light curves
described in Table~\ref{tableSimulations}. Each set contains
$2500$ simulated EBs as described in \S\ref{subsecSimulate}, and
the key parameters of each set are listed in the upper right
corner of each panel. The leftmost bin contains the sum of all
results with values less than 0.0001. The ability of the method to
accurately assign the masses to the component stars degrades
significantly in the absence of any color information (upper left
panel), but is generally robust against changes in the particular
choice of $w$ or $N$ (see \S\ref{subsecSimulate}).}
\label{figHist}
\end{figure}

In order to summarize the accuracy of the MECI results,
we computed the quadrature sum of the relative differences between
the assumed and derived values for the masses of the two components.
We plot the histograms of these values in Figure~\ref{figHist}.
In each histogram, we identify the value encompassing the region
that contains 90\% of the results.  We call
this range the ``$90^{th}$ percentile error'', and list
it in the final column of Table~\ref{tableSimulations}.

We find that the inclusion of color information significantly
improves the accuracy of the MECI results, lowering the $90^{th}$
percentile error from $30$\% in set (A), to less than $6$\% in
sets (B) and (C). In contrast, changing the value of $w$ from
$100$ in set (C), to $10$ in set (E), results in only a modest
increase of $0.8$\% in the size of the $90^{th}$ percentile error.
This indicates that the results are robust to the particular
choice of $w$.  We note, however, that a value of $w>100$ will
usually provide too little weight to the color information, which
results in poorer accuracy. An extreme example of this is seen in
set (A).

Similarly, MECI is not sensitive to the exact value of the search
grid size.   In particular, decreasing the grid size from $15
\times 15$ in set (C), to $10 \times 10$ in set (F), increases the
$90^{th}$ percentile error only modestly, from $5.8$\% to $6.1$\%.
This stability results from the fact that the $\chi^2_\nu$
function contains a broad minimum, which is well sampled even with
$N=10$ grid points.  We note, however, that this is no longer the
case when considering evolved star systems (e.g.,
\S\ref{subsecLikelihoodScore}), for which a larger number of grid
points is required.

When we decreased the level of the noise injected into the
photometric time series from $0.01$ mag in set (C), to $0.001$ mag
in set (D), the $90^{th}$ percentile error dropped from $5.8$\% to
$4.0$\%. Surprisingly, the tail of the upper end of the error
distribution extends to larger values in set (D). This appears to
be due to the phenomenon discussed in
\S\ref{subsecLikelihoodScore}, whereby the $\chi^2_\nu$ function
occasionally contains many local minima. This problem becomes
acute for eccentric systems, since they have a far more complex
$\chi^2_\nu$ function. Decreasing their noise from $0.01$ mag in
set (G), to $0.001$ mag in set (H), raises the $90^{th}$
percentile error from $8.8$\% to $23$\%. This relatively poor
performance reflects the algorithm's inability to robustly
identify the global minimum under these conditions. In such cases
one must increase the size of the search grid and iteratively
solve for the orbital parameters of the systems, which results in
a significant increase in the computational time.

\begin{deluxetable}{llccclc}
\tabletypesize{\scriptsize}
\tablecaption{Accuracy of MECI mass estimates for simulated systems.}
\tablewidth{0pt}
\tablehead{Set & Noise & Orbit & Search grid & Weighting & Color information & $90^{th}$ percentile error}
\startdata
A & 0.01  mag & circular  & $15 \times 15$ & N/A    & No color information            & 30\%  \\
B & 0.01  mag & circular  & $15 \times 15$ & $w=100$  & $(V-I)_{Cousins}$               & 5.9\% \\
C & 0.01  mag & circular  & $15 \times 15$ & $w=100$  & $(J-H)_{ESO}$ and $(H-K)_{ESO}$ & 5.8\% \\
D & 0.001 mag & circular  & $15 \times 15$ & $w=100$  & $(J-H)_{ESO}$ and $(H-K)_{ESO}$ & 4.0\% \\
E & 0.01  mag & circular  & $15 \times 15$ & $w=\ 10$ & $(J-H)_{ESO}$ and $(H-K)_{ESO}$ & 6.6\% \\
F & 0.01  mag & circular  & $10 \times 10$ & $w=100$  & $(J-H)_{ESO}$ and $(H-K)_{ESO}$ & 6.1\% \\
G & 0.01  mag & eccentric & $15 \times 15$ & $w=100$  & $(J-H)_{ESO}$ and $(H-K)_{ESO}$ & 8.8\% \\
H & 0.001 mag & eccentric & $15 \times 15$ & $w=100$  & $(J-H)_{ESO}$ and $(H-K)_{ESO}$ & 23\%
\enddata
\tablecomments{The parameters of the 8 distinct sets of
simulated EB light curves that we generated and subsequently
analyzed with MECI.  The rightmost column lists the range of the
quadrature sum of the fractional errors on the masses which
encompasses 90\% of the solutions (see Figure~\ref{figHist}), which
we take to be indicative of the accuracy of MECI under the
specified conditions.}
\label{tableSimulations}
\end{deluxetable}

\subsection{Limitations}
\label{secPitfalls}

A significant degeneracy results for light curves in which two
distinct eclipses are not apparent.  For such systems, two
distinct possibilities exist, namely that either the EB consists
of two twin components with an orbital period $P$, or that the EB
consists of two stars with very disparate sizes (such that the
secondary eclipse is not discernible), with an orbital period $2\,
P$.  It is often necessary to flag such systems and conduct
analyses with both possible values for the orbital periods.
Distinguishing which of these possibilities is the correct
solution is challenging, but in some instances there are clues.
One such clue is a variable light curve plateau that results from
the mutual tidal distortions, which in turn might indicate the
true orbital period (twice that of the observed modulation).  A
second possibility is a red excess in the system color indicating
a low-mass secondary.  Of course, follow-up spectroscopic
observations can readily resolve this degeneracy, either by
indicating the presence of two components of similar brightness,
or through a direct determination of the orbital period.

We note that MECI employs a simplified model for the generation of
the light curves (DEBiL), which can bring about additional
complications when applied to systems in which our assumptions
(see \S\ref{subsecStellarParams}) do not hold. For example, our
model ignores the effect of third light, from either a physically
associated star or a chance superposition, which reduces the
apparent depths of the eclipses and may contaminate the estimate
of the system color. Furthermore, we have ignored reflection
effects, which can raise the light curve plateau at times
immediately preceding or following eclipses. Finally, tidal
distortions will increase the apparent system brightness at
orbital quadrature, which can serve to increase the apparent depth
of the eclipses. In order for MECI to be able to properly handle
these cases, its light curve generator must be replace with a more
sophisticated one (e.g., WD or EBOP), which will likely make MECI
significantly more computationally expensive.

\section{Conclusions}
\label{secConclusions}

We have described a method for identifying an EB's components
using only its photometric light curve and combined colors. By
utilizing theoretical isochrones and limb darkening coefficients,
this method greatly reduces the EB parameter space over which one
needs to search. Using this approach, we can quickly estimate the
masses, radii and absolute magnitudes of the components, without
spectroscopic data. We described an implementation of this method,
which enables the systematic analyses of datasets consisting of
photometric time series of large numbers of stars, such as those
produced by OGLE, MACHO, TrES, HAT, and many others.  Such
techniques are expected to grow in importance with the next
generation surveys, such as Pan-STARRS \citep{Kaiser02} and LSST
\citep{Tyson02}.  In a future publication, we shall describe a
specific application of these codes, namely to search for low-mass
eclipsing binaries in the TrES dataset.

\section*{Acknowledgments}

We would like to thank Guillermo Torres and Tsevi Mazeh
for many useful discussions, and we
would like to thank Sarah Dykstra for her editorial assistance.

\chapter{Identification, Classifications, and Absolute Properties of 773 Eclipsing Binaries Found in the TrES Survey
\label{chapter5}}

\title{Identification, Classifications, and Absolute Properties of\\ 773 Eclipsing Binaries Found in the TrES Survey}

J.~Devor, D.~Charbonneau, F.~T.~O'Donovan, G.~Mandushev, \& G.~Torres 2008,\\
\emph{The Astronomical Journal}, {\bf 135}, 850$-$877

\section*{Abstract}

In recent years, we have witnessed an explosion of photometric
time-series data, collected for the purpose of finding a small
number of rare sources, such as transiting extrasolar planets and
gravitational microlenses. Once combed, these data are often set
aside, and are not further searched for the many other variable
sources that they undoubtedly contain. To this end, we describe a
pipeline that is designed to systematically analyze such data,
while requiring minimal user interaction. We ran our pipeline on a
subset of the Trans-Atlantic Exoplanet Survey dataset, and used it
to identify and model 773 eclipsing binary systems. For each
system we conducted a joint analysis of its light curve, colors,
and theoretical isochrones. This analysis provided us with
estimates of the binary's absolute physical properties, including
the masses and ages of their stellar components, as well as their
physical separations and distances. We identified three types of
eclipsing binaries that are of particular interest and merit
further observations. The first category includes 11 low-mass
candidates, which may assist current efforts to explain the
discrepancies between the observation and the models of stars at
the bottom of the main-sequence. The other two categories include
34 binaries with eccentric orbits, and 20 binaries with abnormal
light curves. Finally, this uniform catalog enabled us to identify
a number of relations that provide further constraints on binary
population models and tidal circularization theory.

\section{Introduction}

Since the mid-1990s, there has been an explosion of large-scale
photometric variability surveys. The search for gravitational
microlensing events, which were predicted by \citet{Paczynski86},
motivated the first wave of surveys [e.g., OGLE:
\citet{Udalski94}; EROS: \citet{Beaulieu95}; DUO: \citet{Alard97};
MACHO: \citet{Alcock98}]. Encouraged by their success, additional
surveys, searching for gamma-ray bursts [e.g., ROTSE:
\citet{Akerlof00}] and general photometric variabilities [e.g.,
ASAS: \citet{Pojmanski97}] soon followed.

Shortly thereafter, with the discovery of the first transiting
extrasolar planet \citep{Charbonneau00, Henry00, Mazeh00}, a
second wave of photometric surveys ensued [e.g., OGLE-III:
\citet{Udalski03}; TrES: \citet{Alonso04}; HAT: \citet{Bakos04};
SuperWASP: \citet{Christian06}; XO: \citet{McCullough06}; for
 a review, see \citet{Charbonneau07}]. Each
of these projects involved intensive efforts to locate a few
proverbial ``needles'' hidden in a very large data haystack. With
few exceptions, once the needles were found, thus fulfilling the
survey's original purpose, the many gigabytes of photometric light
curves (LCs) collected were not made use of in any other way. In
this paper, we demonstrate how one can extract a great deal more
information from these survey datasets, with comparably little
additional effort, using automated pipelines. To this end, we have
made all the software tools described in this paper freely
available (see web links listed in \S\ref{secMethod}), and they
are designed to be used with any LC dataset.

In the upcoming decade, a third wave of ultra-large ground-based
synoptic surveys [e.g., Pan-STARRS: \citet{Kaiser02}; LSST:
\citet{Tyson02}], and ultra-sensitive space-based surveys [e.g.,
KEPLER: \citet{Borucki97}; COROT: \citet{Baglin98}; GAIA:
\citet{Gilmore98}] are expected to come online. These surveys are
designed to produce photometric datasets that will dwarf all
preceding efforts. To make any efficient use of such large
quantities of data, it will become imperative to have in place a
large infrastructure of automated pipelines for performing even
the most casual data mining query.

In this paper, we focus exclusively on the identification and
analysis of eclipsing binary (EB) systems. EBs provide favorable
targets, as they are abundant and can be well modeled using
existing modeling programs [e.g., WD: \citet{Wilson71}; EBOP:
\citet{Popper81b}]. Once modeled, EBs can provide a wealth of
useful astrophysical information, including constraints on binary
component mass distributions, mass-radius-luminosity relations,
and theories describing tidal circularization and synchronization.
These findings, in turn, will likely have a direct impact on our
understanding of star formation, stellar structure, and stellar
dynamics. These physical distributions of close binaries may even
help solve open questions relating to the progenitors of Type Ia
supernovae \citep{Iben84}. In additional to these, EBs can be used
as tools; both as distance indicators \citep{Stebbing10,
Paczynski97} and as sensitive detectors for tertiary companions
via eclipse timing \citep{Deeg00, Holman05, Agol05}.

In order to transform such large quantities of data into useful
information, one must construct a robust and computationally
efficient automated pipeline. Each step along the pipeline will
either measure some property of the LC, or filter out LCs that do
not belong, so as to reduce the congestion in the following, more
computationally intensive steps. One can achieve substantial gains
in speed by dividing the data into subsets, and processing them in
parallel on multiple CPUs. The bottlenecks of the analysis are the
steps that require user interaction. In our pipeline, we reduce
user interaction to essentially yes/no decisions regarding the
success of the EB models, and eliminate any need for interaction
in all but two stages. We feel that this level of interaction
provides good quality control, while minimizing its detrimental
subjective effects.

The data that we analyzed originate from ten fields of the
Trans-atlantic Exoplanet Survey [TrES ; \citet{Alonso04}].
\citet{Alonso04}. TrES employs a network of three automated
telescopes to survey $6^\circ \times 6^\circ$ fields of view. To
avoid potential systematic noise, we used the data from only one
telescope, Sleuth, located at the Palomar Observatory in Southern
California \citep{ODonovan04}. This telescope has a 10 cm physical
aperture and a photometric aperture of radius of 30". The number
of LCs in each field ranges from 10,405 to 26,495 (see Table
\ref{tableFieldsObs}), for a total of 185,445 LCs. The LCs consist
of $\sim$2000 $r$-band photometric measurements at a 9 minute
cadence. These measurements were created by binning the
image-subtraction results of five consecutive 90 second
observations, thus improving their non-systematic photometric
noise. As a result $\sim$16\% of the LCs have an RMS $<$1\%, and
$\sim$38\% of the LCs have an RMS $<$2\% (see Table
\ref{tableFieldsYield}). The calibration of TrES images,
identification of stars therein, extraction, and decorrelation of
the LCs is described elsewhere \citep{Dunham04, Mandushev05,
ODonovan06, ODonovan07}. TrES is currently an active survey that
is continuously observing new fields, though for this paper we
have limited ourselves to these ten fields.

\begin{deluxetable}{cccccccc}
\tabletypesize{\tiny}
\rotate
\tablecaption{Observational parameters of the TrES fields}
\tablewidth{0pt}
\tablehead{\colhead{Field} &
           \colhead{Constellation} &
           \colhead{\begin{tabular}{c} $\alpha$\\ (J2000)\tablenotemark{a}\end{tabular}} &
           \colhead{\begin{tabular}{c} $\delta$\\ (J2000)\end{tabular}} &
           \colhead{\begin{tabular}{c} Galactic\\ coordinates (l,b)\end{tabular}} &
           \colhead{\begin{tabular}{c} Starting\\ epoch (HJD)\end{tabular}} &
           \colhead{\begin{tabular}{c} Ending\\ epoch (HJD)\end{tabular}} &
           \colhead{\begin{tabular}{c} Duration\\ (days)\end{tabular}}}
\startdata
And0& Andromeda      & 01 09 30.1255& +47 14 30.453& (126.11, -015.52)& 2452878.9& 2452934.9& 56.0\\
Cas0& Cassiopeia     & 00 39 09.8941& +49 21 16.519& (120.88, -013.47)& 2453250.8& 2453304.6& 53.8\\
CrB0& Corona Borealis& 16 01 02.6616& +33 18 12.634& (053.49, +048.92)& 2453493.8& 2453536.8& 43.0\\
Cyg1& Cygnus         & 20 01 21.5633& +50 06 16.902& (084.49, +010.28)& 2453170.7& 2453250.0& 79.3\\
Dra0& Draco          & 16 45 17.8177& +56 46 54.686& (085.68, +039.53)& 2453093.8& 2453163.0& 69.2\\
Her0& Hercules       & 16 49 14.2185& +45 58 59.963& (071.61, +039.96)& 2452769.9& 2452822.0& 52.1\\
Lyr1& Lyra           & 19 01 26.3713& +46 56 05.325& (077.15, +017.86)& 2453541.8& 2453616.7& 74.9\\
Per1& Perseus        & 03 41 07.8581& +37 34 48.712& (156.37, -014.04)& 2453312.8& 2453402.8& 90.0\\
Tau0& Taurus         & 04 20 21.2157& +27 21 02.713& (169.83, -015.94)& 2453702.7& 2453770.9& 68.2\\
UMa0& Ursa Major     & 09 52 06.3560& +54 03 51.596& (160.87, +047.70)& 2453402.9& 2453487.8& 84.9\\
\enddata
\tablenotetext{a}{ICRS 2000.0 coordinates of the guide star, which is located at the center of the field of view.}
\label{tableFieldsObs}
\end{deluxetable}

\begin{deluxetable}{ccccccc}
\tabletypesize{\tiny}
\rotate
\tablecaption{The number of sources and yield of the TrES fields}
\tablewidth{0pt}
\tablehead{\colhead{Field} &
           \colhead{\begin{tabular}{c} Number\\ of LCs\end{tabular}} &
           \colhead{\begin{tabular}{c} Number of observations\\ in each LC\end{tabular}} &
           \colhead{\begin{tabular}{c} Fraction\\ RMS $<$ 1\%\end{tabular}} &
           \colhead{\begin{tabular}{c} Fraction\\ RMS $<$ 2\%\end{tabular}} &
           \colhead{\begin{tabular}{c} Found\\ EBs\end{tabular}} &
           \colhead{\begin{tabular}{c} EB discovery\\ yield\end{tabular}}}
\startdata
And0& 26495& 2357& 16.5\%& 40.4\%& 111& 0.42\%\\
Cas0& 22615& 2069& 11.0\%& 38.2\%& 119& 0.53\%\\
CrB0& 18954& 1287& 11.0\%& 22.4\%&  28& 0.15\%\\
Cyg1& 17439& 3256& 30.3\%& 65.7\%& 125& 0.72\%\\
Dra0& 15227& 2000& 11.8\%& 26.4\%&  42& 0.28\%\\
Her0& 15916&  974& 16.8\%& 35.0\%&  28& 0.18\%\\
Lyr1& 22964& 2815& 19.4\%& 49.0\%& 135& 0.59\%\\
Per1& 20988& 1647& 15.9\%& 38.4\%&  93& 0.44\%\\
Tau0& 14442& 1171& 13.1\%& 32.5\%&  68& 0.47\%\\
UMa0& 10405& 1343& 13.6\%& 29.5\%&  24& 0.23\%\\
\enddata
\label{tableFieldsYield}
\end{deluxetable}

\section{Method}
\label{secMethod}

The pipeline we have developed is an extended version of the
pipeline described by \citet{Devor05}. At the heart of this
analysis lie two computational routines that we have described in
earlier papers: the Detached Eclipsing Binary Light curve
fitter\footnote{The DEBiL source code, utilities, and running
example files are available online at:\newline
http://www.cfa.harvard.edu/$\sim$jdevor/DEBiL.html.} [DEBiL ;
\citet{Devor05}], and the Method for Eclipsing Component
Identification\footnote{The MECI source code and running examples
are available online at:\newline
http://www.cfa.harvard.edu/$\sim$jdevor/MECI.html.} [MECI ;
\citet{Devor06a, Devor06b}]. DEBiL fits each LC to a
$\it{geometric}$ model of a detached EB (steps 3 and 5 below).
This model consists of two luminous, limb-darkened spheres that
orbit in a Newtonian two-body orbit. MECI restricts the DEBiL fit
along theoretical isochrones, and is thus able to create a
$\it{physical}$ model of each EB (step 9). This second model
describes the masses and absolute magnitudes of the EB's stellar
components, which are then used to determine the EB's distance and
absolute separation.

The pipeline consists of ten steps. We elaborate on each of these
steps below.

\begin{enumerate}
\item Determine the period.
\item If a distinct secondary eclipse is not observed, add an entry with twice the period.
\item Fit the orbital parameters with DEBiL.
\item Fine-tune the period using eclipse timing.
\item Refine the orbital parameters with DEBiL using the revised period.
\item Remove contaminated LCs.
\item Visually assess the quality of the EB models.
\item Match the LC sources with external databases.
\item Estimate the absolute physical properties of the binary components using MECI.
\item Classify the resulting systems using both automatic and manual criteria.
\end{enumerate}

We use the same filtering criteria as described in
\citet{Devor05}, both for removing LCs that are not periodic (step
1) and then for removing non-EB LCs (step 3). Together, these
automated filters remove approximately 97\% of the input LCs. In
addition to these filters, we perform stringent manual inspections
(steps 7 and 10) whereby we removed all the LCs we were not
confident were EBs. These inspections ultimately removed
approximately 86\% of the remaining LCs. Thus only 1 out of every
240 input LCs, were included in the final catalog.

In step 1, we use both the Box-fitting Least Squares (BLS) period
finder \citep{Kovacs02}, and a version of the analysis of
variances (AoV) period finder \citep{SchwarzenbergCzerny89,
SchwarzenbergCzerny96} to identify the periodic LCs within the
dataset and to measure their periods. In our AoV implementation,
we scan periods from 0.1 days up to the duration of each LC. We
then select the period that minimizes the variance of a linear fit
within eight phase bins. We remove all systems with weak
periodicities [see \citet{Devor05} for details], and with one
exception (T-Lyr1-14413), all the systems whose optimal period is
found to be longer than half their LC duration. In this way we are
able to filter out many of the non-periodic variables.

\begin{figure}
\includegraphics[width=5in]{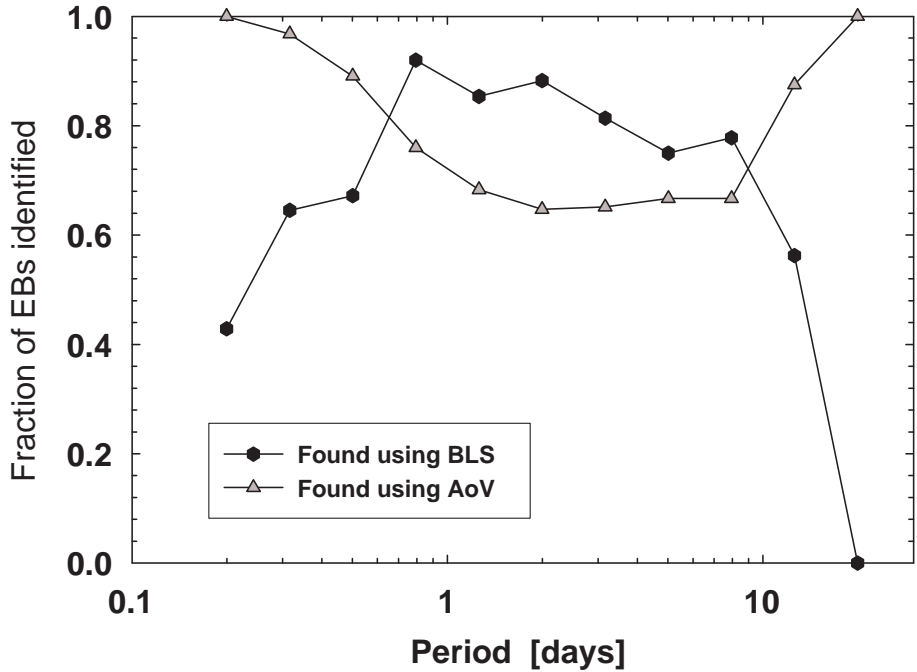}
\caption{The fraction of the EBs in the final catalog found using
the BLS algorithm and the AoV algorithm. The number
of EBs in each bin is shown in Figure \ref{figPeriodDistrib}. The
BLS method excelled at identifying EBs with short-duration eclipses
(compared to the orbital period), which predominately occur at periods
 $> 0.75$ days. The AoV method fared better with EBs that have
long-duration eclipses, which predominately occur in sub-day periods. The AoV
method also does well with EBs with period longer than $10$ days,
which may be dominated by giant-giant binaries \citep{Derekas07}, and
so also have broad eclipses. This plot demonstrates the importance
of using multiple independent methods of identifying EB, otherwise
the results will have a significant selection effect that may
bias any statistical results.}
\label{periodDiffFrac}
\end{figure}

The AoV algorithm is most effective in identifying the periods of
LCs with long duration features, such as semi-detached EBs and
pulsating stars. The BLS algorithm, in contrast, is effective at
identifying periodic systems whose features span only a brief
portion of the period, such as detached EBs and transiting planets
(see Figure \ref{periodDiffFrac}). However, the BLS algorithm is
easily fooled by outlier data points, identifying them as short
duration features. For this reason, the BLS algorithm has a
significantly higher rate of false positives than AoV, especially
for long periods, which have only a few cycles over the duration
of the observations. Therefore, we limit the search range of the
BLS algorithm to periods shorter than 12 days, although as Figure
\ref{periodDiffFrac} illustrates, its efficiency at locating EBs
rapidly declines at periods greater than 10 days.

In step 2, we address the ambiguity between EBs with identical
components in a circular orbit, and EBs with extremely disparate
components. The phased LC of EBs with identical components
contains two identical eclipses, whereas the phased LC of EBs with
disparate components will have a secondary eclipse below the
photometric noise level. These two cases are degenerate, since
doubling the period of a disparate system will result in an LC
that looks like an equal-component system. In the pipeline, we
handle this problem by doubling such entries; one with the period
found in step (1), and another with twice that period. Both of
these entries proceed through the pipeline independently. In many
cases, after additional processing by the following steps, one of
these entries will emerge as being far less likely than the other
(see Appendix \ref{appendixSingleEclipse}), at which point it is
removed. But in cases where photometry alone cannot determine
which is correct, one needs to perform spectroscopic follow-up to
break the ambiguity. In particular, a double-lined spectrum would
support the equal-component hypothesis.

Step 3 is performed using DEBiL, which fits the fractional radii
($r_{1,2}$) and observed magnitudes ($mag_{1,2}$) of the EB's
stellar components, their orbital inclination ($i$) and
eccentricity ($e$), and their epoch ($t_0$) and argument of
periastron ($\omega$). DEBiL first produces an initial guess for
these parameters, and then iteratively improves the fit using the
downhill simplex method \citep{Nelder65} with simulated annealing
\citep{Kirkpatrick83, Press92}.

\begin{figure}
\includegraphics[width=5in]{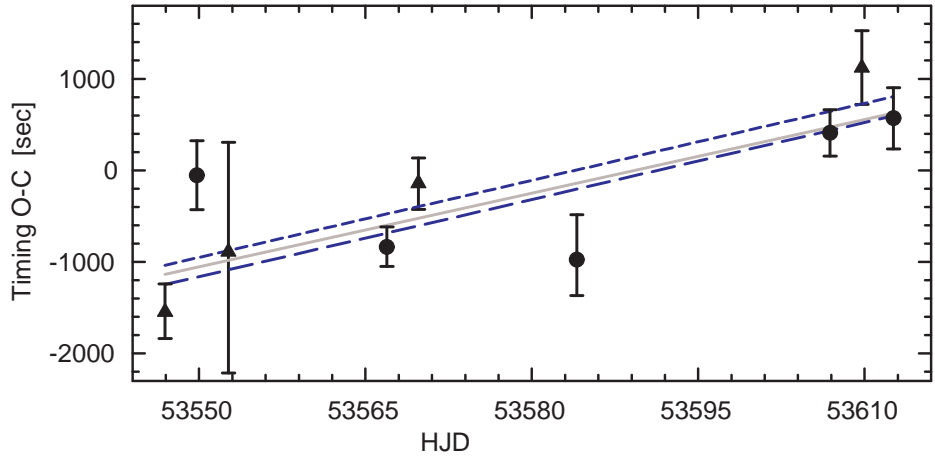}
\caption{An eclipse timing plot produced
in step 4, showing the $O-C$ residuals of the primary eclipses
(circles) and the secondary eclipses (triangles). Here, T-Lyr1-14962
is shown with an assumed period of $5.710660\: days$, as measured
with an AoV periodogram. The slope of the residuals indicates that
the assumed period is inaccurate. The gray solid line is predicted
by the best circular-orbit model, whereas the dashed lines are
predicted by the best eccentric-orbit model (compare to Figure
\ref{figTimingVariationsEcc}). After correction, we get a
fine-tuned period of $5.712516\: days$. This 0.03\% correction is
small but significant in that without having had this correction,
the eclipses would have smeared out and widened.}
\label{figTimingVariations}
\end{figure}

In step 4, we fine-tune the period ($P$) using a method based on
eclipse timing\footnote{The source code and running examples are
available online at\newline
http://www.cfa.harvard.edu/$\sim$jdevor/Timing.html.}, which we
describe below. In order to produce an accurate EB model in step
9, it is necessary to know the system's period with greater
accuracy than that produced in step 1. If we neglect to
fine-tune the period, the eclipses may be out of phase with
respect to one another, and so the phased eclipses will appear
broadened. Our timing method employs the DEBiL model produced in
step (3), and uses it to find the difference between the observed
and calculated ($O-C$) eclipse epochs. This is done by minimizing
the chi-squared fit of the model to the data points in each
eclipse, while varying only the model's epoch of periastron. When
the period estimate is off by a small quantity ($\Delta P$), the
$O-C$ difference increases by $\Delta P$ each period. This change
in the $O-C$ over time can be measured from the slope of the
linear regression, which is expected to equal $\Delta P / P$. Thus
measuring such an $O-C$ slope will yield the desired period
correction (see Figure \ref{figTimingVariations}).

\begin{figure}
\includegraphics[width=5in]{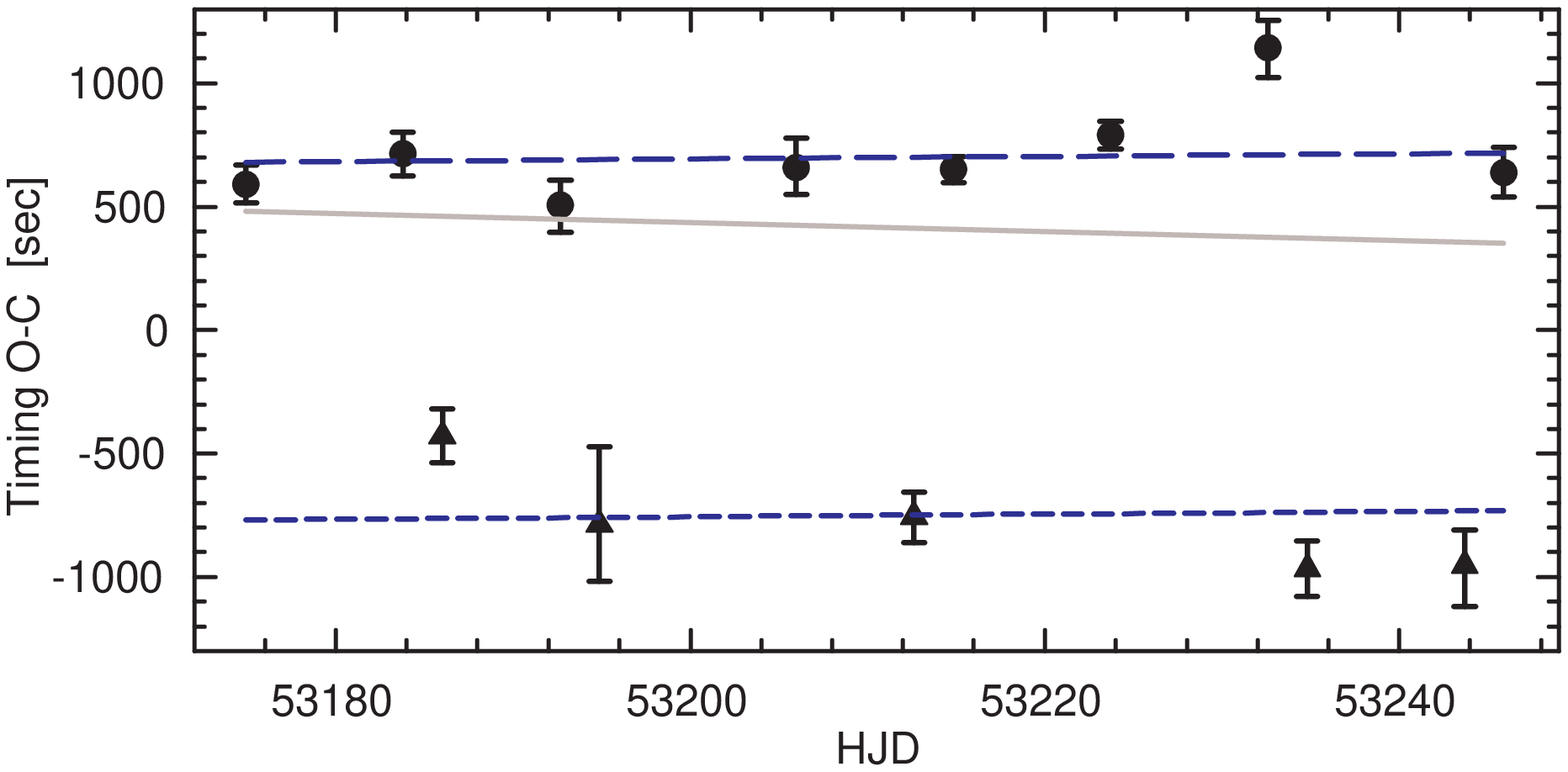}
\caption{An eclipse timing plot for T-Cyg1-01373, with an assumed
period of $4.436013\: days$. In contrast to Figure
\ref{figTimingVariations}, the slope here is consistent with zero,
thus indicating that the period does not need to be fine-tuned.
However, the $O-C$ offset between the primary (circles) and
secondary (triangle) eclipses is significant ($1449\: seconds$),
indicating that this EB has an eccentric orbit. The reduced
chi-squared of the best circular-orbit model (gray solid line) is
$\chi^2_\nu = 12.9$, while the reduced chi-squared of the best
eccentric-orbit model (dashed lines) is $\chi^2_\nu = 0.95$.
Applying the $O-C$ timing offset to equation \ref{eqOmC} provides
a lower limit to the binary's orbital eccentricity: $e \geq | e
\cos \omega | \simeq 0.00594$.} \label{figTimingVariationsEcc}
\end{figure}

If the EB has an eccentric orbit, the primary and secondary
eclipse will separate on the $O-C$ plot, and form two parallel
lines with a vertical offset of $\Delta t$ (see Figure
\ref{figTimingVariationsEcc}). We measure this offset and use it
as a sensitive method to detect orbital eccentricities. In
particular, the value of $\Delta t$ constrains $e \cos \omega$,
which in turn provides a lower limit for the system's eccentricity
\citep{Tsesevich73}:

\begin{equation}
\label{eqOmC}
e \cos \omega \simeq \frac{\pi}{2} \frac{\Delta t}{P}\ .
\end{equation}

This formula assumes an orbital inclination of $i=90^\circ$,
making it a good approximation for eclipsing binaries. We use this
method, in combination with DEBiL, to identify the eccentric EBs
in the catalog (see Table \ref{tableEccentric}). However, in cases
where the eclipse timing measures $|e \cos \omega| < 0.005$, or
when the eccentricity is consistent with zero, we assume that the
EB is non-eccentric, and model it using a circular orbit. We
further discuss the physics of these systems in
\S\ref{subsecEccentricEBs}.

Step 5 is identical to step 3, except that it uses the revised
period from step 4. This step provides an improved fit to the
LCs, as evidenced by an improved chi-squared value in over 70\% of
the cases.

In step 6, we locate and remove non-EB sources that seem to be
periodic due to photometric contamination by true EBs. Such
contaminations result from overlapping point spread functions
(PSF) that cause each source to partially blend into the other.
These cases can be easily identified with a program that scans
through pairs of targets\footnote{We ran a brute force scan, which
required O($N^2$) iterations. But by employing a data structure
that can restrict the scan to nearby pairs, it is possible to
perform this scan in only O($N$) iterations, assuming that such
pairs are rare.}, and selects those that both have similar
periods (see description below) and are separated by an angle that
is smaller than twice the PSF. We found 14 such pairs, all of
which were separated by less than $41"$, which is well within
twice the TrES PSF ($60"$), while the remaining pairs with similar
periods were separated by over $450"$. Upon inspection, all 14 of
the pairs we found had similar eclipse shapes, indicating that we
had no false positives. Between each pair, we identify the LC with
shallower eclipses (in magnitudes) as being contaminated and
remove it from the catalog.

We define periods as being similar if the difference between them
is smaller than their combined uncertainty. We estimate the period
uncertainty using the relation: $\varepsilon_P \propto P^2 / T$,
where $T$ is the time interval between the initial and the final
observations. One arrives at this relation by noticing that when
phasing the LC, the effect of any perturbation from the true
period will grow linearly with the number of periods in the LC
(see step 4). This amplified effect will become evident once it
reaches some fraction of the period itself, in other words, when
$\varepsilon_P (T / P) \propto P$. A typical TrES LC with a
revised period will have a proportionality constant of
approximately $1/1000$. In order to avoid missing contaminated
pairs (false negatives), we adopt in this step the extremely
liberal proportionality constant of unity.

In step 7, we conduct a visual inspection of all the LC fits.
Most EBs were successfully modeled and were included into the
catalog as is. About $1\%$ of the LCs analyzed had misidentified
periods, as a result of failures of the period-finding method of
step 1. In most of these cases, the period finder indicated
either a harmonic of the true period or a rational multiple of a
solar or sidereal day. In such cases, we use an interactive
periodogram\footnote{LC, created by Grzegorz Pojmanski.} to find
the correct period and then reprocess the LCs through the
pipeline. Some entries were misidentified at step 2 as being
ambiguous, even though they have a detectable secondary eclipse or
have slightly unequal eclipses. In these cases, the erroneous
doubled entry was removed. Lastly, some of the EBs were not fit
sufficiently well with DEBiL in step 5. These cases were
typically due to clustered outlier data points, systematic noise,
or severe activity of a stellar component (e.g., flares or spots),
which caused DEBiL to produce erroneous initial model parameters.
These cases were typically handled by having DEBiL produce the
initial model parameters from a more smoothed version of the LC.

In step 8, we match each system, through its coordinates, with the
corresponding source in the Two Micron All Sky Survey catalog
[2MASS ; \citep{Skrutskie06}]. This was done to obtain both
accurate target positions and observational magnitudes. These
magnitude measurements are then used to derive the colors of each
EB, which are incorporated into the MECI analysis, as well as to
estimate the EB's distance modulus (step 9). To this end, 2MASS
provides a unique combination of high astrometric accuracy
($\sim$0.1") together with high photometric accuracy ($\sim$0.015
mag) at multiple near-infrared bands, all while maintaining a
decent spacial resolution ($\sim$3"). By employing these
near-infrared bands, we both inherently reduce the detrimental
effects of stellar reddening, and are able to correct for much of
the remaining extinction by fitting for the Galactic interstellar
absorption.

In order to use the measurements from the 2MASS custom $J$, $H$,
and $K_s$ filters, we converted them to the equivalent ESO-filter
values so that they could be compared to the isochrone table
values used in the MECI analysis. This conversion was done using
approximate linear transformations \citep{Carpenter01}. However,
the colors of three EBs (T-And0-10336, T-Cyg1-02304, and
T-Per1-05205) were so anomalous that they did not permit a
reasonable model solution; thus, we chose not to include any color
information in their MECI analyses.

\begin{figure}
\includegraphics[width=3.5in]{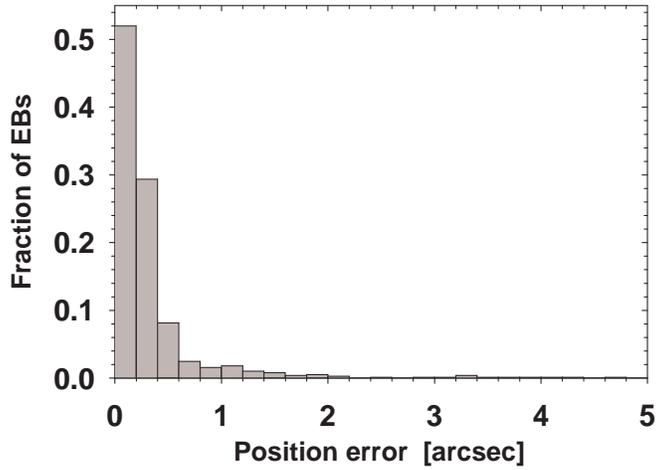}
\caption{The distribution of the catalog position errors when
matching targets to the proper motion databases. In some cases,
the position errors are dominated by the
motion of the EB during the intervening years.}
\label{figPosErr}
\end{figure}

In addition to its brightness, we also look up each EB's proper
motion. Although proper motion is not required for any of the
pipeline analyses, it provides a useful verification for low-mass
candidates (see \S\ref{subsecLowMassEBs}). These systems are
expected to have large proper motions, since they must be nearby
to be observable in this magnitude-limited survey. The most
extreme such case in the catalog is CM Draconis (T-Dra0-01363),
which has a proper motion of over 1300 ${\rm mas\,yr^{-1}}$ \citep{Salim03}, and
is probably the lowest mass system in our catalog. To this end, we
match each system to the Second U.S. Naval Observatory CCD
Astrograph Catalog [UCAC release 2.4 ; \citet{Zacharias04}]. When
there is no match with UCAC, we use the more comprehensive but
less accurate U.S. Naval Observatory photographic sky survey
[USNO-B release 1.0 ; \citet{Monet03}]. These matches are made
using the more accurate aforementioned adopted 2MASS coordinates.
However, because of their increased observational depth, and the
fact that some high-proper motion targets are expected to have
moved multiple arcseconds in the intervening decades, we chose to
match each target to the brightest ($R$-band) source within 7.5".
It should be noted that the position of CM Draconis shifted by
more than 22" and had to be matched manually, though 90\% of the
matches were separated by less than 0.6", and 98\% were separated
by less than 2" (see Figure \ref{figPosErr}).

The proper motions garnered from these databases can be combined
with distance estimates ($D$), to calculate the absolute
transverse velocity ($v_{tr}$) of a given EB:

\begin{equation}
v_{tr} \simeq 4.741\: {\rm km\,s^{-1}} \left(\frac{PM}{1\: {\rm mas\,yr^{-1}}}\right)\left(\frac{D}{1\: {\rm kpc}}\right),
\end{equation}

where $PM$ is the system's angular proper motion. In the catalog
we list the right ascension and declination components
($PM_{\alpha}$ and $PM_{\delta}$, respectively), so as to allow
one to compute the system's direction of motion in the sky. The
value of $PM$ can be computed from its components, using: $PM^2 =
PM_{\delta}^2 + PM_{\alpha}^2 \cos^2 \delta$, where $\delta$ is
the system's declination. When applying this formula, one should
be aware that the $\cos \delta$ coefficient is generally folded
into $PM_{\alpha}$. We follow this convention as well in our
catalog.

\begin{figure}
\includegraphics[width=3.5in]{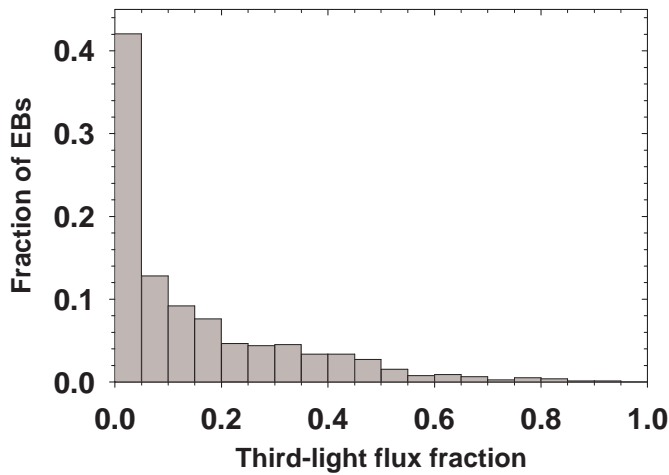}
\caption{The distribution of the $R$-band third-light flux fraction in
the catalog LCs. This fraction was calculated by summing
the fluxes of all the USNO-B sources within 30" of the target, excluding the
target, and dividing this value by the total flux within 30", including the
target. The resulting fraction ranges from 0 to 1.}
\label{figBlending}
\end{figure}

Finally, we incorporate the USNO-B photographic $B$- and
$R$-magnitude measurements into our catalog to provide a rough
estimate of the optical brightness of each target. USNO-B lists
two independent measurements in each of these filter; however, in
some cases one or both of these measurements failed. When both
measurements are available, we average them for improved accuracy.
However, each measurement has a large photometric uncertainty of
$\sim$0.3 mag; thus, even these averaged values will have errors
that are over an order of magnitude larger than the photometric
measurements of 2MASS. For this reason, and because of the
increased effect of stellar reddening, we chose not to incorporate
these data into the MECI analysis. However, USNO-B's high spacial
resolution ($\sim$1") enabled us to detect many sources that
blended with our targets in the TrES exposures. By summing the
$R$-band fluxes of all the USNO-B sources within 30" of each
target, we estimated the fraction of third-light included in each
LC (see Figure \ref{figBlending}). Note that this measure provides
only a lower bound to the true third-light fraction, as some EBs
are expected to have additional close hierarchical components that
would not be resolved by USNO-B. For most of the catalog targets,
the third-light flux fraction was found to be small ($<$10\%). We
therefore conclude that stellar blending will usually have only a
minor effect on the MECI analysis results; however, users should
be aware of the potential biases in the calculated properties of
highly blended targets. Though it was not applied to this catalog,
in principle, given a third-light flux fraction at a
well-determined LC phase, one could correct for the effects of
blending.

\begin{figure}[H]
\centerline{
\begin{tabular}{c@{\hspace{2pc}}c}
\includegraphics[width=2.4in]{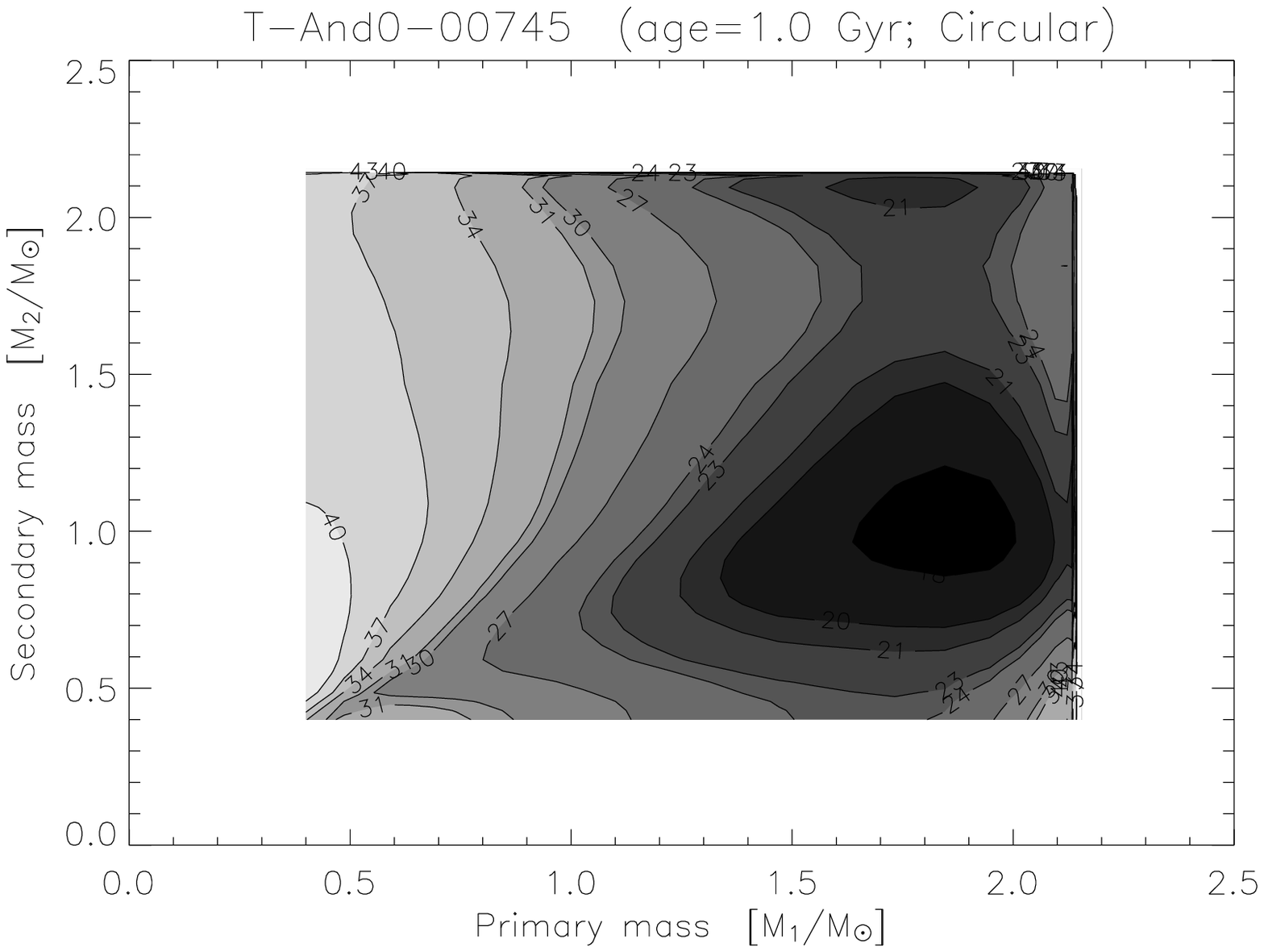} &
\includegraphics[width=2.4in]{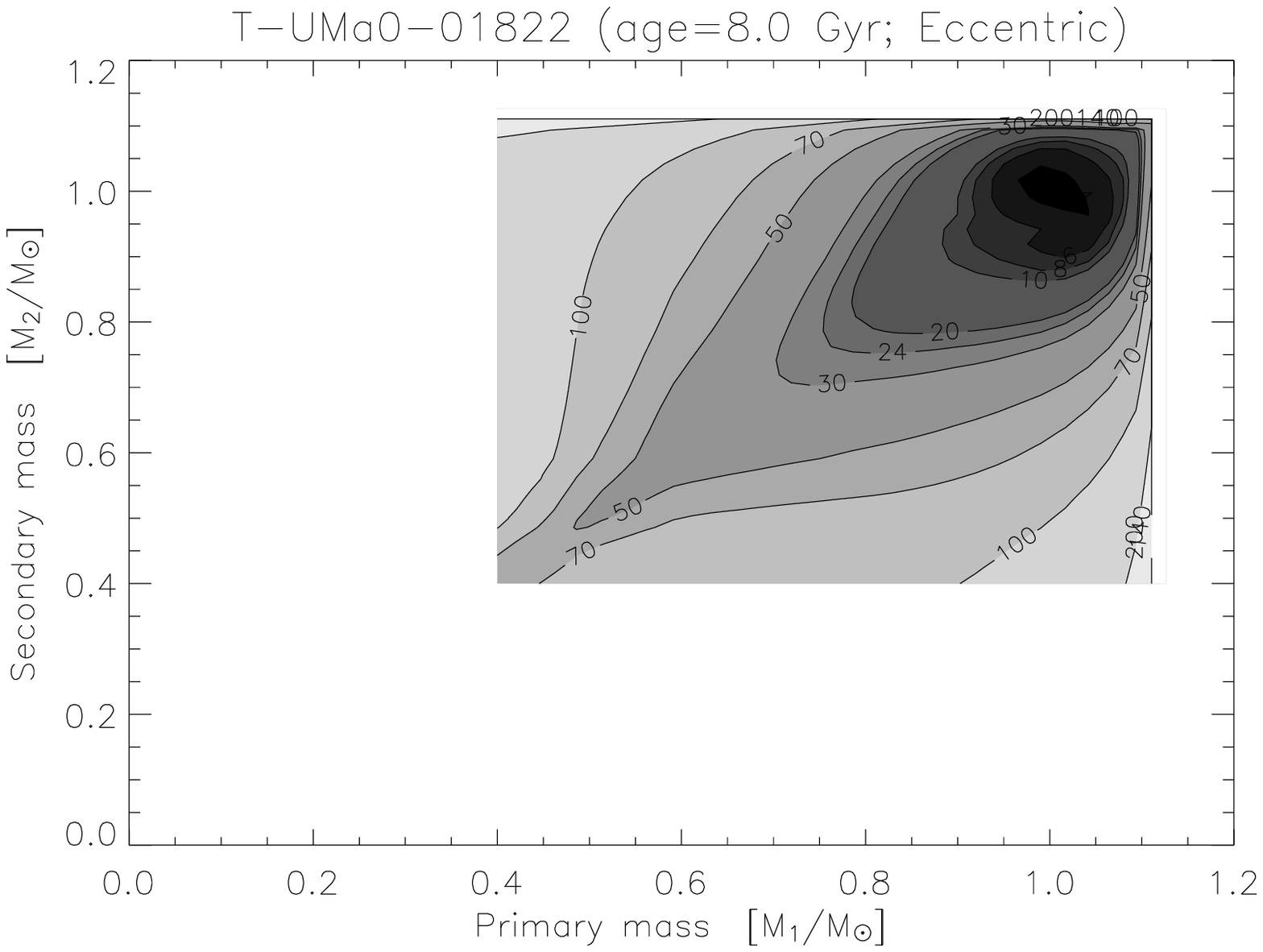} \\
a. T-And0-00745: A circular-orbit EB & b. T-UMa0-01822: An eccentric-orbit EB
\end{tabular}}
\caption{MECI likelihood contour plots of a typical circular-orbit EB and
a typical eccentric-orbit EB.
There is no significant difference in the way MECI handles
these cases, and both usually have a single contour
minimum. The plots shown here have the ages set to the values
that produced the lowest MECI minima.}
\label{figMECI1}
\end{figure}

\begin{figure}[H]
\centerline{
\begin{tabular}{c@{\hspace{2pc}}c}
\includegraphics[width=2.4in]{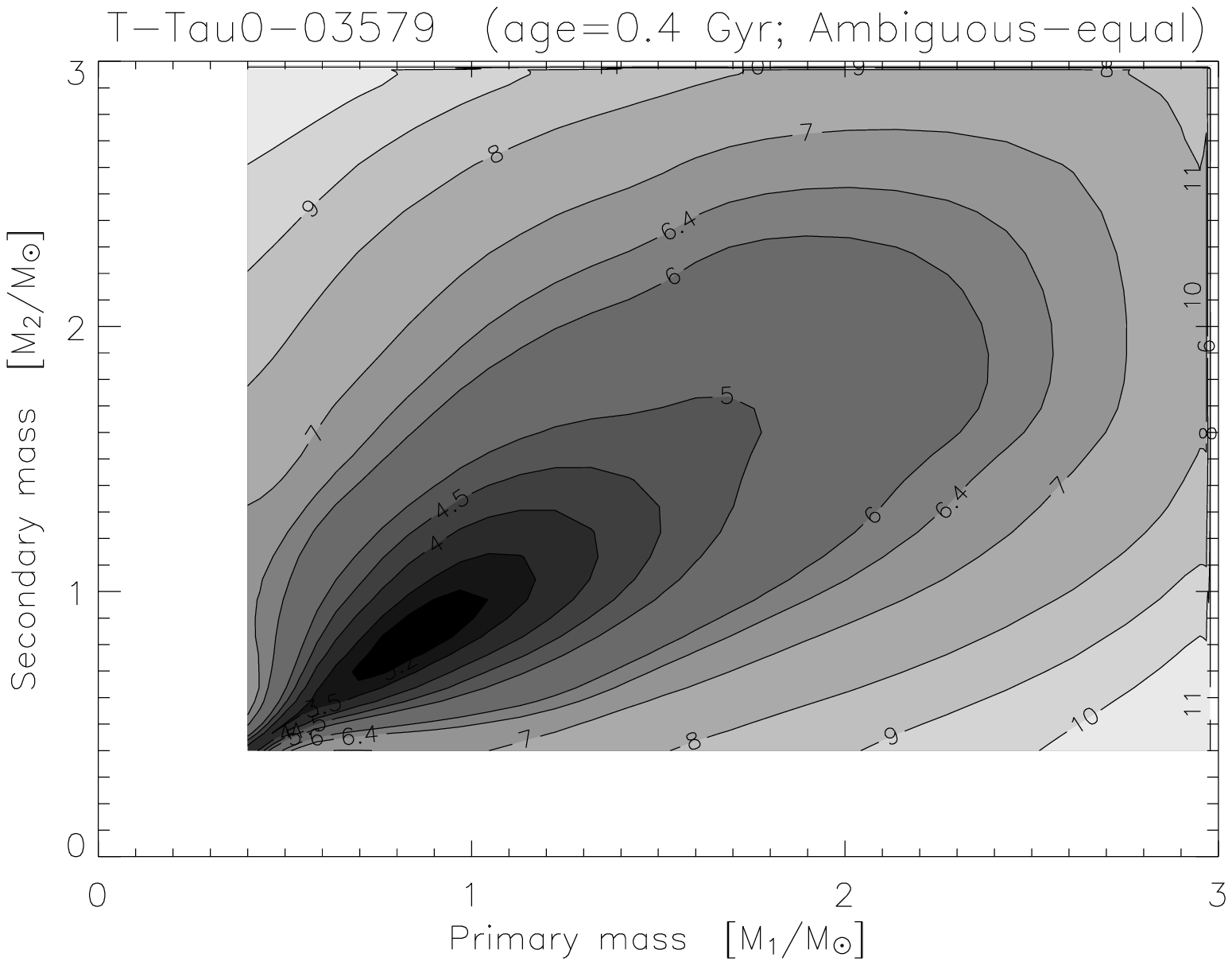} &
\includegraphics[width=2.4in]{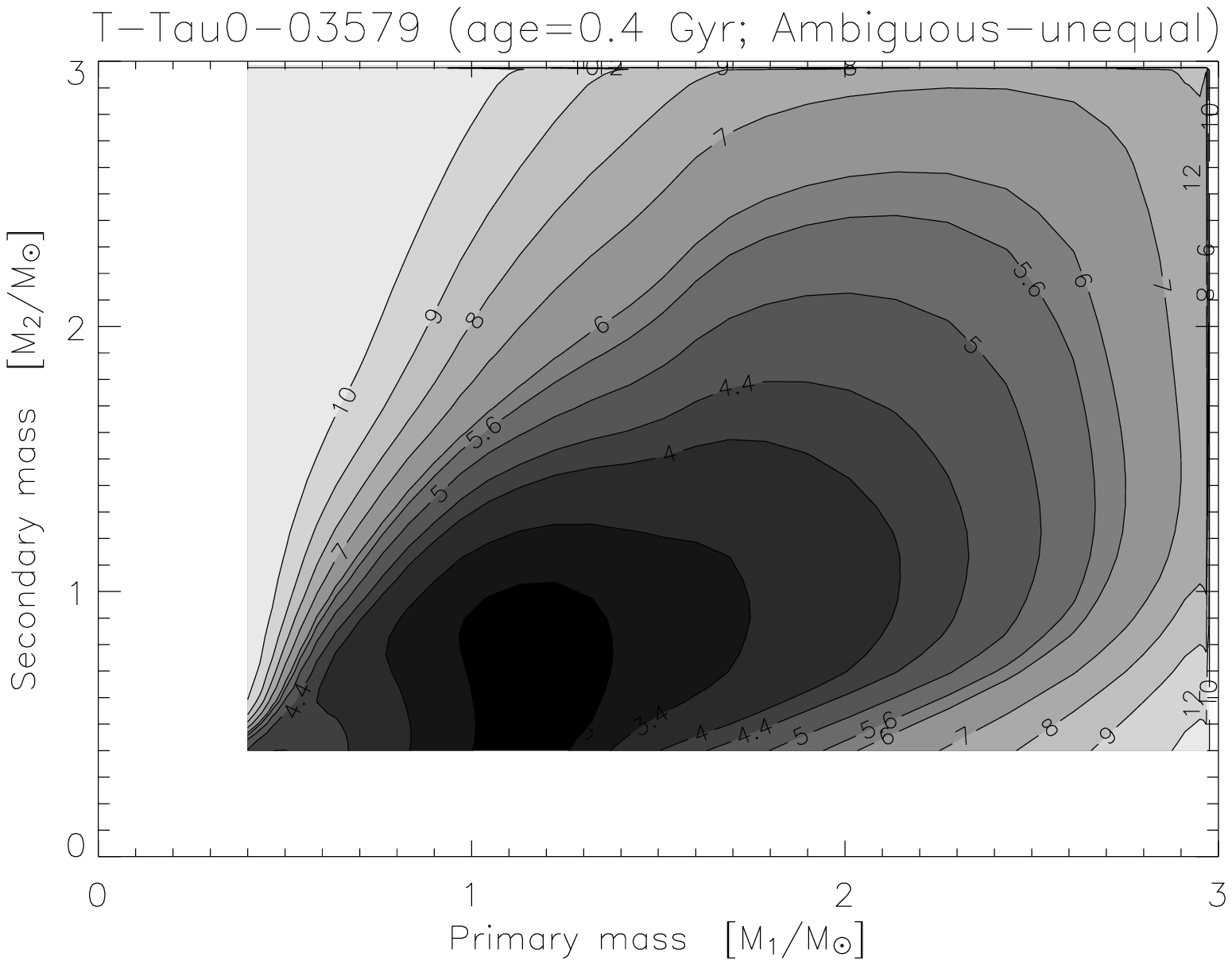} \\
a. T-Tau0-03579: Assuming equal components & b. T-Tau0-03579: Assuming unequal components
\end{tabular}}
\caption{MECI likelihood contour plots of a typical ambiguous EB
(T-Tau0-03579). These plots show the effect of assuming that the
binary components are equal (left) or unequal (right). Note that
the equal-component solution will have a nearly symmetric contour
around the diagonal, while the unequal-component solution can
provide only an upper limit to the secondary component's mass,
in this case $M_2 \lesssim 1\:M_{\sun}$. The plots shown here have
the ages set to the values that produced the lowest MECI minima.}
\label{figMECI2}
\end{figure}

In step 9, we analyze the LCs with MECI. We refer the reader to
the full description of this method in \citet{Devor06a, Devor06b},
and provide here only a brief outline. Given an observed EB LC and
out-of-eclipse colors, MECI will iterate through a range of values
for the EB age and the masses of its two components. By looking up
their radii and luminosities in theoretical isochrone tables, MECI
simulates the expected LC and combined colors, and selects the
model that best matches the observations, as measured by the
chi-squared statistic. Or, more concisely, MECI searches the ($M_1,
M_2, age$)-parameter space for the chi-squared global minimum of
each EB. Figures \ref{figMECI1} and \ref{figMECI2} show
constant-age slices through such a parameter space. Once found,
the curvature of the global minimum along the parameter space axes
is used to determine the uncertainties of the corresponding
parameters.

The MECI analysis makes two important assumptions. The first is
that EB stellar components are coeval, which has been shown to
generally hold for close binaries \citep{Claret02}. When this
assumption is violated, MECI will often not be able to find an EB
model that successfully reproduces the LC eclipses. Such systems,
which may be of interest in their own right, make up $\sim$3\% of
the catalog and are further discussed later in this section. The
second assumption is that there is no significant reddening, or
third-light blended into the observations (i.e. from a photometric
binary or hierarchical triple). Such blending in the LC will make
the eclipses shallower, which produces an effect very similar to
that of the EB having a grazing orbit. Thus, it will cause the
measured orbital inclination to be erroneous, although it should
rarely otherwise affect the results of the MECI analysis
significantly. However, the MECI analysis is sensitive to color
biases caused by stellar reddening and blending.

We reduce both these biases by incorporating 2MASS colors (see
step 8), which are both less suspectable to reddening than optical
colors, and suffer from significantly less blending than TrES, as
the radius of the 2MASS photometric aperture is $\sim$20 times
smaller than that of TrES. We then attempt to further mitigate
this problem by analyzing each EB twice, using different relative
LC/color information weighting values [see \citet{Devor06b} for
further details]. We first run MECI with the default weighting
value ($w = 10$), and then run MECI again with an increased LC
weighting ($w = 100$) thereby decreasing the relative color
weighting. Finally, we adopt the solution that has a smaller
reduced chi-squared. Typically, the results of the two MECI
analyses are very similar, indicating that the observed colors are
consistent with those predicted by the theoretical isochrones.
In such cases, the color information provides an important
constraint, which significantly reduces the parameter
uncertainties. However, when there is a significant color bias,
the default model will not fit the observed data as well as the
model that uses a reduced weighting of the color information. In
such a case, the reduced color information model, which has a
smaller chi-squared, is adopted. Following this procedure, we find
that in $\sim$9\% of our EBs, the reduced color information model
provided a better fit, indicating that while significant
color-bias is uncommon, it is a source of error that should not be
ignored.

By default, we had MECI use the Yonsei-Yale \citep{Yi01, Kim02}
isochrone tables of solar metallicity stars. Although they
successfully describe stars in a wide range of masses, these
tables become increasingly inaccurate for low-mass stars, as the
stars become increasingly convective. For this reason we
re-analyze EBs for which both components were found to have masses
below $0.75M_{\sun}$, using instead the \citet{Baraffe98}
isochrone tables, assuming a convective mixing length equal to the
pressure scale height. Our EB models also take into account the
effects of the limb darkening of each of the stellar components.
To this end, we employ the ATLAS \citep{Kurucz92} and PHOENIX
\citep{Claret98, Claret00} tables of quadratic limb darkening
coefficients.

As previously mentioned, once we know the absolute properties of
an EB system, we are able to estimate its distance
\citep{Stebbing10, Paczynski97}, and thus such systems can be
considered standard candles. We use the extinction coefficients of
\citet{Cox00}, assuming the standard Galactic ISM optical
parameter, $R_V = 3.1$, to create the following system:

\begin{eqnarray}
mag_J - Mag_J & = & \Delta Mag + 0.282 \cdot A(V)\\
mag_H - Mag_H & = & \Delta Mag + 0.176 \cdot A(V)\\
mag_K - Mag_K & = & \Delta Mag + 0.108 \cdot A(V)
\end{eqnarray}

Where $\Delta Mag$ is the extinction-corrected distance modulus,
and $A(V)$ is the $V-mag$ absorption due to Galactic interstellar
extinction. The estimated distance can then be solved using: $D =
10pc \cdot 10^{\Delta Mag / 5}$. Because we have three equations
for only two unknowns, we adopt the solution that minimizes the
sum of the squares of the residuals. In some cases we remove one
of the bands as being an outlier (i.e. if it would have resulted
in a negative absorption), after which we are still able to solve
the systems. But in cases where we need to remove two bands, we
set $A(V) = 0$ in order to solve for the distance modulus.
Although this method has a typical uncertainty of 10\% to 20\%, it
can be applied to EBs that are far more distant and dim than are
accessible in other methods, such as parallax measurement. It can
be used to map broad features of the Galaxy, and identify binaries
that are in the Galactic halo. This method can also be applied to
a clustered group of EBs, whereby averaging their distances will
reduce the distance uncertainty to the cluster as the inverse
square root of the number of EBs measured.

In step 10, we perform a final quality check for the EB model
fits, and classify them into seven groups.

\renewcommand{\theenumi}{\Roman{enumi}}
\begin{enumerate}
\item Eccentric: EBs with unequally-spaced eclipses
\item Circular: EBs with equally-spaced but distinct eclipses
\item Ambiguous-unequal: EBs with undetected secondary eclipses
\item Ambiguous-equal: EB with equally-spaced and indistinguishable eclipses
\item Inverted: detached EBs that are not successfully modeled by MECI
\item Roche-lobe-filling: non-detached EBs that are filling at least one Roche-lobe
\item Abnormal: EBs with atypical out-of-eclipse distortions
\end{enumerate}

We list the model parameters for the EBs of groups I-IV in the
electronic version of this catalog (see full description in
Appendix \ref{appendixCatalogDescription}). The EBs of groups
V-VII could not be well modeled by MECI; therefore, we list only
their coordinates and periods, so that they can be followed up.

\begin{figure}
\includegraphics[width=5in]{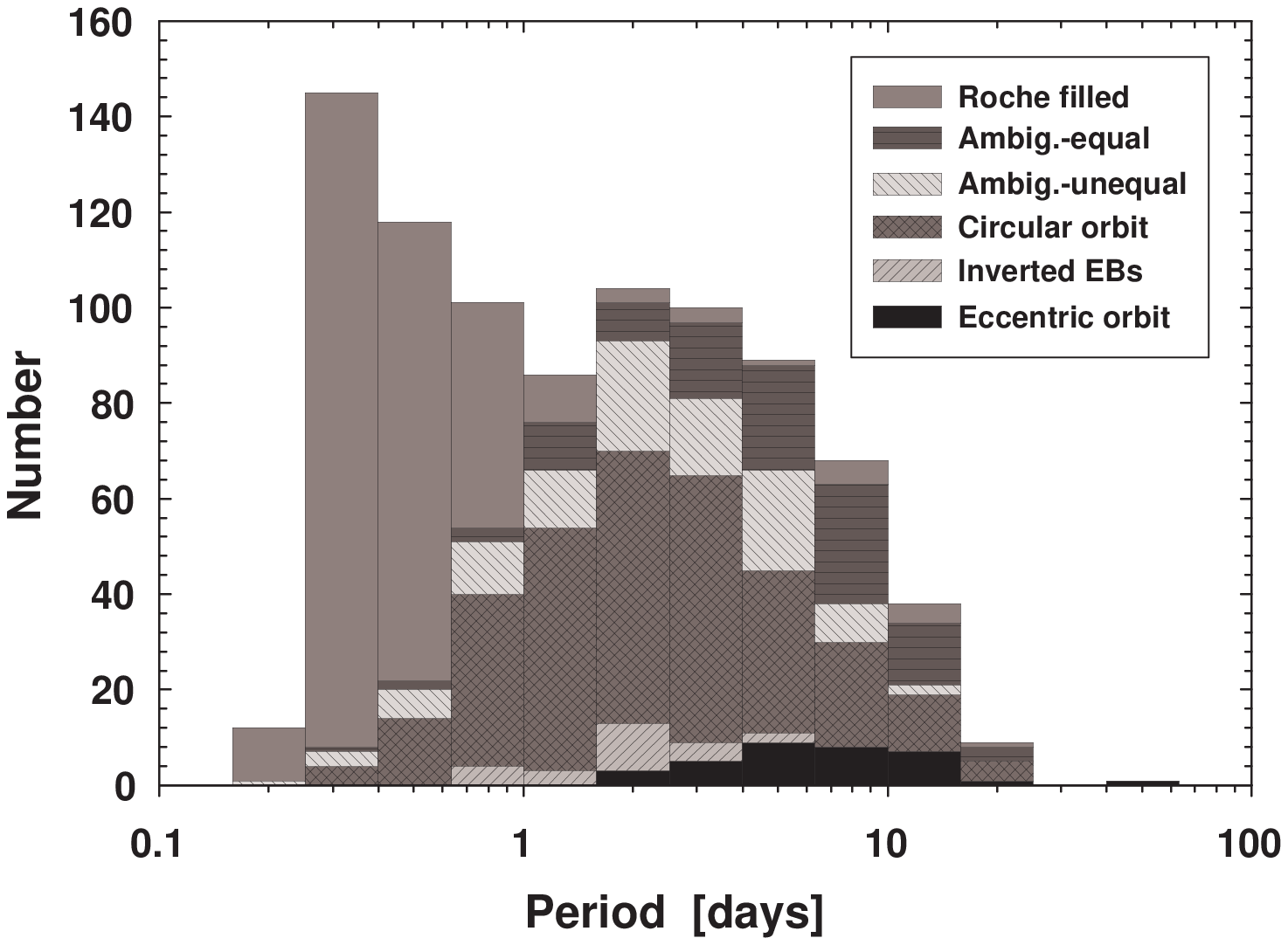}
\caption{The EB orbital period distribution within the catalog.
Each bin is subdivided to show the number of binaries belonging to
each of the classification groups described in \S\ref{secMethod}.
We do not include here five unclassified abnormal systems (see
Table \ref{tableAbnormal}). The remaining abnormal systems were
all classified as having circular orbits (group [II]) and are
included with them. Note that the ambiguous-equal and
ambiguous-unequal entries represent the same stars, with entries
in the former group having double the period of the latter. Note
also how the Roche-lobe-filling EBs dominate the sub-day bins, and
have a long tail stretching well above 10 day periods.
Furthermore, the circular-orbit EBs have a period distribution
peak of at $\sim$2 days, while the eccentric orbit EBs peak at
$\sim$5 days. This is likely due to the orbital circularization
that occurs preferentially in short-period systems (see also
Figures \ref{figPeriodEcc} and \ref{figPeriodM1}).}
\label{figPeriodDistrib}
\end{figure}

Figure \ref{figPeriodDistrib} illustrates the period distribution
of these groups. Note, however, that both the orbital geometry of
EBs (eclipse probability $\propto P^{-2/3}$), and the limited
duration of the TrES survey data ($\leq$90 days ; varies from
field to field ; see Table \ref{tableFieldsObs}), act to suppress
the detection of binaries with longer periods. An added
complication for single-telescope surveys is that about half of
the EBs with periods close to an integer number of days will not
be detectable, as they eclipse only during the daytime. This EB
distribution is consistent with the far deeper OGLE II field
catalog \citep{Devor05}, where the long tail of Roche-lobe-filling
systems has recently been explained by \citet{Derekas07} as being
the result of a strong selection toward detecting eclipsing giant
stars.

\begin{figure}
\includegraphics[width=5in]{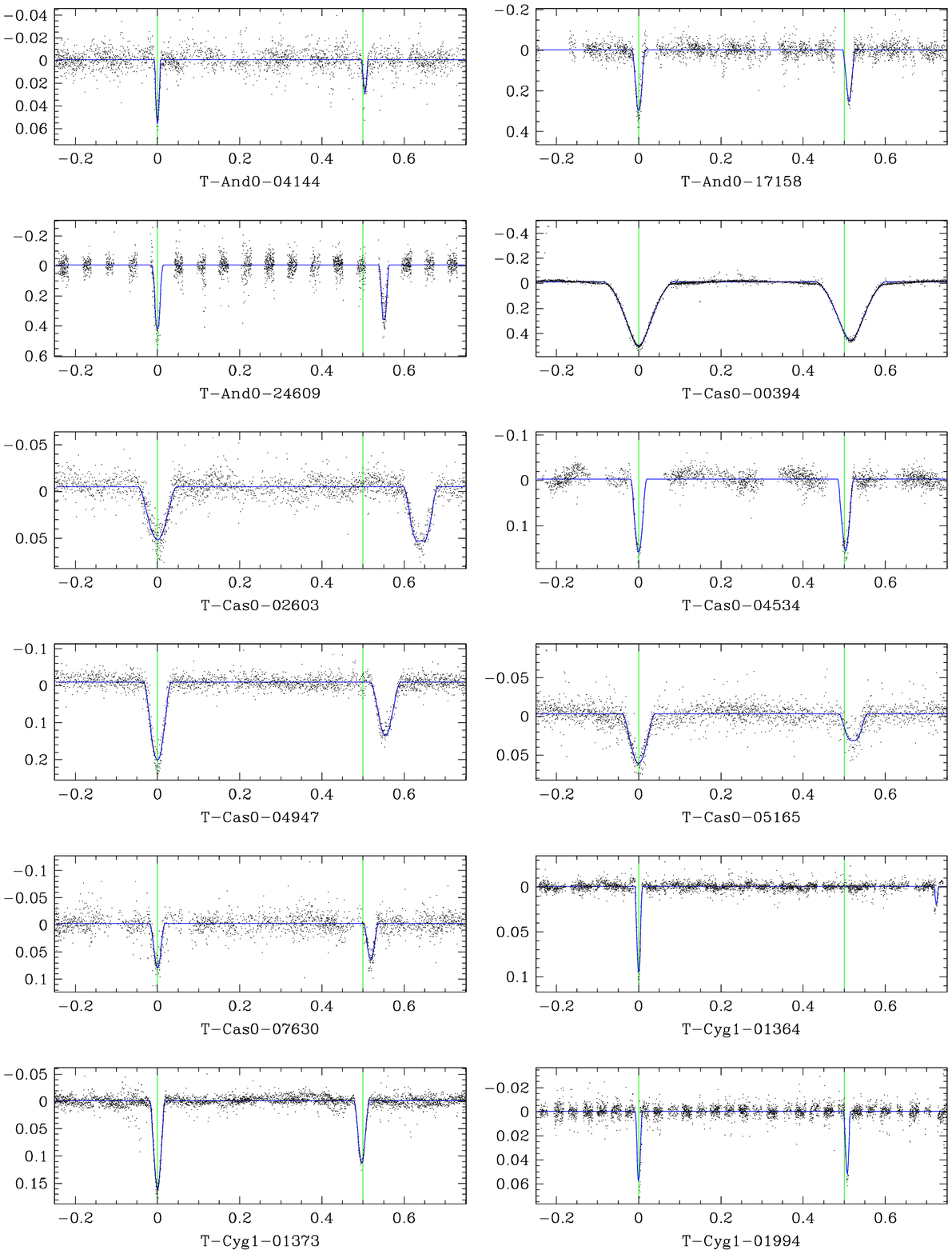}
\caption{Eccentric EBs (panel 1).
Note how the secondary eclipse is not at phase 0.5, as it would be in circular orbit EBs.}
\label{figEcc1}
\end{figure}

\begin{figure}
\includegraphics[width=5in]{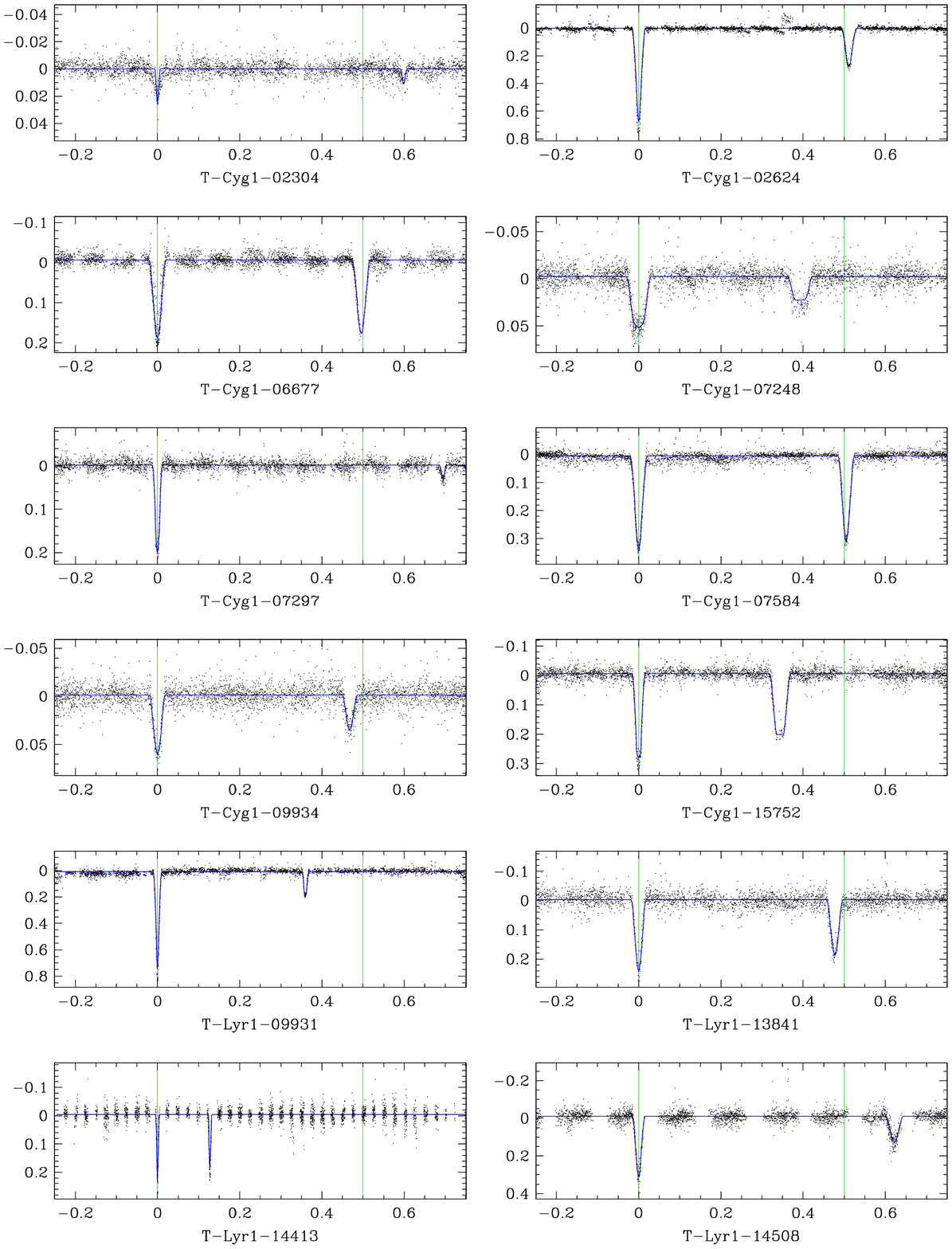}
\caption{Eccentric EBs (panel 2).}
\label{figEcc2}
\end{figure}

\begin{figure}
\includegraphics[width=5in]{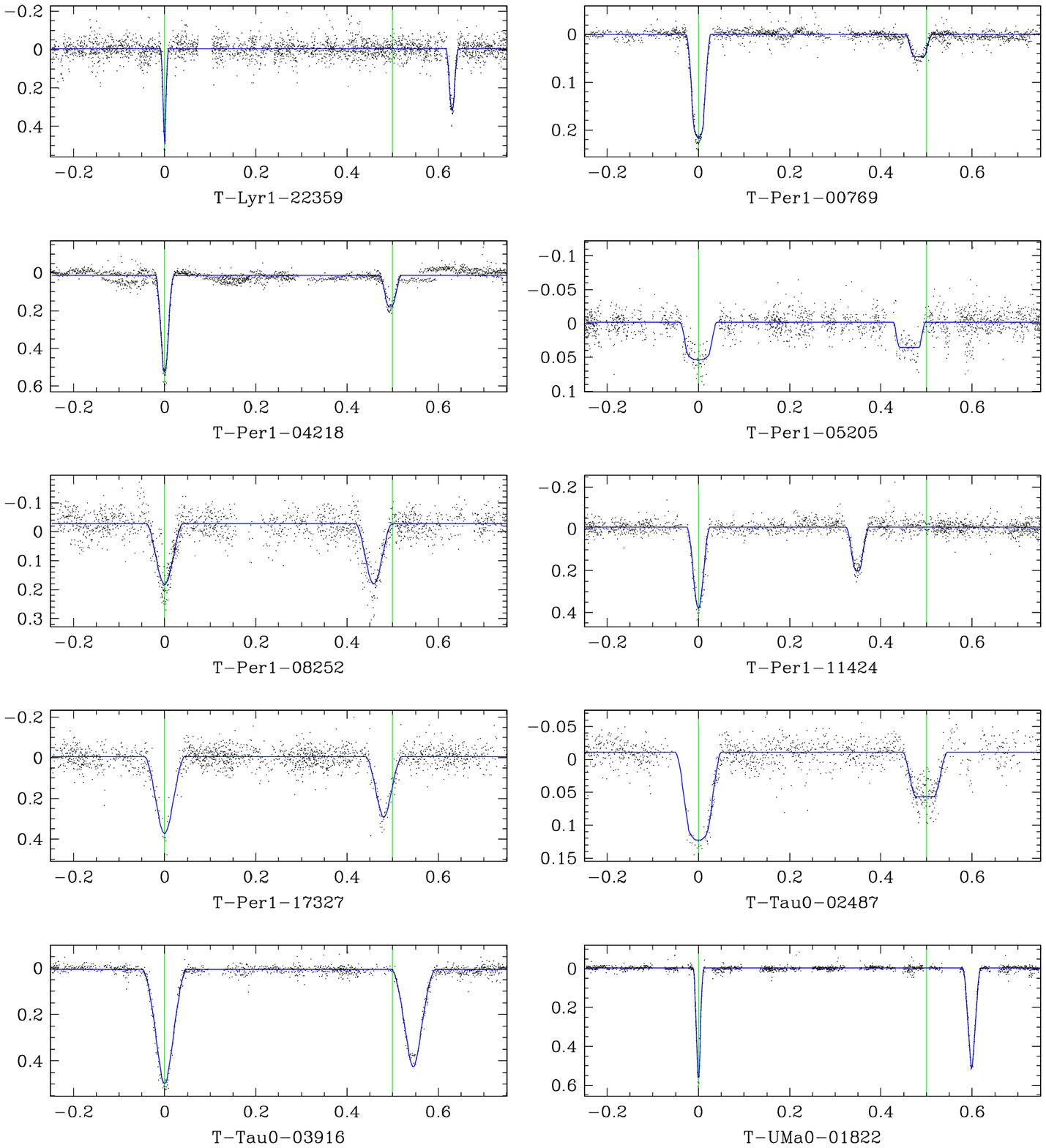}
\caption{Eccentric EBs (panel 3).}
\label{figEcc3}
\end{figure}

\begin{figure}
\includegraphics[width=5in]{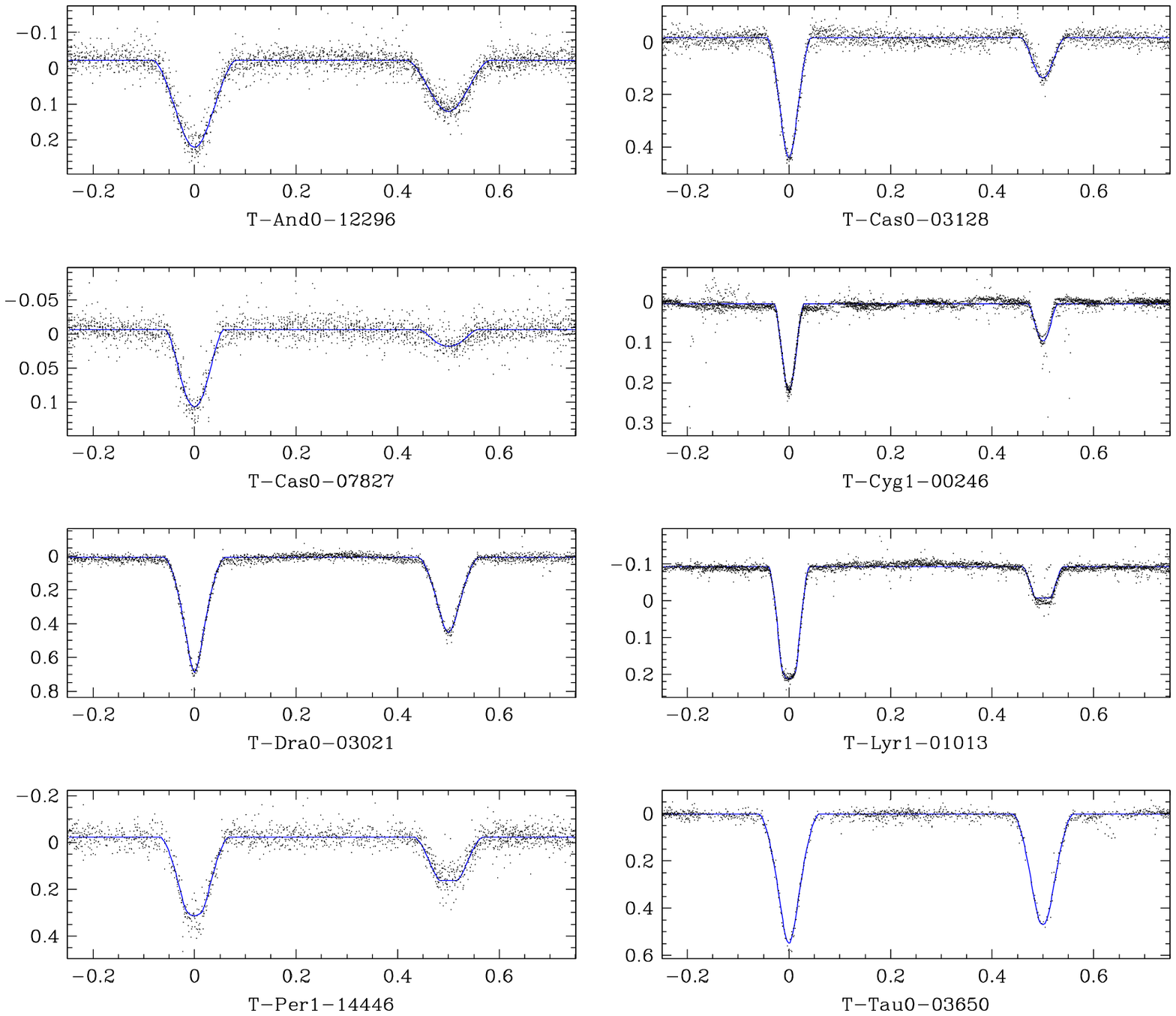}
\caption{Examples of unambiguous
EBs with circular orbits, with their best-fit MECI models (solid line).}
\label{figE0Cat}
\end{figure}

Group [I] contains the eccentric EBs identified in step (4) as
having centers of eclipse that are separated by a duration
significantly different from half an orbital period (see Figures
\ref{figEcc1}- \ref{figEcc3}). This criterion
is sufficient for demonstrating eccentricity, but not necessary,
since we miss systems for which $\cos \omega \simeq 0$ (see
equation \ref{eqOmC}). Fortunately, we are able to detect
eccentricities in well-detached EBs with $|e \cos \omega| \geq
0.005$, using eclipse timing. Therefore, assuming that $\omega$ is
uniformly distributed, we are approximately 67\% complete for $e
= 0.01$, and over 92\% complete for $e = 0.04$. In
principle, it would be possible to be 100\% complete for these
systems by measuring the differences in their eclipse durations;
however, this measurement is known to be unreliable \citep{Etzel91}
and so would likely contaminate this group with false positives.
Group [II] consists of all circular-orbit EBs that were
successfully fit by a single MECI model (see Figure
\ref{figE0Cat}).

\begin{figure}
\includegraphics[width=5in]{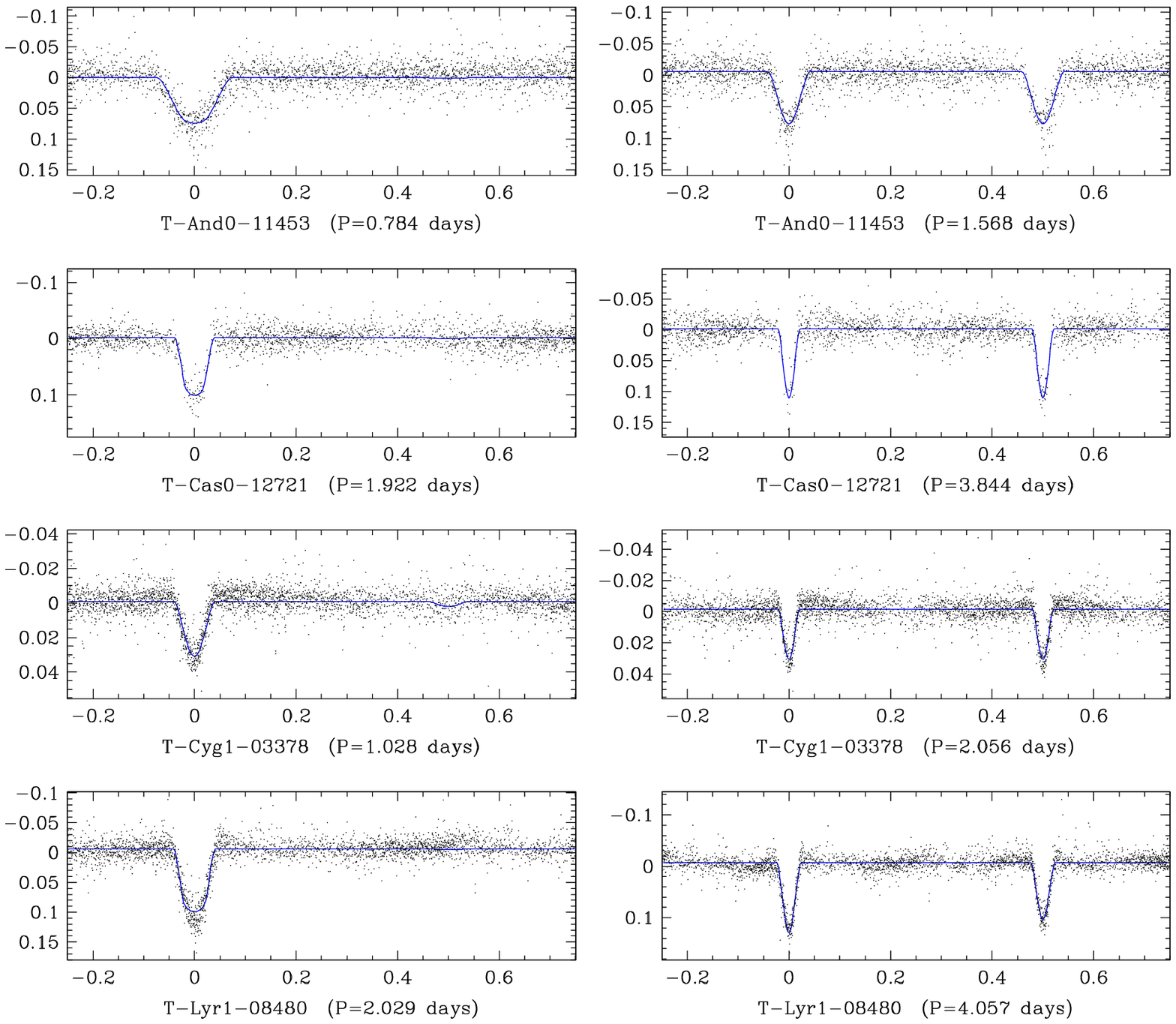}
\caption{Examples of ambiguous EBs. Left
column: assuming very unequal components. Right column: assuming
approximately equal components with double the period.}
\label{figAmbigCat}
\end{figure}

EBs with only one detectable eclipse can potentially be modeled in
two alternative ways\footnote{There is in fact a third
possibility. A highly eccentric binary could be oriented in such a
way that one component is eclipsed near its periastron, while its
orbital plain is not sufficiently inclined to produce a second
eclipse near its apoastron. For example, when $\cos \omega \simeq
0$ and $\cos i \simeq r_1 + r_2$. This possibility, however, is
expected to be very rare, especially in datasets such as this one
that contain very few binaries with large eccentricities.}. One
way is to assume very unequal stellar components, which have a
very shallow undetected secondary eclipse (group [III]). Since we
cannot estimate the eccentricity of such systems, we assume that
they have circular orbits. The other way is to assume that the
period at hand is twice the correct value, and that the components
are nearly equal (group [IV]). The entries of such ambiguous LCs
were doubled in step (2), so that these two solutions would be
independently processed through the pipeline (see Figure
\ref{figAmbigCat}). Therefore, these two groups have a one-to-one
correspondence between them, although only one entry of each pair
can be correct. Resolving this ambiguity may not always be
possible without spectroscopic data. In some cases, we were able
to resolve this ambiguity using either a morphological or a
physical approach. The morphological approach consists of manually
examining the LCs of group [IV] for any asymmetries in the two
eclipses (e.g., width, depth, or shape), or in the two plateaux
between the eclipses (e.g., perturbations due to tidal effects,
reflections, or the ``O'Connell effect''). The physical approach
consists of applying our understanding of stellar evolution in
order to exclude entries that cannot be explained through any
coeval star pairing (see Appendix \ref{appendixSingleEclipse}).
Either way, once one of the two models has been eliminated, the
other model is moved into group [II] and is adopted as a
non-ambiguous solution. It is interesting to note that when
analyzing the two models with MECI, the equal-component solution
(group [IV]) has masses approximately equal to the primary
component of the unequal-component solution (group [III]). The
mass of the unequal-component solution's secondary component will
typically be the smallest value listed in the isochrone table, as
this configuration will produce the least detectable secondary
eclipse.

\begin{figure}
\includegraphics[width=5in]{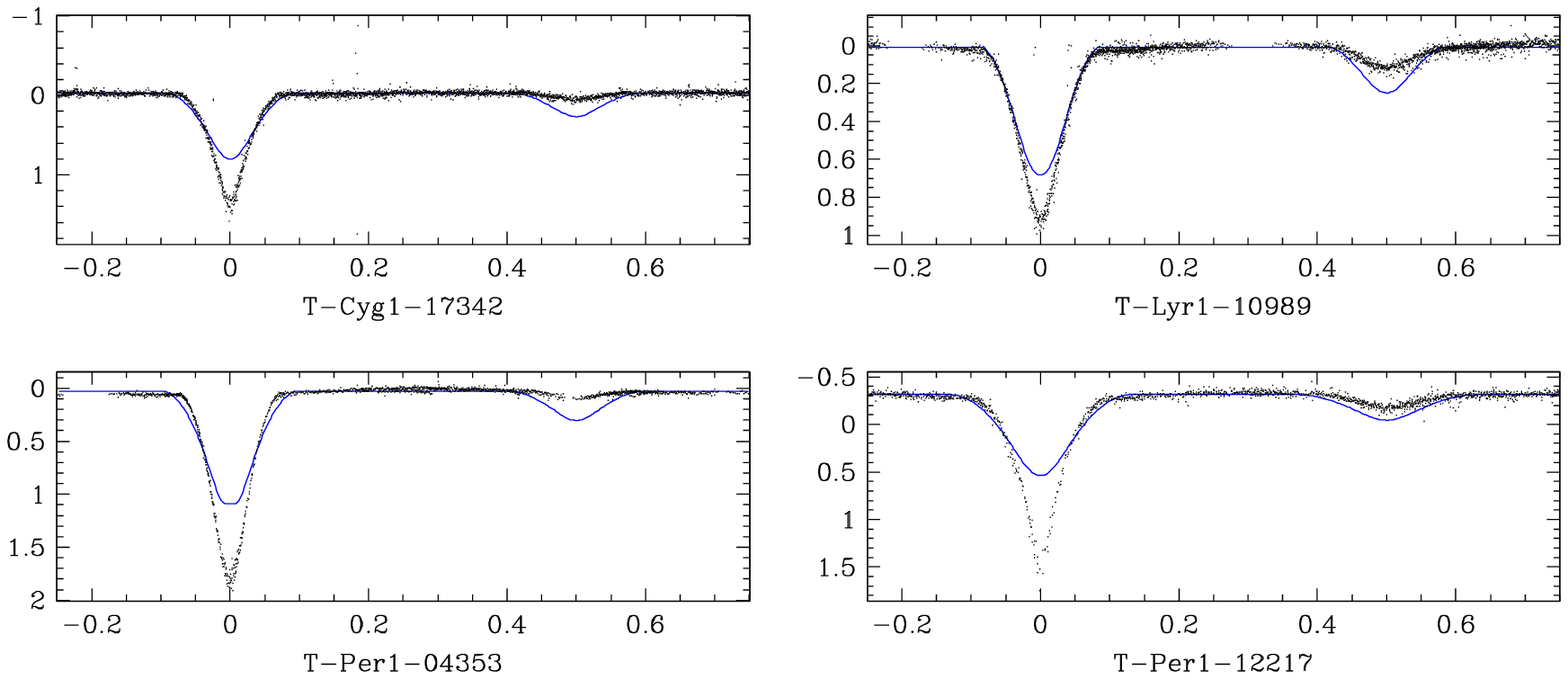}
\caption{Examples of EBs classified as inverted EBs. We included the
unsuccessful best-fit MECI model (solid curve) as an approximate reference to
illustrate the LC of a corresponding binary that has had no mass transfer.
Note how the model LC is unable to achieve a sufficiently deep primary
eclipse, while producing a secondary eclipse that is too deep.}
\label{figInvertedCat}
\end{figure}

Group [V] consists of detached EBs that cannot be modeled by two
coeval stellar components. As mentioned earlier, we can reject the
single-eclipse solution for EBs with sufficiently deep eclipses
(see Appendix \ref{appendixSingleEclipse}). This argument can be
further extended to cases where we can detect both eclipses in the
LC, but where one is far shallower than the other. In some cases,
no two coeval main-sequence components will reproduce such an LC,
but unlike the previous case, since both eclipses are seen, we
cannot conclude that the period needs to be doubled. Such systems
are likely to have had mass transfer from a sub-giant component
onto a main-sequence component through Roche-lobe overflow, to the
point where currently the main-sequence component has become
significantly more massive and brighter than it was originally
\citep{Crawford55}. This process will cause the components to
effectively behave as non-coeval stars, even though they have in
fact the same chronological age. In extreme cases, the originally
lower-mass main-sequence component can become more massive than
the sub-giant, and thus swap their original primary/secondary
designations, so that the main-sequence component is now the
primary component. We call such systems ``inverted'' EBs, and
place them into group [V] (see Figure \ref{figInvertedCat}). This
phenomenon is often referred to in the literature as the ``Algol
paradox,'' though we chose not to adopt this term so as to avoid
confusing it with the term ``Algol-type EB'' (EA), which is
defined by the General Catalogue of Variable Stars [GCVS ;
\citep{Kukarkin48, Samus06}] as being the class of all
well-detached EBs.

\begin{figure}
\includegraphics[width=5in]{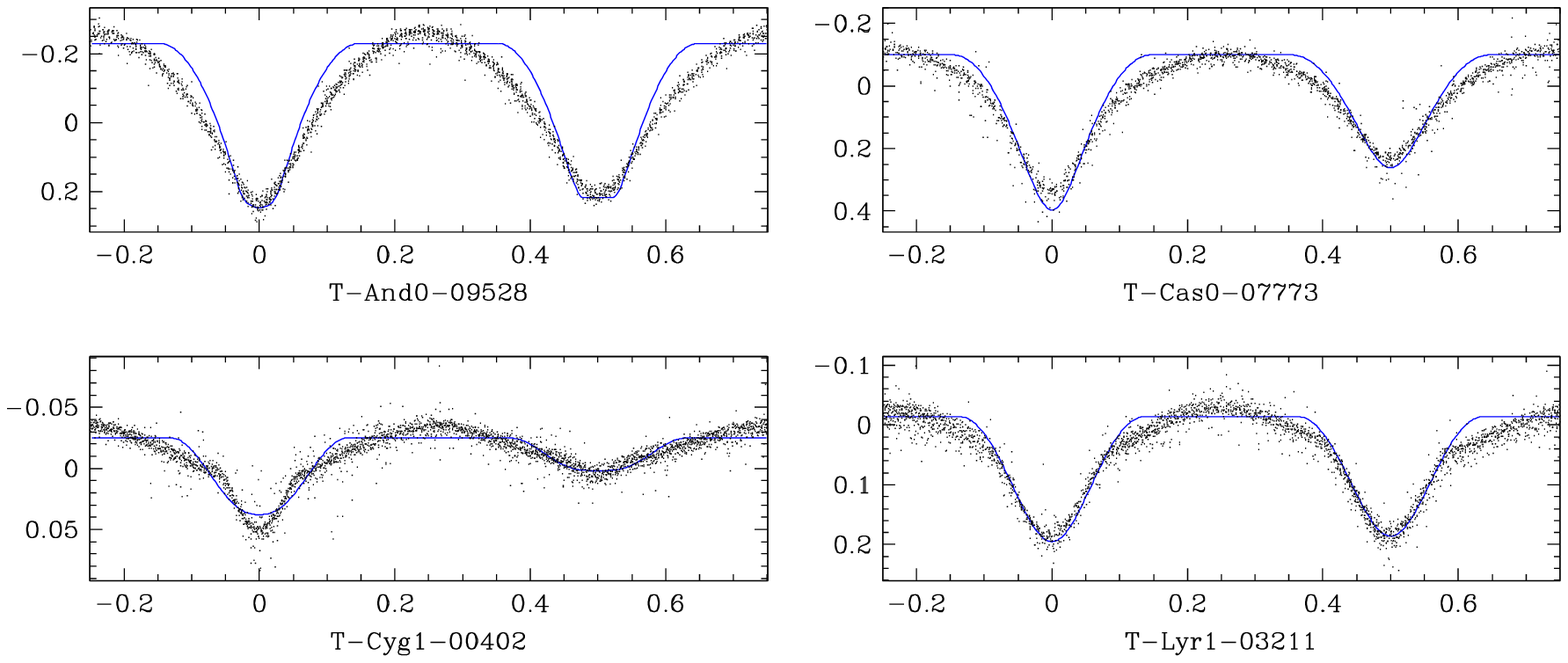}
\caption{Examples of EBs that are assumed to have filled at least one of their Roche-lobes.
We included, for illustration purposes only, their best-fit MECI models (solid line).
These models were not adopted since they neglect tidal distortions, reflections, and gravity-
darkening effects, and so produce a poor fit to the data.}
\label{figRocheFilledCat}
\end{figure}

\begin{figure}
\includegraphics[width=5in]{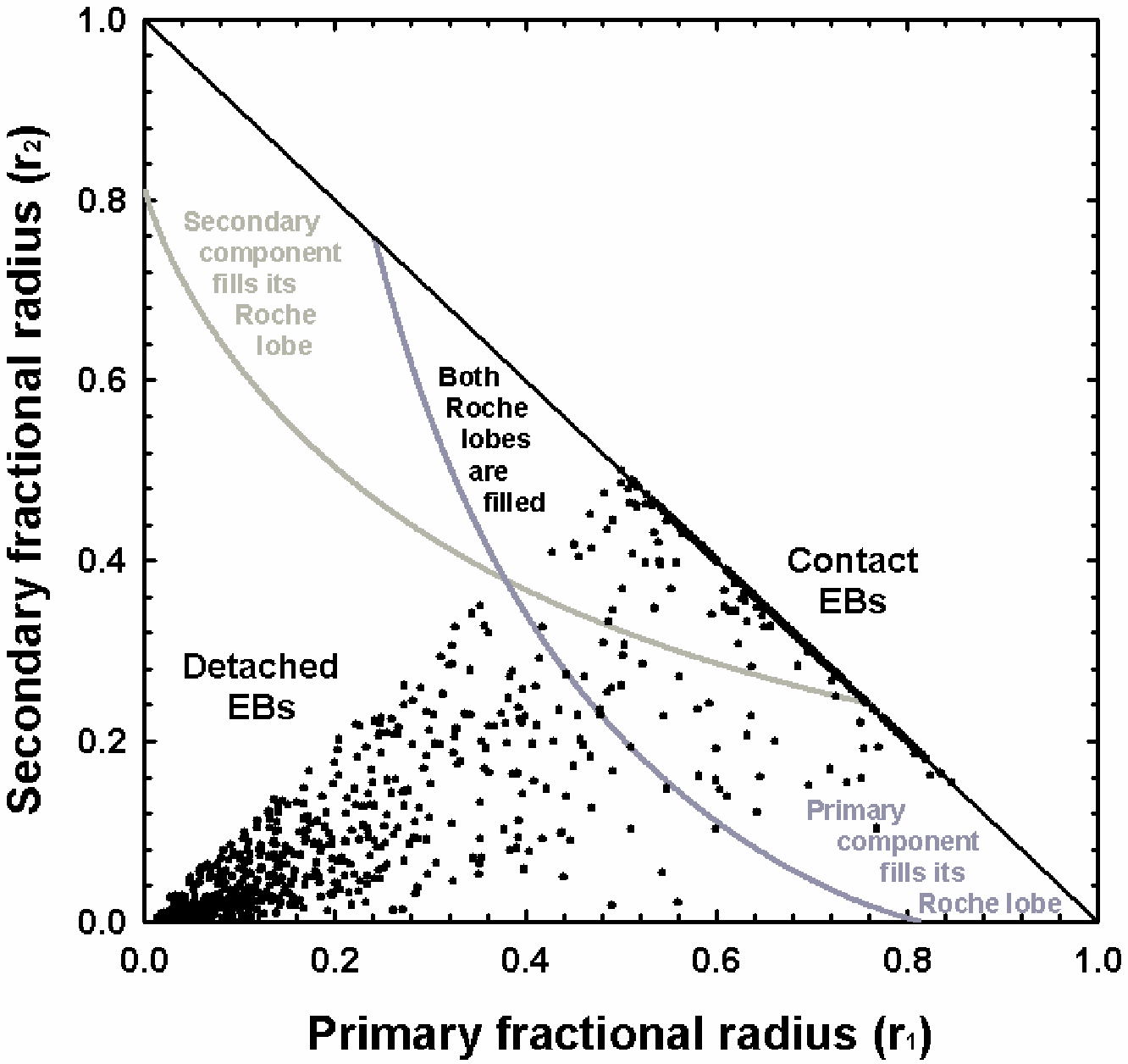}
\caption{The criterion applied in
equations \ref{eqRoche1} and \ref{eqRoche2} to determine whether one or both the EB
components have filled their Roche-lobe, and thus need to be
placed into group [VI]. This categorization is similar to the one performed
for the OGLE~II dataset in \S\ref{secResults} (see Figure~\ref{figR1R2}),
although the threshold criteria here are slightly revised.}
\label{figRoche}
\end{figure}

Group [VI] contains the EBs that have at least one component
filling its Roche-lobe (see Figure \ref{figRocheFilledCat}). Such
system cannot be well fit by either DEBiL or MECI since they
assume that the binary components are detached, and so neglect
tidal and rotational distortions, gravity darkening, and
reflection effects. These systems must be separated from the rest
of the catalog since their resulting best-fit models will be poor
and therefore their evaluated physical attributes will likely be
erroneous. In a similar fashion to \citet{Tamuz06}, we detect
these systems automatically by applying the \citet{Eggleton83}
approximation for the Roche-lobe radius, and place in group [VI]
all the systems for which at least one of the EB components has
filled its Roche-lobe (see Figure \ref{figRoche}), that is, if
either one of the following two inequalities occurs:

\begin{eqnarray}
\label{eqRoche1}
r_1 &>& \frac{0.49\: q^{-2/3}}{0.6\: q^{-2/3} + \ln \left(1 + q^{-1/3}\right)} {\rm \ \ \ or }\\
\label{eqRoche2}
r_2 &>& \frac{0.49\: q^{2/3}}{0.6\: q^{2/3} + \ln \left(1 + q^{1/3}\right)}\ ,
\end{eqnarray}

where $q = M_2 / M_1$ is the EB components' mass ratio. Since we
expect non-detached EBs to be biased toward evolving, higher-mass
stellar components, we estimated $q$ using the early-type
mass-radius power law relation found in binaries \citep{Gorda98}:
$q \simeq (r_2/r_1)^{1.534}$. Although in principle, we could have
estimated $q$ directly from the EB component masses resulting from
the MECI analysis, we chose not to, since as stated above, the
analysis of such systems is inaccurate. The analytic approximation
we used, though crude, proved to be remarkably robust, as we found
only five false negatives and no false positives when visually
inspecting the LCs. We found many more false positives/negatives
when using the alarm criteria suggested by \citet{Devor05} or
\citet{Mazeh06}, both of which attempt to identify bad model fits
by evaluating the temporal correlations of the model's residuals.

\begin{figure}
\includegraphics[width=5in]{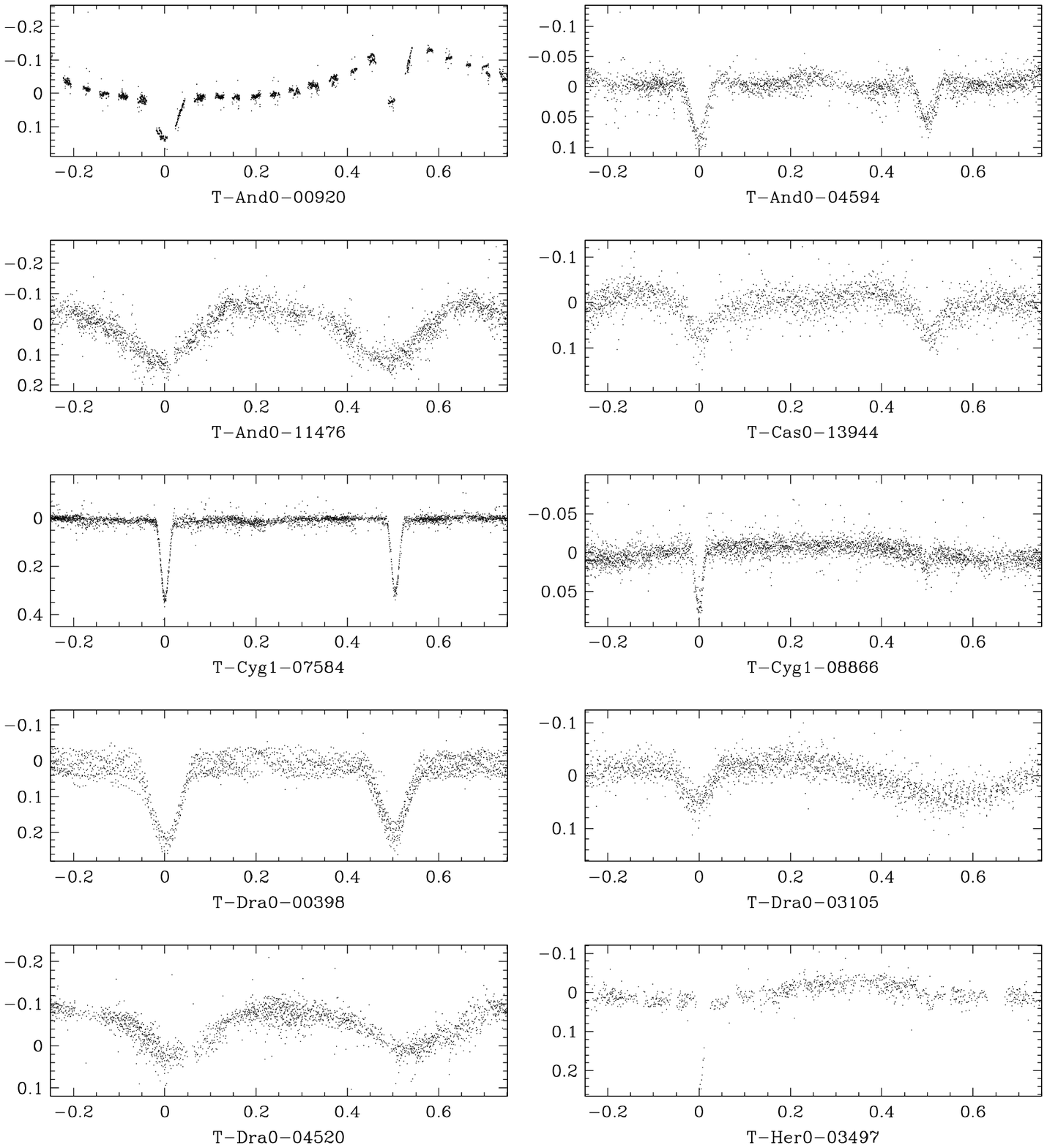}
\caption{LCs of abnormal EBs (panel 1).}
\label{figAbnormalEBs1}
\end{figure}

\begin{figure}
\includegraphics[width=5in]{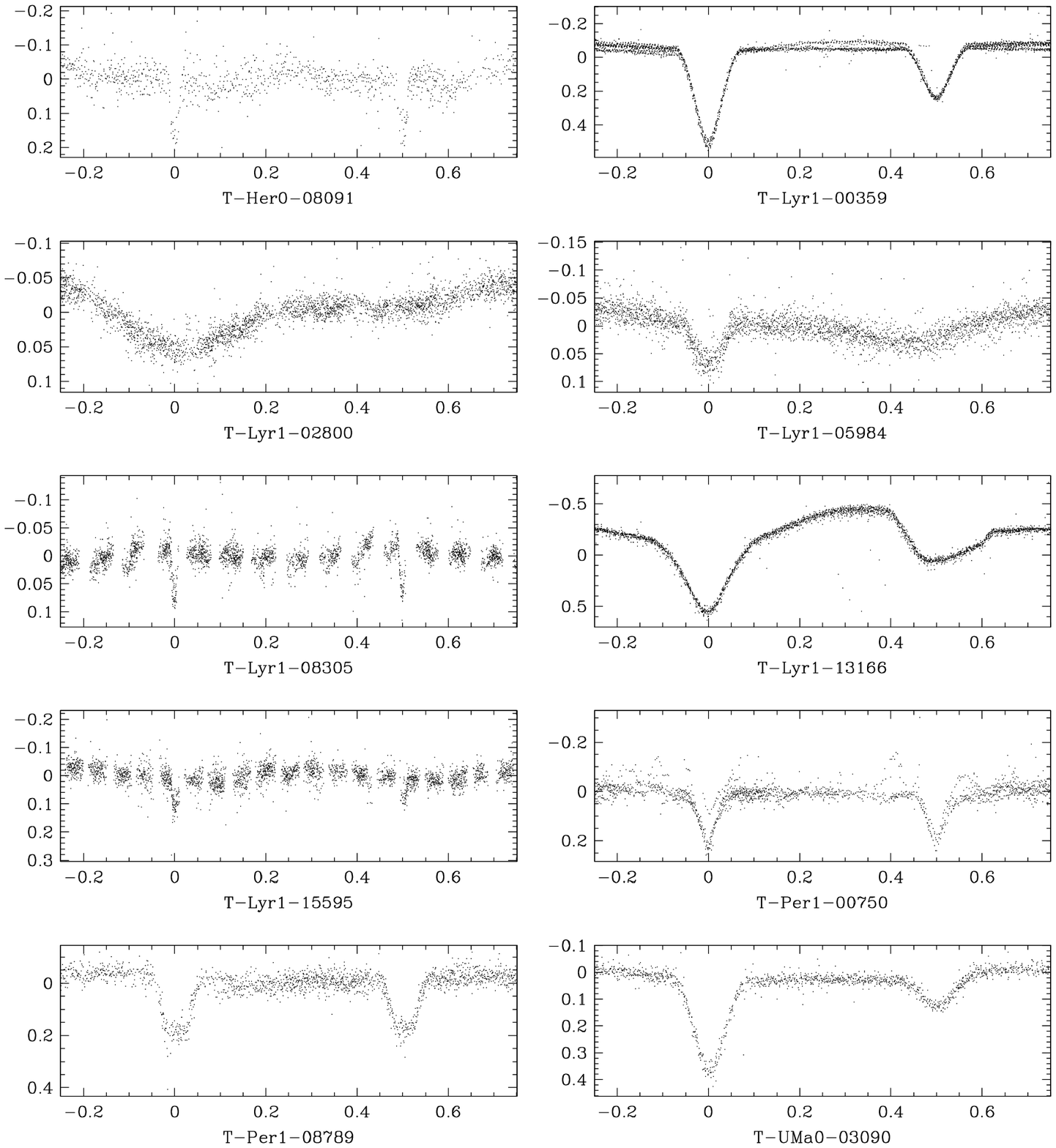}
\caption{LCs of abnormal EBs (panel 2).}
\label{figAbnormalEBs2}
\end{figure}

Finally, group [VII] contains systems visually identified as EBs
(i.e. having LCs with periodic flux dips), yet having atypical LC
perturbations that indicate the existence of additional physical
phenomena (see Figures \ref{figAbnormalEBs1} and
\ref{figAbnormalEBs2}). For lack of a better descriptor, we call
such systems ``abnormal'' (see further information in
\S\ref{subsecAbnomalEBs}). This group is different from the
previous six in that we cannot automate their classification, and
their selection is thus inherently subjective. In 15 of the 20
systems, we were able to approximately model the LCs, and included
them in one of the aforementioned groups. In these cases, users
should be aware that these models may be biased by the phenomenon
that brought about their LC distortion.

\section{Results}

\begin{figure}
\includegraphics[width=5in]{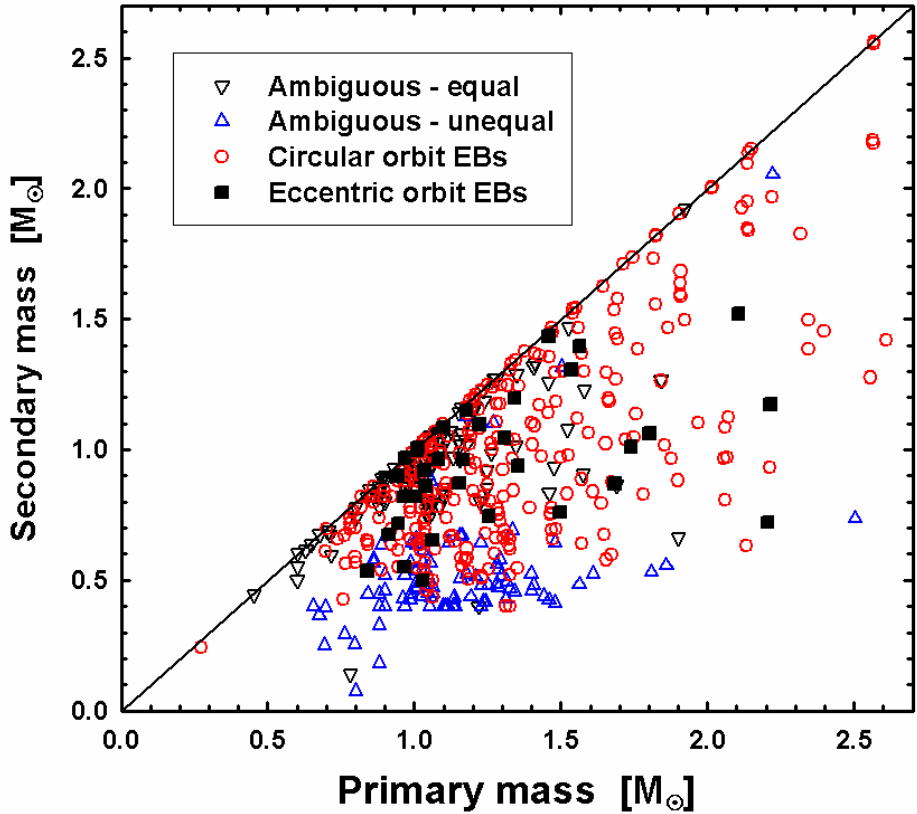}
\caption{The mass-mass relation for the detached EBs of the TrES
dataset. Each category is represented by a different symbol. Note
that the ambiguous EBs are plotted twice, where only one of the
solutions can be correct. Note also that the equal-component
solutions are clustered along the diagonal, while the
unequal-component solutions with $M_1 > 0.75\:M_{\sun}$ are
clustered along the minimum available mass of the Yonsei-Yale
isochrones ($0.4 \:M_{\sun}$). Some of the ambiguous solutions
deviate from these clusters due to poor constraints on the
secondary eclipse, which brings about a large uncertainty.
Finally, note the sparsity of EBs populating the low-mass corner
of this plot ($M_{1,2} < 0.75\:M_{\sun}$). These systems, whose
importance is outlined in \S\ref{subsecLowMassEBs}, were modeled
using the Baraffe isochrones. CM Draconis (T-Dra0-01363) clearly
sets itself apart, being the lowest-mass binary in the catalog
(circle at bottom left).}
\label{figMassMass}
\end{figure}

We identified and classified a total of 773 EBs\footnote{The
observed LCs, fitted models, and model residuals of each of these
EBs are shown at \newline
http://www.cfa.harvard.edu/$\sim$jdevor/Catalog.html.}. These
systems consisted of 734 EBs with circular orbits, 34 detached EBs
with eccentric orbits (group [I] ; Table \ref{tableEccentric}),
and 5 unclassified abnormal EBs (group [VII] ; Table
\ref{tableAbnormal}). We marked 15 of the detached EBs with
circular orbits as also being abnormal. Of the 734 EBs with
circular orbits, we classify 290 as unambiguous detached EBs
(group [II] ; Table \ref{tableCircular}), 103 as ambiguous
detached EBs, for which we could not determine photometrically if
they consisted of equal or disparate components (groups [III] and
[IV] ; Table \ref{tableAmbig}), 23 as inverted EBs (group [V] ;
Table \ref{tableInverted}), and 318 as non-detached (group [VI] ;
Table \ref{tableFillRoche}). With the exception of the abnormal
EBs, which were selected by eye, we use an automated method to
classify each of these groups (see \S\ref{secMethod} for details).
Our mass estimates for the primary and secondary components are
plotted in Figure \ref{figMassMass}.

\begin{figure}
\includegraphics[width=5in]{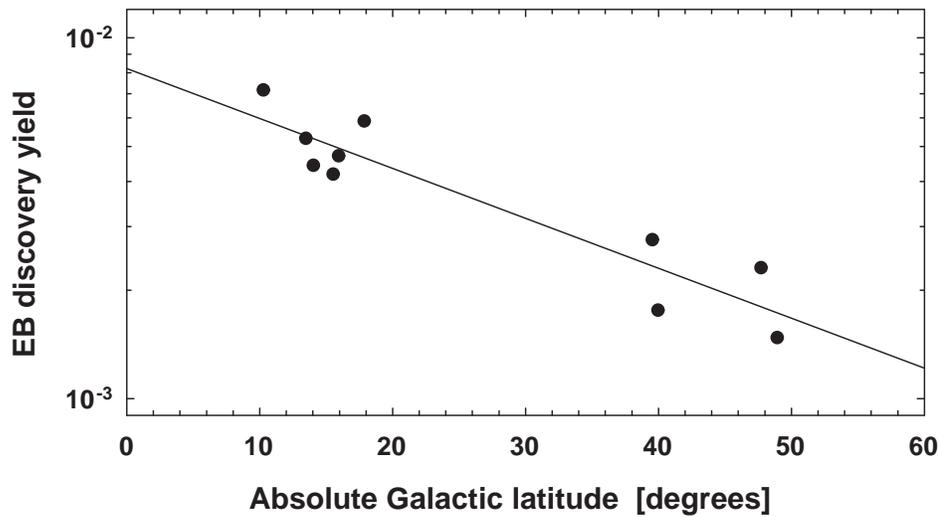}
\caption{The relation between the EB
discovery yield (the fraction of LCs found to be EBs) and the
absolute value of the Galactic latitude, or $|b|$, for the ten TrES
fields used in this catalog (see Tables \ref{tableFieldsObs} and
\ref{tableFieldsYield}). The solid line is the linear regression
of the log of the EB discovery yield ($r^2 = 0.867$). Some of the
residual scatter can be explained as being due to differences in
the duration of observations in each field. By including the
duration in a bi-linear regression, we get a substantially improved
fit ($r^2 = 0.911$).}
\label{figDiscoveryYield}
\end{figure}

The EB discovery yield (the fraction of LCs found to be EBs),
varies greatly from field to field, ranging from 0.72\% for
Cygnus, to 0.15\% for Corona Borealis (see Table
\ref{tableFieldsYield}). This variation is strongly correlated
with Galactic latitude, where fields near the Galactic plane have
larger discovery yields than those that are farther from it (see
Figure \ref{figDiscoveryYield}). This effect is likely due to the
fact that fields closer to the Galactic plane contain a higher
fraction of early-type stars. These early-type stars are both
physically larger, making them more likely to be eclipsed, and are
more luminous, which causes them to produce brighter and less noisy LCs,
thereby enabling the detection of EBs with shallower eclipses.
Furthermore, much of the residual scatter can be attributed to the
variation in the observed duration of each field (see Table
\ref{tableFieldsObs}). That is, we find additional EBs, with
longer periods, in fields that were observed for a longer
duration.

Currently, 88 of the cataloged EBs (11\%) appear in either the
International Variable Star Index\footnote{Maintained by the
American Association of Variable Star Observers (AAVSO).} (VSX),
or in the SIMBAD\footnote{Maintained by the Centre de Donn\'{e}es
astronomiques de Strasbourg (CDS).} astronomical database (Table
\ref{tableSIMBAD}). However, only 49 systems (6\%) have been
identified as being variable. Not surprisingly, with few
exceptions, these targets were among the brightest sources of the
catalog. Using only photometry, it is often notoriously difficult
to distinguish non-detached EBs from pulsating variables that vary
sinusoidally in time, such as type-C RR Lyrae. Furthermore,
unevenly spotted stars may also cause false positive
identifications, especially in surveys with shorter durations.
Ultimately, spectroscopic follow-up will always be necessary to
confirm the identification of such variables.

We highlight three groups of EBs as potentially having special
importance as test beds for current theory. For more accurate
properties, these EBs will likely need to be followed-up both
photometrically and spectroscopically. The brightness of these EBs
will considerably facilitate their follow-up.

\newpage

\begin{deluxetable}{ccccccccccc}
\tabletypesize{\tiny}
\rotate
\tablecaption{Eccentric EBs}
\tablewidth{0pt}
\tablehead{\colhead{Object} & \colhead{$\alpha$ (J2000)} & \colhead{$\delta$ (J2000)} & \colhead{Period $[days]$\tablenotemark{a}} & \colhead{$|e \cos\omega|_{timing}$\tablenotemark{b}} & \colhead{$|e \cos\omega|_{adopted}$\tablenotemark{c}} & \colhead{$e$\tablenotemark{d}} & \colhead{$M_1/M_{\sun}$} & \colhead{$M_2/M_{\sun}$}& \colhead{age $[Gyr]$} & \colhead{$t_{circ}$ $[Gyr]$\tablenotemark{e}}}
\startdata
T-And0-04144& 01 17 35.247& 49 46 16.97&  7.869& 0.0072& 0.0068& $0.14^{+0.08}_{-0.08}$  & 0.84 (-1)\tablenotemark{f}      & 0.54 (-1)      & 10.0 (-3)     & 140\\
T-And0-17158& 01 10 09.143& 48 18 19.68& 11.415& 0.0182& 0.0180& $0.038^{+0.12}_{-0.02}$ & 1.03 (-1)      & 0.92 (-1)      & 10.0 (-3)     & 370\\
T-And0-24609& 00 58 29.826& 49 25 08.88& 17.997& 0.0794& 0.0799& $0.10^{+0.10}_{-0.02}$  & 1.22 $\pm$ 0.10& 1.10 $\pm$ 0.30& 5.4 $\pm$ 12.0& 6400\\
T-Cas0-00394& 00 32 51.608& 49 19 39.36&  1.746& 0.0235& 0.0242& $0.024^{+0.03}_{-0.001}$& 1.46 $\pm$ 0.01& 1.44 $\pm$ 0.01& 3.4 $\pm$ 0.3 & 260\\
T-Cas0-02603& 00 47 08.610& 50 37 19.32&  2.217& 0.2098& 0.2143& $0.25^{+0.14}_{-0.04}$  & 1.25 $\pm$ 0.01& 0.75 $\pm$ 0.04& 5.4 $\pm$ 5.0 & 0.26\\
T-Cas0-04534& 00 31 04.585& 51 52 10.88&  6.909& 0.0057& 0.0048& $0.014^{+0.03}_{-0.01}$ & 1.17 $\pm$ 0.04& 0.96 $\pm$ 0.15& 6.4 $\pm$ 6.9 & 29\\
T-Cas0-04947& 00 47 10.336& 50 45 12.36&  3.285& 0.0845& 0.0845& $0.10^{+0.04}_{-0.02}$  & 1.04 (-1)      & 0.86 (-1)      & 10.0 (-3)     & 0.53\\
T-Cas0-05165& 00 43 59.256& 51 14 00.07&  2.359& 0.0311& 0.0327& $0.15^{+0.08}_{-0.08}$  & 1.50 $\pm$ 0.21& 0.76 $\pm$ 0.17& 2.7 $\pm$ 2.9 & 0.34\\
T-Cas0-07630& 00 37 23.347& 47 19 20.68&  5.869& 0.0200& 0.0298& $0.038^{+0.15}_{-0.008}$& 1.15 $\pm$ 0.12& 0.87 $\pm$ 0.34& 5.9 $\pm$ 9.7 & 13\\
T-Cyg1-01364& 20 09 38.211& 49 05 08.02& 12.233&   N/A & 0.3254& $0.53^{+0.04}_{-0.04}$  & 1.03 $\pm$ 0.18& 0.50 $\pm$ 0.09& 0.4 $\pm$ 1.2 & 1100\\
T-Cyg1-01373& 19 55 44.105& 52 13 34.61&  4.436& 0.0059& 0.0054& $0.010^{+0.02}_{-0.005}$& 0.97 (-1)      & 0.82 (-1)      & 10.0 (-3)     & 3.0\\
T-Cyg1-01994& 20 03 03.111& 52 42 04.17& 14.482&   N/A & 0.0107& $0.15^{+0.15}_{-0.14}$  & 1.80 (-1)      & 1.06 (-1)      & 0.20 (-2)     & 2300\\
T-Cyg1-02304& 20 02 04.388& 47 34 14.75&  5.596& 0.1549& 0.1529& $0.23^{+0.10}_{-0.08}$  & 2.20 $\pm$ 1.28& 0.72 $\pm$ 0.41& 0.7 $\pm$ 4.8 & 46\\
T-Cyg1-02624& 19 59 25.926& 52 23 59.91& 11.608& 0.0172& 0.0172& $0.068^{+0.03}_{-0.03}$ & 2.11 $\pm$ 0.05& 1.52 $\pm$ 0.03& 0.3 $\pm$ 0.1 & $10^7$\\
T-Cyg1-06677& 20 07 25.526& 52 22 00.54&  6.512& 0.0077& 0.0069& $0.062^{+0.03}_{-0.03}$ & 1.54 $\pm$ 0.20& 1.31 $\pm$ 0.22& 1.6 $\pm$ 1.9 & $10^6$\\
T-Cyg1-07248& 19 54 45.937& 50 24 05.32&  6.058& 0.1674& 0.1681& $0.17^{+0.07}_{-0.001}$ & 1.68 $\pm$ 0.01& 0.87 $\pm$ 0.20& 2.0 $\pm$ 2.1 & 33\\
T-Cyg1-07297& 20 10 46.910& 49 09 29.42& 11.613& 0.3019& 0.3010& $0.38^{+0.08}_{-0.08}$  & 0.97 (-1)      & 0.55 (-1)      & 10.0 (-3)     & 830\\
T-Cyg1-07584& 19 58 58.012& 47 38 19.26&  4.925& 0.0074& 0.0074& $0.022^{+0.08}_{-0.01}$ & 0.94 (-1)      & 0.90 (-1)      & 10.0 (-3)     & 4.7\\
T-Cyg1-09934& 20 10 44.209& 51 07 51.77&  4.549& 0.0505& 0.0501& $0.11^{+0.06}_{-0.06}$  & 1.35 $\pm$ 0.64& 0.94 $\pm$ 0.41& 3.5 $\pm$ 5.6 & 5.6\\
T-Cyg1-15752& 20 13 52.454& 50 52 23.12&  9.372& 0.2402& 0.2402& $0.35^{+0.05}_{-0.05}$  & 1.31 $\pm$ 0.04& 1.05 $\pm$ 0.11& 3.6 $\pm$ 4.9 & 230\\
T-Lyr1-09931& 18 59 08.441& 48 36 00.04& 11.632& 0.2207& 0.2209& $0.25^{+0.04}_{-0.03}$  & 0.91 $\pm$ 0.09& 0.67 $\pm$ 0.08& 2.7 $\pm$ 3.3 & 730\\
T-Lyr1-13841& 19 06 26.558& 48 28 47.04&  6.640& 0.0362& 0.0362& $0.075^{+0.11}_{-0.04}$ & 1.01 $\pm$ 0.27& 1.01 $\pm$ 0.24& 8.7 $\pm$ 13.1 & 19\\
T-Lyr1-14413& 19 03 41.143& 47 36 55.78& 39.861& 0.5922& 0.6240& $0.64^{+0.006}_{-0.006}$& 1.08 $\pm$ 0.34& 0.96 $\pm$ 0.26& 6.4 $\pm$ 18.9 & $10^5$\\
T-Lyr1-14508& 18 57 40.271& 48 40 51.28&  8.050& 0.1861& 0.1862& $0.31^{+0.16}_{-0.12}$  & 1.34 $\pm$ 0.28& 1.20 $\pm$ 0.78& 2.9 $\pm$ 8.4 & 220\\
T-Lyr1-22359& 19 10 54.290& 49 26 06.95& 12.319& 0.1990& 0.1984& $0.33^{+0.05}_{-0.05}$  & 0.97 $\pm$ 0.48& 0.97 $\pm$ 0.46& 6.9 $\pm$ 29.3 & 550\\
T-Per1-00769& 03 31 43.915& 36 31 52.36&  3.648& 0.0248& 0.0263& $0.055^{+0.05}_{-0.03}$ & 1.06 $\pm$ 0.01& 0.65 $\pm$ 0.03& 7.6 $\pm$ 2.1 & 1.4\\
T-Per1-04218& 03 35 33.667& 40 00 49.18&  4.070& 0.0072& 0.0079& $0.10^{+0.19}_{-0.09}$  & 0.94 (-1)      & 0.72 (-1)      & 10.0 (-3)     & 2.4\\
T-Per1-05205& 03 34 19.432& 39 32 44.41&  8.472& 0.0558& 0.0592& $0.095^{+0.11}_{-0.04}$ & 2.22 $\pm$ 0.01& 1.17 $\pm$ 0.28& 0.9 $\pm$ 1.7 & 210\\
T-Per1-08252& 03 52 00.670& 40 03 47.73&  4.457& 0.0656& 0.0645& $0.065^{+0.06}_{-0.001}$& 1.56 $\pm$ 0.01& 1.40 $\pm$ 0.34& 2.4 $\pm$ 2.5 & $10^5$\\
T-Per1-11424& 03 47 56.473& 37 31 31.83&  4.247& 0.2403& 0.2404& $0.24^{+0.02}_{-0.006}$ & 1.01 (-1)      & 0.82 (-1)      & 10.0 (-3)     & 2.3\\
T-Per1-17327& 03 40 45.644& 34 47 57.26&  3.946& 0.0332& 0.0305& $0.069^{+0.25}_{-0.04}$ & 1.10 $\pm$ 0.02& 1.09 $\pm$ 0.09& 8.4 $\pm$ 16.4& 1.2\\
T-Tau0-02487& 04 21 55.933& 25 35 49.28&  2.826& 0.0125& 0.0054& $0.014^{+0.005}_{-0.005}$&1.74 $\pm$ 0.07& 1.01 $\pm$ 0.08& 1.6 $\pm$ 0.7 & 0.39\\
T-Tau0-03916& 04 23 37.351& 25 46 36.00&  3.217& 0.0713& 0.0706& $0.071^{+0.02}_{-0.004}$& 1.18 $\pm$ 0.01& 1.15 $\pm$ 0.03& 6.0 $\pm$ 4.4 & 0.56\\
T-UMa0-01822& 09 53 37.710& 52 45 44.72&  9.551& 0.1502& 0.1503& $0.31^{+0.02}_{-0.02}$  & 1.01 $\pm$ 0.04& 1.00 $\pm$ 0.05& 8.3 $\pm$ 4.8 & 130\\
\enddata
\tablenotetext{a}{The full precision of the measured period is
listed in the electronic version of the catalog, together with its
uncertainty and the epoch of the center of eclipse (see appendix
\ref{appendixCatalogDescription}).}
\tablenotetext{b}{Measurements
made using the eclipse timing of step (4). Although these values
are approximations, they do not suffer from nearly as much
numerical error as the DEBiL measurement, and are therefore
usually accurate. ``N/A'' marks LCs for which there were too few
eclipses to be able to apply the timing method.}
\tablenotetext{c}{The adopted value is a combination of the values
measured with the timing method and with DEBiL.}
\tablenotetext{d}{The uncertainties of the eccentricities are
non-Gaussian, since they have a strict lower bound ($e \geq |e
\cos\omega|$). We truncated the quoted lower uncertainties at this
value, though even at this truncated value the real uncertainty is
beyond $1\sigma$.}
\tablenotetext{e}{For each EB, the circularization timescales of both components
were estimated using Equation \ref{eqCircTime}, then these values were combined
as described in \S\ref{subsecEccentricEBs}.}
\tablenotetext{f}{When the most likely model is
at the edge of the parameter space, MECI is not able to bound the
solution, and therefore cannot estimate the uncertainties. We mark
(-3) when the upper limit was reached, (-2) when the lower limit
was reached, and (-1) if one of the other parameters is at its
limit.}
\label{tableEccentric}
\end{deluxetable}

\newpage

\begin{deluxetable}{ccccccl}
\tabletypesize{\tiny}
\rotate
\tablecaption{Abnormal EBs}
\tablewidth{0pt}
\tablehead{
\colhead{Object} &
\colhead{$\alpha$ (J2000)} &
\colhead{$\delta$ (J2000)} &
\colhead{\begin{tabular}{c} Period\\ (days) \end{tabular}} &
\colhead{\begin{tabular}{c} Classified\\ in catalog?\end{tabular}} &
\colhead{\begin{tabular}{c} In SIMBAD/VSX?\\ (see Table \ref{tableSIMBAD}) \end{tabular}} &
\colhead{Notes}}
\startdata
T-And0-00920& 01 17 30.677& 47 03 31.61&24.073& no & no & Large asymmetric reflection ($0.1$ mag) \ offset eclipse\\
T-And0-04594& 01 16 10.713& 48 52 18.97& 3.910& yes& no & Spots / active\\
T-And0-11476& 01 07 32.106& 45 55 44.93& 6.380& yes& no & Tilted plateaux (spots?)\\
T-Cas0-13944& 00 29 48.990& 50 49 54.06& 1.739& yes& no & Irregular eclipse depths\\
T-Cyg1-07584& 19 58 58.012& 47 38 19.26& 4.925& yes& no & Large persistent spot\\
T-Cyg1-08866& 20 08 36.448& 49 29 35.79& 2.876& yes& no & Offset eclipse\tablenotemark{a}\\
T-Dra0-00398& 16 57 33.875& 59 31 51.98& 1.046& yes& yes& Active (has $0.2$ mag fluctuations with periods of a few hours)\\
T-Dra0-03105& 16 23 02.558& 59 27 23.44& 0.485& no & yes& Unequal eclipses\tablenotemark{b} / semi-detached\\
T-Dra0-04520& 16 49 57.960& 56 26 45.56& 3.113& yes& no & Tilted plateaux (spots?)\\
T-Her0-03497& 16 52 28.391& 44 51 29.63& 7.853& yes& no & Unequal plateaux\tablenotemark{c}\\
T-Her0-08091& 16 51 52.608& 47 01 47.98& 2.694& yes& no & Offset eclipse\\
T-Lyr1-00359& 19 15 33.695& 44 37 01.30& 1.062& yes& yes& Large recurring spots ($\sim0.05$ mag)\\
T-Lyr1-02800& 19 08 18.809& 47 12 48.16& 4.876& no & no & Semi-detached / unequal plateaux (spots?)\\
T-Lyr1-05984& 18 53 50.481& 45 33 20.90& 1.470& no & no & Unequal eclipses\tablenotemark{b} / semi-detached\\
T-Lyr1-08305& 18 56 43.798& 48 07 02.86&14.081& yes& no & Large asymmetric reflection ($0.05$ mag) ; offset eclipse\\
T-Lyr1-13166& 19 02 28.120& 46 58 57.75& 0.310& no & no & Unequal plateaux ; misshapen eclipse (persistent spot?)\\
T-Lyr1-15595& 19 06 05.267& 49 04 08.95& 9.477& yes& no & Offset eclipse\\
T-Per1-00750& 03 47 45.543& 35 00 37.08& 1.929& yes& yes& Spots / active\\
T-Per1-08789& 03 54 33.282& 39 07 41.53& 2.645& yes& no & Tilted plateaux\\
T-UMa0-03090& 10 08 52.180& 52 45 52.49& 0.538& yes& yes& Unequal plateaux\\
\enddata
\tablenotetext{a}{Even when the LC plateaux are
not flat, due to tidal distortion or reflections, the system's
mirror symmetry normally guarantees that the eclipses will occur
during a plateau minimum or maximum. When, as in these cases, the
eclipses are significantly offset from the plateau minima/maxima
we can conclude that some mechanism, perhaps severe tidal lag, is
breaking the system's symmetry.}
\tablenotetext{b}{Might not be an EB. This LC could be due to
non-sinusoidal pulsations.}
\tablenotetext{c}{The two LC plateaux
between the eclipses, have a significantly different mean
magnitude. This may be due to one or both components being tidally
locked, and having a persistent spot or surface temperature
variation at specific longitudes.}
\label{tableAbnormal}
\end{deluxetable}

\begin{deluxetable}{cccccccccc}
\tabletypesize{\tiny}
\rotate
\tablecaption{Circular EBs (first 20)}
\tablewidth{0pt}
\tablehead{\colhead{Object} & \colhead{$\alpha$ (J2000)} & \colhead{$\delta$ (J2000)} & \colhead{Period $[days]$} & \colhead{$M_1/M_{\sun}$} & \colhead{$M_2/M_{\sun}$}& \colhead{age $[Gyr]$} &
\colhead{\begin{tabular}{c} Proper motion \\ source catalog \end{tabular}} & \colhead{\begin{tabular}{c} $PM_{\alpha}$ \\ $[MAS/year]$ \end{tabular}} & \colhead{\begin{tabular}{c} $PM_{\delta}$\\ $[MAS/year]$\end{tabular}}}
\startdata
T-And0-00194& 01 20 12.816& 48 36 41.36&  2.145&  2.07 $\pm$ 0.02& 0.97 $\pm$ 0.02& 0.59 $\pm$ 0.12& UCAC& 28.4& -12.2\\
T-And0-00459& 01 11 24.845& 46 57 49.44&  3.655&  1.20 $\pm$ 0.01& 1.19 $\pm$ 0.01& 5.35 $\pm$ 1.13& UCAC& -1.6& -20.6\\
T-And0-00745& 01 03 45.076& 44 50 41.14&  2.851&  1.86 $\pm$ 0.23& 1.02 $\pm$ 0.22& 1.04 $\pm$ 0.72& UCAC& -6.6& -4.8\\
T-And0-01461& 01 06 15.353& 45 08 25.66&  5.613&  1.47 $\pm$ 0.01& 1.45 $\pm$ 0.08& 2.76 $\pm$ 2.72& UCAC& -11.4& 2.8\\
T-And0-01554& 01 17 04.999& 45 54 06.20&  1.316&  0.90    (-1)\tablenotemark{a}   & 0.84     (-1)  & 10.00    (-3)  & UCAC& -44.6& -40.8\\
T-And0-01597& 01 10 32.071& 46 49 53.18&  3.503&  1.55 $\pm$ 0.03& 1.54 $\pm$ 0.01& 2.37 $\pm$ 0.76& UCAC& 2.9& -5.5\\
T-And0-02462& 01 18 00.594& 49 27 12.47&  3.069&  1.97 $\pm$ 0.69& 1.10 $\pm$ 1.31& 1.02 $\pm$ 1.58& UCAC& 5.8& -1.1\\
T-And0-02699& 01 06 44.813& 47 31 08.61&  1.759&  1.18 $\pm$ 0.02& 0.53 $\pm$ 0.07& 5.21 $\pm$ 3.37& UCAC& 0.2& -6.8\\
T-And0-02798& 01 21 18.345& 48 48 05.63&  2.860&  1.04 $\pm$ 0.10& 0.65 $\pm$ 0.13& 6.14 $\pm$ 9.51& UCAC& 6.3& -8.1\\
T-And0-03526& 01 20 17.451& 47 39 23.32&  1.536&  1.04 $\pm$ 0.02& 0.84 $\pm$ 0.02& 6.29 $\pm$ 2.37& UCAC& 17.9& -11.1\\
T-And0-04046& 00 55 20.157& 47 44 53.20&  3.916&  1.30 $\pm$ 0.09& 1.25 $\pm$ 0.12& 3.10 $\pm$ 4.31& UCAC& -3.8& -7.3\\
T-And0-04594& 01 16 10.713& 48 52 18.97&  3.910&  1.05    (-1)   & 0.82    (-1)   & 10.00    (-3)  & UCAC& 1.5& -1.9\\
T-And0-04829& 01 15 15.228& 47 45 58.97&  0.678&  0.99    (-1)   & 0.92    (-1)   & 10.00    (-3)  & UCAC& -23.8& 44.4\\
T-And0-05241& 00 56 34.679& 46 37 02.91&  1.454&  1.56 $\pm$ 0.01& 1.47 $\pm$ 0.31& 2.69 $\pm$ 7.01& UCAC& -4.5& -0.5\\
T-And0-05375& 01 10 58.225& 49 52 48.69&  1.640&  2.13    (-1)   & 1.85    (-2)   & 1.00    (-1)   & UCAC& -6.3& 0.1\\
T-And0-05794& 01 12 11.763& 47 32 30.94&  1.053&  2.06 $\pm$ 0.19& 1.08 $\pm$ 0.54& 1.08 $\pm$ 1.92& UCAC& -0.4& -1.4\\
T-And0-06039& 01 23 37.548& 48 25 37.73&  4.923&  1.22 $\pm$ 0.05& 1.08 $\pm$ 0.31& 5.33 $\pm$ 7.17& UCAC& -2.5& -5.0\\
T-And0-06340& 01 01 55.269& 49 18 38.23&  5.437&  1.33    (-1)   & 0.40    (-2)   & 4.00    (-1)   & UCAC& 0.3& -2.9\\
T-And0-06538& 01 20 58.907& 49 29 08.89& 18.669&  1.33 $\pm$ 0.15& 0.97 $\pm$ 0.17& 3.38 $\pm$ 3.45& UCAC& 1.1& -6.8\\
T-And0-06632& 01 22 36.840& 47 52 53.29&  1.669&  1.69 $\pm$ 0.01& 1.45 $\pm$ 0.24& 2.21 $\pm$ 1.02& UCAC& -7.2& -7.6\\
\enddata
\label{tableCircular}
\tablenotetext{a}{When the most likely model is at the edge of the parameter space, MECI is not
able to bound the solution, and therefore cannot estimate the uncertainties. We mark (-3) when
the upper limit was reached, (-2) when the lower limit was reached, and (-1) if one of the other
parameter is at its limit.}
\end{deluxetable}

\begin{deluxetable}{cccccccc}
\tabletypesize{\tiny}
\rotate
\tablecaption{Ambiguous EBs (first 10)}
\tablewidth{0pt}
\tablehead{\colhead{Ver.} & \colhead{Object} & \colhead{$\alpha$ (J2000)} & \colhead{$\delta$ (J2000)} & \colhead{Period $[days]$} & \colhead{$M_1/M_{\sun}$} & \colhead{$M_2/M_{\sun}$}& \colhead{age $[Gyr]$}}
\startdata
A& T-And0-00657& 01 06 06.159& 47 31 59.37&  6.725&  2.50       (-1)& 0.74       (-1)& 0.20       (-2)\\
B& T-And0-00657& 01 06 06.159& 47 31 59.37& 13.456&  1.92       (-1)\tablenotemark{c}& 1.92       (-1)& 0.20       (-2)\\
A& T-And0-01203& 01 03 34.745& 48 32 39.27&  3.505&  1.86 $\pm$ 0.09& 0.56 $\pm$ 0.10& 0.89 $\pm$ 0.83\\
B& T-And0-01203& 01 03 34.745& 48 32 39.27&  7.011&  1.90 $\pm$ 0.12& 0.66 $\pm$ 0.19& 0.80 $\pm$ 1.13\\
A& T-And0-06017& 01 12 48.217& 49 58 07.16&  2.543&  1.40 $\pm$ 0.35& 0.52 $\pm$ 0.77& 3.49 $\pm$ 4.28\\
B& T-And0-06017& 01 12 48.217& 49 58 07.16&  5.085&  1.18 $\pm$ 0.71& 1.12 $\pm$ 0.85& 3.12 $\pm$ 11.15\\
A& T-And0-06500& 01 25 56.083& 49 23 31.74&  5.337&  0.97 $\pm$ 0.20& 0.49 $\pm$ 0.53& 7.71 $\pm$ 16.33\\
B& T-And0-06500& 01 25 56.083& 49 23 31.74& 10.674&  1.01 $\pm$ 0.30& 0.93 $\pm$ 0.45& 0.74 $\pm$ 1.60\\
A& T-And0-06680& 00 55 48.153& 45 02 48.57&  4.551&  1.16 $\pm$ 0.04& 0.51 $\pm$ 0.20& 6.09 $\pm$ 8.91\\
B& T-And0-06680& 00 55 48.153& 45 02 48.57&  9.104&  1.16 $\pm$ 0.09& 0.96 $\pm$ 0.29& 6.24 $\pm$ 10.78\\
A& T-And0-08053& 01 13 59.402& 45 51 43.43&  4.116&  1.14       (-1)& 0.40       (-2)& 6.00       (-1)\\
B& T-And0-08053& 01 13 59.402& 45 51 43.43&  8.231&  1.09 $\pm$ 0.55& 1.05 $\pm$ 0.64& 3.22 $\pm$ 16.37\\
A& T-And0-08417& 01 01 39.041& 45 03 32.98&  2.053&  1.01       (-1)& 0.47       (-1)& 10.00       (-3)\\
B& T-And0-08417& 01 01 39.041& 45 03 32.98&  4.106&  1.01       (-1)& 0.90       (-1)& 10.00       (-3)\\
A& T-And0-09365& 01 01 00.459& 45 14 24.77&  1.887&  1.05 $\pm$ 0.03& 0.43 $\pm$ 0.39& 8.74 $\pm$ 16.06\\
B& T-And0-09365& 01 01 00.459& 45 14 24.77&  3.774&  1.05 $\pm$ 0.05& 0.93 $\pm$ 0.52& 9.47 $\pm$ 23.55\\
A& T-And0-10518& 01 07 44.417& 48 44 58.11&  0.194&  0.90       (-1)& 0.40       (-2)& 0.40       (-1)\\
B& T-And0-10518& 01 07 44.417& 48 44 58.11&  0.387&  0.45 $\pm$ 0.27& 0.45 $\pm$ 0.28& 0.27 $\pm$ 0.54\\
A& T-And0-11453& 01 05 42.744& 44 54 02.26&  0.784&  1.12       (-1)& 0.40       (-2)& 7.00       (-1)\\
B& T-And0-11453& 01 05 42.744& 44 54 02.26&  1.568&  1.02 $\pm$ 0.43& 1.01 $\pm$ 0.32& 8.81 $\pm$ 14.54\\
\enddata
\tablenotetext{A}{Unequal eclipse model, assuming an unseen secondary eclipse.}
\tablenotetext{B}{Equal eclipse model, with double the period of the unequal model.}
\tablenotetext{c}{When the most likely model is at the edge of the parameter space, MECI is not
able to bound the solution, and therefore cannot estimate the uncertainties. We mark (-3) when
the upper limit was reached, (-2) when the lower limit was reached, and (-1) if one of the other
parameter is at its limit.}
\label{tableAmbig}
\end{deluxetable}

\begin{deluxetable}{cccc}
\tabletypesize{\tiny}
\rotate
\tablecaption{Inverted EBs}
\tablewidth{0pt}
\tablehead{\colhead{Object} & \colhead{$\alpha$ (J2000)} & \colhead{$\delta$ (J2000)} & \colhead{Period $[days]$}}
\startdata
T-And0-13653& 00 59 57.881& 45 03 41.53&  3.342\\
T-Cas0-02069& 00 49 17.959& 50 39 02.92&  2.830\\
T-Cas0-03012& 00 45 41.832& 51 01 35.40&  1.108\\
T-Cas0-04618& 00 46 22.661& 50 39 17.57&  2.798\\
T-Cas0-07780& 00 34 18.779& 52 00 35.72&  1.852\\
T-Cas0-19045& 00 21 44.707& 50 32 29.55&  0.785\\
T-Cas0-19668& 00 48 01.342& 47 06 11.58&  1.848\\
T-Cas0-21651& 00 26 34.895& 46 38 42.69&  1.155\\
T-Cyg1-01956& 19 53 29.106& 47 48 49.86&  2.045\\
T-Cyg1-02929& 20 11 57.009& 48 07 03.59&  4.263\\
T-Cyg1-17342& 19 49 54.197& 50 53 28.08&  2.220\\
T-Her0-05469& 16 54 51.245& 43 20 35.89&  0.899\\
T-Lyr1-04431& 19 12 16.047& 49 42 23.58&  0.903\\
T-Lyr1-05887& 18 52 10.489& 47 48 16.67&  1.802\\
T-Lyr1-07179& 18 49 14.039& 45 24 38.61&  1.323\\
T-Lyr1-10989& 19 06 22.791& 45 41 53.82&  2.015\\
T-Lyr1-11067& 18 52 53.489& 47 51 26.58&  2.241\\
T-Per1-04353& 03 45 04.887& 37 47 15.91&  2.953\\
T-Per1-06993& 03 40 59.668& 39 12 35.90&  2.125\\
T-Per1-09366& 03 49 20.305& 39 55 41.97&  2.374\\
T-Per1-12217& 03 28 59.454& 37 37 42.14&  1.690\\
T-Tau0-00686& 04 07 13.870& 29 18 32.44&  5.361\\
T-UMa0-00127& 09 38 06.716& 56 01 07.32&  0.687\\
\enddata
\label{tableInverted}
\end{deluxetable}

\begin{deluxetable}{cccc}
\tabletypesize{\tiny}
\rotate
\tablecaption{EBs that fill at least one of their Roche-lobes (first 20)}
\tablewidth{0pt}
\tablehead{\colhead{Object} & \colhead{$\alpha$ (J2000)} & \colhead{$\delta$ (J2000)} & \colhead{Period $[days]$}}
\startdata
T-And0-03774& 00 59 01.029& 46 47 17.08&  1.362\\
T-And0-04813& 01 16 37.880& 47 33 23.43&  0.552\\
T-And0-05140& 01 03 22.258& 44 56 24.31&  0.981\\
T-And0-05153& 01 18 48.278& 49 39 36.86&  0.492\\
T-And0-05343& 00 52 55.122& 48 01 37.68&  0.824\\
T-And0-07638& 01 09 27.871& 49 20 33.81&  0.403\\
T-And0-07892& 00 56 15.567& 48 39 10.73&  0.380\\
T-And0-08330& 01 19 15.949& 48 00 17.45&  0.630\\
T-And0-08652& 00 56 58.855& 49 05 05.00&  0.335\\
T-And0-09528& 01 22 09.328& 47 14 29.86&  0.918\\
T-And0-10071& 01 14 50.412& 49 17 46.28&  0.387\\
T-And0-10206& 00 55 55.724& 49 49 46.56&  0.859\\
T-And0-10511& 01 19 16.430& 47 07 46.27&  0.563\\
T-And0-10722& 01 04 03.859& 48 37 13.04&  1.062\\
T-And0-11354& 01 18 05.168& 46 10 14.66&  0.331\\
T-And0-11476& 01 07 32.106& 45 55 44.93&  6.380\\
T-And0-11599& 01 09 28.113& 46 18 24.85&  0.280\\
T-And0-11617& 01 07 28.020& 45 22 40.35&  0.503\\
T-And0-12453& 01 17 12.316& 46 42 35.43&  0.448\\
T-And0-12769& 00 52 58.164& 44 44 11.26&  0.325\\
\enddata
\label{tableFillRoche}
\end{deluxetable}

\begin{deluxetable}{cccccll}
\tabletypesize{\tiny}
\rotate
\tablecaption{EBs that appear in either the VSX or the SIMBAD databases}
\tablewidth{0pt}
\tablehead{\colhead{Category} & \colhead{Object} & \colhead{$\alpha$ (J2000)} & \colhead{$\delta$ (J2000)} & \colhead{Spectral type} & \colhead{Classification} & \colhead{Identifiers}}
\startdata
Circular  & T-And0-00194& 01 20 12.816& 48 36 41.36& A5  & Star                  & BD+47 378 ; GSC 03269-00662 ; SAO 37126 ; AG+48 143\\
          &             &             &            &     &                       & PPM 43886 ; TYC 3269-662-1\\
Circular  & T-And0-00459& 01 11 24.845& 46 57 49.44& F8  & EB of Algol type      & CO And ; GSC 03268-00398 ; TYC 3268-398-1 ; BD+46 281 ; BV 74\\
Ambiguous & T-And0-00657& 01 06 06.159& 47 31 59.37& K0  & Star                  & BD+46 254 ; GSC 03267-01349 ; TYC 3267-1349-1 ; AG+47 120 ; PPM 43637\\
Circular  & T-And0-00745& 01 03 45.076& 44 50 41.14&     & Star                  & TYC 2811-470-1 ; GSC 02811-00470\\
Ambiguous & T-And0-01203& 01 03 34.745& 48 32 39.27&     & Star                  & TYC 3267-1176-1 ; GSC 03267-01176\\
Circular  & T-And0-04046& 00 55 20.157& 47 44 53.20&     & Star                  & GPM 13.833991+47.748193\\
Roche-fill& T-And0-05153& 01 18 48.278& 49 39 36.86&     & EB of W UMa type      & QW And\\
Roche-fill& T-And0-05343& 00 52 55.122& 48 01 37.68&     & Star                  & GPM 13.232700+48.019757\\
Roche-fill& T-And0-07892& 00 56 15.567& 48 39 10.73&     & EB                    & NSVS 3757820\\
Circular  & T-And0-23792& 00 54 09.254& 47 45 19.91&     & Star                  & GPM 13.538629+47.755510\\
Roche-fill& T-Cas0-00170& 00 53 37.847& 48 43 33.83&     & Star                  & TYC 3266-195-1 ; GSC 03266-00195\\
Eccentric & T-Cas0-00394& 00 32 51.608& 49 19 39.36& B3  & EB of $\beta$ Lyr type& V381 Cas ; BD+48 162 ; BV 179\\
Roche-fill& T-Cas0-00430& 00 40 06.247& 50 14 15.64& K4  & EB of W UMa type      & V523 Cas ; GSC 03257-00167 ; WR 16 ; CSV 5867\\
          &             &             &            &     &                       & 1RXS J004005.0+501414 ; TYC 3257-167-1\\
Circular  & T-Cas0-00640& 00 47 06.277& 48 31 13.14&     & Star                  & TYC 3266-765-1 ; GSC 03266-00765\\
Circular  & T-Cas0-00792& 00 48 26.554& 51 35 02.52&     & Star                  & TYC 3274-664-1 ; GSC 03274-00664\\
Roche-fill& T-Cas0-02013& 00 40 46.427& 46 56 57.41&     & Star                  & TYC 3253-1767-1 ; GSC 03253-01767\\
Inverted  & T-Cas0-02069& 00 49 17.959& 50 39 02.92&     & EB                    & V385 Cas\\
Roche-fill& T-Cas0-08802& 00 51 32.351& 47 16 42.57&     & Star                  & GPM 12.884787+47.278540\\
Roche-fill& T-CrB0-00654& 16 00 14.507& 35 12 31.56&     & EB of W UMa type      & AS CrB ; GSC 02579-01125 ; NSVS 7847829\\
          &             &             &            &     &                       & ROTSE1 J160014.54+351228.4\\
Roche-fill& T-CrB0-00705& 15 55 51.838& 33 11 00.39&     & EB of W UMa type      & ROTSE1 J155551.87+331100.5\\
Roche-fill& T-CrB0-01589& 16 10 09.313& 35 57 30.57&     & Variable of $\delta$ Sct type& ROTSE1 J161009.33+355730.8\\
Roche-fill& T-CrB0-01605& 16 00 58.472& 34 18 54.34&     &EB of W UMa or RR Lyr-C& NSVS 7848126 ; ROTSE1 J160058.45+341854.5\\
Roche-fill& T-CrB0-04254& 16 09 19.589& 35 32 11.48&     & EB of W UMa type      & ROTSE1 J160919.62+353210.8\\
Circular  & T-Cyg1-00246& 19 44 01.777& 50 13 57.42&     & Star                  & TYC 3565-643-1 ; GSC 03565-00643\\
Roche-fill& T-Cyg1-00402& 19 54 39.939& 50 36 41.91&     & Star                  & TYC 3566-606-1 ; GSC 03566-00606\\
Ambiguous & T-Cyg1-01385& 20 15 21.936& 48 17 14.14&     & Star                  & TYC 3576-2035-1 ; GSC 03576-02035\\
Circular  & T-Cyg1-01627& 19 45 20.426& 51 35 07.22&     & Star                  & TYC 3569-1752-1 ; GSC 03569-01752\\
Roche-fill& T-Cyg1-04652& 20 07 07.305& 50 34 01.34&     &EB of W UMa type       & GSC 03567-01035\\
Roche-fill& T-Cyg1-04852& 19 51 59.208& 50 05 29.61&     &EB of W UMa type       & NSVS 5645908\\
Circular  & T-Cyg1-09274& 20 16 06.814& 51 56 26.07&     & EB of W UMa type      & V1189 Cyg ; CSV 8488 ; GSC 03584-01600 ; SON 7885\\
Roche-fill& T-Cyg1-11279& 19 59 53.377& 49 23 27.86&     & X-ray source          & 1RXS J195954.0+492318\\
Roche-fill& T-Cyg1-12518& 19 58 15.339& 48 32 15.79&     & Variable star         & Mis V1132\\
Roche-fill& T-Cyg1-14514& 19 48 05.077& 52 51 16.25&     &EB of W UMa or RR Lyr-C& V997 Cyg ; GSC 03935-02233 ; ROTSE1 J194804.79+525117.6 ; SON 7839\\
Ambiguous & T-Dra0-00240& 17 03 52.919& 57 21 55.54&     & Star                  & TYC 3894-898-1  ; GSC 03894-00898\\
Ambiguous & T-Dra0-00358& 16 45 38.339& 54 31 32.02&     & Star                  & TYC 3879-2689-1 ; GSC 03879-02689\\
Circular  & T-Dra0-00398& 16 57 33.875& 59 31 51.98&     & EB of Algol type/X-ray source& RX J1657.5+5931 ; 1RXS J165733.5+593156\\
          &             &             &            &     &                       & VSX J165733.8+593151 ; GSC 03898-00272\\
Roche-fill& T-Dra0-00405& 16 27 49.103& 58 50 23.30&     & Star                  & TYC 3884-1488-1 ; GSC 03884-01488\\
Roche-fill& T-Dra0-00959& 16 27 44.159& 56 45 59.30&     & EB of W UMa type/X-ray source& NSVS 2827877 ; 1RXS J162743.9+564557\\
Circular  & T-Dra0-01363& 16 34 20.417& 57 09 48.95&M4.5V& EB of BY Dra type     & CM Dra ; CSI+57-16335 1 ; LSPM J1634+5709 ; G 225-67 ; G 226-16\\
          &             &             &            &     &High proper-motion Star& IDS 16326+5721 A ; [RHG95] 2616 ; SBC7 580 ; CCDM J16343+5710A\\
          &             &             &            &     &                       & GJ 630.1 A ; LP 101-15 ; IDS 16325+5721 A ; [GKL99] 324 ; LHS 421\\
          &             &             &            &     &                       & 2MASS J16342040+5709439 ; CCABS 108 ; CABS 134 ; GEN\# +9.80225067\\
          &             &             &            &     &                       & RX J1634.3+5709 ; 1RXH J163421.2+570941 ; 1RXS J163421.2+570933\\
          &             &             &            &     &                       & PM 16335+5715 ; USNO 168 ; USNO-B1.0 1471-00307615 ; NLTT 43148\\
Roche-fill& T-Dra0-01346& 16 52 12.345& 57 43 31.70&     & EB of Algol type      & BPS BS 16080-0095 ; VSX J165212.3+574331 ; GSC 03885-00583\\
Roche-fill& T-Dra0-02224& 16 30 01.408& 54 45 55.80&     & Star                  & BPS BS 16084-0159\\
Circular  & T-Dra0-03021& 17 01 03.618& 55 14 54.70&     & EB of Algol type      & VSX J170103.5+551455 ; GSC 03890-01216 \\
Abnormal  & T-Dra0-03105& 16 23 02.558& 59 27 23.44&     & X-ray source          & 1RXS J162303.6+592717\\
Roche-fill& T-Dra0-05259& 16 41 48.751& 56 22 34.40&     & EB of W UMa type      & VSX J164148.7+562234 ; GSC 03882-02264 ; USNO-B1.0 1463-0278621\\
Ambiguous & T-Her0-00274& 17 00 51.150& 45 25 35.94&     & Star                  & TYC 3501-2245-1 ; GSC 03501-02245\\
Roche-fill& T-Her0-01086& 16 48 15.539& 44 44 28.73&     & EB of W UMa type      & GSC 03082-00896 ; NSVS 5252572 ; 1RXS J164817.3+444430\\
Roche-fill& T-Her0-03579& 16 35 47.390& 45 24 58.19&     & EB of W UMa type      & GSC 03499-01631\\
Inverted  & T-Her0-05469& 16 54 51.245& 43 20 35.89&     & EB                    & V747 Her ; SVS 2066\\
Circular  & T-Lyr1-00359& 19 15 33.695& 44 37 01.30& G0V & EB                    & V2277 Cyg ; GSC 03133-01149 ; ROTSE1 J191533.92+443704.9\\
          &             &             &            &     & X-ray source          & BD+44 3087 ; ILF1+44 155 ; 1RXS J191533.7+443704\\
Circular  & T-Lyr1-00687& 18 55 27.911& 47 13 41.76&     & Star                  & TYC 3544-1392-1 ; GSC 03544-01392\\
Circular  & T-Lyr1-01013& 18 55 03.963& 47 49 08.39&     & Star                  & TYC 3544-2565-1 ; GSC 03544-02565\\
Circular  & T-Lyr1-01439& 19 06 13.439& 46 57 26.42&     & Star                  & TYC 3545-2716-1 ; GSC 03545-02716\\
Circular  & T-Lyr1-02109& 18 57 35.415& 45 07 44.10&     & Cepheid variable star & ROTSE1 J185735.99+450752.5\\
Roche-fill& T-Lyr1-02166& 19 05 07.448& 46 15 07.51&     & X-ray source          & 1RXS J190504.8+461512\\
Roche-fill& T-Lyr1-03173& 18 59 45.531& 47 20 07.34&     & EB of W UMa type      & ROTSE1 J185945.43+472007.0\\
Roche-fill& T-Lyr1-03211& 18 45 56.939& 47 19 09.54&     & EB of W UMa type/X-ray source& ROTSE1 J184556.86+471914.4 ; 1RXS J184557.9+471906\\
Roche-fill& T-Lyr1-03270& 18 57 33.098& 48 05 22.49&     & EB of W UMa type      & ROTSE1 J185733.12+480522.5\\
Roche-fill& T-Lyr1-03783& 18 50 12.684& 45 35 44.05&     & Star                  & GPM 282.552858+45.595521\\
Inverted  & T-Lyr1-04431& 19 12 16.047& 49 42 23.58&     & EB of Algol type      & NSV 11822 ; GSC 03550-01770 ; NSVS 5578839 ; SON 9371\\
Roche-fill& T-Lyr1-05706& 18 47 57.211& 44 38 11.30&     & EB of W UMa type      & ROTSE1 J184757.18+443810.8\\
Inverted  & T-Lyr1-05887& 18 52 10.489& 47 48 16.67&     & EB of Algol type      & WX Dra ; AN 24.1925\\
Roche-fill& T-Lyr1-06583& 18 52 26.837& 44 55 20.86&     & EB                    & ROTSE1 J185226.53+445527.8\\
Inverted  & T-Lyr1-07179& 18 49 14.039& 45 24 38.61&     & Star                  & GPM 282.308454+45.410868\\
Roche-fill& T-Lyr1-08406& 18 50 06.942& 45 41 05.95&     & Star                  & GPM 282.528833+45.685035\\
Roche-fill& T-Lyr1-10276& 18 46 55.088& 45 00 52.27&     & EB of W UMa type      & V596 Lyr ; GPM 281.729421+45.014635 ; GSC 03540-00085\\
          &             &             &            &     &                       & ROTSE1 J184654.98+450054.7\\
Inverted  & T-Lyr1-10989& 19 06 22.791& 45 41 53.82&     & EB of Algol type      & V512 Lyr ; SON 10931\\
Roche-fill& T-Lyr1-11226& 18 45 21.748& 45 53 28.79&     & EB of W UMa type or $\delta$ Sct & V594 Lyr ; GPM 281.340617+45.891326 ; GSC 03540-01842\\
          &             &             &            &     &                       & ROTSE1 J184522.47+455321.0\\
Roche-fill& T-Lyr1-12772& 18 52 25.096& 44 55 40.23&     & EB of W UMa type      & ROTSE1 J185226.53+445527.8\\
Abnormal  & T-Lyr1-13166& 19 02 28.120& 46 58 57.75& F9V & EB                    & V361 Lyr ; SON 9349\\
Roche-fill& T-Per1-00328& 03 41 57.108& 39 07 29.60& G5  & EB of Algol type      & HD 275743 ; BD+38 787 ; GSC 02863-00755 ; TYC 2863-755-1\\
Circular  & T-Per1-00459& 03 34 57.745& 39 33 18.70& G5  & Star                  & HD 275547 ; GSC 02866-01995 ; TYC 2866-1995-1\\
Circular  & T-Per1-00750& 03 47 45.543& 35 00 37.08&     &Double or multiple star& TYC 2364-2327-1 ; GSC 02364-02327 ; CCDM J03478+3501BC\\
          &             &             &            &     &                       & ADS 2771 BC ; BD+34 732B ; CSI+34 732 2 ; NSV 1302\\
Roche-fill& T-Per1-00974& 03 34 43.738& 38 40 22.22&  A  & Star                  & HD 275481\\
Circular  & T-Per1-01218& 03 42 33.165& 39 06 03.63&  A  & EB                    & HU Per ; HD 275742 ; SVS 922\\
Roche-fill& T-Per1-01482& 03 48 45.999& 35 14 10.05& F0  & Star                  & HD 279025\\
Circular  & T-Per1-02597& 03 44 32.202& 39 59 34.94& K4V & T Tau type Star       & [LH98] 94 ; 1RXS J034432.1+395937 ; 1SWASP J034433.95+395948.0\\
Inverted  & T-Per1-04353& 03 45 04.887& 37 47 15.91&     & EB of Algol type      & HV Per ; SVS 368 ; P 107\\
Roche-fill& T-Tau0-00397& 04 30 09.466& 25 32 27.05& A3  & EB of $\beta$ Lyr type& GW Tau ; SVS 1421 ; HD 283709 ; ASAS 043009+2532.4\\
Inverted  & T-Tau0-00686& 04 07 13.870& 29 18 32.44&     & EB of Algol type      & IL Tau ; SON 9543\\
Roche-fill& T-Tau0-00781& 04 12 51.218& 24 41 44.26& G9  &Eruptive/T Tau-type Star& V1198 Tau ; NPM2+24.0013 ; 1RXS J041250.9+244201\\
          &             &             &            &     &                       & GSC 01819-00498 ; RX J0412.8+2442 ; [WKS96] 14\\
Roche-fill& T-Tau0-01262& 04 16 28.109& 28 07 35.81& K7V &Variable Star of Orion Type& V1068 Tau ; EM StHA 25 ; JH 165 ; EM LkCa 4\\
          &             &             &            &     &                       & HBC 370 ; ASAS 041628+2807.6\\
Roche-fill& T-Tau0-01715& 04 19 26.260& 28 26 14.30& K7V &T Tau-type Star/X-ray source& V819 Tau ; HBC 378 ; NAME WK X-Ray 1 ; 1E 0416.3+2830\\
          &             &             &            &     &                       & IRAS C04162+2819 ; TAP 27 ; [MWF83] P1 ; WK81 1\\
          &             &             &            &     &                       & 1RXS J041926.1+282612 ; X 04163+283\\
Roche-fill& T-Tau0-06463& 04 07 27.415& 27 51 06.36&     & EB of W UMa type      & V1022 Tau ; HV 6199 ; NSV 1464\\
Inverted  & T-UMa0-00127& 09 38 06.716& 56 01 07.32& A2V & EB of Algol type      & VV UMa ; GEN\# +0.05601395 ; HIP 47279 ; TYC 3810-1290-1\\
          &             &             &            &     &                       & GSC 03810-01290 ; SBC7 384 ; GCRV 6211 ; BD+56 1395\\
          &             &             &            &     &                       & HIC 47279 ; SVS 770 ; AAVSO 0931+56\\
Circular  & T-UMa0-00222& 10 07 18.023& 56 12 37.12& A0  & Star                  & HD 237866 ; GSC 03818-00504 ; SAO 27524 ; AG+56 778 ; HIC 49581\\
          &             &             &            &     &                       & BD+56 1432 ; HIP 49581 ; YZ 56 6209 ; TYC 3818-504-1\\
Roche-fill& T-UMa0-01701& 10 03 02.856& 55 47 53.34&     & X-ray source          & RX J100303.4+554752 ; [PTV98] H22 ; [PTV98] P29\\
Circular  & T-UMa0-03090& 10 08 52.180& 52 45 52.49& K2e & Star                  & GSC 03815-01151 ; RIXOS 229-302 ; RX J100851.6+524553\\
Roche-fill& T-UMa0-03108& 10 04 16.780& 54 12 02.83&     & EB of W UMa type      & NSVS 2532137
\enddata
\label{tableSIMBAD}
\end{deluxetable}

\subsection{Low-Mass EBs}
\label{subsecLowMassEBs}

\begin{figure}
\includegraphics[width=5in]{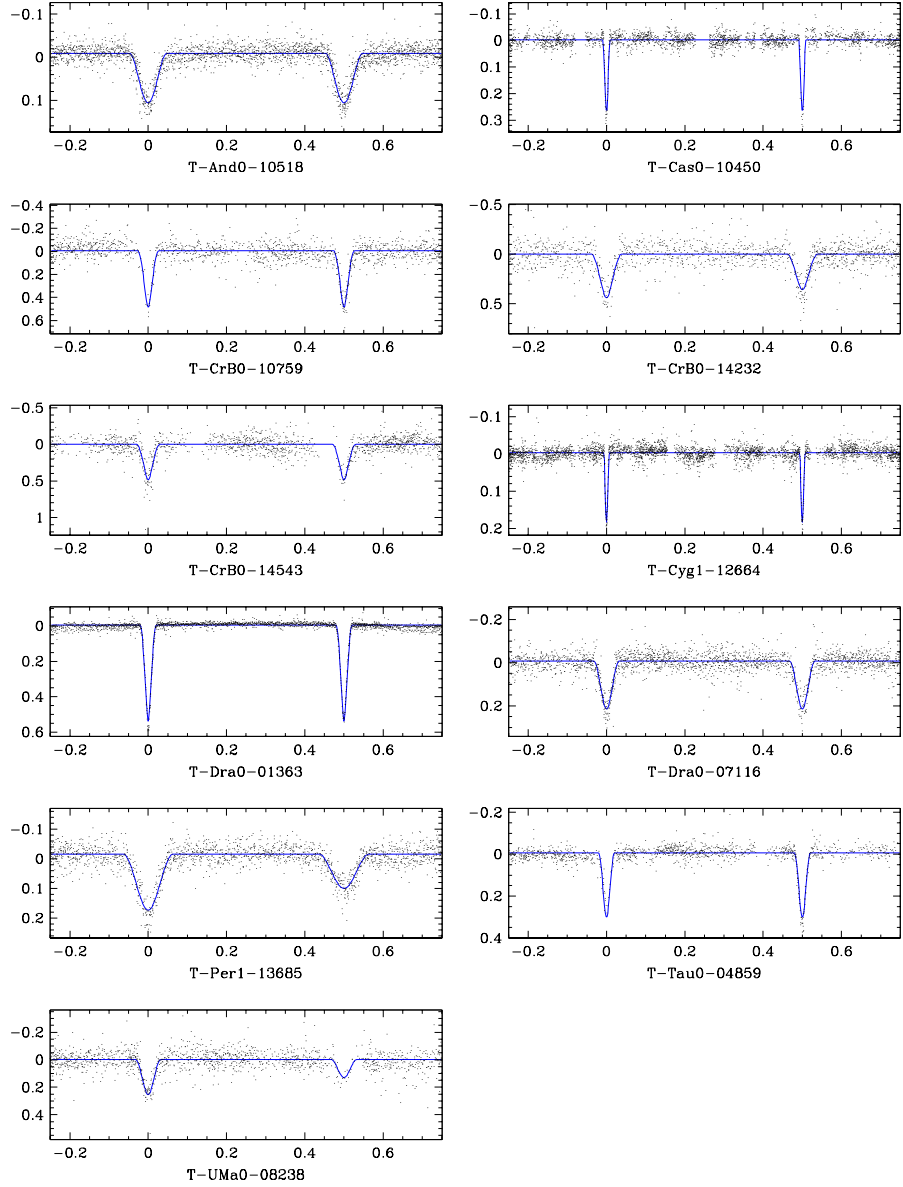}
\caption{Low-mass candidates ($M_1 < 0.75\: M_{\sun}$), with their best-fit MECI models (solid line).}
\label{figLowMassCat}
\end{figure}

The first group consists of 11 low-mass EB candidates, including
10 newly discovered EBs with either K or M-dwarf stellar
components. Our criteria for selecting these binaries were that
they be well-detached, and that both components have estimated
masses below $0.75M_{\sun}$ (see Table \ref{tableLowMass} and Figure
\ref{figLowMassCat}). Currently, only seven such detached low-mass EBs
have been confirmed [YY~Gem: \citep{Kron52, Torres02}; CM~Dra:
\citep{Lacy77a, Metcalfe96}; CU~Cnc: \citep{Delfosse99, Ribas03};
T-Her0-07621: \citep{Creevey05}; GU~Boo: \citep{LopezMorales05};
NSVS01031772: \citep{LopezMorales06}; and UNSW-TR-2:
\citep{Young06}].

Despite a great deal of work that has been done to understand the
structure of low-mass stars [e.g., \citet{Chabrier00}], models
continue to underestimate their radii by as much as $15$\%
\citep{Lacy77b, Torres02, Creevey05, Ribas06}, a significant
discrepancy considering that for solar-type stars the agreement
with the observations is typically within $1-2$\%
\citep{Andersen91, Andersen98}. In recent years, an intriguing
hypothesis has been put forward that strong magnetic fields may
have bloated these stars through chromospheric activity
\citep{Ribas06, Torres06, LopezMorales07, Chabrier07}.
Furthermore, \citet{Torres06} find that such bloating occurs even
for stars with nearly a solar mass, and suggest that this effect
may also be due to magnetically induced convective disruption. In
either case, these radius discrepancies should diminish for widely
separated binaries with long periods, as they become
non-synchronous and thus rotate slower, which according to dynamo
theory would reduce the strength of their magnetic fields.

Unfortunately, the small number of well-characterized low-mass EBs
makes it difficult to provide strong observational constraints to
theory. Despite the fact that such stars make up the majority of
the Galactic stellar population, their intrinsic faintness renders
them extremely rare objects in magnitude-limited surveys. In
addition, once found, their low flux severely limits the ability
to observe their spectra with both sufficiently high resolution
and a high signal-to-noise ratio. To this end, the fact that the
TrES survey was made with small-aperture telescopes is a great
advantage, as any low-mass EB candidate found is guaranteed to be
bright, and thus requires only moderate-aperture telescopes for
their follow-up. Thus we propose multi-epoch spectroscopic study
of the systems listed here, in order to confirm their low mass and
to estimate their physical properties with an accuracy sufficient
to test models of stellar structure. Moreover, two of our
candidates (T-Cyg1-12664 and T-Cas0-10450), if they are in fact
ambiguous-equal (group [IV]), have periods greater than 8 days,
making them prime targets for testing the aforementioned
magnetic-bloating hypothesis.

\begin{deluxetable}{lcccccccccc}
\tabletypesize{\tiny}
\rotate
\tablecaption{Low-mass EB candidates  ($M_{1,2} < 0.75M_{\sun}$ ; sorted by mass)}
\tablewidth{0pt}
\tablehead{\colhead{Category} & \colhead{Object} & \colhead{$\alpha$ (J2000)} & \colhead{$\delta$ (J2000)} & \colhead{Period $[days]$} & \colhead{$M_1/M_{\sun}$} & \colhead{$M_2/M_{\sun}$} & \colhead{age $[Gyr]$} &
 \colhead{\begin{tabular}{c} Proper motion \\ source catalog\tablenotemark{a} \end{tabular}} & \colhead{\begin{tabular}{c} $PM_{\alpha}$ \\ mas/year \end{tabular}} & \colhead{\begin{tabular}{c} $PM_{\delta}$ \\ mas/year \end{tabular}}}
\startdata
Circular& T-Dra0-01363\tablenotemark{b}& 16 34 20.417& 57 09 48.95&  1.268&   0.27 $\pm$ 0.02&  0.24 $\pm$ 0.03&  1.6 $\pm$ 1.6  & \citet{Salim03} & -1121& 1186\\
AmbigEq\tablenotemark{c}&  T-And0-10518& 01 07 44.417& 48 44 58.11&  0.387&   0.45 $\pm$ 0.27&  0.45 $\pm$ 0.28&  0.3 $\pm$ 0.5  & UCAC  &  2.7& -2.0\\
AmbigEq&  T-Cyg1-12664& 19 51 39.824& 48 19 55.38&  8.257&   0.50 $\pm$ 0.20&  0.48 $\pm$ 0.19&  0.3 $\pm$ 0.4  & USNO-B&  -18&  -6\\
AmbigEq&  T-CrB0-14232& 16 10 22.495& 33 57 52.33&  0.971&   0.60 $\pm$ 0.24&  0.55 $\pm$ 0.29&  4.4 $\pm$ 8.8  & UCAC  &-15.2& -24.2\\
AmbigEq&  T-CrB0-14543& 15 57 45.926& 33 56 07.28&  1.506&   0.60 (-1)\tablenotemark{d}      &  0.60 (-1)      &  0.2 (-2)       & UCAC&  -13.9&  13.3\\
Circular& T-Per1-13685& 03 53 51.217& 37 03 16.73&  0.384&   0.60 (-1)      &  0.50 (-1)      &  10.0 (-3)      & UCAC  &-24.1&-15.9\\
AmbigEq&  T-CrB0-10759& 15 52 18.455& 30 35 32.13&  1.901&   0.63 $\pm$ 0.24&  0.62 $\pm$ 0.21&  7.3 $\pm$ 49.6 & UCAC  &  3.6&-19.4\\
AmbigEq&  T-UMa0-08238& 10 09 25.384& 53 57 01.31&  1.250&   0.69 $\pm$ 0.54&  0.61 $\pm$ 0.51&  4.1 $\pm$ 15.0 & USNO-B&    6&   -4\\
AmbigEq&  T-Cas0-10450& 00 29 16.288& 50 27 38.58&  8.656&   0.71 $\pm$ 0.21&  0.67 $\pm$ 0.20&  0.3 $\pm$ 0.4  & UCAC  & -3.1& -4.2\\
AmbigEq&  T-Dra0-07116& 17 02 53.025& 55 07 47.44&  1.369&   0.71 $\pm$ 0.22&  0.69 $\pm$ 0.22&  2.1 $\pm$ 3.6  & USNO-B&   -2&  -16\\
Circular& T-Tau0-04859& 04 08 11.608& 24 51 10.18&  3.068&   0.74 $\pm$ 0.10&  0.66 $\pm$ 0.10&  8.8 $\pm$ 14.8 & UCAC  &  3.4& -8.0\\
\enddata
\tablenotetext{a}{Where possible, we used the more accurate UCAC
catalog, otherwise we reverted to the USNO-B catalog. Since they
are dim and nearby, we expect most of the low-mass binaries to
have comparably large proper motions.}
\tablenotetext{b}{This binary is CM Draconis, which has been extensively
studied and found to have a masses of $M_1 = 0.2307 \pm 0.0010 M_{\sun}$ and
$M_2 = 0.2136 \pm 0.0010 M_{\sun}$ \citep{Lacy77a, Metcalfe96}. For consistency,
we listed the MECI results, which are off by less than $0.04 M_{\sun}$
($\sim1.5\sigma$). We also adopted an alternate proper motion
estimate, as its USNO-B values seems to be erroneous, probably due
to its very high angular velocity.}
\tablenotetext{c}{For clarity
we list for the ambiguous systems, only the solution with
approximately equal components. But it is likely that at least a
few of the ambiguous systems may be unequal, with half the period.
Such cases can be identified as single-line spectroscopic
binaries, with the secondary component being no larger than a few
$0.1M_{\sun}$.}
\tablenotetext{d}{When the most likely model is at
the edge of the parameter space, MECI is not able to bound the
solution, and therefore cannot estimate the uncertainties. We mark
(-3) when the upper limit was reached, (-2) when the lower limit
was reached, and (-1) if one of the other parameter is at its
limit.}
\label{tableLowMass}
\end{deluxetable}

\subsection{Eccentric EBs}
\label{subsecEccentricEBs}

The second group of EBs consists of 34 binaries with eccentric
orbits (see Table \ref{tableEccentric}, and Figures \ref{figEcc1}-\ref{figEcc3}). We were able to reliably
measure values of $|e \cos \omega|$ as low as $\sim0.005$ by using
the eclipse timing technique (see \S\ref{secMethod} and Figure
\ref{figTimingVariationsEcc}). Since this measure provides a lower
limit to the eccentricity, it is well suited to identify eccentric
EBs, even though the actual value of the eccentricity may be
uncertain. As mentioned earlier, in an effort to avoid
false-positives, we do not include in this group EBs whose eclipse
timing measures $|e \cos \omega| < 0.005$, or EBs with an
eccentricity consistent with zero.

Our interest in these eccentric binaries stems from their
potential to constrain tidal circularization theory
\citep{Darwin1879}. This theory describes how the eccentricity of
a binary orbit decays over time due to tidal dissipation, with a
characteristic timescale ($t_{circ}$) that is a function of the
components' stellar structure and orbital separation. As long as
the components' stellar structure remains unchanged, the orbital
eccentricity is expected to decay approximately exponentially over
time [$e \propto \exp (-t/t_{circ})$]. However, once the
components evolve off the main-sequence, this timescale may vary
considerably \citep{Zahn89}. Thus, to understand the
circularization history of binaries with circularization
timescales similar to or larger than their evolutionary
timescales, one must integrate over the evolutionary tracks of
both stellar components.

Three alternative tidal dissipation mechanisms have been proposed:
dynamical tides \citep{Zahn75, Zahn77}, equilibrium tides
\citep{Zahn77, Hut81}, and hydrodynamics \citep{Tassoul88}.
Despite its long period of development, the inherent difficulty of
observing tidal dissipation has prevented definitive conclusions.
\citet{Zahn89} add a further complication by maintaining that most
of the orbital circularization process takes place at the
beginning of the Hayashi phase, and that the eccentricity of a
binary should then remain nearly constant throughout its lifetime
on the main-sequence.

Observational tests of these tidal circularization theories,
whereby $t_{circ}$ is measured statistically in coeval stellar
populations, have so far proved inconclusive. \citet{North03}
found that short-period binaries in both the Large and Small
Magellanic Clouds seem to have been circularized in agreement with
the theory of dynamical tides. However, \citet{Meibom05} show that,
with the exception of the Hyades, the stars in the clusters that
they observed were considerably more circularized than any of the
known dissipation mechanisms would predict. Furthermore, they find
with a high degree of certainty, that older clusters are more
circularized than younger ones, thereby contradicting the Hayashi
phase circularization model.

\begin{figure}
\includegraphics[width=5in]{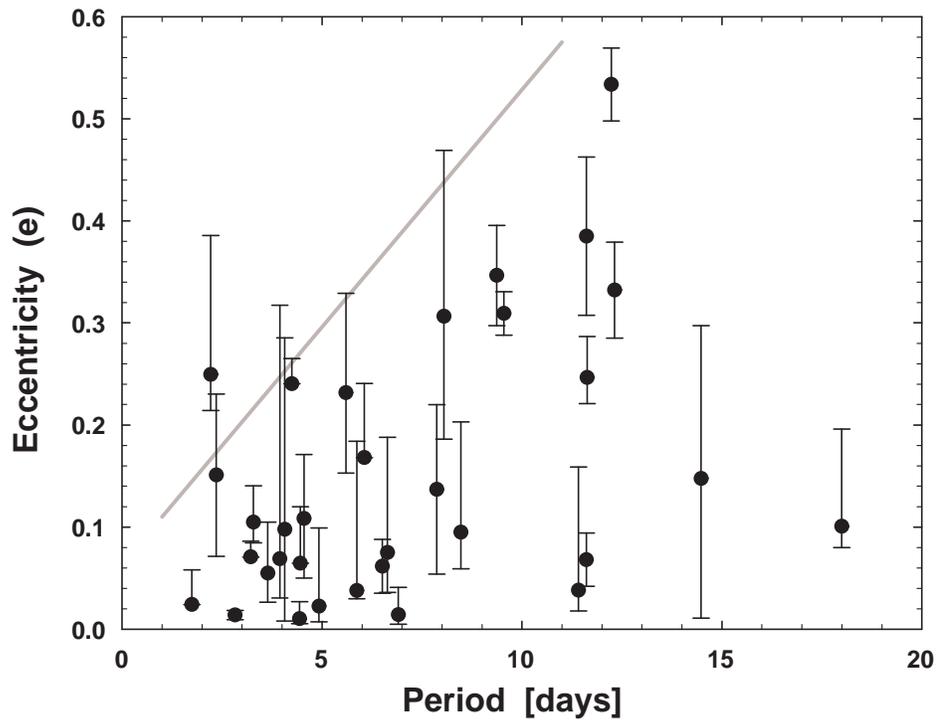}
\caption{The period-eccentricity
relation. The lower ends of the error bars were truncated, where
needed, by the measured lower limit, $|e \cos \omega|$. Note the
lack of eccentric short-period systems. The diagonal line is
provided to guide the eye.}
\label{figPeriodEcc}
\end{figure}

Encouraged by the statistical effect of circularization that can
be seen in our catalog (Figure \ref{figPeriodEcc}), we further
estimated $t_{circ}$ for each of the eccentric systems as follows.
\citet{Zahn77, Zahn78} provides an estimate for the orbital
circularization timescale due to turbulent dissipation in stars
possessing a convective envelope, assuming that corotation has
been achieved:

\begin{equation}
t_{circ} = \frac{1}{21q(1+q)k_2} \left(\frac{MR^2}{L}\right)^{1/3}\left(\frac{a}{R}\right)^8
\end{equation}

where $M, R, L$ are the star's mass, radius, and luminosity, and
$k_2$ is the apsidal motion constant of the star, which is
determined by its internal structure and dynamics.

More massive stars, which do not have a convective envelope but
rather develop a radiative envelope, are thought to circularize
their orbit using radiative damping \citep{Zahn75, Claret97}. This
is a far slower mechanism, whose circularization timescale can be
estimated by:

\begin{equation}
t_{circ} = \frac{2}{21 q (1+q)^{11/6} E_2} \left(\frac{R^3}{GM}\right)^{1/2} \left(\frac{a}{R}\right)^{21/2}
\end{equation}

where $E_2$ is the tidal torque constant of the star, and $G$ is
the universal gravitational constant. We can greatly simplify
these expressions by applying Kepler's law [$a^3 =
GM(1+q)(P/{2\pi})^2$], and adopt the \citet{Cox00} power law
approximations for the main-sequence mass-radius and
mass-luminosity relations. For the convective envelope case, we
adopt the late-type mass-radius relation ($M < 1.3 M_{\sun}$), and
for the radiative envelope case we adopt the early-type
mass-radius relation ($M \geq 1.3 M_{\sun}$), thus arriving at:

\begin{equation}
\label{eqCircTime}
t_{circ} \simeq \cases{
0.53 Myr \; (k_2 / 0.005)^{-1}q^{-1}(1+q)^{5/3} \left(P / day\right)^{16/3} \left(M / M_{\sun} \right)^{-4.99}, \ M < 1.3 M_{\sun}\cr
1370 Myr \; (E_2 / 10^{-8})^{-1}q^{-1}(1+q)^{5/3} \left(P / day\right)^7 \left(M / M_{\sun} \right)^{-2.76}, \ M \geq 1.3 M_{\sun} \cr}
\end{equation}

Determining the values of $k_2$ and $E_2$ is the most difficult
part of this exercise, since their values are a function of the
detailed structure and dynamics of the given star, which in turn
changes significantly as the star evolves \citep{Claret97,
Claret02}. In our calculation, we estimate these values by
interpolating published theoretical tables [$k_2$: \citet{Zahn94},
$E_2$: \citet{Zahn75, Claret97}]. Since both stellar components
contribute to the circularization process, the combined
circularization timescale becomes $t_{circ} = 1/(t_{circ,1}^{-1} +
t_{circ,2}^{-1})$, where the subscripts $1$ and $2$ refer to the
primary and secondary binary components \citep{Claret97}. In Table
\ref{tableEccentric}, we list the combined circularization
timescale for each of the eccentric EBs we identify.

The value of $t_{circ}$ for most of the eccentric systems (21 of
34) is larger than the Hubble time, indicating that no significant
circularization is expected to have taken place since they settled
on the main-sequence. About a quarter of the eccentric systems (8
of 34) have a $t_{circ}$ smaller than the Hubble time but larger
than $1\:Gyr$. While circularization is underway, the fact that
they are still eccentric is consistent with theoretical
expectations. The remaining systems (5 of 34) all have $t_{circ} <
1\:Gyr$, have periods less than 3.3 days, and unless they are
extremely young, require an explanation for their eccentric
orbits. Two of these EBs (T-Tau0-02487 and T-Tau0-03916) are
located near the star-forming regions of Taurus, supporting the
hypothesis that they are indeed young. However, this hypothesis
does not seem to be adequate for T-Cas0-02603, which has a period
of only 2.2 days and $t_{circ} \simeq 0.26\:Gyr$, while possessing
a large eccentricity of $e \simeq 0.25$. An alternative
explanation is that some of these binaries were once further
apart, having larger orbital periods, and thus larger
circularization timescales. These systems may have been involved
in a comparably recent interaction with a third star (a collision
or near miss), or have been influenced by repeated resonant
perturbations of a tertiary companion.

\begin{figure}
\includegraphics[width=5in]{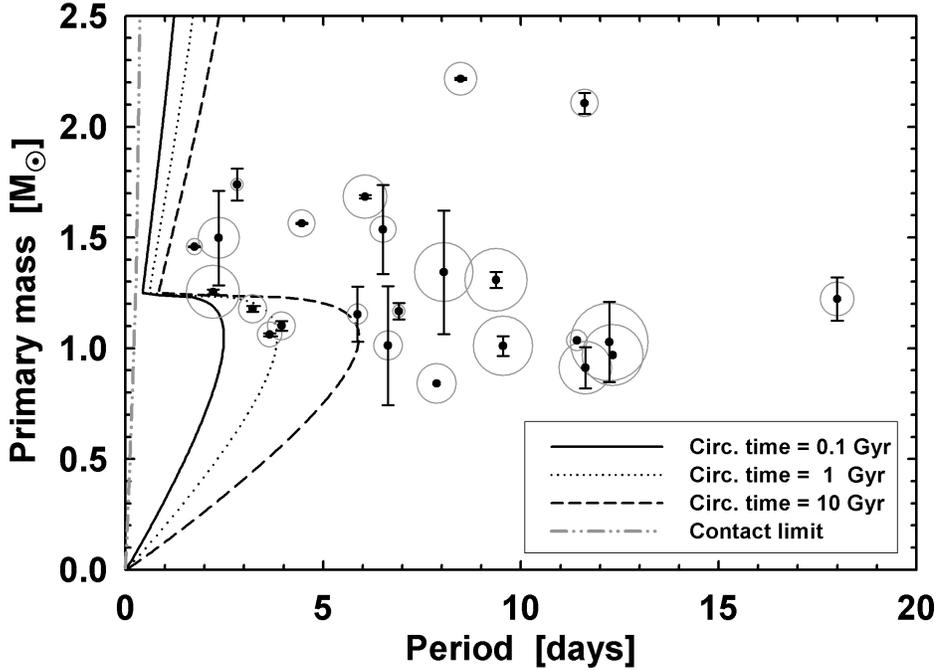}
\caption{The period-primary mass relation for eccentric EBs. We
included all systems with well-determined masses. The area of the
gray circles is proportional to the EB's eccentricity. All the
curves are theoretical boundaries, assuming that the binary
components are both on the main-sequence and have equal masses
($q=1$). The left-most dot-dash line demarcates the binary contact
limit, and the remaining curves mark systems with increasing
circularization time (see equation \ref{eqCircTime}). Note the
abrupt increase in the circularization time for systems more
massive than $\sim1.25\:M_{\sun}$, at which point the stellar
convective envelope becomes radiative, and thus far less efficient
at tidal dissipation.} \label{figPeriodM1}
\end{figure}

Finally, we would like to draw the reader's attention to our
shortest-period eccentric EB, T-Cas0-00394, whose period is a mere
1.7 days. Notably, this system is entirely consistent with theory,
since its mass falls in a precarious gap, where the stellar
envelopes of its components are no longer convective, yet their
radiative envelopes are not sufficiently extended to produce
significant tidal drag (see Figure \ref{figPeriodM1}).

\subsection{Abnormal EBs}
\label{subsecAbnomalEBs}

The third group of EBs consists of 20 abnormal systems (see Table
\ref{tableAbnormal}, and Figures \ref{figAbnormalEBs1} and
\ref{figAbnormalEBs2}). While possessing the distinctive
characteristics of EBs, these LCs stood out during manual
inspection for a variety of reasons. These systems underline the
difficulty of fully automating any LC pipeline, as any such system
will inevitably need to recognize atypical EBs that were not
encountered before.

The LCs we listed can be loosely classified into groups according
to the way they deviate from a simple EB model. A few cases
exhibited pulsation-like fluctuations that were not synchronized
with the EB period (shorter-period: T-Dra0-00398, longer-period:
T-Lyr1-00359, T-Per1-00750). These fluctuations may be due either
to the activity of an EB component, or to a third star whose light
is blended with the binary. In principle, one can identify the
active star by examining the amplitude of the fluctuations during
the eclipses. If the fluctuations originate from one of the
components, their observed amplitude will be reduced when the
component is being eclipsed. In such a case, if the fluctuations
are due to pulsations, they can further provide independent
constraints to the stellar properties through astro-seismological
models \citep{Mkrtichian04}. To identify such fluctuating EBs one
must subtract the fitted EB model from the LC, and evaluate the
residuals [e.g., \citet{Pilecki07}]. When the fluctuation period is
fixed, one can simply search the residual LC using a periodogram,
as was done in step (1) of our pipeline (see \S\ref{secMethod}).
However, when the fluctuation period varies (i.e. non-coherent),
as in the aforementioned LCs, one must employ alternative methods,
since simply phasing their LC will not produce any discernable
structure. For LCs with long-period fluctuations, one can directly
search the residuals for time dependencies, while for LCs with
short-period fluctuations one can search the residuals for
non-Gaussian distributions. However, in practice these
measurements will likely not be robust, as there are many
instrumental effects that can produce false positives. Thus, we
employ a search for auto-correlations in the residual time series,
which overcomes most instrumental effects, while providing a
reliable indicator for many types of pseudo-periodic fluctuations.

The remaining systems had LC distortions that appear to be
synchronized with the orbital period. The source of these
fluctuations is likely due to long-lasting surface inhomogeneities
on one or both of the rotationally synchronized components. When
the LC has brief periodic episodes of darkening (T-And0-11476,
T-Cas0-13944, T-Cyg1-07584, T-Dra0-04520), they can usually be
explained as stable star spots, but brief periodic episodes of
brightening (T-And0-04594, T-Her0-08091), which may indicate the
presence of stable hot-spots, are more difficult to interpret.
This phenomenon is especially puzzling in the aforementioned two
cases, in which the brightening episodes are briefer than one
would expect from a persistent surface feature and repeat at the
middle of both plateaux.

When the two plateaux of an LC are not flat, they are usually
symmetric about the center of the eclipses. This is due to the
physical mirror symmetry about the line intersecting the binary
components' centers. When the axis of symmetry does not coincide
with the center of eclipse (T-And0-00920, T-Cyg1-08866,
T-Dra0-03105, T-Lyr1-07584, T-Lyr1-15595), a phenomenon we term
``eclipse offset,'' we conclude that this symmetry must somehow be
broken. This may occur if the EB components are not rotationally
synchronized, or have a substantial tidal lag. Another form of
this asymmetry can appear as an amplitude difference between the
two LC plateaux (T-Her0-03497, T-Lyr1-13166, T-Per1-08789,
T-UMa0-03090). This phenomenon, which was originally called the
``periastron effect'' and has since been renamed the ``O'Connell
effect,'' has been known for over a century, and has been
extensively studied [e.g., \citet{OConnell51, Milone86}]. Classical
hypotheses suggest an uneven distribution of circumstellar
material orbiting with the binary \citep{Struve48} or surrounding
the stars \citep{Mergentaler50}, either of which could induce a
preferential $H^-$ absorption on one side. \citet{Binnendijk60}
was the first of many to suggest that this asymmetry is due to
subluminous regions of the stellar surface (i.e. star spots).
However, this explanation also requires the stars to be
rotationally synchronized, and for the spots to be stable over the
duration of the observations. Alternative models abound, including
a hot spot on one side of a component brought about through mass
transfer from the other component, persistent star spots created
by an off-axis magnetic field, and circumstellar material being
captured by the components and heating one side of both stars
\citep{Liu03}. As with many phenomena that have multiple possible
models, the true answer may involve a combination of a number of
these mechanisms, and will likely vary from system to system
\citep{Davidge84}.

Finally, a few particularly unusual LCs (T-Dra0-03105,
T-Lyr1-05984) display a very large difference between their
eclipse durations. Although a moderate difference could be
explained by an eccentric orbit, such extreme eccentricities in
systems with such short orbital periods (0.5 and 1.5 days) are
highly unlikely.

\section{Conclusions}

We presented a catalog of 773 eclipsing binaries found in ten
fields of the TrES survey, identified and analyzed using an
automated pipeline. We described the pipeline we used to identify
and model them. The pipeline was designed to be mostly automated,
with manual inspections taking place only once the vast majority
of non-EB LCs had been automatically filtered out. At the final
stage of the pipeline, we classified the EBs into seven groups:
eccentric, circular, ambiguous-equal, ambiguous-unequal, inverted,
Roche-lobe-filling, and abnormal. The former four groups were all
successfully modeled with our model fitting program. However, the
latter three groups possessed significant additional physical
phenomena (tidal distortions, mass-transfer, and surface
activity), which did not conform to the simple detached-EB model
we employed.

We highlighted three groups of binaries, which may be of
particular interest and warrant follow-up observations. These
groups are: low-mass EBs, EBs with eccentric orbits, and abnormal
EBs. The low-mass EBs (both components $< 0.75\:M_{\sun}$) allow
one to probe the mass-radius relation at the bottom of the
main-sequence. Only seven such EBs have previously been confirmed,
and the physical properties of many of them are inconsistent with
current theoretical models. Our group of ten new candidates will
likely provide considerable additional constraints to the models,
and the discovery of two long-period systems could help confirm a
recent hypothesis that this inconsistency is due to stellar
magnetic activity. The eccentric-orbit EBs may help confirm and
constrain tidal circularization theory, as many of them have
comparably short circularization timescales. We demonstrated that,
as one would predict from the theory, the shortest-period systems
fall within a narrow range of masses, in which their stellar
envelopes cease to be convective yet their envelopes are not
extended enough to produce significant tidal drag. The abnormal
EBs seem to show a plethora of effects that are indicative of
asymmetries, stellar activity, persistent hot and cold spots, and
a host of other physical phenomena. Some of these systems may
require dedicated study to be properly understood.

In the future, as LC datasets continue to grow, it will become
increasingly necessary to use such automated pipelines to identify
rare and interesting targets. Such systematic searches promise a
wealth of data that can be used to test and constrain theories in
regions of their parameter space that were previously
inaccessible. Furthermore, even once the physics of ``vanilla''
systems has been solved, more complex cases will emerge to
challenge us to achieve a better understanding of how stars form,
evolve, and interact.

\section*{Acknowledgments}

We would like to thank Tsevi Mazeh for many useful discussions, as
well as S{\o}ren Meibom for his repeated help. We would also like
to thank Sarah Dykstra for her continuous support throughout the
preparation of this paper. We are grateful to the staff of the
Palomar Observatory for their assistance in operating the Sleuth
instrument, and we acknowledge support from NASA through grant
NNG05GJ29G issued through the Origins of Solar Systems Program.
This research has made use of NASA's Astrophysics Data System
Bibliographic Services, the SIMBAD database, operated at CDS,
Strasbourg, and the VSX database, which was created by Christopher
Watson for the AAVSO. This publication also utilizes data products
from 2MASS, which is a joint project of the University of
Massachusetts and the Infrared Processing and Analysis
Center/California Institute of Technology, funded by NASA and NSF.
Finally, we would like to thank the anonymous referee for very
insightful comments and suggestions, which significantly improved
this manuscript.

\section{Appendix - Rejecting Single-Eclipse EB Models}
\label{appendixSingleEclipse}

An EB LC comprising a deep eclipse and a very shallow eclipse,
can occur in one of two ways. Either the secondary component is
luminous but extremely small (e.g., a white dwarf observed in UV),
thus producing a shallow primary eclipse, or the secondary
component is comparably large but extremely dim, thus producing a
shallow secondary eclipse. The first case, though possible [e.g.,
\citet{Maxted04}], is extremely rare, and will have a signature
``flat bottom'' to the eclipse. We have not encountered such an LC
in our dataset. The second case will have a rounded eclipse
bottom, due to the primary component's limb darkening. Assuming
this latter contingency, in which the secondary component is dark
in comparison to the primary component, we can place a lower bound
on its radius ($R_2$):

\begin{equation}
R_2 \geq R_1 \sqrt{1 - 10^{-0.4 \Delta mag_1}}\ ,
\end{equation}

where $R_1$ is the radius of the primary component, and $\Delta
mag_1$ is the magnitude depth of the primary eclipse. Thus, if the
eclipse is very deep, the size of the secondary component must
approach the size of the primary component. However, coeval
short-period detached EBs with components of similar sizes yet
disparate luminosities are expected to be very rare, assuming that they
follow normal stellar evolution. Therefore, if only one eclipse is
detected, and it is both rounded and sufficiently deep, we may
conclude that this configuration entry is likely to be incorrect,
and that the correct configuration has double the orbital period
and produces two equal eclipses. Only when we cannot apply such a
period-doubling solution (i.e. when the secondary eclipse is
detectable), do we resort to questioning our assumption of normal
stellar evolution (see classification group V, described in
\S\ref{secMethod}).

\section{Appendix - Description of the Catalog Fields}
\label{appendixCatalogDescription}

Due to the large size of the catalog, we were only able to list
small excerpts of it in the body of this paper. Readers interested
in viewing the catalog in its entirety can download it
electronically. Note that although the catalog lists 773 unique
systems, each of the 103 ambiguous EBs appears in both possible
configurations (see \S\ref{secMethod}), raising the total number
of catalog entries to 876. Below, we briefly describe the
catalog's 38 columns. The column units, if any, are listed in
square brackets.

\renewcommand{\theenumi}{\arabic{enumi}}
\begin{enumerate}
\item $Category$-- the EB's classification (see \S\ref{secMethod}).
\item $Binary\ name$-- the EB's designation, which is composed of its TrES field (see Table \ref{tableFieldsObs}) and index.
\item $\alpha$-- the EB's right ascension (J2000).
\item $\delta$-- the EB's declination (J2000).
\item $Period$ [days]-- the EB's orbital period.
\item $Period \ uncertainty$ [days] -- the uncertainty in the EB's orbital period.
\item $Mass_1$ [$M_{\sun}$]-- the mass of the EB's primary (more massive) component.
\item $Mass_1\ uncertainty$ [$M_{\sun}$]-- the uncertainty in the primary component's mass.
\item $Mass_2$ [$M_{\sun}$]-- the mass of the EB's secondary (less massive) component.
\item $Mass_2\ uncertainty$ [$M_{\sun}$]-- the uncertainty in the secondary component's mass.
\item $Age$ [Gyr] -- the age of the EB (assumed to be coeval).
\item $Age\ uncertainty$ [Gyr] -- the uncertainty in the EB's age.
\item $Score$-- a weighted reduced $\chi^2$ of the MECI model fit [see \citet{Devor06b} for further details].
\item $Isochrone\ source$-- isochrone tables used [Y2: \citet{Kim02}, or Baraffe: \citet{Baraffe98}].
\item $Color\ weighting$-- the relative weight ($w$) of the LC fit, compared to the color fit [see \citet{Devor06b} for further details].
\item $PM\ source$-- the database that provided the proper motion measurement [UCAC: \citet{Zacharias04}, USNO-B: \citet{Monet03}, or Salim03: \citet{Salim03}].
\item $PM_{\alpha}$ [${\rm mas\,yr^{-1}}$]-- the right ascension component of the EB's proper motion.
\item $PM_{\delta}$ [${\rm mas\,yr^{-1}}$] -- the declination component of the EB's proper motion.
\item $Location\ error$ [arcsec]-- the distance between our listed location (columns 3 and 4) and the location listed by the proper motion database.
\item $mag_B$-- the USNO-B $B$-band observational magnitude of the EB (average of both magnitude measurements, if available).
\item $mag_R$-- the USNO-B $R$-band observational magnitude of the EB (average of both magnitude measurements, if available).
\item $Third-light\ fraction$-- the fraction of third-light flux ($R$-band) blended into the LC (i.e. the flux within 30", excluding the target, divided by the total flux within 30").
\item $mag_J$-- the 2MASS observational $J$-band magnitude of the EB, converted to ESO $J$-band.
\item $mag_H$-- the 2MASS observational $H$-band magnitude of the EB, converted to ESO $H$-band.
\item $mag_K$-- the 2MASS observational $K_s$-band magnitude of the EB, converted to ESO $K$-band.
\item $Mag_J$-- the absolute ESO $J$-band magnitude of the EB listed in the isochrone tables.
\item $Mag_H$-- the absolute ESO $H$-band magnitude of the EB listed in the isochrone tables.
\item $Mag_K$-- the absolute ESO $K$-band magnitude of the EB listed in the isochrone tables.
\item $Distance$ [pc]-- the distance to the EB, as calculated from the extinction-corrected distance modulus.
\item $A(V)$-- the EB's V-mag absorption due to Galactic interstellar extinction (assuming $R_V = 3.1$).
\item $\sin(i)$-- the sine of the EB's orbital inclination.
\item $|e \cos(\omega)|$-- a robust lower limit for the EB's eccentricity (see equation \ref{eqOmC}).
\item $Eccentricity$-- the orbital eccentricity of the EB.
\item $Eccentricity\ uncertainty$-- the uncertainty in the orbital eccentricity of the EB.
\item $\Delta mag_1$-- the $r$-band primary (deeper) eclipse depth in magnitudes.
\item $Epoch_1$-- the Heliocentric Julian date (HJD) at the center of a primary eclipse, minus 2,400,000.
\item $\Delta mag_2$-- the $r$-band secondary (shallower) eclipse depth in magnitudes.
\item $Epoch_2$-- the Heliocentric Julian date (HJD) at the center of a secondary eclipse, minus 2,400,000.
\end{enumerate}

Note that the values of the uncertainties (columns 6, 8 10, 12, and
34), were calculated by measuring the curvature of the
parameter space $\chi^2$ contour, near its minimum. This method
implicitly assumes a Gaussian distribution of the parameter
likelihood. If the likelihood distribution not Gaussian, but
rather has a flattened (boxy) distribution, then the computed
uncertainty becomes large. In extreme cases, the estimated formal
uncertainty can be larger than the measurement itself.

\chapter{T-Lyr1-17236: A Long-Period Low-Mass Eclipsing Binary
\label{chapter6}}

\def\kms{\ifmmode{\rm km\thinspace s^{-1}}\else km\thinspace s$^{-1}$\fi}

\title{T-Lyr1-17236: A Long-Period Low-Mass Eclipsing Binary}

J.~Devor, D.~Charbonneau, G.~Torres, C.~H.~Blake, R.~J.~White,
M.~Rabus, F.~T.~O'Donovan, G.~Mandushev, G.~{\'A}.~Bakos,
G.~F\H{u}r{\'e}sz, \& A.~Szentgyorgyi\\
\emph{The Astrophysical Journal}, {\bf 687}, 1253$-$1263

\section*{Abstract}

We describe the discovery of a 0.68+0.52~$M_{\sun}$ eclipsing
binary (EB) with an 8.4-day orbital period, found through a
systematic search of 10 fields of the Trans-atlantic Exoplanet
Survey (TrES). Such long-period low-mass EBs constitute critical
test cases for resolving the long standing discrepancy between the
theoretical and observational mass-radius relations at the bottom
of the main-sequence. It has been suggested that this discrepancy
may be related to strong stellar magnetic fields, which are not
properly accounted for in current theoretical models. All
previously well-characterized low-mass main-sequence EBs have
periods of a few days or less, and their components are therefore
expected to be rotating rapidly as a result of tidal
synchronization, thus generating strong magnetic fields. In
contrast, the binary system described here has a period that is
more than 3 times longer than previously characterized low-mass
main-sequence EBs, and its components rotate relatively slowly. It
is therefore expected to have a weaker magnetic field and to
better match the assumptions of theoretical stellar models. Our
follow-up observations of this EB yield preliminary stellar
properties that suggest it is indeed consistent with current
models. If further observations confirm a low level of activity in
this system, these determinations would provide support for the
hypothesis that the mass-radius discrepancy is at least partly due
to magnetic activity.

\section{Introduction}
\label{ch6_secIntro}

\begin{figure}
\includegraphics[width=5in]{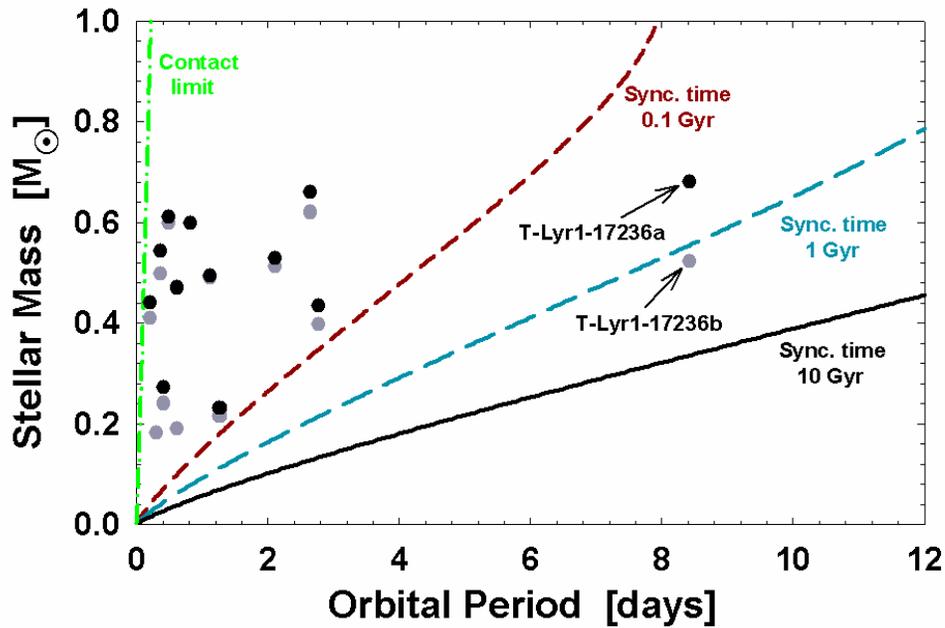}
\caption{The predicted synchronization timescales
due to turbulent dissipation \citep{Zahn77, Zahn94} for well
characterized low-mass EBs from Table~\ref{ch6_tablePreviousEBs}. The
lines trace constant synchronization timescales of binary
components for which $q=1$ (see \S\ref{ch6_secPhysical} for further
details on this calculation). The black circles indicate primary
components and the gray circles indicate secondary components.
Note that in some cases the primary and secondary symbols nearly
overlap.}
\label{ch6_figTsync}
\end{figure}

Despite a great deal of work that has been done to understand the
structure of low-mass ($<$ 0.8 $M_{\sun}$) main-sequence stars
\citep[e.g.,][]{Chabrier00}, models continue to underestimate
their radii by as much as 15\% \citep{Lacy77b, Torres02, Ribas06}.
This is a significant discrepancy, considering that for solar-type
stars the agreement with the observations is typically within
1\%--2\% \citep{Andersen91, Andersen98}. In recent years an
intriguing hypothesis has been put forward, suggesting that strong
magnetic fields may have bloated these stars, either through
chromospheric activity \citep[e.g.,][]{Ribas06, Torres06,
LopezMorales07, Chabrier07} or through magnetically induced
convective disruption \citep{Torres06}. Such strong magnetic
fields are expected to be formed by the dynamo mechanism of
rapidly rotating stars.\footnote{Dynamo theory predicts that this
mechanism operates only in partially convective stars. However,
the strong magnetic activity observed in fully convective low-mass
stars indicates that they also possess a mechanism for generating
strong magnetic fields \citep[see][and references
therein]{Browning07}.} To test this hypothesis, one needs to
measure both the masses and radii of low-mass stars, which thus
far can be done most accurately with eclipsing binary (EB)
systems. However, all well characterized low-mass main-sequence
EBs have orbital periods shorter than 3 days (see
Table~\ref{ch6_tablePreviousEBs}) and are therefore expected to
have synchronization timescales shorter than $\sim$100 Myr
\citep[][see Figure~\ref{ch6_figTsync} and further description in
\S\,\ref{ch6_secPhysical}]{Zahn77, Zahn94}. As a result of these
short periods and synchronization timescales, the rotations of
these binary components are expected to have accelerated to the
point that they now match the rapid angular velocity of their
orbits. With such rapid rotations, these binary components could
have a wide range of dynamo-induced magnetic field strengths. To
better constrain current stellar models, we set out to find
systems with slowly rotating components. Such systems would
presumably have comparably weak magnetic fields, thus being more
consistent with the model assumptions. Furthermore, by comparing
the mass-radius relations of binary components with
well-determined levels of magnetic activity, one could test
various magnetic disruption models.

We note here that in addition to EB analysis, long-baseline
optical interferometry has also been used recently to measure the
radii of nearby low-mass stars \citep{Lane01, Segransan03,
Berger06}.  While these stars are single and are therefore
expected to rotate slowly, their masses can only be estimated
through empirical mass-luminosity relations or other indirect
methods. Those determinations are thus less fundamental, in a
sense, and arguably of lesser value for accurately constraining
stellar models and testing the magnetic disruption hypothesis.

\begin{deluxetable}{lcl}
\tabletypesize{\tiny}
\tablecaption{Periods of Well-characterized
Main-Sequence EBs with Both Component Masses below 0.8 $M_{\sun}$}
\tablewidth{0pt}
\tablehead{\colhead{Name} & \colhead{Period (days)} & \colhead{Citation}}
\startdata
OGLE~BW5~V38\tablenotemark{a} & 0.198 & \citet{Maceroni04}\\
RR~Cae\tablenotemark{b}     & 0.304 & \citet{Maxted07}\\
NSVS01031772 & 0.368 & \citet{LopezMorales06}\\
SDSS-MEB-1   & 0.407 & \citet{Blake07}\\
GU~Boo       & 0.489 & \citet{LopezMorales05}\\
2MASS~J04463285+1901432 & 0.619 & \citet{Hebb06}\\
YY~Gem       & 0.814 & \citet{Kron52, Torres02}\\
T-Her0-07621 & 1.121 & \citet{Creevey05}\\
CM~Dra       & 1.268 & \citet{Lacy77a, Metcalfe96}\\
UNSW-TR-2    & 2.117 & \citet{Young06}\\
2MASS~J01542930+0053266 & 2.639 & \citet{Becker08}\\
CU~Cnc       & 2.771 & \citet{Delfosse99, Ribas03}\\
\enddata
\label{ch6_tablePreviousEBs}
\tablenotetext{a}{This binary might
not be detached, as its components seem to be undergoing
significant mutual heating and tidal interactions due to their
proximity ($a=1.355 \pm 0.066 R_{\sun}$).}
\tablenotetext{b}{This
is an unusual case of an EB containing a white dwarf (primary) and
an M dwarf (secondary). As such, the primary component is likely
to have transferred mass to the secondary component, and perhaps
even enveloped it during the red giant phase of its evolution.}
\end{deluxetable}

\section{Initial Photometric Observations}
\label{ch6_secInitial}

T-Lyr1-17236 was first identified as a likely low-mass EB
candidate in the \cite{Devor08} catalog, following a systematic
analysis of the light curves (LCs) within 10 fields of the
Trans-atlantic Exoplanet Survey \citep[TrES;][]{Alonso04}. TrES
employs a network of three automated telescopes to survey
$6\arcdeg \times 6\arcdeg$ fields of view. To avoid potential
systematic noise, we performed our initial search using data from
only one telescope, Sleuth, located at the Palomar Observatory in
Southern California \citep{ODonovan04}, and we combined additional
data at subsequent follow-up stages. Sleuth has a 10-cm physical
aperture and a photometric aperture radius of 30\arcsec. The
number of LCs in each field ranges from 10,405 to 26,495, for a
total of 185,445 LCs.  The LCs consist of $\sim$2000 Sloan
$r$-band photometric measurements binned to a 9-minute cadence.
The calibration of the TrES images, the identification of stars
therein, and the extraction and decorrelation of the LCs are
described elsewhere \citep{Dunham04, Mandushev05, ODonovan06,
ODonovan07}.

An automated pipeline was used to identify and characterize the
EBs among the TrES LCs. This pipeline has been described in detail
in a previous paper \citep{Devor08}. At the heart of this analysis
lie two computational tools: the Detached Eclipsing Binary Light
curve fitter\footnote{The DEBiL source code, utilities, and
example files are available at\newline
http://www.cfa.harvard.edu/$\sim$jdevor/DEBiL.html}
\citep[DEBiL;][] {Devor05}, and the Method for Eclipsing Component
Identification\footnote{The MECI source code and running examples
are available at\newline
http://www.cfa.harvard.edu/$\sim$jdevor/MECI.html}
\citep[MECI;][]{Devor06a, Devor06b}. DEBiL fits each LC to a
$\it{geometric}$ model of a detached EB that consists of two
luminous, limb-darkened spheres that describe a Newtonian two-body
orbit. MECI then incorporated some of the DEBiL results, and
together with 2MASS color information \citep{Skrutskie06}, refit
each LC to a $\it{physical}$ model that is constrained by the
solar metallicity Yonsei-Yale theoretical isochrones \citep{Yi01,
Kim02}. Thus, using only photometric data, the DEBiL/MECI pipeline
provided initial estimates of the absolute physical properties of
each EB. These estimates were then used to locate promising
candidates for follow-up.

\begin{figure}
\includegraphics[width=5in]{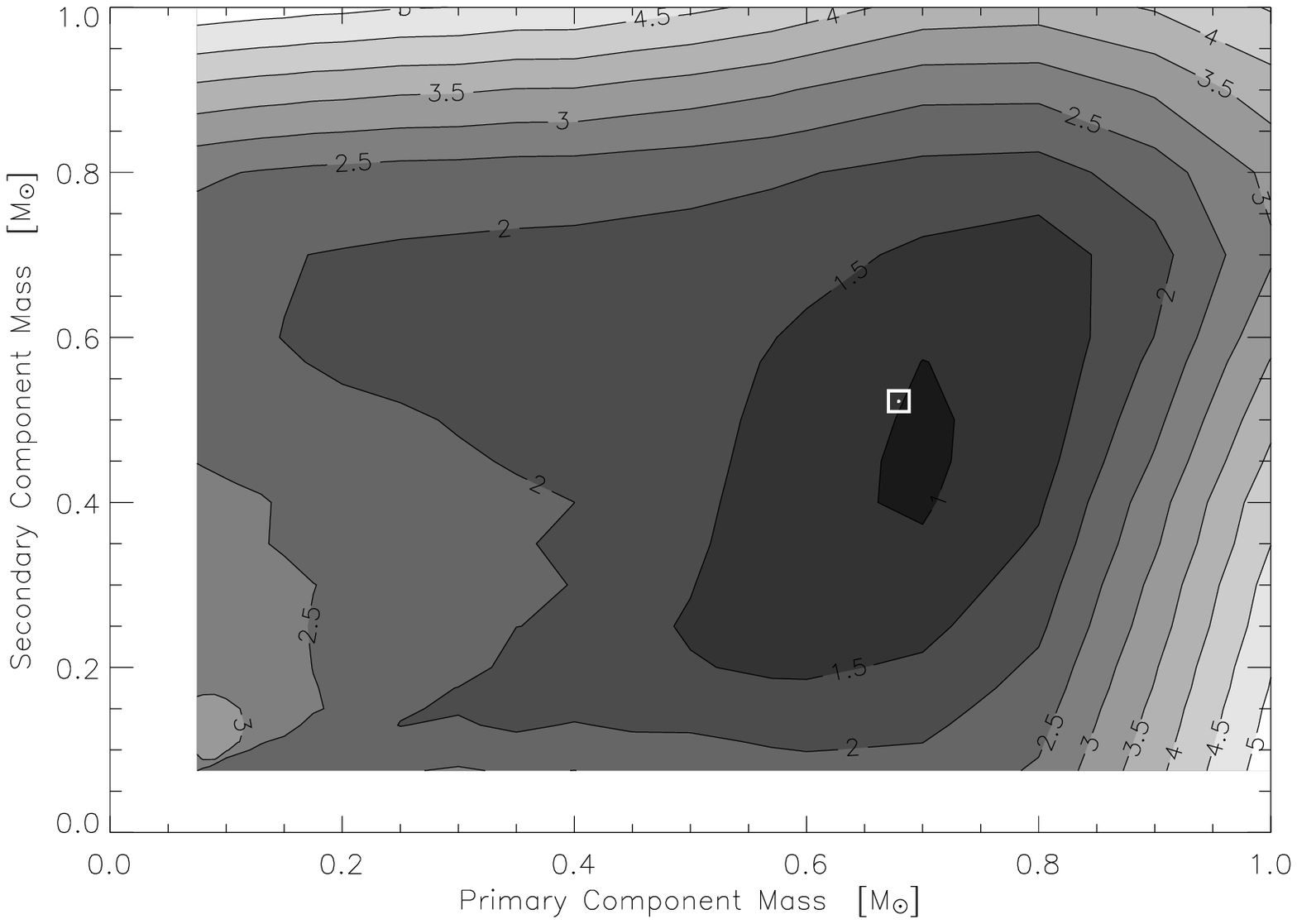}
\caption{Mass-mass likelihood plot for T-Lyr1-17236 created with
MECI, using the \cite{Baraffe98} isochrones for an age of 2.5~Gyr.
This analysis incorporated the $r$-band LC and the 2MASS colors of
the target. The contour lines indicate the weighted reduced
$\chi^2$ values of each component mass pairing, using $w=10$
\citep{Devor06b}. The white point indicates our final mass
estimate from this paper, and the white square approximates our
current mass uncertainties.} \label{ch6_figMECI}
\end{figure}

Using this pipeline a total of 773 EBs were identified within the
TrES dataset. Of these, 427 EBs were both detached and had small
out-of-eclipse distortions, thereby enabling the DEBiL/MECI
pipeline to estimate their component masses. These results,
together with many other properties, are listed for each EB in an
online catalog\footnote{Available at
http://www.cfa.harvard.edu/$\sim$jdevor/Catalog.html}
\citep{Devor08}. Of these characterized EBs, we then identified a
handful of promising long-period low-mass candidates and chose
one, T-Lyr1-17236 ($\alpha_{2000.0}=19^h07^m16.621^s,\
\delta_{2000.0}=+46\arcdeg39'53.21'',\ P=8.429441 \pm 0.000033$
days\,; see Table~\ref{ch6_tableObjectParam} for additional
information), for further follow-up and analysis. As with all of
our low-mass candidates, we repeated the MECI analysis using the
\cite{Baraffe98} solar-metallicity isochrones (with a
mixing-length parameter of $\alpha_{\rm ML} = 1.0$), which are
more accurate than the Yonsei-Yale isochrones in this regime. The
resulting MECI mass-mass likelihood contour plot of T-Lyr1-17236
is shown in Figure~\ref{ch6_figMECI}.  Since the MECI analysis
incorporates data from theoretical stellar models, we cannot use
it to constrain stellar models. Rather, once we identified the
candidate, we followed it up photometrically and
spectroscopically, and used only these follow-up data to derive
the binary's absolute properties.

\begin{deluxetable}{llc}
\tabletypesize{\tiny}
\tablecaption{Catalog Information for T-Lyr1-17236}
\tablewidth{0pt}
\tablehead{\colhead{Source Catalog} & \colhead{Parameter} & \colhead{Value}}
\startdata
2MASS\tablenotemark{a}\ \ \ \ \ & $\alpha$ (J2000.0) & $19^h07^m16.621^s$\\
2MASS & $\delta$ (J2000.0)      & $+46\arcdeg39'53.21''$\\
USNO-B\tablenotemark{b} & $B$ mag & 16.11 $\pm$ 0.2\\
GSC2.3\tablenotemark{c} & $V$ mag & 14.37 $\pm$ 0.28\\
USNO-B & $R$ mag                  & 14.41 $\pm$ 0.2\\
CMC14\tablenotemark{d}  & $r'$ mag & 14.073 $\pm$ 0.029\\
2MASS  & $J$ mag  & 12.019 $\pm$ 0.015\\
2MASS & $H$ mag         & 11.399 $\pm$ 0.015\\
2MASS & $K_s$ mag       & 11.235 $\pm$ 0.015\\
USNO-B & $\mu_\alpha$ (${\rm mas\,yr^{-1}}$) &  $-$2 $\pm$ 3\\
USNO-B & $\mu_\delta$ (${\rm mas\,yr^{-1}}$) & $-$28 $\pm$ 2\\
2MASS & Identification    & 19071662+4639532\\
CMC14 & Identification    & 190716.6+463953\\
GSC2.3 & Identification   & N2EH033540\\
USNO-B & Identification   & 1366-0314305
\enddata
\label{ch6_tableObjectParam}
\tablenotetext{a}{Two Micron All Sky
Survey \citep{Skrutskie06}.}
\tablenotetext{b}{U.S. Naval
Observatory photographic sky survey \citep{Monet03}.}
\tablenotetext{c}{Guide Star Catalog ver. 2.3.2
\citep{Morrison01}.}
\tablenotetext{d}{Carlsberg Meridian Catalog
14 \citep{Evans02}.}
\end{deluxetable}

\section{Follow-up Photometric Observations}

In order to characterize T-Lyr1-17236 we combined photometric data
from four telescopes: (1) Sleuth and (2) the Planet Search Survey
Telescope \citep[PSST;][]{Dunham04} of the TrES network, (3) the
Instituto de Astrof\'{\i}sica de Canarias telescope
\citep[IAC80;][]{Galan87}, and (4) the Hungarian Automated
Telescope Network \citep[HATNet;][]{Bakos04}. With the exception
of the IAC80, we obtained our photometric data from archived
survey datasets that were intended for locating exoplanets.

\begin{figure}
\includegraphics[width=5in]{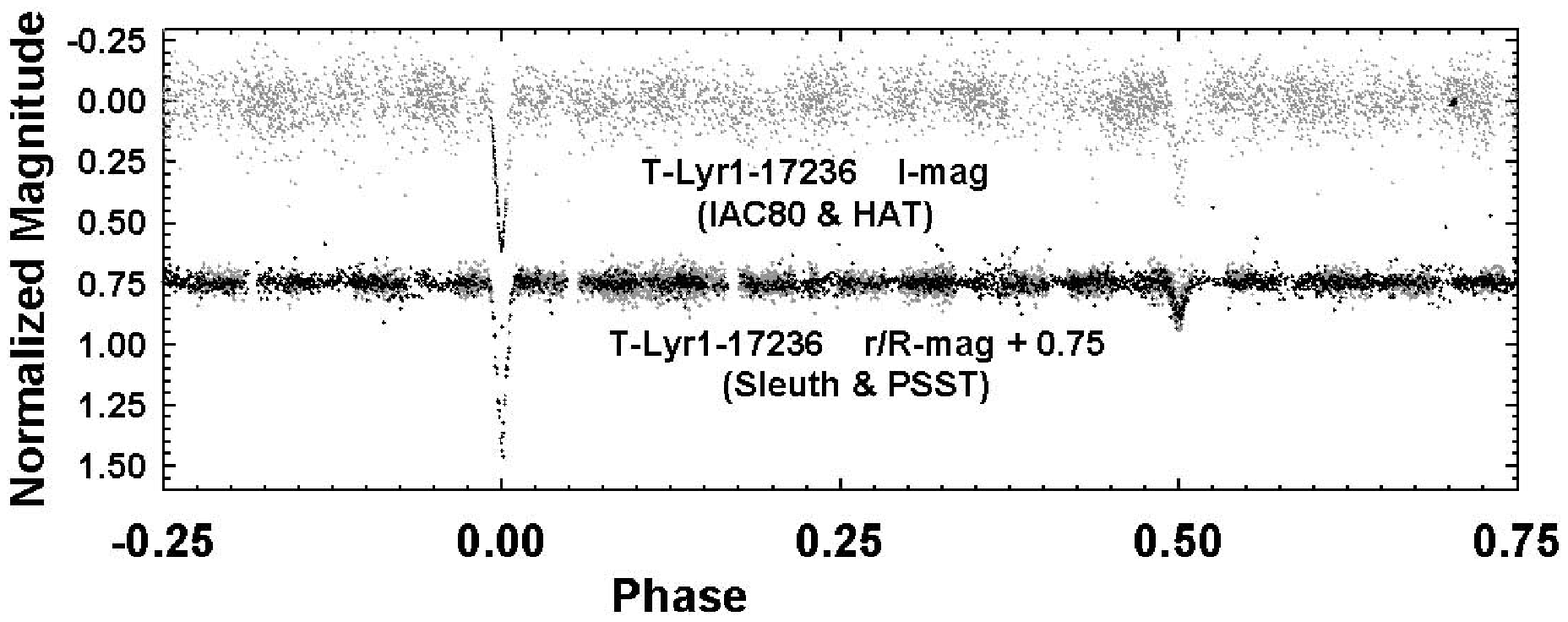}
\caption{Phased LCs of T-Lyr1-17236. The top curve is from the
IAC80 (black symbols) and HATNet (gray symbols) telescopes, both
of which observed in the $I$-band. Note the tight cluster of IAC80
observations near phase 0.7; these points determine the IAC80 LC
zero point. The bottom curve is from the Sleuth (black symbols)
and PSST (gray symbols) telescopes, which observe, respectively,
in the $r$-band and $R$-band. The secondary eclipse is about twice
as deep in the $I$-band as it is in the $r$- or $R$-bands,
indicating that the secondary component is significantly redder
and therefore cooler than the primary.} \label{ch6_figLC_lyr17236}
\end{figure}

\begin{figure}
\includegraphics[width=5in]{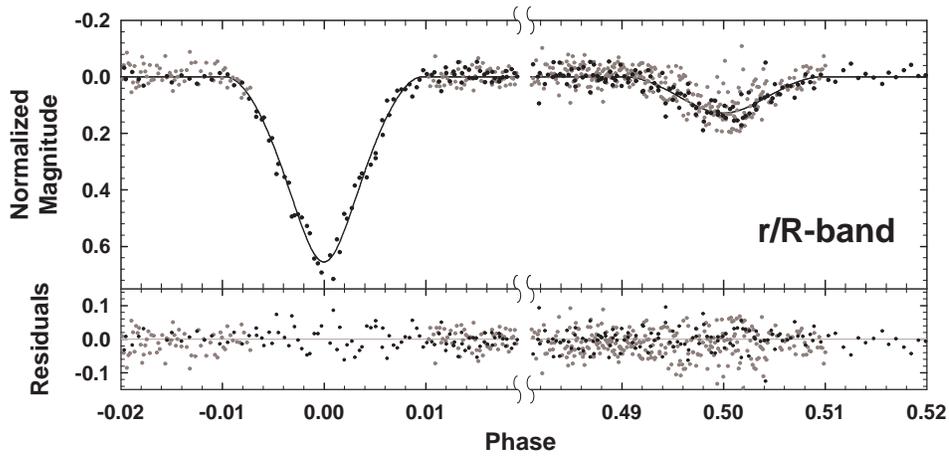}
\caption{Enlargement of the eclipse phases in the
LC of T-Lyr1-17236, as recorded by the Sleuth (black symbols) and PSST
(gray symbols) telescopes ($r$-band and $R$-band, respectively). The solid
line shows the best-fit JKTEBOP model, for which the residuals are
displayed at the bottom.}
\label{ch6_figEclipse_r}
\end{figure}

As part of the TrES network (see \S\,\ref{ch6_secInitial}), Sleuth
and PSST are operated similarly. However, PSST, which is located
at the Lowell Observatory in Arizona, observes in the Johnson
$R$-band whereas Sleuth observes in the Sloan $r$-band (see
Figures~\ref{ch6_figLC_lyr17236} and \ref{ch6_figEclipse_r}).
Furthermore, PSST has a 20\arcsec\ photometric aperture radius
compared to Sleuth's 30\arcsec\ radius, which provides PSST with a
higher resolving power than Sleuth. However, the smaller aperture
of PSST also causes it to have noisier photometry, with an RMS of
0.031 mag for T-Lyr1-17236, compared to the Sleuth photometry,
which has an RMS of 0.028 mag. Though these differences are small,
they would have affected our analysis. We therefore chose not to
use the PSST data for fitting the photometric model, although we
did use them to improve the determination of the orbital period
and the epoch of eclipse (see \S\,\ref{ch6_secAnalysis}).

\begin{figure}
\includegraphics[width=5in]{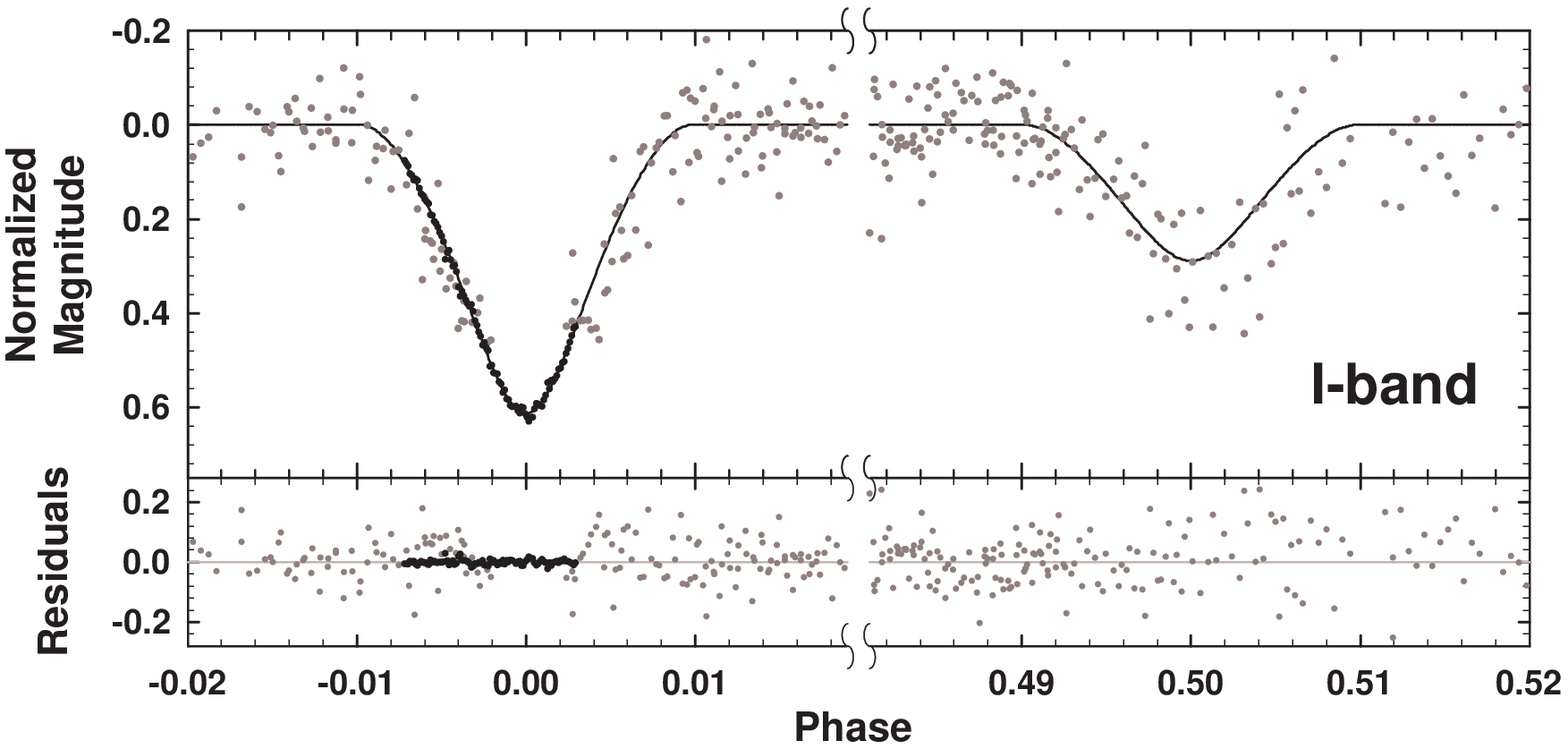}
\caption{Enlargement of the eclipse phases in the
LC of T-Lyr1-17236, as recorded by the IAC80 (black symbols) and the
HATNet (gray symbols) telescopes ($I$-band). The solid line shows the
best-fit JKTEBOP model, for which the residuals are displayed at
the bottom.}
\label{ch6_figEclipse_I}
\end{figure}

In an effort to better constrain the eclipses of T-Lyr1-17236, we
obtained data from the IAC80, an 82-cm aperture telescope with a
$14' \times 14'$ field of view, located at the Observatorio del
Teide in the Canary Islands. We produced an $I$-band LC at a
1.3-minute cadence using the 1024$\times$1024-pixel Tromso CCD
Photometer (TCP), resulting in 0.008 mag RMS photometry for
T-Lyr1-17236. Unfortunately, we were only able to observe a
primary eclipse with the IAC80. We therefore incorporated archival
HATNet observations so as to provide coverage of the secondary
eclipse in a similar bandpass (see Figures~\ref{ch6_figLC_lyr17236}
and \ref{ch6_figEclipse_I}).

HATNet is a network of six 11-cm aperture, fully-automated
telescopes (HATs) located at the F. L. Whipple Observatory in
Arizona and the Submillimeter Array site atop Mauna Kea, Hawaii.
The HATs have an $8\arcdeg \times 8\arcdeg$ field of view, a
response that peaks in the $I$-band, and they operate at a
5.5-minute cadence. To reduce the photometric noise, the HAT point
spread function (PSF) is broadened to an $\sim$15\arcsec\ aperture
radius through microstepping \citep{Bakos02}. Even so, the HATNet
photometric RMS for T-Lyr1-17236 was comparably large, at 0.084
mag. Nevertheless, to provide more complete coverage of the
primary and secondary eclipses in the $I$-band, we combined the
IAC80 observations with data from HAT-7 (Whipple Observatory) and
HAT-8 (Mauna Kea). Due to the very different characteristics of
these two systems, however, we chose not to adopt any of the model
parameters derived from these data, and we only used these results
as an independent confirmation of the Sleuth $r$-band LC analysis.

\section{Spectroscopic Observations}
\label{ch6_secSpecObs}

T-Lyr1-17236 was observed spectroscopically with two instruments:
the Near-Infrared Spectrometer \citep[NIRSPEC;][]{McLean98,
McLean00} at the W.~M.~Keck Observatory in Hawaii, and the
Tillinghast Reflector Echelle Spectrograph
\citep[TRES;][]{Szentgyorgyi07}, installed on the 1.5~m
Tillinghast telescope at the F.\ L.\ Whipple Observatory in
Arizona.

NIRSPEC was operated using a 3-pixel slit (0.432\arcsec) and an N7
blocking filter, thus producing a spectral resolving power of
\textsf{R} $= \lambda/\Delta\lambda \simeq 25,\!000$. The duration
of the exposures, which ranged from 420 to 900 seconds, was
adjusted according to observing conditions. The spectra were
gathered in two consecutive nods, producing a total of five
NIRSPEC nod pairs.  The nods of each pair were then subtracted one
from the other, removing much of the sky emission. We extracted
the spectra of both nods using the optimal extraction procedure
outlined in \citet{Horne86}, and then co-added the two resulting
one-dimensional spectra. We calibrated the wavelengths of the
resulting spectrum using its atmospheric telluric features, and
then corrected for both the telluric absorption and the blaze of
the spectrograph by dividing this spectrum by the spectrum of an
A0V-type star (HR~5511). Finally, we cross-correlated each
spectrum with the spectrum of an M0.5V template star (GJ~182). To
this end, we used a single NIRSPEC order (2290--2320~nm), which is
within the $K$-band, and has a scale of $0.0336\:$nm\,pixel$^{-1}$
at its center. This order covers the CO(2-0) band head, which
includes a rich forest of R-branch transition lines, as well as
many telluric absorption features due to methane in the Earth's
atmosphere. The advantages offered by this spectral region and the
details of the instrument setup are described in \citet{Blake08}.

TRES is a high-resolution fiber-fed optical echelle spectrograph
designed to cover a large range of wavelengths (390--934~nm) in 51
orders. We employed the medium-size fiber (2.3\arcsec) so as to
cover the full stellar PSF, while providing a spectral resolving
power of \textsf{R} $\simeq 47,\!000$. Following each of our three
900--1000 second exposures, the TRES data were read from a
4638$\times$1090-pixel CCD, which we set to a 2$\times$2 binning
mode for a more rapid read-out. We then used a dedicated IRAF
toolset to process and extract 51 spectral orders simultaneously,
ultimately producing 2319 data points along each order. The IRAF
processing of the TRES data involved merging the mosaic FITS
files, removing cosmic ray hits, flattening fringing effects, and
then extracting the orders. We wavelength-calibrated the TRES
spectra using Thorium-Argon (ThAr) exposures, and then corrected
the telluric absorption and spectroscopic blazing by dividing each
spectrum by a TRES spectrum of a rapidly-rotating B0IV-type star
(HR~264). Though TRES produces 51 spectral orders, we used only 4
of them, covering wavelengths of 665--720~nm (similar to the
$R$-band), and with a post-binning scale of
$\sim$0.0065$\:$nm\,pixel$^{-1}$. These orders contain a diverse
array of absorption features, including those of TiO, Fe$\:$I,
Ca$\:$I, Ni$\:$I, and Cr$\:$I. We limited ourselves to these
orders because at shorter wavelengths there was insufficient flux
from our red target, while at longer wavelengths the spectra were
dominated by telluric absorption features, produced largely by
terrestrial O$_2$ and H$_2$O. We cross-correlated these four
orders with the corresponding orders of an M1.5V template star
(GJ~15A, also known as GX~And~A) and averaged their
cross-correlation functions. We repeated this final calculation
using the \citet{Zucker03} maximum-likelihood method, which
reproduced our results to within a fraction of their
uncertainties, although with slightly larger errors.\footnote{The
\citet{Zucker03} method is more accurate than simple
cross-correlation averaging for large N. However, because it takes
the absolute value of the correlation, it loses some information
and effectively increases the noise baseline. This increased noise
will negate its advantage when combining a small number of
correlations, as is the case in our TRES analysis ($N = 4$).}

In total, we produced five radial velocity (RV) measurements of
each component with NIRSPEC and three with TRES. In all cases we
were able to measure the RVs of both binary components by
employing a cross-correlation method that transforms the spectra
to Fourier-space using the Lomb-Scargle algorithm \citep{Press92}.
This method allowed us to cross-correlate spectra with arbitrary
sampling, without having to interpolate or resample them onto an
equidistant grid. We then multiplied the Fourier-transformed
target and template spectra, inverse-Fourier-transformed the
product, and normalized it. Since the resulting two peaks in the
cross-correlation functions were always well separated, we were
able to fit each with a parabola, and thus measure their offsets
and widths. The uncertainties of these RVs are somewhat difficult
to determine with our procedures, but tests indicate that they are
approximately 1.0 and 1.4~\kms\ for the primary and secondary in
our NIRSPEC spectra, and about 0.5 and 1.2~\kms\ in our TRES
spectra. These internal errors are adopted below in the
spectroscopic analysis, but they have relatively little effect on
the results. Finally, the RVs were transformed to the barycentric
frame, and the TRES RV measurements were further offset by
$-$2.82~\kms\ in order to place them on the same reference frame
as the NIRSPEC measurements, which were obtained with a different
template (GJ~182). This offset was determined by including it as
an additional free parameter in the Keplerian RV model (see
\S\,\ref{ch6_secAnalysis}). Once the offset was determined, we
held its value fixed in all subsequent analyses. The final
velocities are listed in Table~\ref{ch6_tableLyr17236RVs} and
include this offset. Note that these listed RVs are all relative
to GJ~182, for which \cite{Montes01} have measured the value
$+32.4 \pm 1.0$~\kms.

\begin{deluxetable}{ccccll}
\tabletypesize{\tiny}
\tablecaption{Radial Velocity Measurements for T-Lyr1-17236 in the Barycentric Frame Relative to GJ~182}
\tablewidth{0pt}
\tablehead{\colhead{Epoch (BJD)} &
\colhead{\begin{tabular}{c} Primary RV\\ (\kms) \end{tabular}} &
\colhead{\begin{tabular}{c} Secondary RV\\ (\kms) \end{tabular}} &
\colhead{\begin{tabular}{c} Exposure Time \\ (sec) \end{tabular}}
& \colhead{Template} & \colhead{Instrument}}
\startdata
$2,\!453,\!927.9400$   &  \phn$-$2.87  &  $-$45.24  &  480 & GJ 182  &  NIRSPEC\\
$2,\!453,\!930.9258$   &  $-$68.09  &  \phs38.85    &  900 & GJ 182  &  NIRSPEC\\
$2,\!453,\!946.8846$   &  $-$64.26  &  \phs36.53    &  600 & GJ 182  &  NIRSPEC\\
$2,\!453,\!948.9100$   &  $-$43.45  &  \phs\phn7.03    &  420 & GJ 182  &  NIRSPEC\\
$2,\!454,\!312.7985$   &  \phs\phn7.66    &  $-$57.68  &  480 & GJ 182  &  NIRSPEC\\
$2,\!454,\!372.6179$   &  \phs23.99    &  $-$80.14  &  900 & GJ 15A  &  TRES\\
$2,\!454,\!377.6382$   &  $-$68.03  &  \phs40.10    & 1000 & GJ 15A  &  TRES\\
$2,\!454,\!377.6624$   &  $-$67.97  &  \phs39.73    & 1000 & GJ 15A  &
TRES
\enddata
\label{ch6_tableLyr17236RVs}
\end{deluxetable}

\section{Orbital Analysis}
\label{ch6_secAnalysis}

\begin{figure}
\includegraphics[width=5in]{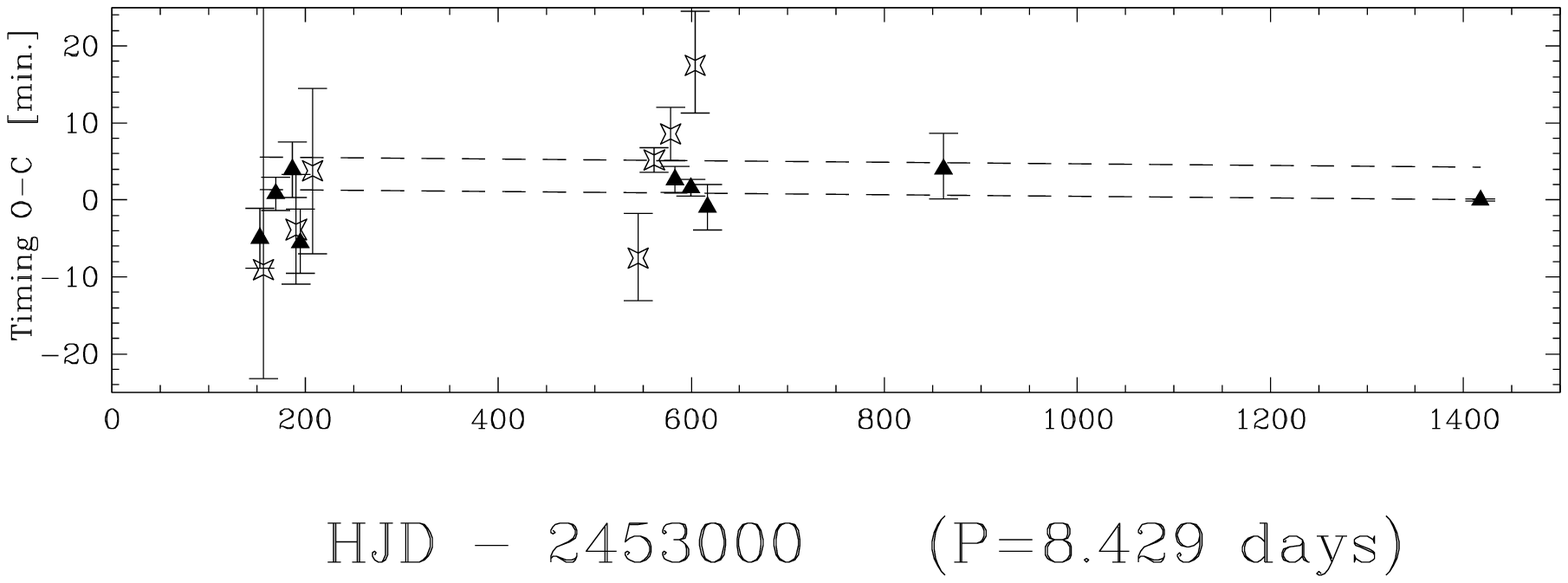}
\caption{Eclipse timing ($O\!-\!C$) measurements
of T-Lyr1-17236. The triangles indicate primary eclipses,
and the four-pointed stars indicate secondary eclipses. The large
error bars are generally due to eclipses that are constrained by
only a few observations, or for which only the ingress or egress
was observed.  The cluster of points at the very left (HJD $<
2,\!453,\!300$) are measurements from HATNet, the single data
point at HJD $2,\!454,\!417$ is from the IAC80, and the remaining
data are from Sleuth and PSST. The two parallel dashed lines
indicate the expected $O\!-\!C$ location of the primary (bottom)
and secondary (top) eclipses, in the best-fit eccentric model
($|e\cos\omega| \simeq 0.0005$). This eccentric model provides
only a very small improvement in the fit compared to the circular
model (F-test: $\chi^2_{\nu,circ} / \chi^2_{\nu,ecc} \simeq 1.29$,
indicating a $p \simeq 0.33$ significance).}
\label{ch6_figTiming}
\end{figure}

We began our analysis by determining the orbital period ($P$) and
the epoch of primary eclipse ($t_0$) and constraining the
eccentricity ($e$) of T-Lyr1-17236 through eclipse timing. The
times of eclipse determined from our photometric observations are
listed in Table~\ref{ch6_tableLyr17236Timing}. Since our data span
3.5 years, we were able to determine the period to an accuracy of
3 seconds (see Table~\ref{ch6_tablePhotoParam}). To estimate the
binary's eccentricity, we first measured the observed minus
calculated ($O\!-\!C$) timing difference between the primary and
secondary eclipses in all available LCs, which provided an upper
bound of $|e\cos\omega| \lesssim 0.0008$, where $\omega$ is the
argument of periastron (see Figure~\ref{ch6_figTiming}). Though
$\omega$ and $e$ cannot be determined separately in this way, this
result indicates that the orbit of T-Lyr1-17236 is likely to be
circular or very nearly so. This conclusion is further supported
by a weaker upper limit of $|e\sin\omega| \lesssim 0.06$, obtained
through preliminary LC model fitting (see below). Theoretical
estimates \citep{Zahn77, Zahn78, Zahn94} of this binary suggest a
circularization timescale of $t_{\rm circ} \simeq 390$~Gyr
\citep[see also][]{Devor08}. Being many times the age of the
binary, this long timescale suggests that T-Lyr1-17236 formed in a
circular orbit. However, this timescale value is an instantaneous
estimate for the current epoch, and is likely to have been
significantly different in the past \citep[see][and references
therein]{Zahn89, Mazeh08}. Therefore, it is quite possible that
the binary circularized while it was in the pre-main-sequence;
however, to the extent that this theory is correct, it is unlikely
to have circularized once it settled on the main-sequence.

\begin{deluxetable}{cccc}
\tabletypesize{\tiny}
\tablecaption{Eclipse Timings Measured for T-Lyr1-17236}
\tablewidth{0pt}
\tablehead{\colhead{Eclipse Type} & \colhead{Epoch (HJD)} & \colhead{O-C (sec)} & \colhead{Data Source}}
\startdata
Primary    & $2,\!453,\!152.96121$ &   $-299^{+232}_{-236}$  & HATNet\\
Secondary  & $2,\!453,\!157.17593$ &   $-546^{+6868}_{-849}$ & HATNet\\
Primary    & $2,\!453,\!169.82009$ &   $48^{+126}_{-131}$ & HATNet\\
Primary    & $2,\!453,\!186.67897$ &   $237^{+214}_{-221}$ & HATNet\\
Secondary  & $2,\!453,\!190.89369$ &   $-231^{+431}_{-423}$ & HATNet\\
Primary    & $2,\!453,\!195.10841$ &   $-333^{+263}_{-238}$ & HATNet\\
Secondary  & $2,\!453,\!207.75258$ &   $225^{+642}_{-648}$ & HATNet\\
Secondary  & $2,\!453,\!544.93022$ &   $-452^{+346}_{-332}$ & Sleuth \\
Secondary  & $2,\!453,\!561.78910$ &   $312^{+97}_{-98}$ & Sleuth + PSST\\
Secondary  & $2,\!453,\!578.64798$ &   $515^{+206}_{-208}$ & Sleuth\\
Primary    & $2,\!453,\!582.86270$ &   $159^{+99}_{-98}$ & Sleuth\\
Primary    & $2,\!453,\!599.72158$ &   $94^{+64}_{-64}$ & Sleuth\\
Secondary  & $2,\!453,\!603.93630$ &   $1047^{+424}_{-371}$ & Sleuth + PSST\\
Primary    & $2,\!453,\!616.58046$ &   $-57^{+175}_{-175}$ & Sleuth\\
Primary    & $2,\!453,\!861.03425$ &   $238^{+280}_{-233}$ & PSST\\
Primary    & $2,\!454,\!417.37736$ &   $-1^{+10}_{-10}$ & IAC80
\enddata
\label{ch6_tableLyr17236Timing}
\end{deluxetable}

\begin{deluxetable}{lcc}
\tabletypesize{\tiny}
\tablecaption{Photometric Parameters of T-Lyr1-17236}
\tablewidth{0pt}
\tablehead{\colhead{Parameter} & \colhead{Symbol} & \colhead{Value}}
\startdata
Period (days)               & $P$   & 8.429441 $\pm$ 0.000033\\
Epoch of eclipse (HJD)      & $t_0$ & 2453700.87725 $\pm$ 0.00041\\
Primary fractional radius   & $r_A$ & 0.0342 $\pm$ 0.0023\\
Secondary fractional radius & $r_B$ & 0.0283 $\pm$ 0.0028\\
Orbital inclination [deg]   & $i$   & 89.02 $\pm$ 0.26\\
Eccentricity                & $e$   & 0.0 (fixed)\\
Sum of fractional radii     & $r_A+r_B$ & 0.06256 $\pm$ 0.00095\\
Ratio of radii ($R_B/R_A$)  & $k$   & 0.83 $\pm$ 0.15\\
Light ratio ($r$-band)     & $L_B/L_A$ &   0.173 $\pm$ 0.073\\
Surface brightness ratio ($r$-band) & $J_B/J_A$ & 0.2525 $\pm$
0.0099
\enddata
\label{ch6_tablePhotoParam}
\end{deluxetable}

\begin{figure}
\includegraphics[width=5in]{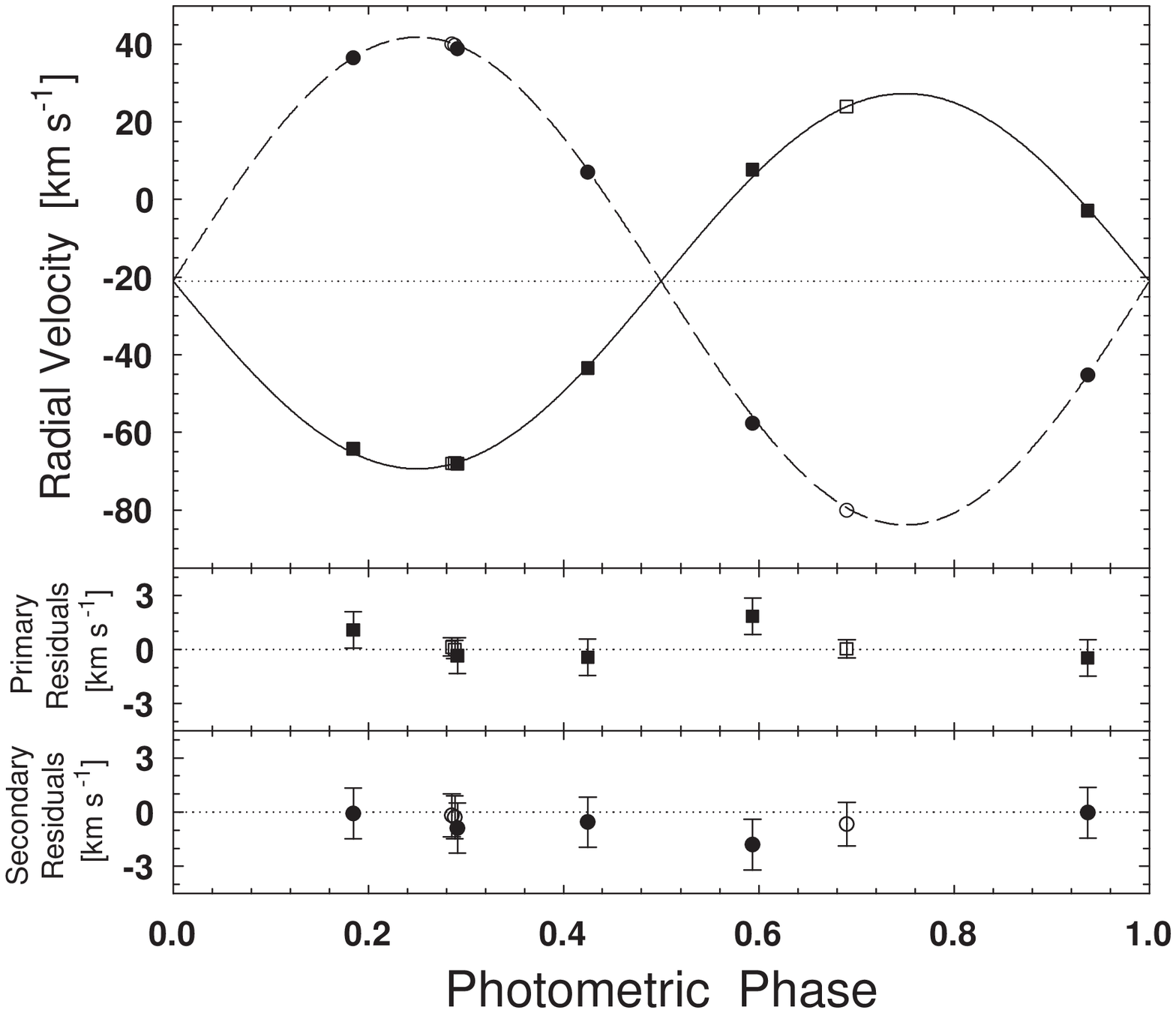}
\caption{RV measurements of
T-Lyr1-17236, relative to GJ~182, shown as a function of orbital
phase. The velocities of the primary component are represented
with squares, and those of the secondary with circles. The filled
symbols correspond to data taken with NIRSPEC, and the open
symbols represent TRES measurements. Residuals from the model fit
are shown in the bottom panels for the primary and secondary components.}
\label{ch6_figPhasedRV}
\end{figure}

A Keplerian model was fit to the RVs to determine the elements of
the spectroscopic orbit of T-Lyr1-17236. We assumed the
eccentricity to be zero based on the evidence above and the lack
of any indications to the contrary from preliminary spectroscopic
solutions. The period and $t_0$ were held fixed at the values
determined above. We solved simultaneously for the velocity
semi-amplitudes of the components ($K_{A,B}$) and the RV of their
center of mass ($V_\gamma$). The results are shown graphically in
Figure~\ref{ch6_figPhasedRV}, and the elements are listed in
Table~\ref{ch6_tableSpecParam}. The minimum masses $M_{A,B} \sin^3
i$ are formally determined to better than 2\%. However, because of
the small number of observations ($N = 8$), the possibility of
systematic errors cannot be ruled out and further observations are
encouraged to confirm the accuracy of these results.

\begin{deluxetable}{lcc}
\tabletypesize{\tiny}
\tablecaption{Spectroscopic Parameters of T-Lyr1-17236}
\tablewidth{0pt}
\tablehead{\colhead{Parameter} & \colhead{Symbol} & \colhead{Value}}
\startdata
Primary RV semi-amplitude ($\kms$)   & $K_A$          & 48.36 $\pm$ 0.23\phn\\
Secondary RV semi-amplitude ($\kms$) & $K_B$          & 62.86 $\pm$ 0.46\phn\\
Barycentric RV, relative to GJ~182\tablenotemark{a}\ \ ($\kms$) & $V_\gamma$  & $-$21.01 $\pm$ 0.18\phs\\
Binary separation with projection factor ($R_{\sun}$)      & $a \sin i$     & 18.526 $\pm$ 0.083\phn\\
Primary mass with projection factor ($M_{\sun}$)           & $M_A \sin^3 i$ & 0.6792 $\pm$ 0.0107\\
Secondary mass with projection factor ($M_{\sun}$)         & $M_B \sin^3 i$ & 0.5224 $\pm$ 0.0061\\
Mass ratio ($M_B/M_A$)                                     & $q$
& 0.7692 $\pm$ 0.0069
\enddata
\label{ch6_tableSpecParam}
\tablenotetext{a}{\citet{Montes01} list the RV of GJ~182 as $+32.4 \pm 1.0$~\kms.}
\end{deluxetable}

We then proceeded to find the remaining photometric parameters of
T-Lyr1-17236. To this end, we analyzed the Sleuth $r$-band LC
using JKTEBOP \citep{Southworth04a, Southworth04b}, a LC modeling
program based on the EBOP light curve generator \citep{Nelson72,
Etzel81, Popper81b}. We assumed a circular orbit, as before, a mass
ratio of $q = 0.7692$ from the spectroscopic model, and the period
determined above. We solved simultaneously for the orbital
inclination ($i$), the fractional radii ($r_{A,B}$), the central
surface brightness ratio of the secondary in units of the primary
($J$), the time of primary eclipse ($t_0$), and the out-of-eclipse
magnitude (zero point). We estimated the uncertainties of the
fitted parameters by evaluating the distribution generated by 1000
Monte Carlo simulations \citep{Southworth05}.

\begin{figure}
\includegraphics[width=4.5in]{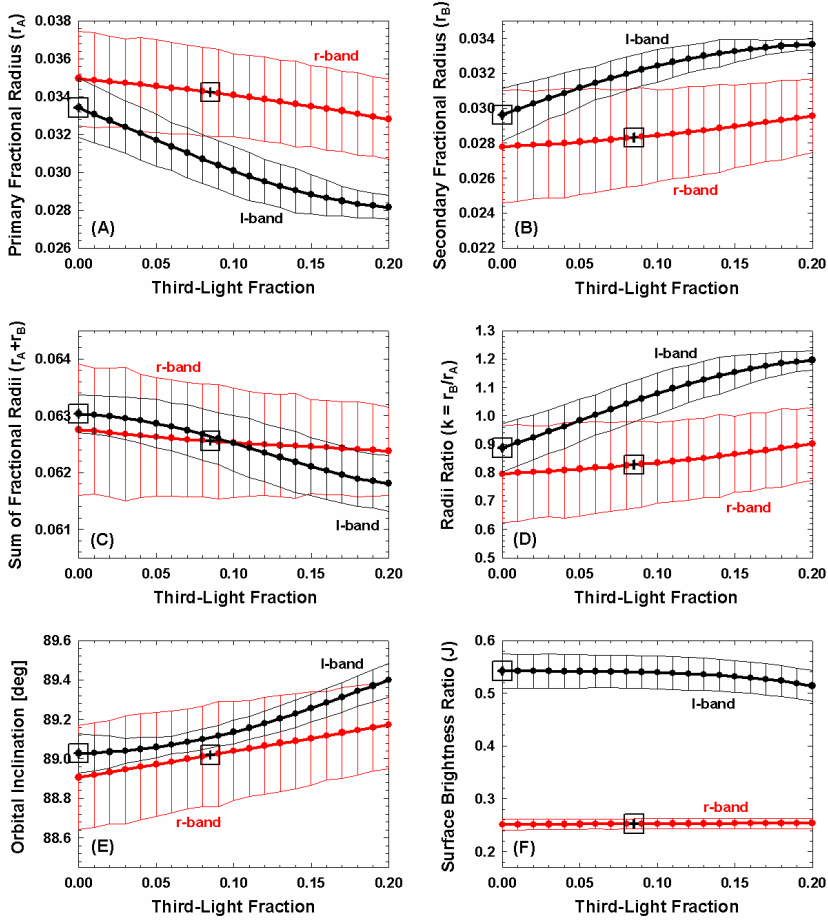}
\caption{JKTEBOP parameter
fits over a range of values for the third light fraction. The
panels show the best-fit values and uncertainties for (A) the
fractional radii of the primary ($r_A$) and (B) secondary ($r_B$)
components; (C) the sum of the fractional radii ($r_A + r_B$) and
(D) the radius ratio ($k = r_B / r_A$); (E) the binary orbital
inclination ($i$); and (F) the central surface brightness ratio
($J$, secondary over primary). Note that in contrast to the other
panels, (F) shows distinct values for the $I$- and $r$-band
LCs. This is expected, since the two components have different
colors, and therefore different relative fluxes through different
filters. In all cases the estimated third light fractions for the
$r$-band and the $I$-band LCs are indicated by squares.}
\label{ch6_fig3light}
\end{figure}

Because of the large photometric aperture of Sleuth, the presence
of significant contamination from the light of additional stars is
a distinct possibility. Unfortunately, due to its degeneracy with
the orbital inclination and the fractional radii, we were not able
to simultaneously determine the fractional third light of the
system ($l_3$). We therefore sequentially refit the LC model
parameters with fixed fractional third-light values ranging from 0
to 0.2 (see Figure~\ref{ch6_fig3light}). We repeated this routine with
the $I$-band IAC80/HATNet LC as well, although these results were
not used because of their larger uncertainties. We obtained an
external estimate of the third-light fraction affecting the Sleuth
observations using the USNO-B catalog \citep{Monet03}, which lists
two dim objects within 30\arcsec\ of T-Lyr1-17236 (USNO-B1.0
1366-0314297 and 1366-0314302). Assuming that these objects are
completely blended into T-Lyr1-17236, we expect an $R$-band
third-light fraction of $l_3=0.085 \pm 0.018$, and we adopted this
value for the $r$-band LC. Fortunately, the fitted parameters are
quite insensitive to third light, so that the uncertainty in $l_3$
only moderately increases their uncertainties.  No objects were
listed within the smaller photometric apertures of either IAC80 or
HATNet, so we conclude that the $I$-band LC should have little or
no third-light contamination. It is important to note that these
third-light estimates assume that there are no further unresolved
luminous objects that are blended with T-Lyr1-17236 (e.g., a
hierarchical tertiary component).  However, the divergence of the
$r$-band and $I$-band solutions at higher third-light fractions
(see Figure~\ref{ch6_fig3light}), and the deep primary eclipse in both
the $r$- and $I$-bands (0.649 and 0.604 mag, respectively),
suggest that if such unresolved objects exist, they are unlikely
to account for more than $\sim$0.1 of the total flux, and
therefore would not bias the fitted results beyond the current
estimated uncertainties. The final results of our LC fits are
given in Table~\ref{ch6_tablePhotoParam}.

\section{Physical Parameters}
\label{ch6_secPhysical}

The fundamental parameters of T-Lyr1-17236, such as their absolute
masses and radii, were derived by combining the results of the
spectroscopic analysis (Table~\ref{ch6_tableSpecParam}) with those
from the photometric analysis (Table~\ref{ch6_tablePhotoParam}). These
and other physical properties are listed in
Table~\ref{ch6_tableSystemParam}. Our estimates of the primary and
secondary component masses, $M_A$ = 0.6795 $\pm$ 0.0107
and $M_B$ = 0.5226 $\pm$ 0.0061 $M_{\sun}$, lead us to infer
spectral types of K5V and M0V, respectively, according to
empirical tables \citep{Cox00}. We are not able to make
independent estimates of the effective temperatures of the stars
from the data in hand. This could be done if, for example, we had
individual color indices based on combined light values and light
ratios in two different bands, but we can only derive a reliable
estimate of the light ratio in the $r$-band. The comparison with
stellar evolution models by \cite{Baraffe98} in
\S\,\ref{ch6_secConclusions} suggests primary and secondary component
temperatures of approximately 4150 and 3700~K, respectively,
although the accuracy of these values is difficult to assess.

\begin{deluxetable}{lccc}
\tabletypesize{\tiny}
\tablecaption{System Parameters of T-Lyr1-17236}
\tablewidth{0pt}
\tablehead{\colhead{Parameter} & \colhead{Symbol} & \colhead{Component A} & \colhead{Component B}}
\startdata
Mass ($M_{\sun}$)          & $M$     & 0.6795 $\pm$ 0.0107 & 0.5226 $\pm$ 0.0061\\
Radius ($R_{\sun}$)        & $R$     & 0.634 $\pm$ 0.043   & 0.525 $\pm$ 0.052\\
Log surface gravity (cgs)  & $\log g$& 4.666 $\pm$ 0.059   & 4.718 $\pm$ 0.086\\
Semimajor axis ($10^6$ km) & $a$     & 5.606 $\pm$ 0.027   & 7.288 $\pm$ 0.053 \\
Maximum rotational velocity\tablenotemark{a}\ \ (\kms) & $v \sin i_r$ & $5.6 \pm 2.0$ & $5.1 \pm 2.3$ \\
Synchronized rotational velocity\tablenotemark{a}\ \ (\kms) & $(v \sin i_r)_{\rm sync}$ & $3.81 \pm 0.26$ & $3.15 \pm 0.31$ \\
Absolute visual magnitude\tablenotemark{b}\ \ (mag) & $M_V$ & 8.03 & 9.67 \\
Bolometric luminosity\tablenotemark{b}\ \ ($L_{\sun}$) & $L$ & 0.110 & 0.039 \\
Effective temperature\tablenotemark{b}\ \ (K)  & $T_{\rm eff}$ & 4150 & 3700 \\
Distance\tablenotemark{b}\ \ (pc)             & $D$ &
\multicolumn{2}{c}{230 $\pm$ 20}
\enddata
\tablenotetext{a}{See description in \S\ref{ch6_secPhysical}.}
\tablenotetext{b}{Inferred using stellar evolution models by
\cite{Baraffe98} assuming solar metallicity and an age of
$2.5$~Gyr.}
\label{ch6_tableSystemParam}
\end{deluxetable}

No trigonometric parallax is available for T-Lyr1-17236.  A rough
distance estimate to the system may be made using the $JHK_s$
brightness measurements in the 2MASS Catalog, collected in
Table~\ref{ch6_tableObjectParam}, along with estimates of the absolute
magnitudes. For these we must rely once again on models. The
Galactic latitude of $+16.8\arcdeg$ suggests the possibility of
some interstellar extinction. From the reddening maps of
\cite{Schlegel98} we infer $E(B-V) \simeq 0.07$ in the direction
of the object (total reddening), which corresponds to extinctions
of $A(J) \simeq 0.061$, $A(H) \simeq 0.038$, and $A(K) \simeq
0.011$, assuming $R_V = 3.1$ \citep{Cox00}. Under the further
assumption that this extinction applies to T-Lyr1-17236, we derive
a mean distance of $230 \pm 20$~pc, after conversion of the
near-infrared magnitudes in the CIT system from the
\cite{Baraffe98} models to the 2MASS system, following
\cite{Carpenter01}. With the proper motion components from the
USNO-B catalog listed in Table~\ref{ch6_tableObjectParam}, the
center-of-mass velocity $V_\gamma$ from the spectroscopic solution
corrected for the velocity of GJ~182 \citep{Montes01}, and the
distance above, we infer space velocity components in the Galactic
frame of ($U$,$V$,$W$) $\simeq$ ($+41$,$+21$,$+2$)~\kms, where $U$
points in the direction of the Galactic center.

Because of the relevance of the rotational velocities of the stars
for the interpretation of the chromospheric activity results of
\S\,\ref{ch6_secActivity}, we have made an effort here to measure the
rotational broadening of both components from the widths of the
cross-correlation functions derived from our TRES spectra. We rely
on the fact that to first order, the width of a cross-correlation
peak is approximately equal to the quadrature sum of the line
broadening of the two spectra. We began our estimation procedure
by finding the effective resolution of the instrument ($\sigma_i$)
in the four TRES orders we used. This was done by auto-correlating
a TRES ThAr spectrum that was taken just before the second
T-Lyr1-17236 observation. We found that the four orders produced
peaks with an average FWHM of $8.90 \pm 0.17$~\kms.  Thus,
assuming that the intrinsic widths of the ThAr emission lines are
negligible compared to the instrumental resolution, we found that
$\sigma_i = 6.29 \pm 0.12$~\kms. This value corresponds to a
spectral resolving power of \textsf{R} $= 47,\!630 \pm 930$, which
is consistent with the TRES specifications. Next, we determined
the intrinsic spectral line broadening of the template star,
GJ~15A ($\sigma_t$). We auto-correlated the template spectrum and
found that it produced peaks with an average FWHM of $9.7 \pm
1.4$~\kms. This value should be equal to $\sqrt{2}(\sigma_i^2 +
\sigma_t^2)^{1/2}$, from which we infer that $\sigma_t = 2.7 \pm
2.5$~\kms. Note that this result is well within the upper bound
provided by \citet{Delfosse98}, following their non-detection of
any rotational broadening in GJ~15A. Using this information, we
can now find the intrinsic spectral line broadening of the
T-Lyr1-17236 components ($\sigma_{A,B}$). The average FWHM of the
primary and secondary peaks, resulting from the cross-correlation
of each observed spectrum of T-Lyr1-17236 against the template,
were measured to be $12.6 \pm 2.0$~\kms\ and $12.0 \pm 2.4$~\kms,
respectively. These widths are expected to be equal to
$[(\sigma_i^2 + \sigma_t^2) + (\sigma_i^2 +
\sigma_{A,B}^2)]^{1/2}$, from which we calculate that $\sigma_A =
8.4 \pm 3.0$ and $\sigma_B = 7.6 \pm 3.8$~\kms.

The rotational profile FWHM expected for a homogeneous stellar
disk is $\sqrt{3}\,v \sin i_r$, where $v$ is the star's equatorial
rotational velocity, and $i_r$ is the inclination of its
rotational axis. Stellar limb darkening, however, will narrow the
rotational profile, thus decreasing the observed FWHM
\citep{Gray92}. Adopting the $R$-band PHOENIX linear limb
darkening coefficients from \cite{Claret98}, we find that the
expected FWHM values for the primary and secondary components of
T-Lyr1-17236 are, respectively, $1.495\,v \sin i_r$ and $1.499\,v
\sin i_r$. Using these results we can set upper bounds to the
components' $v \sin i_r$. These upper bounds represent the
limiting case whereby the spectral line broadening is due entirely
to stellar rotation, and we neglect all other line broadening
mechanisms, such as microturbulence and the Zeeman effect. We thus
determine the maximum rotational velocities of the T-Lyr1-17236
primary and secondary components to be $v \sin i_r = 5.6 \pm
2.0$ and $5.1 \pm 2.3$~\kms, respectively.

An estimate of the timescale for tidal synchronization of the
stars' rotation with their orbital motion may be obtained from
theory following \cite{Zahn77}, and assuming simple power-law
mass-radius-luminosity relations \citep{Cox00}. Thus, for stars
less massive than 1.3~$M_{\sun}$,
\begin{equation}
\label{ch6_eqCircTime}
t_{\rm sync} \simeq 0.00672~{\rm Myr} \;
(k_2 / 0.005)^{-1}q^{-2}(1+q)^2 \left(P / {\rm day}\right)^4
\left(M / M_{\sun} \right)^{-4.82}~,
\end{equation}
where $k_2$ is determined by the structure and dynamics of the
star and can be obtained by interpolating published theoretical
tables \citep{Zahn94}. This calculation leads to timescales of
$t_{\rm sync} \simeq 0.56$ and 1.02~Gyr for the primary and
secondary components of T-Lyr1-17236, respectively, which are much
shorter than the circularization timescale determined in
\S\,\ref{ch6_secAnalysis}. We note that similar to the circularization
timescale, the synchronization timescales estimated above are the
current instantaneous values, and are likely to have changed over
time. The age of the system is undetermined (see
\S\,\ref{ch6_secConclusions}), but assuming its age is at least a few
Gyr, as is typical for field stars, it would not be surprising if
tidal forces between the components had already synchronized their
rotations. This is illustrated in Figure~\ref{ch6_figTsync}, where
T-Lyr1-17236 is shown along with the other systems in
Table~\ref{ch6_tablePreviousEBs} and with curves representing
theoretical estimates of the synchronization timescale as a
function of orbital period.

If we assume that the components are indeed rotationally
synchronized, we can compute their rotational velocities more
accurately using $v_{A,B} = 2\pi R_{A,B}/P$. We thus derive
synchronized velocities of $(v \sin i_r)_{\rm sync} = 3.81 \pm
0.26$ and $3.15 \pm 0.31$~\kms\ for the primary and
secondary components, respectively. These values are slightly
below but still consistent with the maximum rotational velocities
measured above. Thus, observational evidence suggests that the
stars' rotations may well be synchronized with their orbital
motion, although more precise measurements would be needed to
confirm this. Our conclusion from this calculation is that
regardless of whether we assume that the components of
T-Lyr1-17236 are synchronized, their rotational velocities do not
appear to be large.

\section{Chromospheric Activity}
\label{ch6_secActivity}

Our absolute mass and radius determinations for T-Lyr1-17236 offer
the possibility of testing stellar evolution models in the lower
main-sequence, in particular testing the idea that the
discrepancies noted in \S\,\ref{ch6_secIntro} are related to
chromospheric activity and the associated magnetic fields in
systems where the components are rotating relatively rapidly.
Thus, establishing the level of the activity in the system
presented here is of considerable importance. We have shown in
\S\,\ref{ch6_secPhysical} that the relatively long period of
T-Lyr1-17236 ($P \simeq 8.429441$~days) implies that even if the
components are synchronized, their rotational velocities are slow,
and therefore they are not expected to induce a great deal of
chromospheric activity. However, demonstrating that the stars are
indeed inactive requires more direct evidence, given that some
stars of masses similar to these are still found to be quite
active at rotation periods as long as 8 days \citep[see,
e.g.,][]{Pizzolato03}. We present here the constraints available
on the surface activity of T-Lyr1-17236 from its X-ray emission,
optical variations, and spectroscopic indicators.

The present system has no entry in the $\it{ROSAT}$ Faint Source
Catalog \citep{Voges99}, suggesting that the X-ray luminosity,
usually associated with activity, is not strong. Examination of
the original $\it{ROSAT}$ archive images leads to a conservative
upper limit to the X-ray flux of $6.71 \times 10^{-14}$
erg$\thinspace$cm$^{-2}\thinspace$s$^{-1}$ in the energy range
0.1--2.4~keV, and together with information from
Table~\ref{ch6_tableSystemParam}, we infer an upper limit for the
ratio of the X-ray to bolometric luminosity of $\log L_X/L_{\rm
bol} \lesssim -3.13$\,. Values for the four best studied cases of
CM~Dra, YY~Gem, CU~Cnc, and GU~Boo, which are all very active, are
respectively $-3.15$, $-2.88$, $-3.02$, and $-2.90$
\citep[see][]{LopezMorales07}. These are at the level of our limit
or higher, although we do not consider this evidence conclusive.

There are no detectable variations in the $r$-band LC out
of eclipse, within the uncertainties. Such variations would be
expected from activity-related surface features showing
significant contrast with the photospheres. We estimate an upper
limit of $\sim$0.01 mag in $r$ for the night-to-night variations
(see Figure 3). Because the secondary components is significantly
dimmer, it has a weaker variability upper limit of $\sim$0.09 mag.
We note, however, that this evidence for inactivity is not
conclusive either, since the observed photometric variations can
depend significantly on the distribution of spots on the surface.

A number of spectroscopic activity indicators (the Ca~II H and K
lines, H$\alpha$, etc.) should in principle allow a more direct
assessment of the activity level in T-Lyr1-17236.  Unfortunately,
however, the quality of our spectroscopic material in the optical
makes this difficult. The flux in the blue for this very red
system is too low to distinguish the Ca~II H and K lines, and even
at H$\alpha$ the noise is considerable (typical signal-to-noise
ratios at this wavelength are $\sim$12~pixel$^{-1}$). Two of the
three TRES spectra show the H$\alpha$ line in absorption, and the
other appears to show H$\alpha$ in emission. This suggests some
degree of chromospheric activity, although perhaps not at such a
high level as to sustain the emission at all times, as is seen in
other stars. H$\beta$ appears to be in absorption in all three
TRES spectra.

Clearly more spectra with higher signal-to-noise ratios are needed
to better characterize the level of activity, but from the sum of
the evidence above it would not appear that the activity in
T-Lyr1-17236 is as high as in other low-mass EBs studied
previously, thus more closely aligning it with the assumptions of
current standard stellar models. The system may therefore
constitute a useful test case for confirming or refuting the
magnetic disruption hypothesis (see \S\,\ref{ch6_secIntro}), which
predicts that the absolute properties of its slowly rotating
components should match the theoretical models of convective
stars.

\section{Comparison with Models and Conclusions}
\label{ch6_secConclusions}

\begin{figure}
\includegraphics[width=5in]{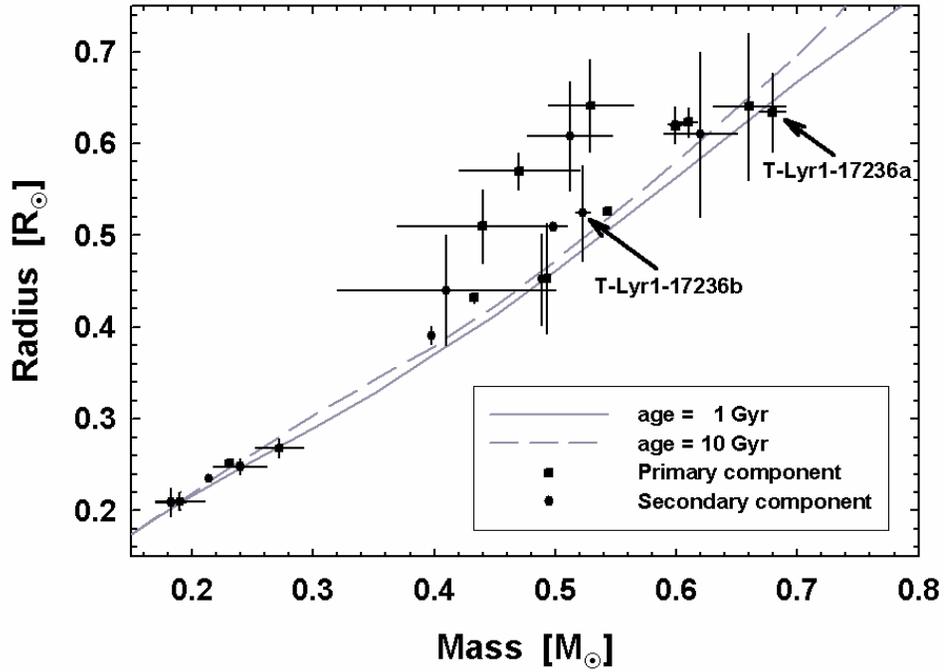}
\caption{Mass-radius diagram for T-Lyr1-17236 and
other low-mass EBs under 0.8~$M_{\sun}$ from
Table~\ref{ch6_tablePreviousEBs}. Theoretical isochrones for solar
metallicity from \cite{Baraffe98} are shown for ages of 1 and 10
Gyr. The components of T-Lyr1-17236 are indicated with arrows.
Most of these binary components (particularly those with smaller
uncertainties) display a systematic offset in which their measured
radii are larger than predicted from models.}
\label{ch6_figIsochrones}
\end{figure}

A comparison with solar-metallicity models by \cite{Baraffe98} for
a mixing-length parameter of $\alpha_{\rm ML} = 1.0$ is presented
in Figure~\ref{ch6_figIsochrones}. Our mass and radius
determinations for T-Lyr1-17236 (see
Table~\ref{ch6_tableSystemParam}) are shown along with those of
the low-mass systems listed in Table~\ref{ch6_tablePreviousEBs}.
The location of the models in this diagram depends only slightly
on age because these stars evolve very slowly. The age of
T-Lyr1-17236 is difficult to establish independently. The space
motions derived in \S\,\ref{ch6_secPhysical} do not associate the
system with any known moving group, and are quite typical of the
thin disk. Thus, all we can say is that it is not likely to be
very old. We display in Figure~\ref{ch6_figIsochrones} two models
for ages of $1$ and $10$~Gyr, which likely bracket the true age of
T-Lyr1-17236. Within the errors, our measurements for the two
components are consistent with the models, which would in
principle support the magnetic disruption hypothesis.
Unfortunately, however, the uncertainties in the radius
measurements ($\sim$7\% and $\sim$10\%) are still large enough
that our statement cannot be made more conclusive. Further
follow-up observations, especially rapid-cadence and precise
photometric measurements during multiple eclipses, should
significantly reduce the uncertainties in the radii and thus
provide far stronger constraints on the theoretical models of
low-mass stars. In addition, higher quality spectroscopic
observations than ours are needed to confirm that the level of
chromospheric activity in the system is relatively low.  If after
such observations, the masses and radii of the T-Lyr1-17236
components remain consistent with the stellar models, then the
magnetic disruption hypothesis will be strengthened. However, if
further observations find that the components of T-Lyr1-17236 are
larger than predicted by current stellar models, as is the case
with most other similar systems investigated in sufficient detail,
then this will provide evidence that additional mechanisms need to
be included in the models of the structure of low-mass
main-sequence stars \citep[see, e.g.,][]{Chabrier07}.

It is important to note here that T-Lyr1-17236 falls within the
field of view of the upcoming NASA $\it{Kepler}$ mission
\citep{Borucki03}. The $\it{Kepler}$ Mission will not return data
for all stars within its field of view; rather, the targets will
be selected by the $\it{Kepler}$ team. We see at least two reasons
why such monitoring of T-Lyr1-17236 would be of significant value.
First, the data would greatly refine the estimates of the physical
parameters of the component stars and might permit a search for
their asteroseismological modes. Second, the data would enable a
search for transits of exoplanets, which are expected to orbit in
the same plane as that defined by the stellar orbits.

Finally, we note that our findings in this paper confirm the
accuracy of the MECI algorithm (see Figure~\ref{ch6_figMECI}), which
can be further used to find additional long-period low-mass EBs,
and indeed a variety of other interesting targets. We have shown
in a recent paper \citep{Devor08} how this can be done with
comparable ease by systematically searching the ever-growing body
of LC survey datasets. We hope that this new approach for locating
rare EBs will motivate additional studies of these vast, largely
untapped datasets, which likely harbor a wealth of information on
the formation, structure, dynamics, and evolution of stars.

\section*{Acknowledgments}

We would like to thank Joel Hartman and Doug Mink for their help
in operating a few of the software analysis tools used for this
paper, and we would like to thank Sarah Dykstra for her editorial
assistance. Valeri Hambaryan provided expert assistance in
examining archival $\it{ROSAT}$ images of T-Lyr1-17236, for which
we are grateful, and we thank the referee for a number of helpful
comments that have improved the paper. G. T. acknowledges partial
support from NSF grant AST-0708229 and NASA's MASSIF SIM Key
Project (BLF57-04). This research has made use of NASA's
Astrophysics Data System Bibliographic Services, as well as the
SIMBAD database operated at CDS, Strasbourg. This publication also
used data products from the Two Micron All Sky Survey, which is a
joint project of the University of Massachusetts and the Infrared
Processing and Analysis Center/California Institute of Technology,
and is funded by NASA and the National Science Foundation. Some of
the data presented herein were obtained at the W. M. Keck
Observatory, which is operated as a scientific partnership among
Caltech, the University of California and NASA. The Observatory
was made possible by the generous financial support of the W. M.
Keck Foundation. The authors wish to recognize and acknowledge the
very significant cultural role and reverence that the summit of
Mauna Kea has always had within the indigenous Hawaiian community.
We are most fortunate to have the opportunity to conduct
observations from this mountain.

\chapter{Conclusions and Future Work
\label{chapter7}}

\title{Conclusions and Future Prospects}

In this thesis, we showed how assembly line astronomical analysis
can be constructed and used. This approach enabled us to produce
orders of magnitude more data than traditional approaches would
have, and with this dramatic increase, we entered a new regime,
where problems once considered difficult become readily
accessible. Such problems include constraining the orbital period
distribution of various binary populations (see Figures
\ref{figPeriod1} and \ref{figPeriodDistrib}), correlating the
masses of the binary components (see Figure \ref{figMassMass}),
and testing tidal circularization theory (see Figure
\ref{figPeriodM1}). Furthermore, we have demonstrated that rare
objects, which would once have required a serendipitous discovery,
can now be found in large numbers through a systematic search. The
bottlenecks of assembly line pipelines are ultimately occur with
those tasks that must be performed manually. However, we have
shown that with the exception of quality control, all the
necessary tasks for the analysis of EBs through their photometric
LCs can be fully automated.

The fastest and most efficient analysis will be of little value if
its results are not reliable. We thus spent considerable effort
demonstrating how each part of our pipeline, namely DEBiL and
MECI, are robust and provide realistic uncertainty estimates.
Though in themselves they are not as accurate as the best manual
methods, when high accuracy measurements are required, one can
follow-up the pipeline results with additional observations, and
manually fine-tune the fitted EB model. This way we are able to
combine both the high-throughput capability of an automated
pipeline with the precision of the traditional approach.

T-Lyr1-17236 was a case in point; this long-period EB was
identified by our pipeline in a systematic search of the TrES
dataset as containing two low-mass components (see Chapter 5). It
was then followed-up photometrically and spectroscopically (see
Table \ref{tableRVsMultiple}), and then reanalyzed manually (see
Chapter 6). The follow-up analysis confirmed the original results
and greatly improved the accuracy of its parameters, thus
establishing T-Lyr1-17236 as the longest-period low-mass EB known
to date. In addition to T-Lyr1-17236, our pipeline identified 11
additional low-mass candidates (see Table \ref{tableLowMass}),
which is remarkable considering that there are only 12 well
characterized main-sequence low-mass EBs currently known (see
Table \ref{ch6_tablePreviousEBs}). Of our 11 candidates, only one
system, T-Dra0-01363, more commonly known as CM Draconis, was
previously known, and was indeed found to be a low-mass EB (0.23 +
0.21 $M_{\sun}$) through the extensive work of \citet{Lacy77a} and
the further parameter refinements of \citet{Metcalfe96}. We
followed-up five of the ten remaining candidates, as well as four
additional low-mass candidates with RV measurements (see Tables
\ref{tableRVsMultiple} and \ref{tableRVsSingle}). Although our
results are preliminary, we were able to confirm the low-mass
status of all the binary candidate with at least three RV
measurements (see \S\ref{secMultipleRVs}). The candidates with
fewer than three RV measurements have very uncertain masses, but
are all consistent with being low-mass (see \S\ref{secSingleRV}).

Considering that they relied only on photometric data, the results
of the MECI analyses were remarkably consistent with the RV
follow-up results. However, we did find a tendency for the MECI
analysis to overestimate the masses of our low-mass binaries. This
bias is probably largely due to the fact that for masses in this
range, the stellar models that MECI incorporates systematically
underestimate the radii of stars. Thus, after inferring a given
EB's fractional radii from its measured eclipse duration, these
stellar models will match it to a mass value that is too large.
Once these theoretical models are further refined, the MECI
analysis should become more accurate.

Though most of the work in this thesis centered on modeling EB LCs
and fitting their parameters, one can reverse this approach and
search for LCs that do not conform to the given model. For
example, in Chapter~5 we located a group of EBs that were flagged
by our pipeline, because they could not be successfully modeled by
MECI due to their very disparate eclipse depths (see Table
\ref{tableInverted} and Figure \ref{figInvertedCat}). We thus
suggested that effectively these systems may not be coeval, as is
assumed by MECI, and that this phenomenon may have come about
through an earlier epoch of mass transfer. Another group that we
identified in Chapter~5 included EBs that have large periodic
perturbations in their LCs (see Table \ref{tableAbnormal}, and
Figures \ref{figAbnormalEBs1} and \ref{figAbnormalEBs2}). Some of
these perturbations may be due to persistent star spots or other
large surface inhomogeneities, however, some perturbations may
require more exotic explanations. We thus conclude that adding an
``other'' category is advisable for any automated pipeline. This
category may at first receive a large number of rejected cases,
however, once these systems are better understood, they can be
further classified until the full taxonomy of the observed
population is revealed. One should be aware, though, that even
with a complete classification of all the LCs there will
inevitably be subtle phenomena that will be overlooked. As an
illustration of this point, we describe two maverick LCs that
would not have been found using our pipeline alone, but rather
required a combination of pipeline results coupled with a
dedicated search. The first, T-Cas0-07656, described in section
\S\ref{secCas07656pulsating}, is an EB that contains a pulsating
component. The second, T-Cyg1-03378, described in section
\S\ref{secCyg03378Timing}, is an EB with large sinusoidal eclipse
timing variations. These two binaries, as well as the aforementioned
low-mass EBs candidates, merit further observation and
investigation. We sincerely hope that others will further pursue
these systems and continue the work begun here.

\section{Low-Mass Candidates with At Least Three RV Measurements}
\label{secMultipleRVs}

Subsequent to the development of our pipeline, we identified a
number of low-mass candidates for further follow-up. From late
2005 until mid 2007, these candidates were observed periodically
using the Near-Infrared Spectrometer [NIRSPEC ; \citet{McLean98,
McLean00}] at the W.~M.~Keck Observatory in Mauna Kea, Hawaii.
Then, in late 2007 they were further observed with the Tillinghast
Reflector Echelle Spectrograph [TRES ; \citet{Szentgyorgyi07}],
installed on the 1.5-meter Tillinghast telescope at the
F.~L.~Whipple Observatory in Arizona. These spectra were processed
and analyzed using the procedure described in Chapter~6, and
converted into RV values. We preferred to make the spectral
observations near the EB's quadrature (i.e., at a phase of $1/4$
or $3/4$) for two related reasons. Firstly, since during this time
each measurement best constrains the RV model, and secondly, since
at this time the spectral lines of the two components are
maximally separated. The latter reason becomes critical when the
components' respective absorption lines overlap, thus lowering the
accuracy of our measurement technique, which simply fits the
absorption line profiles to parabolas. Specifically, such an
overlap would bias the centroid of the two absorption lines
towards one another. One must pay close attention to this issue
when analyzing short-period systems (e.g., T-Tau0-07388; see
\S\ref{subsec_Tau0-07388}), which rotate rapidly and thus have
highly broadened absorption features. Unfortunately, in many cases
we had to compromise and make our observations far from
quadrature. However, we were careful, at the very least, to avoid
making spectral observations during an eclipse, as the partial
occultation of a spinning star can create a significant bias to
its RV measurements, in what is known as the Rossiter-McLaughlin
effect \citep{Rossiter24, McLaughlin24}. To this end we may
estimate the full duration of an EB's eclipse using:

\begin{equation}
\Delta t_{\rm eclipse} = \frac{P}{\pi} \arcsin \left( r_1 + r_2
\right) ~,
\end{equation}

where $P$ is the binary's orbital period and $r_{1,2}$ are its
fraction radii.

In the following subsections, we describe what we know about each
of these EBs, and illustrate their phased LC, their phased RV
measurements, and the results of their MECI analysis (see
Chapter~4). In all cases we modeled these data with circular
orbits, since both the RV data and the photometric LC were found
to be consistent with having no eccentricity. When modeling these
data, we did not distinguish between the sources of the RV data
(i.e. NIRSPEC and TRES), however, we did weight the RV
measurements according to their estimated uncertainties.

By convention, a positive RV value indicates that the object is
receding from the observer. Thus, for a normal main-sequence
binary, we expect the primary component to have a negative
acceleration immediately after the primary eclipse (phase 0), as
it moves out from behind the eclipsing secondary component and
then curves towards the observer. Since the epoch of eclipse
($t_0$) and orbital period ($P$) of the binary are typically well
constrained by its LC, we adopted an RV model with only three free
parameters: the primary and secondary RV amplitudes ($K_{A,B}$),
and the barycentric RV ($V_\gamma$). We were able to fit this
model confidently when we had at least three double-lined RV
measurements, or six observables. Binaries with fewer than three
RV measurements could not be robustly modeled this way, but were
rather fit to a simplified model (see \S\ref{secSingleRV}). The
component RV amplitudes were then used to compute the binary
components' masses ($M_{A,B}$), and the sum of their semimajor
axes ($a$), also known as the binary separation. To this end we
used the following equations [see, e.g., \citet{Batten73}]:

\begin{eqnarray}
a \sin i     & = & P (K_A + K_B)(1-e^2)^{1/2} / 2 \pi \simeq \label{eq_asini}\\
             & \simeq & 0.01977 R_{\sun}\;P_d (k_A + k_B) (1-e^2)^{1/2}, {\rm \ and}\nonumber\\
M_{A,B} \sin^3 i & = & P K_{B,A} (K_A + K_B)^2 (1-e^2)^{3/2} / 2 \pi G \simeq \label{eq_msin3i}\\
             &\simeq&  1.0361 \cdot 10^{-7} M_{\sun}\;P_d k_{B,A} (k_A + k_B)^2 (1-e^2)^{3/2}\nonumber,
\end{eqnarray}

where for convenience, we define the unitless parameters: $P_d
\equiv P/day$, and $k_{A,B} \equiv K_{A,B}/{\rm km\,s^{-1}}$. Note
that since we assume that the orbits are circular, we fix the
eccentricity to $e=0$. The RV projection factor ($\sin i$) can, in
principle, be estimated using second order relativistic effects in
ultra-high precision spectroscopy \citep{Zucker07a}. However,
since our binaries eclipse, we were able to easily derive the
projection factor by modeling the photometric LC and fitting for
the orbital inclination ($i$). To this end, we used the DEBiL
parameter fit, which was able to provide a comparably accurate
estimate, as detached EBs inherently constrain the projection
factor to be within a small interval immediately below unity.

When estimating the uncertainties of $a \sin i$ and $M_{A,B}
\sin^3 i$, we took into account both the primary and secondary RV
amplitude uncertainties, and the asymmetric uncertainty of the
period. In the cases we will describe here, the RV measurements
were often obtained a few years after those of the photometric LC
data. Therefore, even small errors in the orbital period produce
significant shifts in the RV model phase due to the long
extrapolation, and can thus produce a noticeable error in the
measured RV amplitudes. We next estimated each binary component's
spectral classification using its derived mass and empirical
tables of the properties of a range of MK spectral type stars
\citep{Cox00}. Finally, we compared these results with those
produced through the MECI analysis (using $w=100$) and plotted
each binary LC with the LC model of its most likely MECI model.
Because the MECI model has only three free parameters (the
binary's age and the masses of its two components), it will often
produce a LC fit that is worse than the fit that DEBiL or EBOP
would produce. However, since MECI is constrained to use only the
parameters of physically realistic main-sequence stars, if it is
unable to successfully fit a given LC, we will have a strong
indication that either one of our assumptions is erroneous or that
additional effects must be considered.

\begin{deluxetable}{ccccccl}
\tabletypesize{\tiny}
\tablecaption{Measured radial velocities for EBs with at least three RV measurements}
\tablewidth{0pt}
\tablehead{\colhead{Target} & \colhead{Epoch (HJD)} &
\colhead{\begin{tabular}{c} Primary RV\\ ${\rm [km\:s^{-1}]}$ \end{tabular}} &
\colhead{\begin{tabular}{c} Secondary RV\\ ${\rm [km\:s^{-1}]}$ \end{tabular}} &
\colhead{Template} & \colhead{Instrument} & \colhead{Comments}}
\startdata
T-CrB0-10759 & $2,\!453,\!930.816114$   &   47.20  $\pm$  2.0   &   -99.10  $\pm$   2.0  &  GJ 182    &  NIRSPEC & \\
   --        & $2,\!454,\!104.167015$   &   50.53  $\pm$  2.0   &  -110.11  $\pm$   2.0  &  GJ 182    &  NIRSPEC & \\
   --        & $2,\!454,\!311.845306$   &  -60.30  $\pm$  2.0   &     7.93  $\pm$   2.0  &  GJ 182    &  NIRSPEC & \\
T-Cyg1-12664 & $2,\!454,\!311.884551$   &   34.51  $\pm$  3.0   &   -82.48  $\pm$  10.0  &  GJ 182    &  NIRSPEC & \\
   --        & $2,\!454,\!373.642658$   &   28.13  $\pm$  2.0   &  $\cdots$              &  GX And A  &  TRES & offset to match GJ 182\\
   --        & $2,\!454,\!373.655263$   &   28.21  $\pm$  2.0   &  $\cdots$              &  GX And A  &  TRES & offset to match GJ 182\\
   --        & $2,\!454,\!373.667080$   &   29.14  $\pm$  2.0   &  $\cdots$              &  GX And A  &  TRES & offset to match GJ 182\\
   --        & $2,\!454,\!377.686427$   &   23.47  $\pm$  2.0   &  $\cdots$              &  GX And A  &  TRES & offset to match GJ 182\\
   --        & $2,\!454,\!377.678074$   &   24.12  $\pm$  2.0   &  $\cdots$              &  GX And A  &  TRES & offset to match GJ 182\\
   --        & $2,\!454,\!377.710062$   &   24.45  $\pm$  2.0   &  $\cdots$              &  GX And A  &  TRES & offset to match GJ 182\\
T-Tau0-04859 & $2,\!453,\!928.118216$   &  100.17  $\pm$  7.0   &   -27.30  $\pm$   3.0  &  GJ 182    &  NIRSPEC & \\
   --        & $2,\!454,\!024.132604$   &  -21.76  $\pm$  8.0   &   100.89  $\pm$   3.0  &  GJ 182    &  NIRSPEC & \\
   --        & $2,\!454,\!100.852420$   &  -28.44  $\pm$  3.0   &   108.01  $\pm$   3.0  &  GJ 182    &  NIRSPEC & \\
   --        & $2,\!454,\!103.952431$   &  -31.36  $\pm$  3.0   &   108.93  $\pm$   5.0  &  GJ 182    &  NIRSPEC & \\
   --        & $2,\!454,\!377.920886$   &   21.90  $\pm$  3.0   &    52.11  $\pm$   3.0  &  HD 3651   &  TRES & offset to match GJ 182\\
   --        & $2,\!454,\!377.932704$   &   23.37  $\pm$  3.0   &    50.61  $\pm$   3.0  &  HD 3651   &  TRES & offset to match GJ 182\\
   --        & $2,\!454,\!377.944544$   &   25.18  $\pm$  3.0   &    48.33  $\pm$   3.0  &  HD 3651   &  TRES & offset to match GJ 182\\
T-Tau0-07388 & $2,\!453,\!931.117781$   &  -37.68  $\pm$ 25.0   &     2.32  $\pm$  25.0  &  GJ 182    &  NIRSPEC & merged peaks- not used\\
   --        & $2,\!454,\!100.866271$   &  113.33  $\pm$  5.0   &  -150.85  $\pm$   5.0  &  GJ 182    &  NIRSPEC & \\
   --        & $2,\!454,\!102.971832$   & -103.01  $\pm$ 25.0   &   144.20  $\pm$  25.0  &  GJ 182    &  NIRSPEC & uncertain\\
   --        & $2,\!454,\!377.992485$   &   86.42  $\pm$  5.0   &  -118.18  $\pm$  10.0  &  GX And A  &  TRES & offset to match GJ 182\\
   --        & $2,\!454,\!378.001999$   &   92.89  $\pm$  5.0   &  -128.20  $\pm$  10.0  &  GX And A  &  TRES & offset to match GJ 182\\
   --        & $2,\!454,\!378.011501$   &   98.18  $\pm$  5.0   &  -143.22  $\pm$  10.0  &  GX And A  &  TRES & offset to match GJ 182\\
   --        & $2,\!454,\!378.021119$   &  100.66  $\pm$  5.0   &  -136.24  $\pm$   7.5  &  GX And A  &  TRES & offset to match GJ 182\\
   --        & $2,\!454,\!378.030923$   &  107.53  $\pm$  5.0   &  -157.77  $\pm$   7.5  &  GX And A  &  TRES & offset to match GJ 182
\enddata
\label{tableRVsMultiple}
\end{deluxetable}

\newpage
\subsection{T-CrB0-10759}

The nearly equal eclipses of this binary indicate that the
components have similar effective temperatures, and are thus
likely to have similar masses. This conclusion is in agreement
with the fact that the components also show similar RV amplitudes.
The double-line spectra of this binary eliminate the possibility
that the orbital period is half the stated value and that the
secondary component is dark. We thus conclude that this binary
is comprised of 0.60+0.58 $M_{\sun}$ components ($K7+K8$), in
remarkable agreement with the results of its MECI analysis.

\begin{figure}
\includegraphics[width=5in]{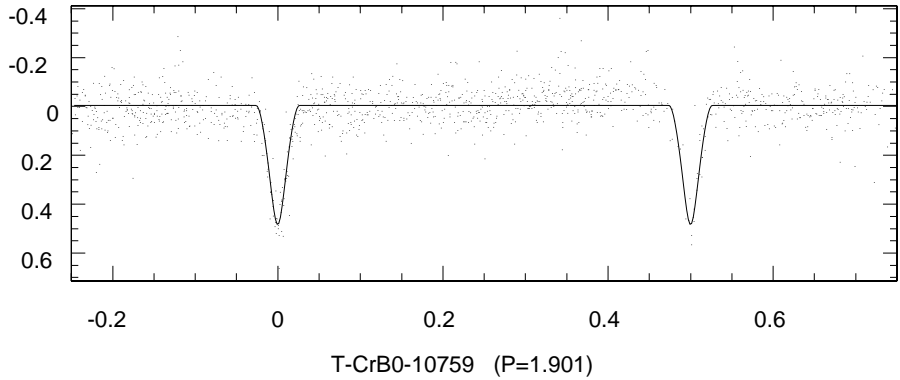}
\caption{The phased TrES light curve ($r$-band), with the best-fit MECI model (solid line).}
\end{figure}

\begin{figure}
\centerline{
\begin{tabular}{cc}
\includegraphics[width=2.72in]{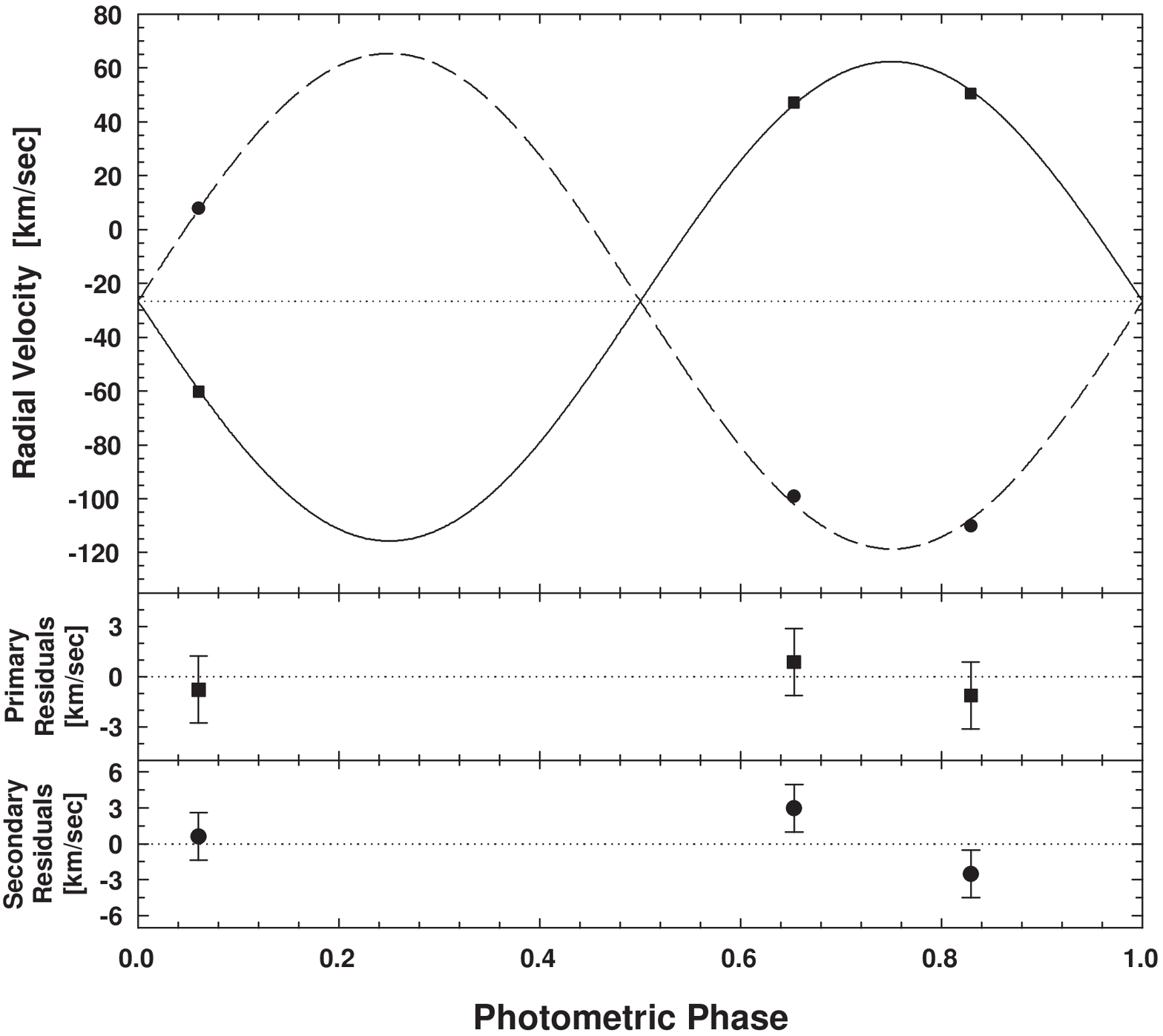} &
\includegraphics[width=3.45in]{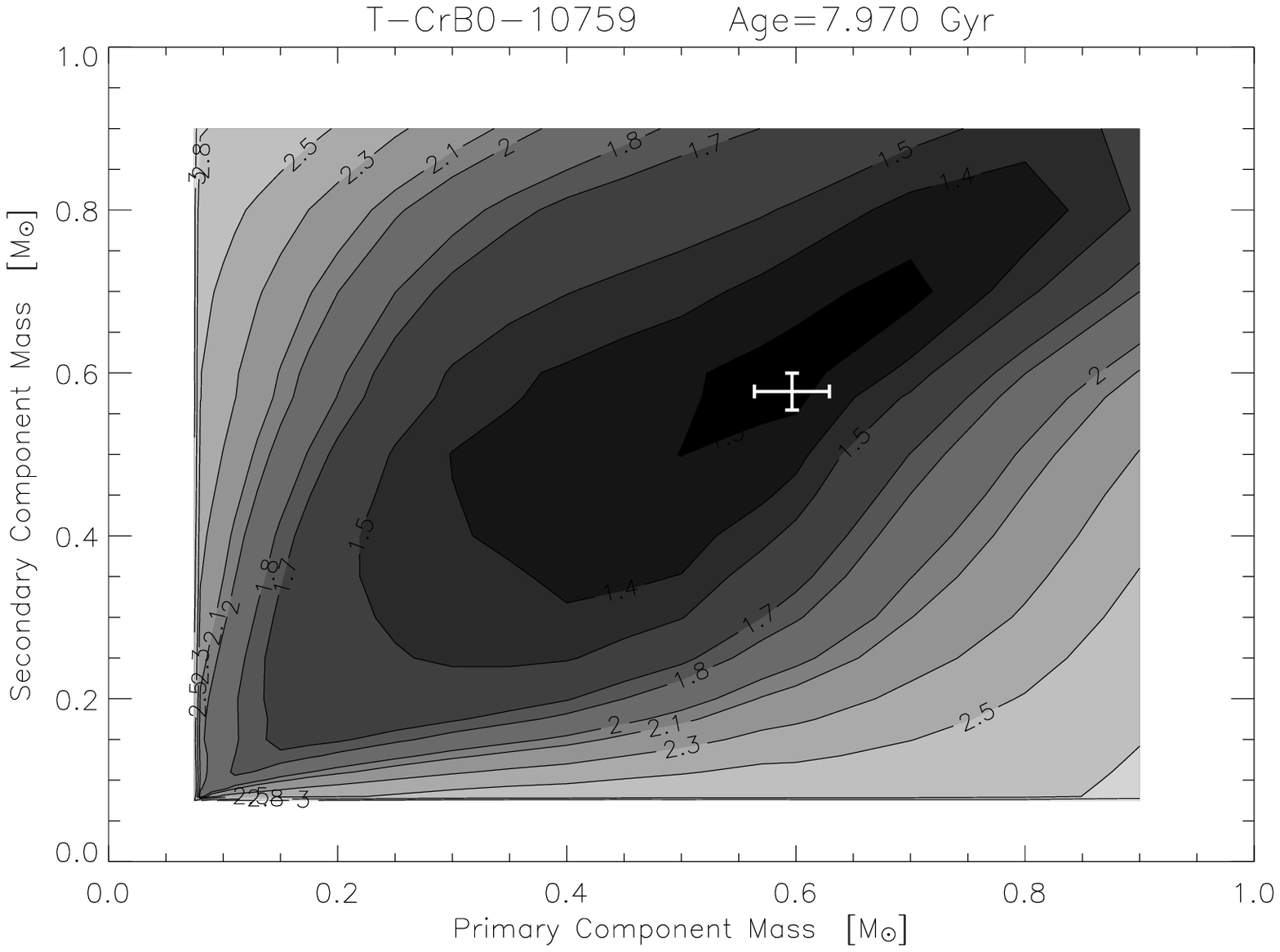}\\
a. Phased RVs with model & b. MECI likelihood contours
\end{tabular}}
\caption{The phased RVs of T-CrB0-10759 are shown with the
best-fit circular orbit model. The square symbols indicate the
primary component's RVs and the solid line illustrate their fitted
model, while the circular symbols indicate the secondary component
RVs and the dashed line illustrates their fitted model. The MECI
likelihood contours are compared with the RV solution (error
bars).}
\end{figure}

\begin{deluxetable}{lc}
\tabletypesize{\tiny}
\tablecaption{Catalog information of T-CrB0-10759}
\tablewidth{0pt}
\tablehead{\colhead{Parameter} & \colhead{Value}}
\startdata
$\alpha$ (J2000)      & 15:52:18.455\\
$\delta$ (J2000)      & 30:35:32.13\\
USNO-B\tablenotemark{a}\ \ \ $B$-mag & 17.495 $\pm$ 0.2\\
GSC2.3\tablenotemark{b}\ \ \ $V$-mag & 15.67 $\pm$ 0.30\\
USNO-B\ \ \ $R$-mag        & 15.185 $\pm$ 0.2\\
CMC14\tablenotemark{c}\ \ \ $r'$-mag & 15.543 $\pm$ 0.021\\
2MASS\tablenotemark{d}\ \ \ $J$-mag & 13.049 $\pm$ 0.015\\
2MASS\ \ \ $H$-mag         & 12.388 $\pm$ 0.015\\
2MASS\ \ \ $K_s$-mag        & 12.160 $\pm$ 0.015\\
UCAC\tablenotemark{e}\ \ \ $\mu_\alpha$ [${\rm mas\,yr^{-1}}$] &  3.6 $\pm$ 5.4\\
UCAC\ \ \  $\mu_\delta$ [${\rm mas\,yr^{-1}}$] & -19.4 $\pm$ 5.3\\
TrES third light\tablenotemark{f}\ \ \ [$R$-mag]\ \ \ \ \ \ \ \ \ \ \ \ \ \ \ \ \ \ & 0.028 $\pm$ 0.006
\enddata
\tablenotetext{a}{U.S. Naval Observatory photographic sky survey \citep{Monet03}.}
\tablenotetext{b}{Guide Star Catalog, version 2.3.2 \citep{Morrison01}.}
\tablenotetext{c}{Carlsberg Meridian Catalog 14 \citep{Evans02}.}
\tablenotetext{d}{Two Micron All Sky Survey catalog \citep{Skrutskie06}.}
\tablenotetext{e}{The Second U.S. Naval Observatory CCD Astrograph Catalog \citep{Zacharias04}.}
\tablenotetext{f}{The fraction of blended light in the TrES LC from resolved USNO-B sources within 30" of the target.}
\end{deluxetable}

\begin{deluxetable}{lcc}
\tabletypesize{\tiny}
\tablecaption{Binary parameters of T-CrB0-10759}
\tablewidth{0pt}
\tablehead{\colhead{Parameter} & \colhead{Symbol} & \colhead{Value}}
\startdata
Orbital Period [days]                                                 & $P$          & $1.901274_{-0.000014}^{+0.000012}$\\
Epoch of eclipse [HJD]                                                & $t_0$        & 2453515.09699 $\pm$ 0.00064\\
Number of light curve data points                                     & $N_{LC}$     & 1287 \\
Number of RV data points                                              & $N_{RV}$     &  3\\
MECI analysis primary mass [$M_{\sun}$]                               & $M_A^{MECI}$ & 0.61 $\pm$ 0.30  \\
MECI analysis secondary mass [$M_{\sun}$]                             & $M_B^{MECI}$ & 0.60 $\pm$ 0.25   \\
MECI analysis binary age [Gyr]                                        & $T^{MECI}$   & 7.4 $\pm$ 58 \\
Primary RV amplitude [${\rm km\,s^{-1}}$]                             & $K_A$        & 89.08 $\pm$ 0.92 \\
Secondary RV amplitude [${\rm km\,s^{-1}}$]                           & $K_B$        & 92.06 $\pm$ 2.22\\
Barycenteric RV, relative to GJ~182 [${\rm km\,s^{-1}}$]              & $V_\gamma$   & -26.65 $\pm$ 0.78\\
DEBiL estimate of the projection factor                               & $\sin i$     & 0.9991 $\pm$ 0.0070\\
Combined semimajor axis with projection factor [$R_{\sun}$]           & $a \sin i$   & $6.809^{+0.040}_{-0.042}$\\
Primary mass with projection factor [$M_{\sun}$]                      & $M_A \sin^3 i$ & $0.595^{+0.030}_{-0.030}$\\
Secondary mass with projection factor [$M_{\sun}$]                    & $M_B \sin^3 i$ & $0.576^{+0.019}_{-0.019}$
\enddata
\vspace*{5in}
\end{deluxetable}

\newpage
\subsection{T-Cyg1-12664}

This binary was originally thought to have an 8.2-day orbital
period, with two equal eclipses and therefore equal components,
because all of the eclipses observed in its TrES ($r$-band) LC are
similar. However, we subsequently observed it to have six
single-lined TRES spectra ($R$-band), and one double-lined NIRSPEC
spectrum ($K$-band) in which the secondary component is very weak.
We therefore revised our model for this EB to one with a 4.1-day
orbital period with an unobserved secondary eclipse, indicating
that the secondary component is far dimmer and cooler than the
primary component. Furthermore, the fact that the secondary
component is unobserved in the optical, but is observed in the
near infrared, supports the assertion that it is very red, and
thus likely to be a cool low-mass star. Since we only have a
single RV measurement for the secondary component, the primary
component's mass remains uncertain, nevertheless we estimate that
this binary is comprised of 0.62+0.32 $M_{\sun}$ components
($K6+M3$), which are smaller that those predicted by its MECI
analysis. As with T-Lyr1-17236, the comparably long orbital period
of this EB may make it a good test case for the magnetic
disruption hypothesis (see Chapter~6). The fact that the MECI
analysis overestimated the components' masses might be due to them
having bloated radii, which could then be explained through
magnetic disruption. Finally, also like T-Lyr1-17236 (see
\S\ref{ch6_secConclusions}), this EB falls within the field of
view of the upcoming NASA $\it{Kepler}$ mission. If it is observed
with this mission, both eclipses would be easily detected and the
high-precision LC would allow for strong constraints on the
fractional radii of both binary components.

\begin{figure}
\includegraphics[width=5in]{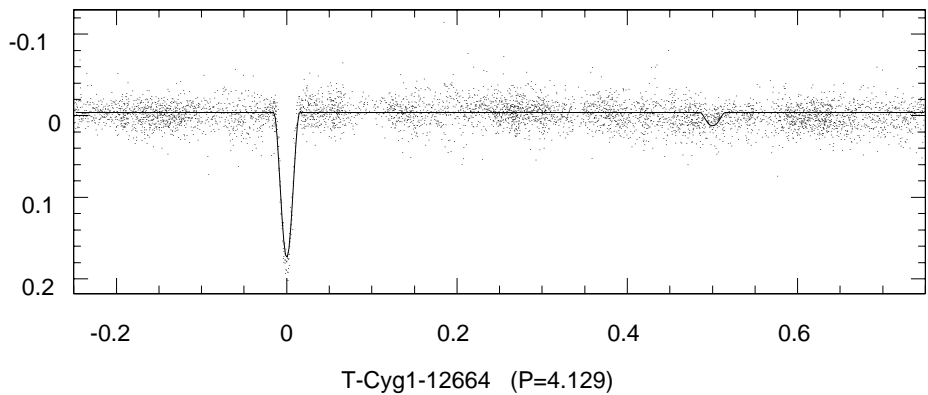}
\caption{The phased TrES light curve ($r$-band), with the best-fit MECI model (solid line).}
\end{figure}

\begin{figure}
\centerline{
\begin{tabular}{cc}
\includegraphics[width=2.72in]{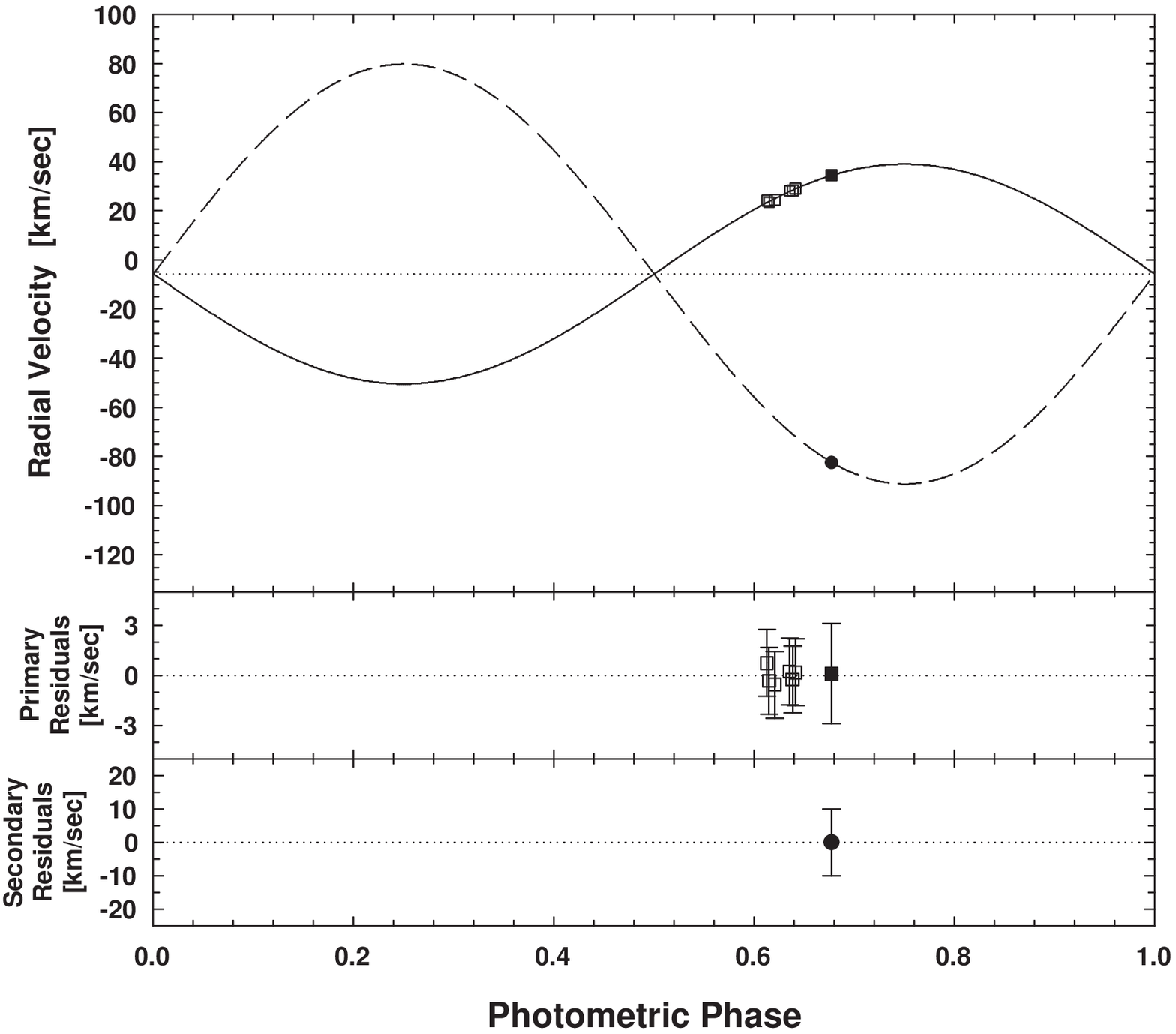} &
\includegraphics[width=3.45in]{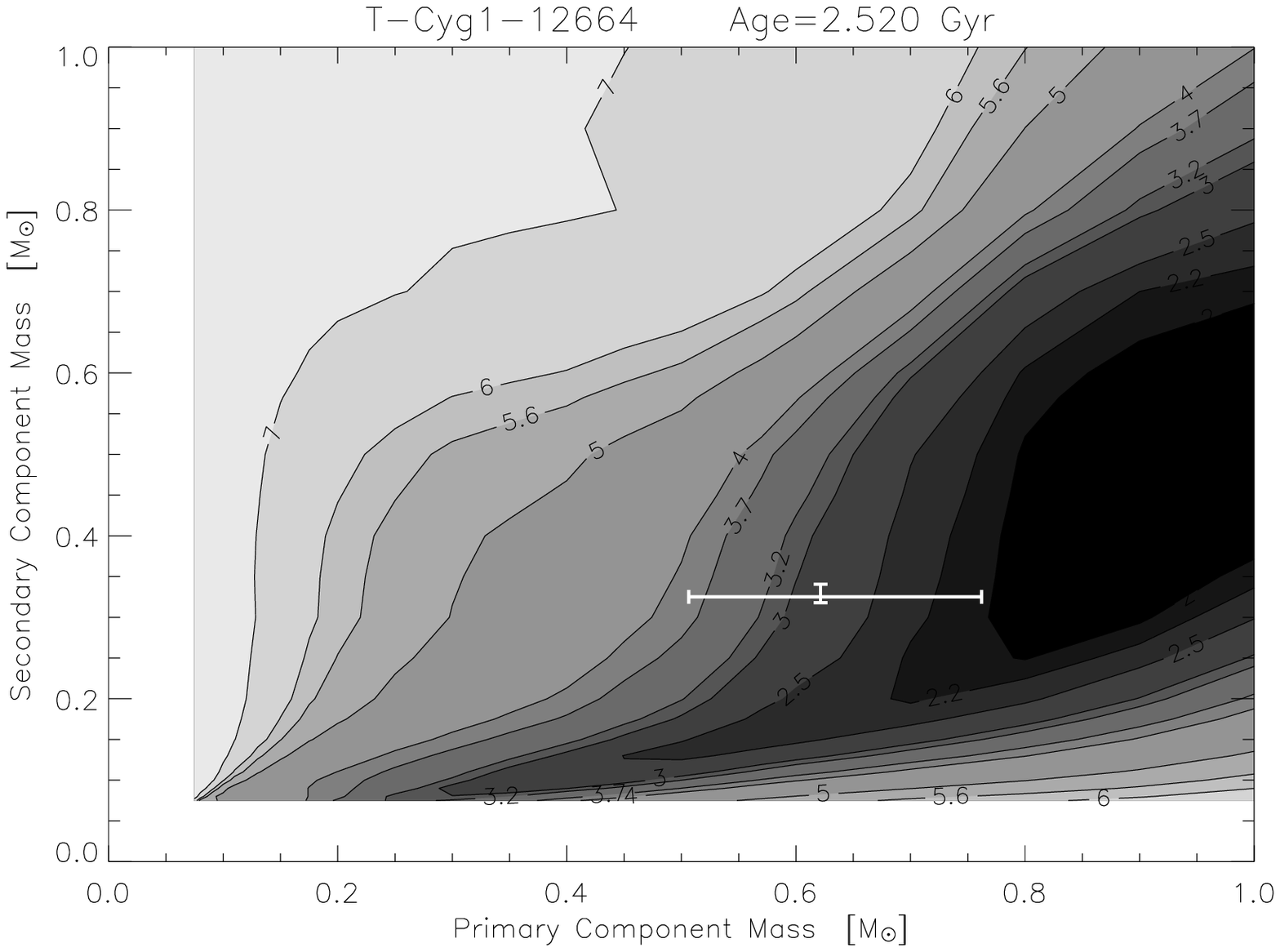}\\
a. Phased RVs with model & b. MECI likelihood contours
\end{tabular}}
\caption{The phased RVs of T-Cyg1-12664 are shown with the
best-fit circular orbit model. The filled symbols indicate
NIRSPEC measurements, while the unfilled symbols indicate TRES
measurements. The square symbols indicate the primary component's
RVs and the solid line is their fitted model, while the circular
symbols indicate the secondary component RVs and the dashed line
is their fitted model. The MECI likelihood contours are compared
with the RV solution (error bars).}
\end{figure}

\begin{deluxetable}{lc}
\tabletypesize{\tiny}
\tablecaption{Catalog information of T-Cyg1-12664}
\tablewidth{0pt}
\tablehead{\colhead{Parameter} & \colhead{Value}}
\startdata
$\alpha$ (J2000)    & 19:51:39.824 \\
$\delta$ (J2000)    & 48:19:55.38\\
USNO-B \ $B$-mag    & 14.240 $\pm$ 0.2\\
GSC2.3 \ $V$-mag    &  13.11 $\pm$ 0.30\\
USNO-B \ $R$-mag    &  13.315 $\pm$ 0.2\\
CMC14 \  $r'$-mag   &  13.024 $\pm$ 0.035\\
2MASS \  $J$-mag    & 11.911  $\pm$ 0.015\\
2MASS \  $H$-mag    & 11.582  $\pm$ 0.015\\
2MASS \  $K_s$-mag  &  11.529 $\pm$ 0.015\\
USNO-B $\mu_\alpha$ [${\rm mas\,yr^{-1}}$] & -18 $\pm$ 6 \\
USNO-B $\mu_\delta$ [${\rm mas\,yr^{-1}}$] &  -6 $\pm$ 2 \\
TrES third light [$R$-mag]    & 0.177 $\pm$ 0.034
\enddata
\end{deluxetable}

\begin{deluxetable}{lcc}
\tabletypesize{\tiny}
\tablecaption{Binary parameters of T-Cyg1-12664}
\tablewidth{0pt}
\tablehead{\colhead{Parameter} & \colhead{Symbol} & \colhead{Value}}
\startdata
Orbital Period [days]                                                 & $P$      & 4.128751 $\pm$ 0.000032 \\
Epoch of eclipse [HJD]                                                & $t_0$     & 2453210.84051 $\pm$ 0.00020\\
Number of light curve data points                                     & $N_{LC}$   &  5280 \\
Number of RV data points                                              & $N_{RV}$   & 7 \\
MECI analysis primary mass [$M_{\sun}$]                               & $M_A^{MECI}$ &  0.91 $\pm$ 0.27    \\
MECI analysis secondary mass [$M_{\sun}$]                             & $M_B^{MECI}$ & 0.41 $\pm$ 0.38  \\
MECI analysis binary age [Gyr]                                        & $T^{MECI}$ & 2.4 $\pm$ 2.4\\
Primary RV amplitude [${\rm km\,s^{-1}}$]                             & $K_A$          &  44.73 $\pm$ 0.23 \\
Secondary RV amplitude [${\rm km\,s^{-1}}$]                           & $K_B$          & -85.48 $\pm$ 1.04\\
Barycenteric RV, relative to GJ~182 [${\rm km\,s^{-1}}$]              & $V_\gamma$     & -5.72 $\pm$ 0.13\\
DEBiL estimate of the projection factor                               & $\sin i$     & 0.9992 $\pm$ 0.0028\\
Combined semimajor axis with projection factor [$R_{\sun}$]           & $a \sin i$   & $10.63^{+0.49}_{-0.39}$\\
Primary mass with projection factor [$M_{\sun}$]                      & $M_A \sin^3 i$ & $0.620^{+0.140}_{-0.115}$\\
Secondary mass with projection factor [$M_{\sun}$]                    & $M_B \sin^3 i$ & $0.324^{+0.015}_{-0.007}$
\enddata
\end{deluxetable}

\newpage
\subsection{T-Tau0-04859}

This systems exhibits clearly disparate eclipses, which eliminates
the possibility that the secondary component is unseen.
Furthermore, the RV measurements indicate that the components are
similar yet not equal, having masses of 0.66+0.62 $M_{\sun}$
($K5+K6$). Though these mass estimates are in agreement with those
derived from its MECI analysis, MECI was not able to reproduce the
large depth of the EB's primary eclipse ($\sim$0.45 mag in
$r$-band) together with the significantly shallower secondary
eclipse ($\sim$0.2 mag in $r$-band). As mentioned earlier, one
explanation for such a phenomenon is that the binary components
are effectively not coeval, bringing about a larger than expected
difference in their surface temperatures. However, it is unlikely
that a $K$-dwarf binary such as this would be old enough to have
evolved off the main-sequence, and therefore it is improbable that
its components underwent mass transfer. An alternative explanation
for this phenomenon is that this binary is very young, and has
therefore not yet reached the main-sequence isochrones used by
MECI. This latter hypothesis is strengthened by the fact that this
candidate resides in the Taurus field, which is known to harbor
many young stars; however, the fact that the eclipse durations are
comparably short, significantly constrains the size of the
components, and therefore limits how young they could be.

\begin{figure}
\includegraphics[width=5in]{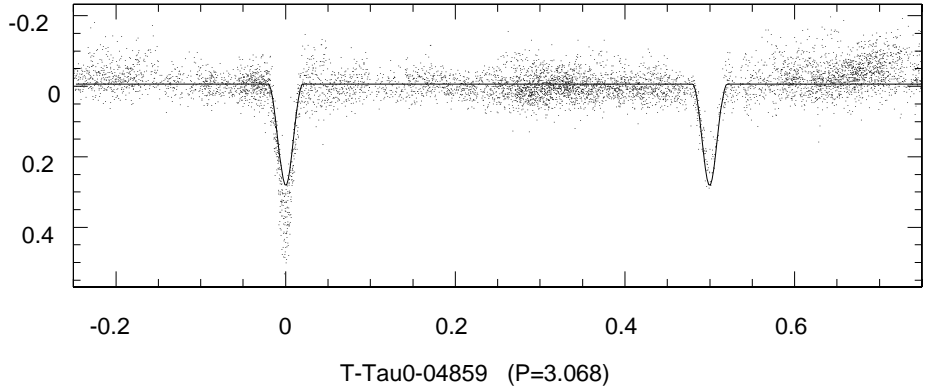}
\caption{The phased TrES light curve ($r$-band), with the best-fit MECI model (solid line).}
\end{figure}

\begin{figure}
\centerline{
\begin{tabular}{cc}
\includegraphics[width=2.72in]{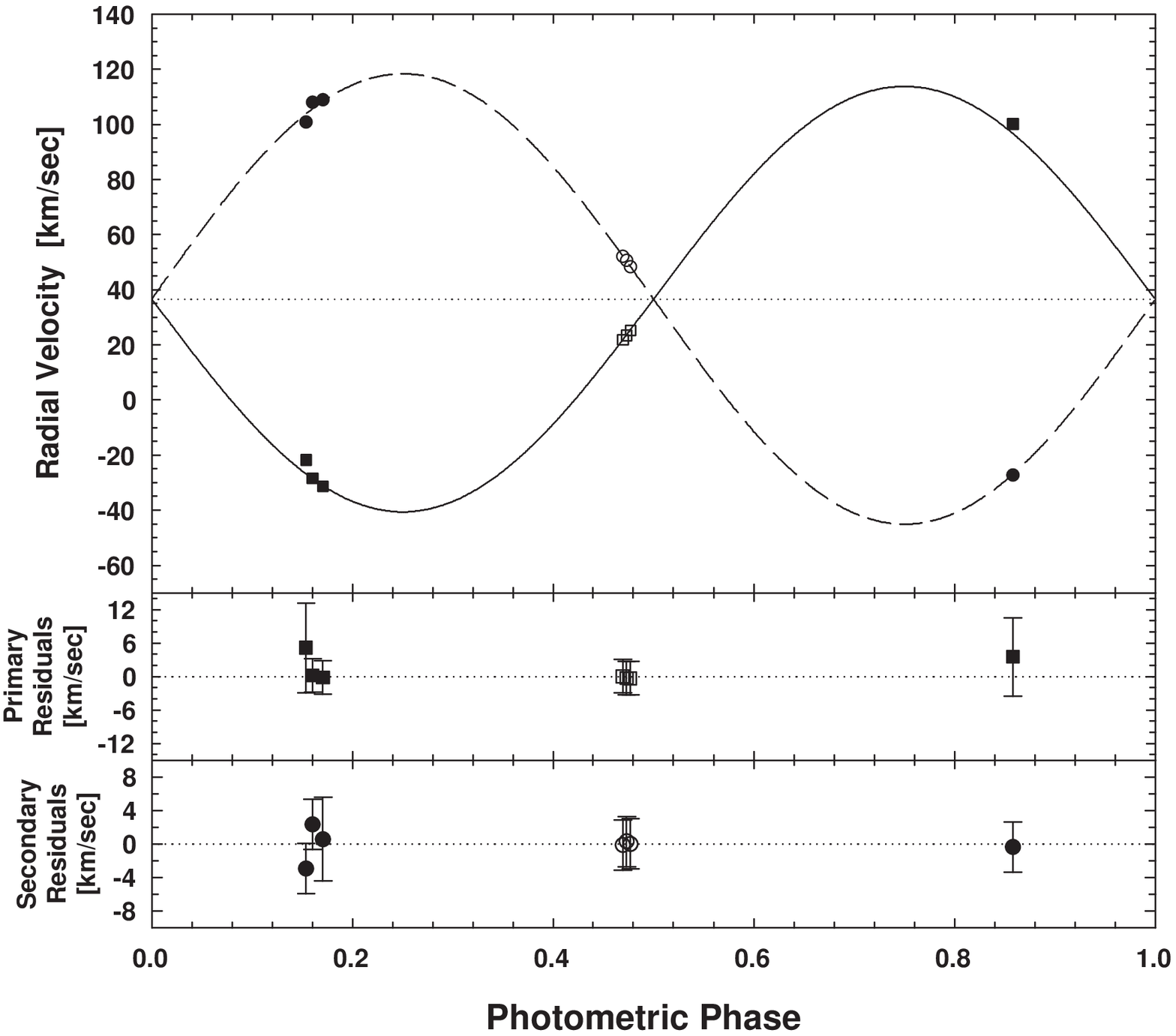} &
\includegraphics[width=3.45in]{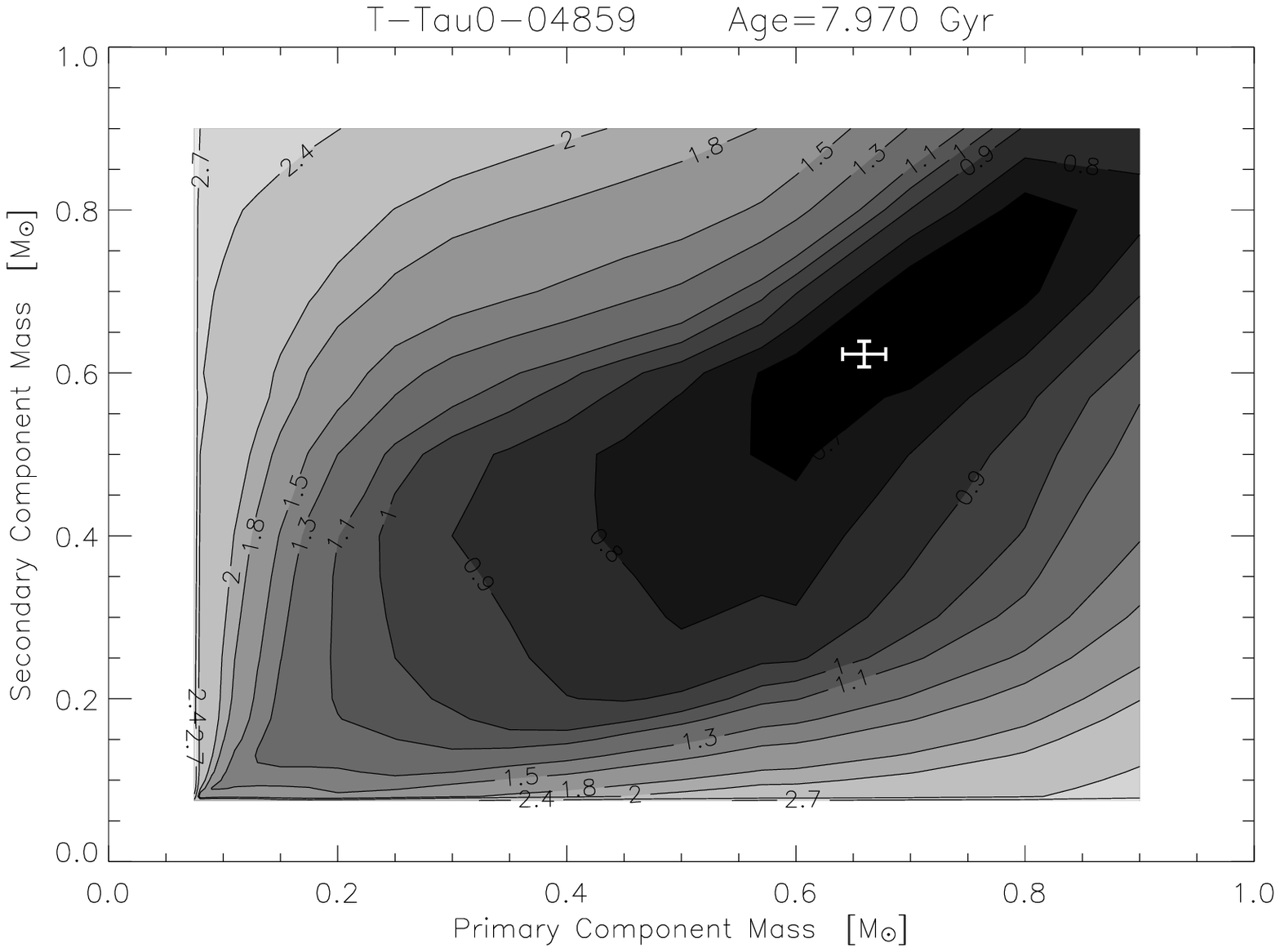}\\
a. Phased RVs with model & b. MECI likelihood contours
\end{tabular}}
\caption{The phased RVs of T-Tau0-04859 are shown with the
best-fit circular orbit model. The filled symbols indicate a
NIRSPEC measurements, while the unfilled symbols indicate TRES
measurements. The square symbols indicate the primary component's
RVs and the solid line is their fitted model, while the circular
symbols indicate the secondary component RVs and the dashed line
is their fitted model. The MECI likelihood contours are compared
with the RV solution (error bars).}
\end{figure}

\begin{deluxetable}{lc}
\tabletypesize{\tiny}
\tablecaption{Catalog information of T-Tau0-04859}
\tablewidth{0pt}
\tablehead{\colhead{Parameter} & \colhead{Value}}
\startdata
$\alpha$ (J2000)    & 04:08:11.608\\
$\delta$ (J2000)    & 24:51:10.18\\
USNO-B \ $B$-mag    & 15.520 $\pm$ 0.2\\
GSC2.3 \ $V$-mag    & 14.05 $\pm$ 0.23\\
USNO-B \ $R$-mag    & 13.495 $\pm$ 0.2\\
CMC14 \  $r'$-mag   & 13.748 $\pm$ 0.022\\
2MASS \  $J$-mag    & 11.748 $\pm$ 0.015\\
2MASS \  $H$-mag    & 11.064 $\pm$ 0.015\\
2MASS \  $K_s$-mag  & 10.869 $\pm$ 0.015\\
UCAC $\mu_\alpha$ [${\rm mas\,yr^{-1}}$] & 3.4 $\pm$ 5.2 \\
UCAC $\mu_\delta$ [${\rm mas\,yr^{-1}}$] & -7.7 $\pm$ 5.2 \\
TrES third light [$R$-mag]       & 0.004 $\pm$ 0.001
\enddata
\end{deluxetable}

\begin{deluxetable}{lcc}
\tabletypesize{\tiny}
\tablecaption{Binary parameters of T-Tau0-04859}
\tablewidth{0pt}
\tablehead{\colhead{Parameter} & \colhead{Symbol} & \colhead{Value}}
\startdata
Orbital Period [days]                                                 & $P$          & $3.068000_{-0.000028}^{+0.000027}$\\
Epoch of eclipse [HJD]                                                & $t_0$        & 2453738.33643 $\pm$ 0.00031 \\
Number of light curve data points                                     & $N_{LC}$     & 6729  \\
Number of RV data points                                              & $N_{RV}$     &   7 \\
MECI analysis primary mass [$M_{\sun}$]                               & $M_A^{MECI}$ &  0.73 $\pm$ 0.24 \\
MECI analysis secondary mass [$M_{\sun}$]                             & $M_B^{MECI}$ & 0.67 $\pm$ 0.24 \\
MECI analysis binary age [Gyr]                                        & $T^{MECI}$ &  8.3 $\pm$ 14.7\\
Primary RV amplitude [${\rm km\,s^{-1}}$]                             & $K_A$          & 77.21 $\pm$ 0.77\\
Secondary RV amplitude [${\rm km\,s^{-1}}$]                           & $K_B$          & 81.73 $\pm$ 1.01\\
Barycenteric RV, relative to GJ~182 [${\rm km\,s^{-1}}$]              & $V_\gamma$     & 36.63 $\pm$ 0.37\\
DEBiL estimate of the projection factor                               & $\sin i$     &  0.9984 $\pm$ 0.0031\\
Combined semimajor axis with projection factor [$R_{\sun}$]           & $a \sin i$   & $9.642^{+0.047}_{-0.047}$\\
Primary mass with projection factor [$M_{\sun}$]                      & $M_A \sin^3 i$ & $0.656^{+0.018}_{-0.018}$\\
Secondary mass with projection factor [$M_{\sun}$]                    & $M_B \sin^3 i$ & $0.620^{+0.015}_{-0.015}$
\enddata
\end{deluxetable}

\newpage
\subsection{T-Tau0-07388}
\label{subsec_Tau0-07388}

 Like T-Tau0-04859, this binary also clearly exhibits
disparate eclipses, which eliminates the possibility that the
secondary component is unseen. However, unlike T-Tau0-04859, the
RV measurements of this binary indicate that the components of
this binary are significantly different, having masses of
0.65+0.47 $M_{\sun}$ ($K6+M1$). The MECI analysis significantly
overestimated these component masses, which suggests that the
components are larger than predicted by stellar models. Such a
size increase could be explained through the magnetic disruption
hypothesis (see Chapter~6), since this binary has a short 0.6-day
orbital period, making it likely that both these components are
spinning rapidly and producing strong magnetic fields. However,
another explanation for the components' large radii is that they
are young and still in the process of forming. As with
T-Tau0-04859, this latter explanation is supported by the fact
that this binary resides in the Taurus field, which is known to
contain star forming regions. Lastly, we note that the LC plateaux
are of unequal brightness, thus indicating an ``O'Connell effect''
(see \S\ref{subsecAbnomalEBs}). Specifically, the out-of-eclipse
$r$-band brightness of the binary is $\sim$0.05 mag dimmer after
the primary eclipse than it was before the eclipse, perhaps
indicating a persistent off-axis hot spot (e.g., reflection) or
cold spot (e.g., gravity darkening) on one of the components.

\begin{figure}
\includegraphics[width=5in]{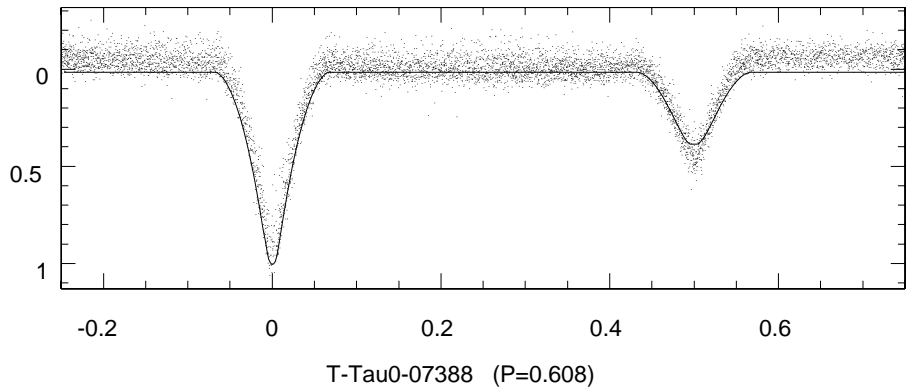}
\caption{The phased TrES light curve ($r$-band), with the best-fit MECI model (solid line).}
\end{figure}

\begin{figure}
\centerline{
\begin{tabular}{cc}
\includegraphics[width=2.72in]{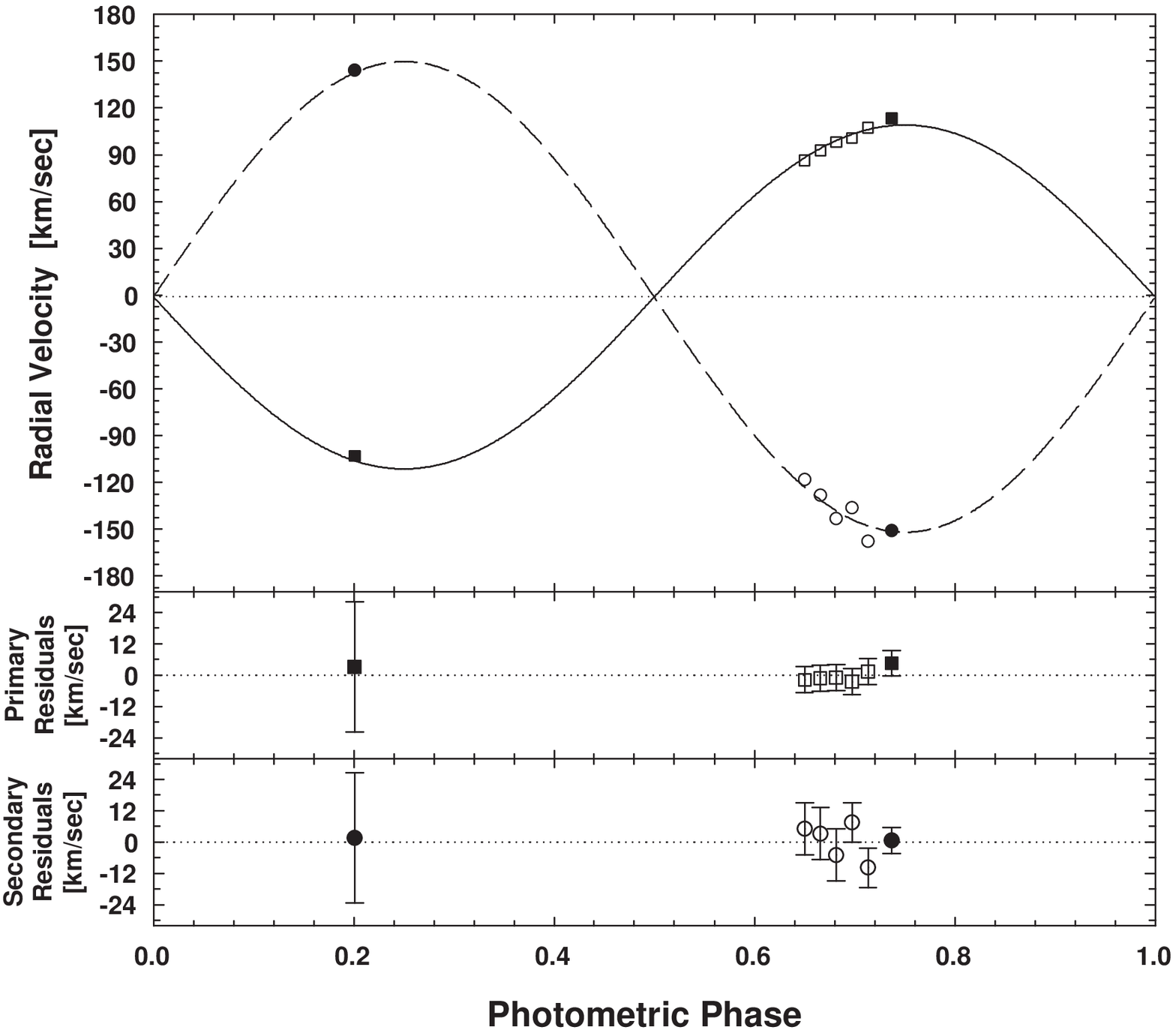} &
\includegraphics[width=3.45in]{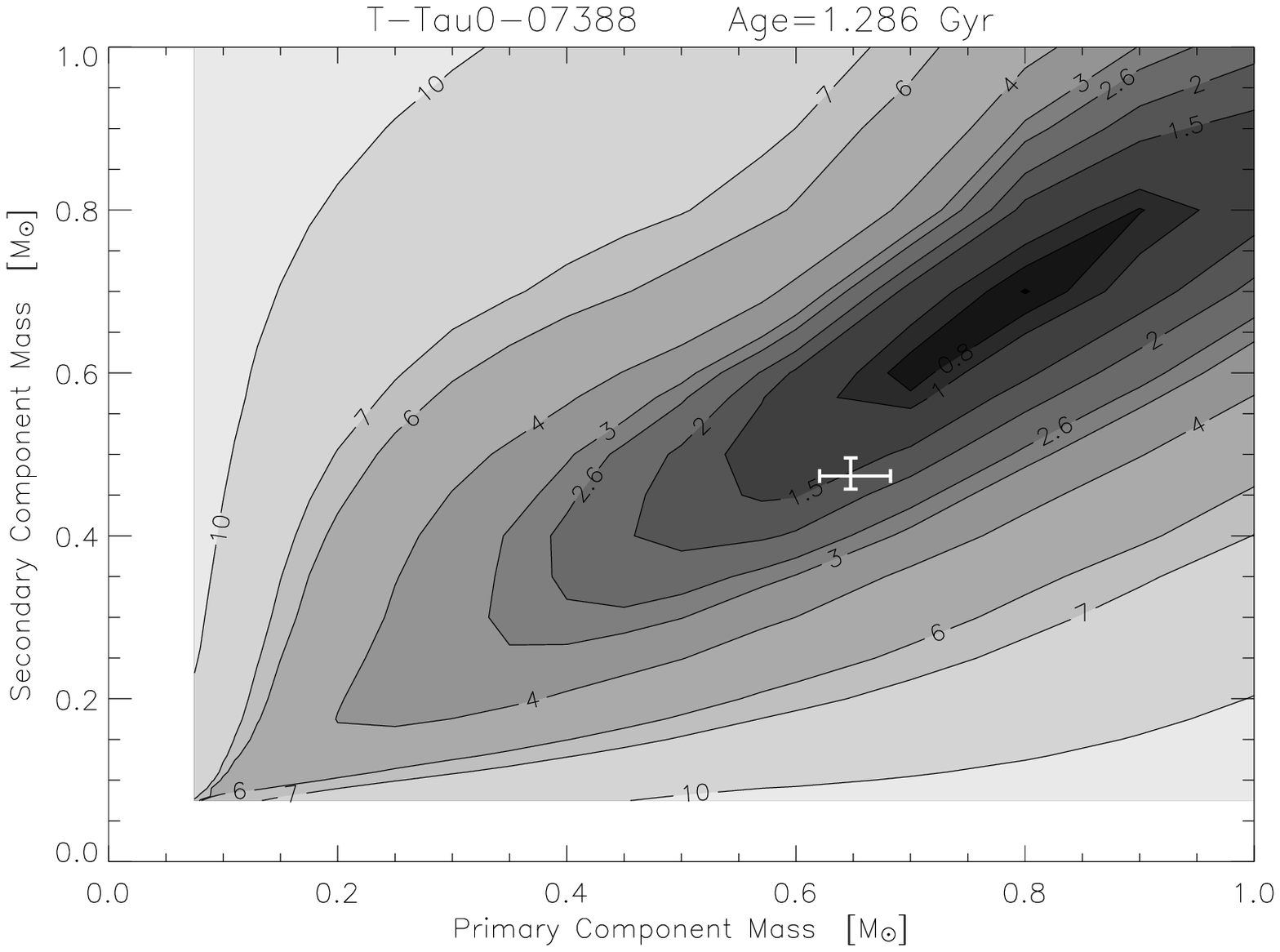}\\
a. Phased RVs with model & b. MECI likelihood contours
\end{tabular}}
\caption{The phased RVs of T-Tau0-07388 are shown with the
best-fit circular orbit model. The filled symbols indicate a
NIRSPEC measurements, while the unfilled symbols indicate TRES
measurements. The square symbols indicate the primary component's
RVs and the solid line is their fitted model, while the circular
symbols indicate the secondary component RVs and the dashed line
is their fitted model. The MECI likelihood contours are compared
with the RV solution (error bars).}
\end{figure}

\begin{deluxetable}{lc}
\tabletypesize{\tiny}
\tablecaption{Catalog information of T-Tau0-07388}
\tablewidth{0pt}
\tablehead{\colhead{Parameter} & \colhead{Value}}
\startdata
$\alpha$ (J2000)   & 04:10:04.977 \\
$\delta$ (J2000)    & 29:31:02.33\\
USNO-B \ $B$-mag    & 16.635 $\pm$ 0.2\\
GSC2.3 \ $V$-mag    & 14.38  $\pm$ 0.26\\
USNO-B \ $R$-mag    & 13.750  $\pm$ 0.2\\
CMC14 \  $r'$-mag   &  14.115 $\pm$ 0.05 \\
2MASS \  $J$-mag    &  11.131 $\pm$ 0.015\\
2MASS \  $H$-mag    & 10.375 $\pm$ 0.015\\
2MASS \  $K_s$-mag  & 10.133 $\pm$ 0.015\\
USNO-B $\mu_\alpha$ [${\rm mas\,yr^{-1}}$] & -36 $\pm$ 2\\
USNO-B $\mu_\delta$ [${\rm mas\,yr^{-1}}$] &  -14 $\pm$ 6\\
TrES third light [$R$-mag]        & 0.031 $\pm$ 0.007
\enddata
\end{deluxetable}

\begin{deluxetable}{lcc}
\tabletypesize{\tiny}
\tablecaption{Binary parameters of T-Tau0-07388}
\tablewidth{0pt}
\tablehead{\colhead{Parameter} & \colhead{Symbol} & \colhead{Value}}
\startdata
Orbital Period [days]                                                 & $P$         & $0.6078486_{-0.0000015}^{+0.0000033}$\\
Epoch of eclipse [HJD]                                                & $t_0$       & 2453736.924811 $\pm$ 0.000080  \\
Number of light curve data points                                     & $N_{LC}$       & 6702 \\
Number of RV data points                                              & $N_{RV}$       & 7 \\
MECI analysis primary mass [$M_{\sun}$]                               & $M_A^{MECI}$ & 0.80 $\pm$ 0.13   \\
MECI analysis secondary mass [$M_{\sun}$]                             & $M_B^{MECI}$ & 0.70 $\pm$ 0.11  \\
MECI analysis binary age [Gyr]                                        & $T^{MECI}$ & 1.4 $\pm$ 1.3 \\
Primary RV amplitude [${\rm km\,s^{-1}}$]                             & $K_A$          & 110.23 $\pm$ 1.06 \\
Secondary RV amplitude [${\rm km\,s^{-1}}$]                           & $K_B$          & 150.83 $\pm$ 2.40 \\
Barycenteric RV, relative to GJ~182 [${\rm km\,s^{-1}}$]              & $V_\gamma$  & -1.10 $\pm$ 1.02\\
DEBiL estimate of the projection factor                               & $\sin i$     &  0.9998 $\pm$ 0.0055\\
Combined semimajor axis with projection factor [$R_{\sun}$]           & $a \sin i$   & $3.137^{+0.040}_{-0.020}$\\
Primary mass with projection factor [$M_{\sun}$]                      & $M_A \sin^3 i$ & $0.647^{+0.033}_{-0.025}$\\
Secondary mass with projection factor [$M_{\sun}$]                    & $M_B \sin^3 i$ & $0.473^{+0.021}_{-0.014}$
\enddata
\end{deluxetable}

\newpage
\section{Low-Mass Candidates with Fewer than Three RV Measurements}
\label{secSingleRV}

The following binaries have only one or two RV measurements, and
their fitted models are therefore speculative at best.
Nevertheless, for completeness we chose to present these systems
as well, together with our best guess as to their properties. Our
analysis of these systems repeated the procedure we outlined in
section \S\ref{secMultipleRVs}. However, because we did not have
sufficient RV data to reliably determine both the primary and
secondary RV amplitudes ($K_{A,B}$), as well as the Barycenteric
RV ($V_\gamma$), as we did previously, we opted instead for a
model of the sum of the components' RVs. This latter model has
only a single free parameter, the sum of the RV amplitudes
($K_A+K_B$), and thus can be fit, in principle, using a single
measurement of the difference between the components' RVs. Note
that, as before, we determine the period ($P$) and epoch of
eclipse ($t_0$) of the binary using the photometric LC.

Once we determined the sum of the RV amplitudes ($K_A+K_B$), we
solve for $a \sin i$ using Equation \ref{eq_asini}, and then solve
for $(M_A+M_B) \sin^3 i$, by summing both instances of Equation
\ref{eq_msin3i} and thus arriving at:

\begin{eqnarray}
(M_A+M_B) \sin^3 i & = & P (K_A + K_B)^3 (1-e^2)^{3/2} / 2 \pi G \simeq \label{eq_mpmsin3i}\\
                   &\simeq&  1.0361 \cdot 10^{-7} M_{\sun}\;P_d (k_A + k_B)^3 (1-e^2)^{3/2}\nonumber.
\end{eqnarray}

As before, we assume that the orbits are circular and fix the
eccentricity to $e=0$, and use the DEBiL parameter fit to estimate
the projection factor ($\sin i$). Since in these cases we only
know the sum of the components' masses, when comparing the RV
results with the MECI analysis, we produce a solid diagonal line
on the mass-mass contour plot, indicating all the mass pairings
that solve Equation \ref{eq_mpmsin3i}. We repeated this
calculation for mass pairings at the 1$\sigma$ uncertainty level,
thus producing two additional diagonals, which we plot as dotted
lines. Though we do not know the component's mass ratio ($q$) for
these EBs, by the fact that at least one observed spectrum was
seen to be double-lined, we may conclude that $q \ga 0.5$ with a
moderate degree of certainty.

\begin{deluxetable}{ccccccl}
\tabletypesize{\tiny}
\tablecaption{Measured radial velocities for EBs with fewer than three RV measurements}
\tablewidth{0pt}
\tablehead{\colhead{Target} & \colhead{Epoch (HJD)} &
\colhead{\begin{tabular}{c} Primary RV\\ ${[\rm km\:s^{-1}]}$
\end{tabular}} & \colhead{\begin{tabular}{c} Secondary RV\\ ${[\rm
km\:s^{-1}]}$ \end{tabular}} & \colhead{Template} &
\colhead{Instrument} & \colhead{Comments}}
\startdata
T-And0-04829 & $2,\!453,\!928.066849$   &   98.89  $\pm$  1.5   &   -29.92  $\pm$  1.5   &  GJ 182     &  NIRSPEC & \\
   --        & $2,\!453,\!929.095220$   &  -77.06  $\pm$  1.5   &   $\cdots$             &  GJ 182     &  NIRSPEC & uncertain\\
T-And0-20382 & $2,\!453,\!931.061037$   &  -24.22  $\pm$  3.0   &   110.72  $\pm$  10.0  &  GJ 182     &  NIRSPEC\\
   --        & $2,\!454,\!377.855671$   &  -47.24  $\pm$ 15.0   &   $\cdots$             &  GX And A   &  TRES     & offset to match GJ 182\\
T-CrB0-14232 & $2,\!453,\!930.846256$   &   67.26  $\pm$  2.0   &  -102.47  $\pm$   2.0  &  GJ 182     &  NIRSPEC & \\
T-Dra0-03021 & $2,\!453,\!929.882148$   &  -25.04  $\pm$  10.0  &    85.84  $\pm$  10.0  &  GJ 182     &  NIRSPEC & \\
   --        & $2,\!454,\!308.878115$   &  -71.31  $\pm$  10.0  &    $\cdots$            &  GJ 182     &  NIRSPEC & \\
T-Dra0-07116 & $2,\!453,\!931.905448$   &   25.89  $\pm$   2.5  &   -47.58  $\pm$  2.5   &  GJ 182     &  NIRSPEC & \\
   --        & $2,\!454,\!308.890984$   &  -77.08  $\pm$   8.0  &    83.63  $\pm$  4.0   &  GJ 182     &  NIRSPEC & \\
\enddata
\label{tableRVsSingle}
\vspace*{5in}
\end{deluxetable}

\newpage
\subsection{T-And0-04829}

This binary was found to have a double-lined spectrum and nearly
identical eclipses, which suggests that the components are likely
similar. The RV model seems to predict components as small as
M-dwarfs, while MECI suggests that the components may be as large
as G-dwarfs. The binary's 2MASS colors are comparably red,
supporting the low-mass hypothesis; however, the eclipse durations
are comparably large, which supports the higher-mass hypothesis.
Consequently, these contradictory results prevent us from arriving
at any firm conclusions.

\begin{figure}
\includegraphics[width=5in]{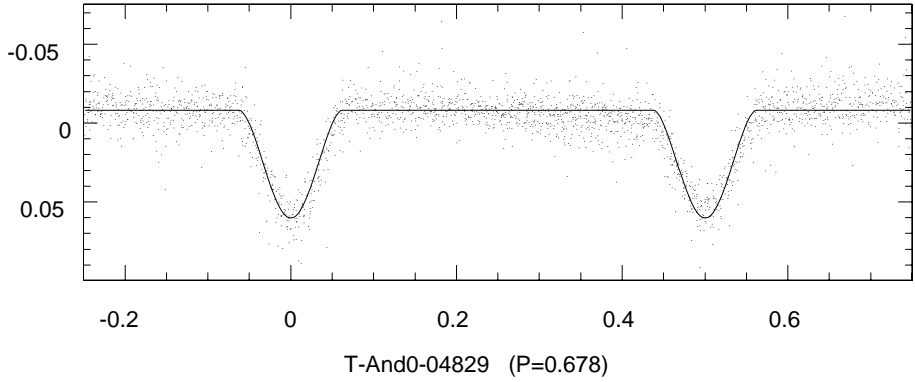}
\caption{The phased TrES light curve ($r$-band), with the best-fit MECI model (solid line).}
\end{figure}

\begin{figure}
\centerline{
\begin{tabular}{cc}
\includegraphics[width=3.1in]{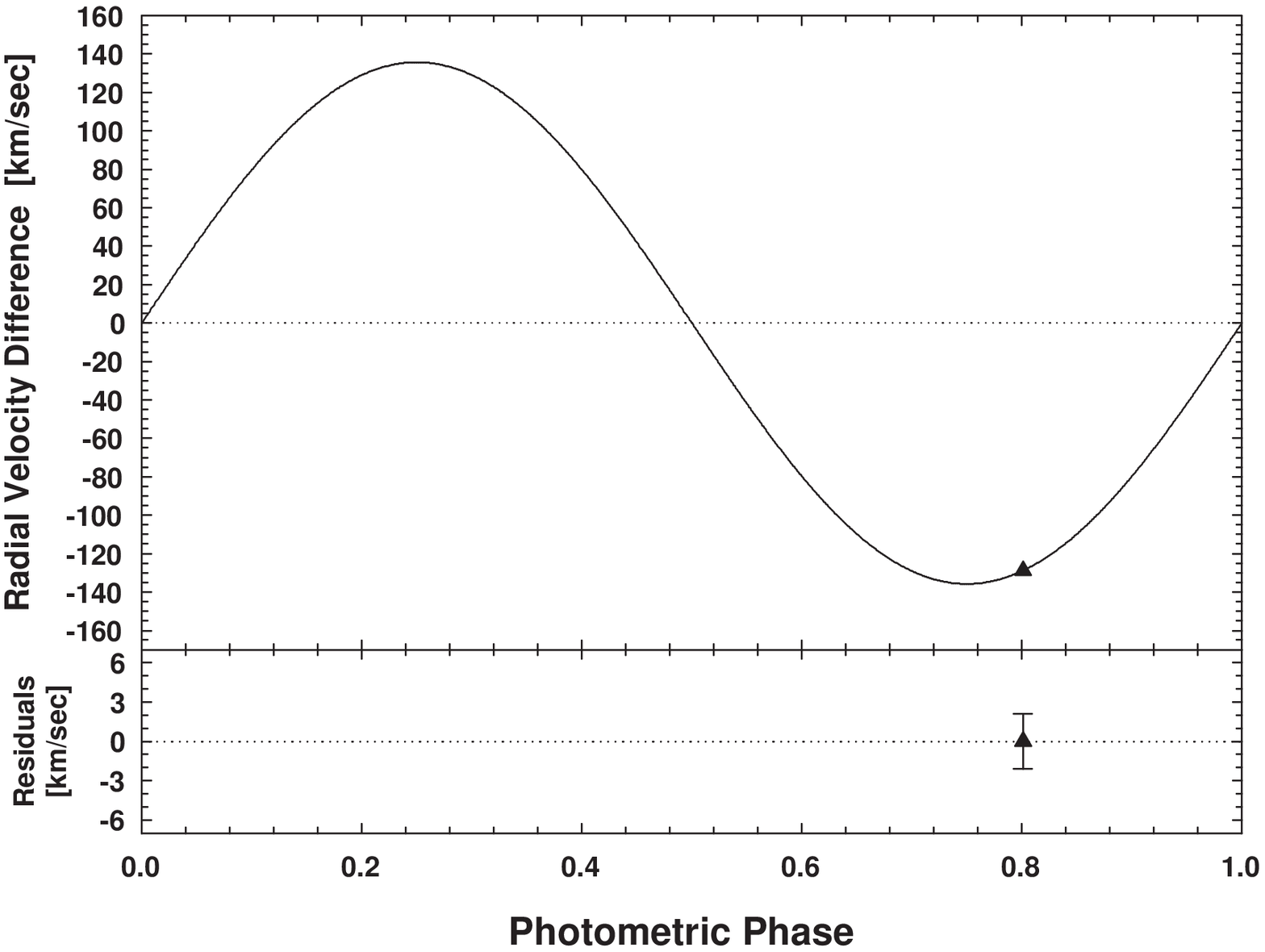} &
\includegraphics[width=3.3in]{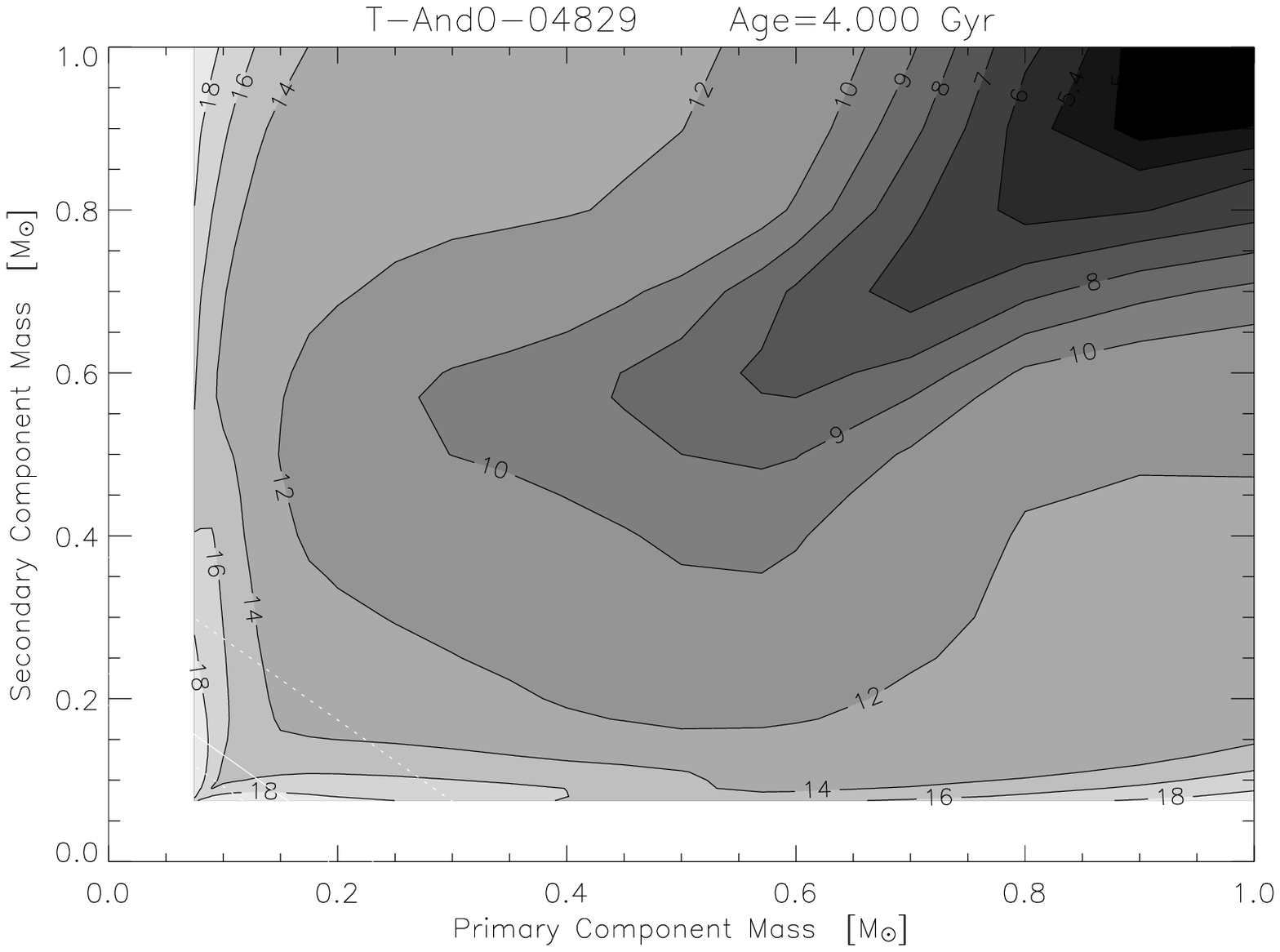}\\
a. Phased RVs with model & b. MECI likelihood contours
\end{tabular}}
\caption{The phased RVs of T-And0-04829 are shown with the best-fit
 circular orbit model to the difference between the component RVs.}
\end{figure}

\begin{deluxetable}{lc}
\tabletypesize{\tiny}
\tablecaption{Catalog information of T-And0-04829}
\tablewidth{0pt}
\tablehead{\colhead{Parameter} & \colhead{Value}}
\startdata
$\alpha$ (J2000)    & 01:15:15.228\\
$\delta$ (J2000)    & 47:45:58.97\\
USNO-B \ $B$-mag    & 13.610 $\pm$ 0.2\\
GSC2.3 \ $V$-mag    & 12.33 $\pm$ 0.29\\
USNO-B \ $R$-mag    & 11.730 $\pm$ 0.2\\
CMC14 \  $r'$-mag   & 12.782 $\pm$ 0.05\\
2MASS \  $J$-mag    &  10.970 $\pm$ 0.015\\
2MASS \  $H$-mag    & 10.339 $\pm$ 0.015\\
2MASS \  $K_s$-mag  & 10.162 $\pm$ 0.015\\
UCAC $\mu_\alpha$ [${\rm mas\,yr^{-1}}$] & -23.8 $\pm$ 4.9\\
UCAC $\mu_\delta$ [${\rm mas\,yr^{-1}}$] & 44.7 $\pm$ 4.9\\
TrES third light [$R$-mag]      & 0.011 $\pm$ 0.003
\enddata
\end{deluxetable}

\begin{deluxetable}{lcc}
\tabletypesize{\tiny}
\tablecaption{Binary parameters of T-And0-04829}
\tablewidth{0pt}
\tablehead{\colhead{Parameter} & \colhead{Symbol} & \colhead{Value}}
\startdata
Orbital Period [days]                                                 & $P$          & 0.6778149 $\pm$ 0.0000036 \\
Epoch of eclipse [HJD]                                                & $t_0$       & 2452906.767874 $\pm$ 0.000081\\
Number of light curve data points                                     & $N_{LC}$       & 2357 \\
Number of RV data points                                              & $N_{RV}$       &  1\\
Relative RV amplitude [${\rm km\,s^{-1}}$]                            & $K_A+K_B$    & 135.74 $\pm$ 2.24\\
DEBiL estimate of the projection factor                               & $\sin i$     &  0.913 $\pm$ 0.021\\
Combined semimajor axis with projection factor [$R_{\sun}$]           & $a \sin i$   & $1.82^{+0.32}_{-0.10}$\\
Mass sum with projection factor [$M_{\sun}$]                          & $(M_A+M_B) \sin^3 i$ & $0.176^{+0.108}_{-0.027}$
\enddata
\end{deluxetable}

\newpage
\subsection{T-And0-20382}

Like T-And0-04829, this binary was found to have a double-lined
spectrum and nearly identical eclipses, which suggests that the
components are likely similar. The RV and MECI analyses are
consistent with one another, and indicate that the components are
K-dwarfs, however both analyses have large uncertainties that
prevent a more specific identification. MECI is limited by poorly
constrained eclipses, while the RV analysis is limited by both a
poorly determined orbital period and a poorly determined RV
measurement.

\begin{figure}
\includegraphics[width=5in]{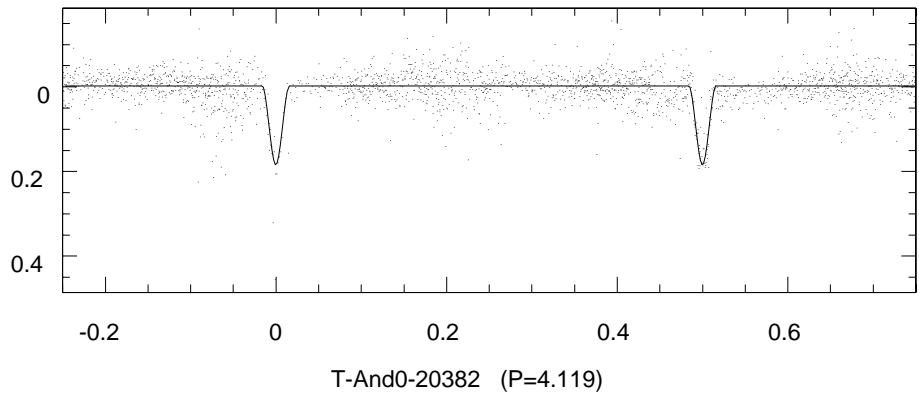}
\caption{The phased TrES light curve ($r$-band), with the best-fit MECI model (solid line).}
\end{figure}

\begin{figure}
\centerline{
\begin{tabular}{cc}
\includegraphics[width=3.1in]{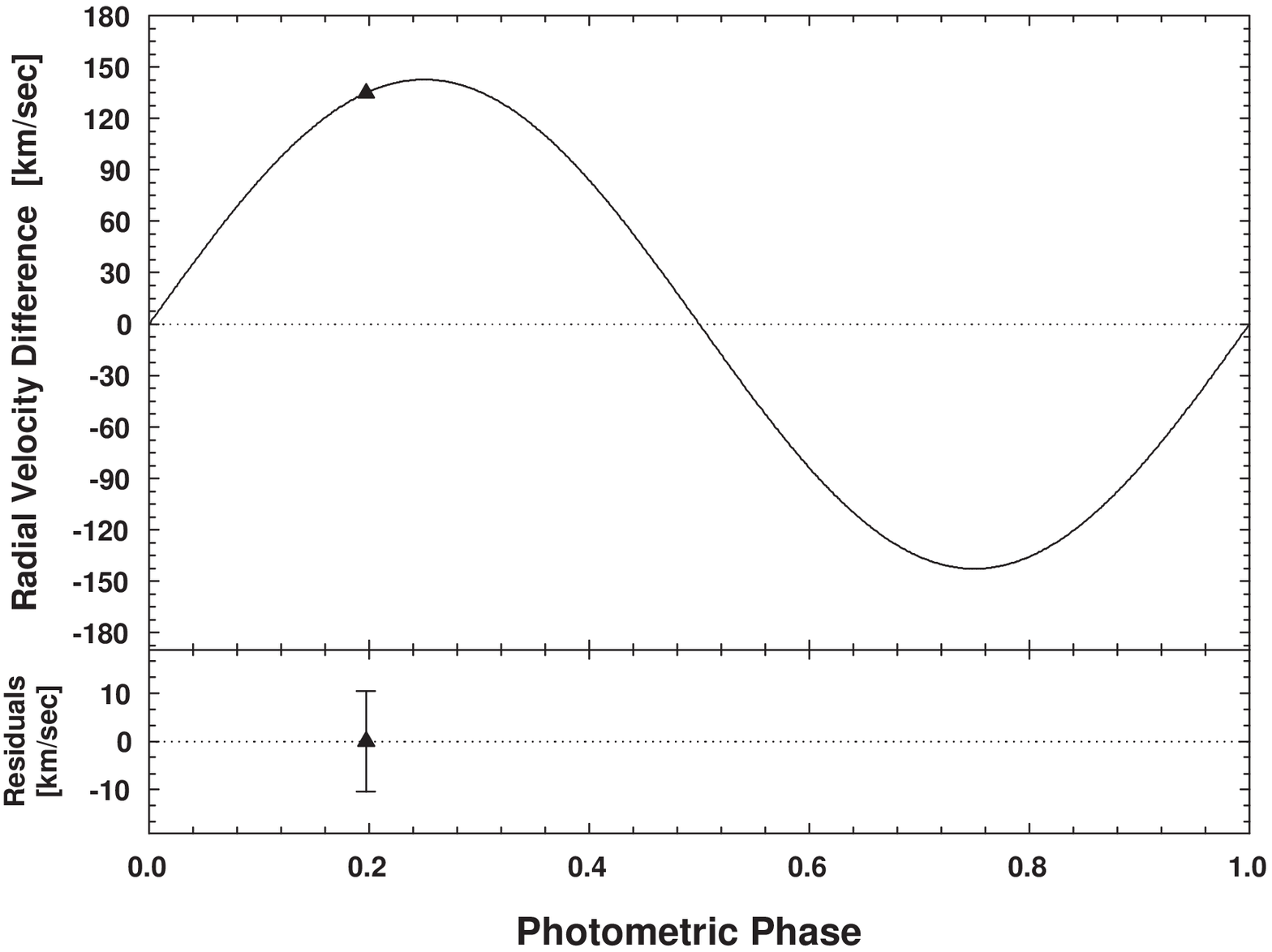} &
\includegraphics[width=3.3in]{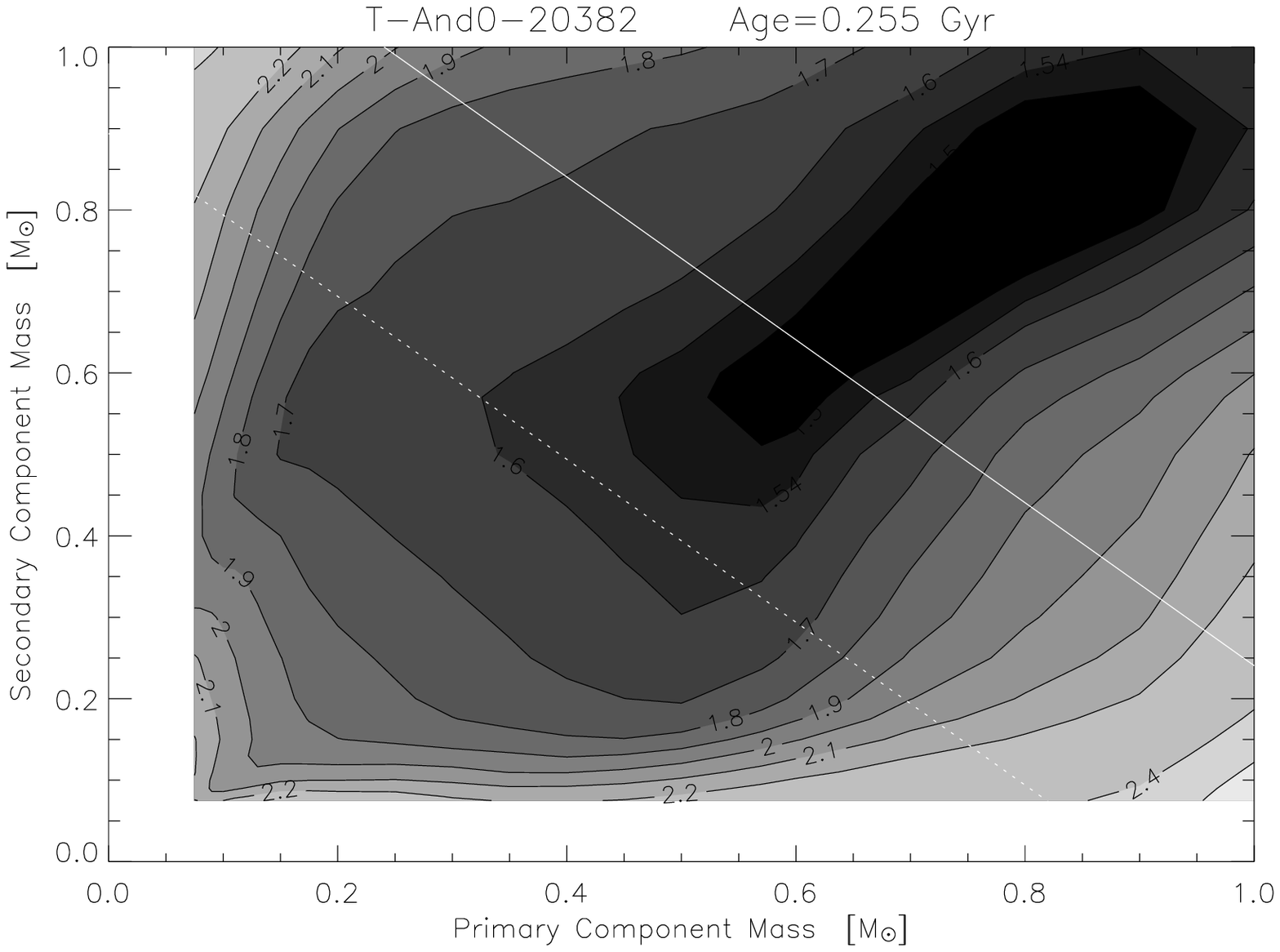}\\
a. Phased RVs with model & b. MECI likelihood contours
\end{tabular}}
\caption{The phased RVs of T-And0-20382 are shown with the best-fit circular orbit model to the difference between the component RVs.}
\end{figure}

\begin{deluxetable}{lc}
\tabletypesize{\tiny}
\tablecaption{Catalog information of T-And0-20382}
\tablewidth{0pt}
\tablehead{\colhead{Parameter} & \colhead{Value}}
\startdata
$\alpha$ (J2000)    & 01:22:08.780\\
$\delta$ (J2000)    & 46:59:55.41\\
USNO-B \ $B$-mag    & 16.190 $\pm$ 0.2\\
GSC2.3 \ $V$-mag    & 14.88  $\pm$ 0.33\\
USNO-B \ $R$-mag    & 14.685 $\pm$ 0.2\\
CMC14 \  $r'$-mag   & 14.859 $\pm$ 0.05\\
2MASS \  $J$-mag    & 13.154 $\pm$ 0.015\\
2MASS \  $H$-mag    & 12.619 $\pm$ 0.015\\
2MASS \  $K_s$-mag  & 12.440 $\pm$ 0.015\\
UCAC $\mu_\alpha$ [${\rm mas\,yr^{-1}}$] &  9.0  $\pm$ 5.7 \\
UCAC $\mu_\delta$ [${\rm mas\,yr^{-1}}$] & -3.5 $\pm$ 5.6\\
TrES third light [$R$-mag]      & 0.333 $\pm$ 0.050
\enddata
\end{deluxetable}

\begin{deluxetable}{lcc}
\tabletypesize{\tiny}
\tablecaption{Binary parameters of T-And0-20382}
\tablewidth{0pt}
\tablehead{\colhead{Parameter} & \colhead{Symbol} & \colhead{Value}}
\startdata
Orbital Period [days]                                                 & $P$       & 4.11975 $\pm$ 0.00031 \\
Epoch of eclipse [HJD]                                                & $t_0$       & 2452949.7657 $\pm$ 0.0046 \\
Number of light curve data points                                     & $N_{LC}$  & 5368 \\
Number of RV data points                                              & $N_{RV}$  & 1 \\
MECI analysis primary mass [$M_{\sun}$]                               & $M_A^{MECI}$ & 0.81 $\pm$ 0.60 \\
MECI analysis secondary mass [$M_{\sun}$]                             & $M_B^{MECI}$ & 0.80 $\pm$ 0.66   \\
MECI analysis binary age [Gyr]                                        & $T^{MECI}$ &  0.25 $\pm$ 0.77\\
Relative RV amplitude [${\rm km\,s^{-1}}$]                            & $K_A+K_B$      & 142.7 $\pm$ 11.0 \\
DEBiL estimate of the projection factor                               & $\sin i$     & 1.0000 $\pm$ 0.0057\\
Combined semimajor axis with projection factor [$R_{\sun}$]           & $a \sin i$   &$11.6^{+10.2}_{-1.1}$ \\
Mass sum with projection factor [$M_{\sun}$]                          & $(M_A+M_B) \sin^3 i$ & $1.2^{+6.9}_{-0.3}$
\enddata
\end{deluxetable}

\newpage
\subsection{T-CrB0-14232}

As with the two previous EBs, this binary was found to have a
double-lined spectrum and nearly identical eclipses, which
suggests that the components are likely similar. As with
T-And0-20382, the RV and MECI analyses are consistent with one
another, though the RV analysis results have large uncertainties
due to poorly determined orbital period and a poorly determined RV
measurement. The MECI results, however, provide stronger mass
constraints, suggesting that the EB components are both late
K-dwarfs.

\begin{figure}
\includegraphics[width=5in]{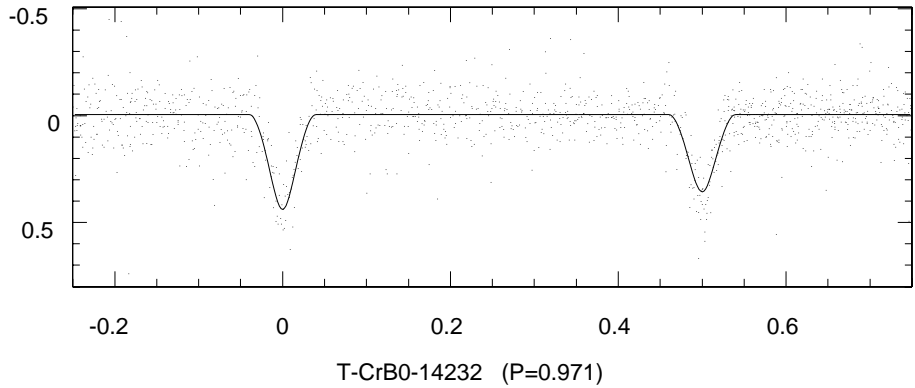}
\caption{The phased TrES light curve ($r$-band), with the best-fit MECI model (solid line).}
\end{figure}

\begin{figure}
\centerline{
\begin{tabular}{cc}
\includegraphics[width=3.1in]{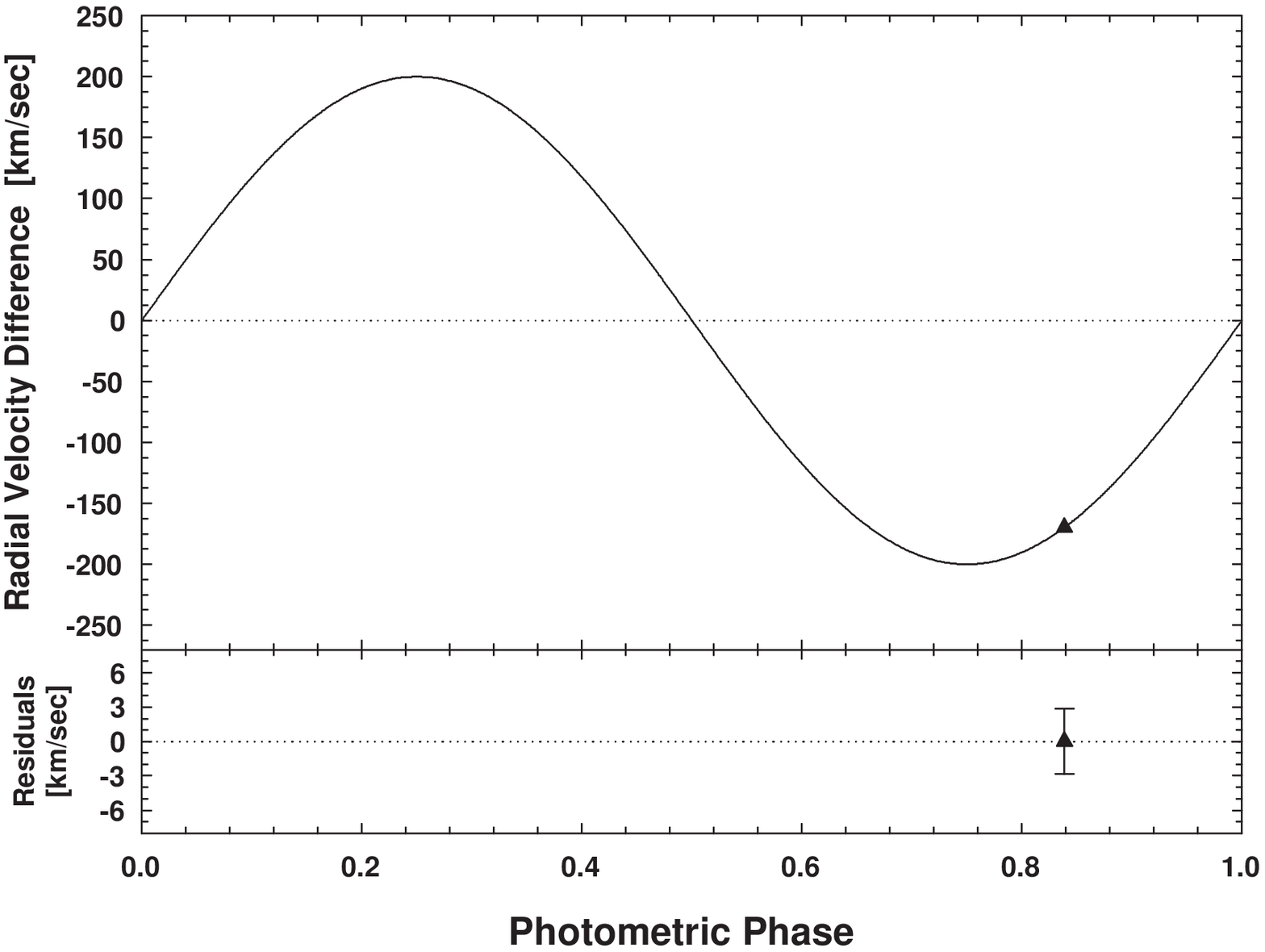} &
\includegraphics[width=3.3in]{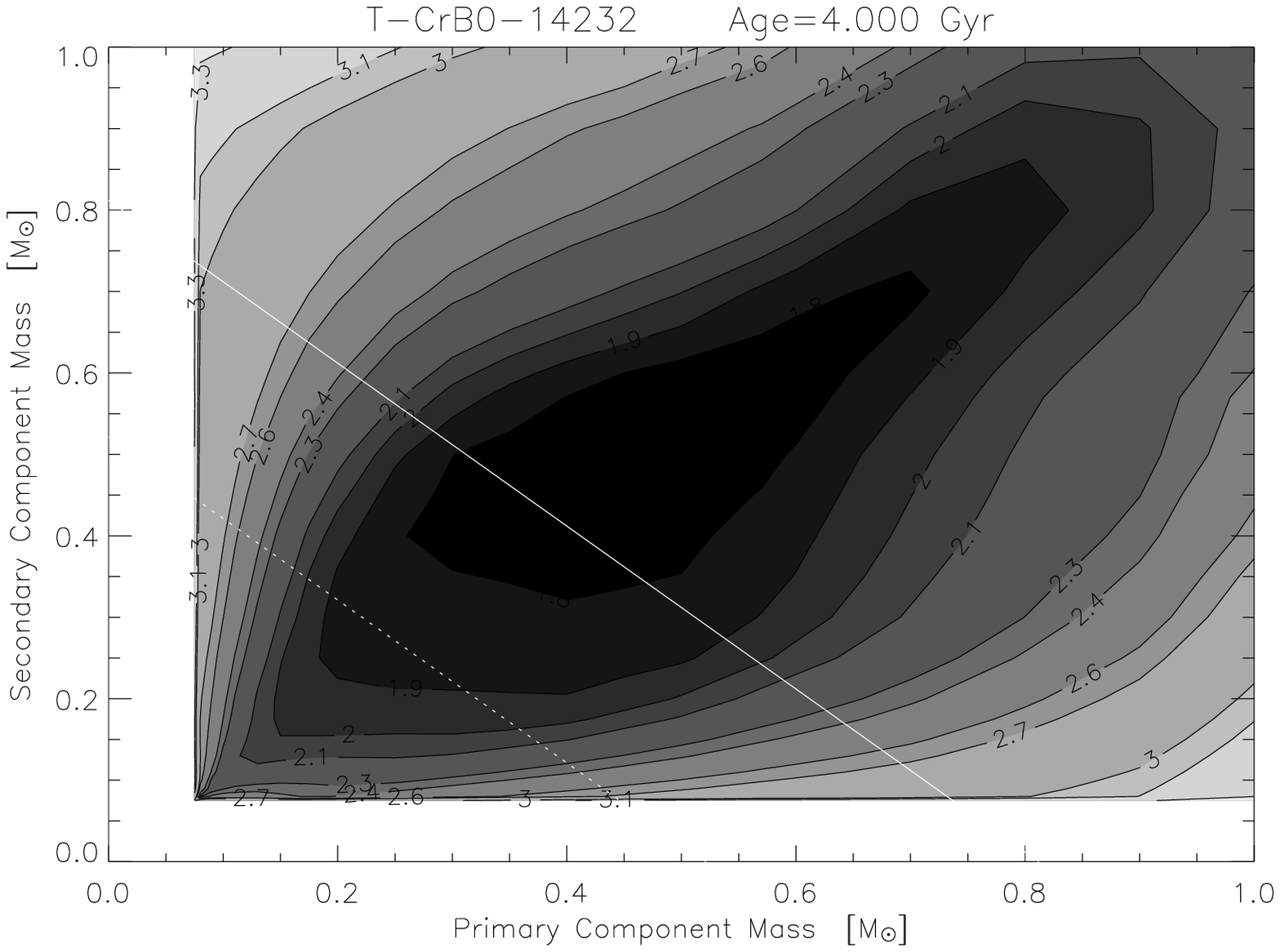}\\
a. Phased RVs with model & b. MECI likelihood contours
\end{tabular}}
\caption{The phased RVs of T-CrB0-14232 are shown with the best-fit circular orbit model to the difference between the component RVs.}
\end{figure}

\begin{deluxetable}{lc}
\tabletypesize{\tiny}
\tablecaption{Catalog information of T-CrB0-14232}
\tablewidth{0pt}
\tablehead{\colhead{Parameter} & \colhead{Value}}
\startdata
$\alpha$ (J2000)    & 16:10:22.495\\
$\delta$ (J2000)    & 33:57:52.33\\
USNO-B \ $B$-mag    & 17.985 $\pm$ 0.2\\
GSC2.3 \ $V$-mag    & 16.27  $\pm$ 0.31\\
USNO-B \ $R$-mag    &  16.170 $\pm$ 0.2\\
CMC14 \  $r'$-mag   &  16.092 $\pm$ 0.01 \\
2MASS \  $J$-mag    & 13.367  $\pm$ 0.015\\
2MASS \  $H$-mag    & 12.757 $\pm$ 0.015\\
2MASS \  $K_s$-mag  & 12.511 $\pm$ 0.015\\
USNO-B $\mu_\alpha$ [${\rm mas\,yr^{-1}}$] & 6  $\pm$ 1\\
USNO-B $\mu_\delta$ [${\rm mas\,yr^{-1}}$] & -4 $\pm$ 3\\
TrES third light [$R$-mag]       & 0.252 $\pm$ 0.039
\enddata
\end{deluxetable}

\begin{deluxetable}{lcc}
\tabletypesize{\tiny}
\tablecaption{Binary parameters of T-CrB0-14232}
\tablewidth{0pt}
\tablehead{\colhead{Parameter} & \colhead{Symbol} & \colhead{Value}}
\startdata
Orbital Period [days]                                                 & $P$      &   0.971305 $\pm$ 0.000038 \\
Epoch of eclipse [HJD]                                                & $t_0$       & 2453514.79889 $\pm$ 0.00055 \\
Number of light curve data points                                     & $N_{LC}$       &  1287\\
Number of RV data points                                              & $N_{RV}$       &  1\\
MECI analysis primary mass [$M_{\sun}$]                               & $M_A^{MECI}$ &  0.59 $\pm$ 0.28   \\
MECI analysis secondary mass [$M_{\sun}$]                             & $M_B^{MECI}$ &  0.56 $\pm$ 0.36  \\
MECI analysis binary age [Gyr]                                        & $T^{MECI}$ & 4.0 $\pm$ 13.4 \\
Relative RV amplitude [${\rm km\,s^{-1}}$]                            & $K_A+K_B$   & 200.00 $\pm$ 3.33 \\
DEBiL estimate of the projection factor                               & $\sin i$     & 0.997 $\pm$ 0.013\\
Combined semimajor axis with projection factor [$R_{\sun}$]           & $a \sin i$   & $3.84^{+1.65}_{-0.52}$\\
Mass sum with projection factor [$M_{\sun}$]                          & $(M_A+M_B) \sin^3 i$ & $0.81^{+1.54}_{-0.29}$
\enddata
\end{deluxetable}

\newpage
\subsection{T-Dra0-03021}

This binary was found to have disparate eclipses, which indicates
that the component are not equal. Though the RV measurement was
uncertain, the high signal-to-noise ratio of this EB's LC allowed
us to determine its orbital period with high accuracy, thus
reducing the RV analysis uncertainties. The RV analysis indicates
that the components are early M-dwarfs, however the binary's 2MASS
colors indicate that it is likely to be a K-dwarf binary.
Furthermore, the eclipse durations are so long that they caused
the MECI analysis to produce mass estimates that are out of range
($>$1$M_{\sun}$). Similar to the case of T-And0-04829, these
contradictory results prevent us from arriving at any firm
conclusion.

\begin{figure}
\includegraphics[width=5in]{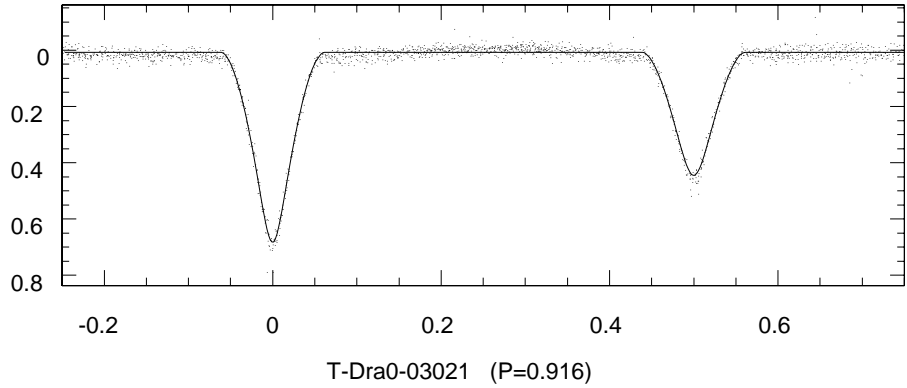}
\caption{The phased TrES light curve ($r$-band), with the best-fit MECI model (solid line).}
\end{figure}

\begin{figure}
\centerline{
\begin{tabular}{cc}
\includegraphics[width=3.1in]{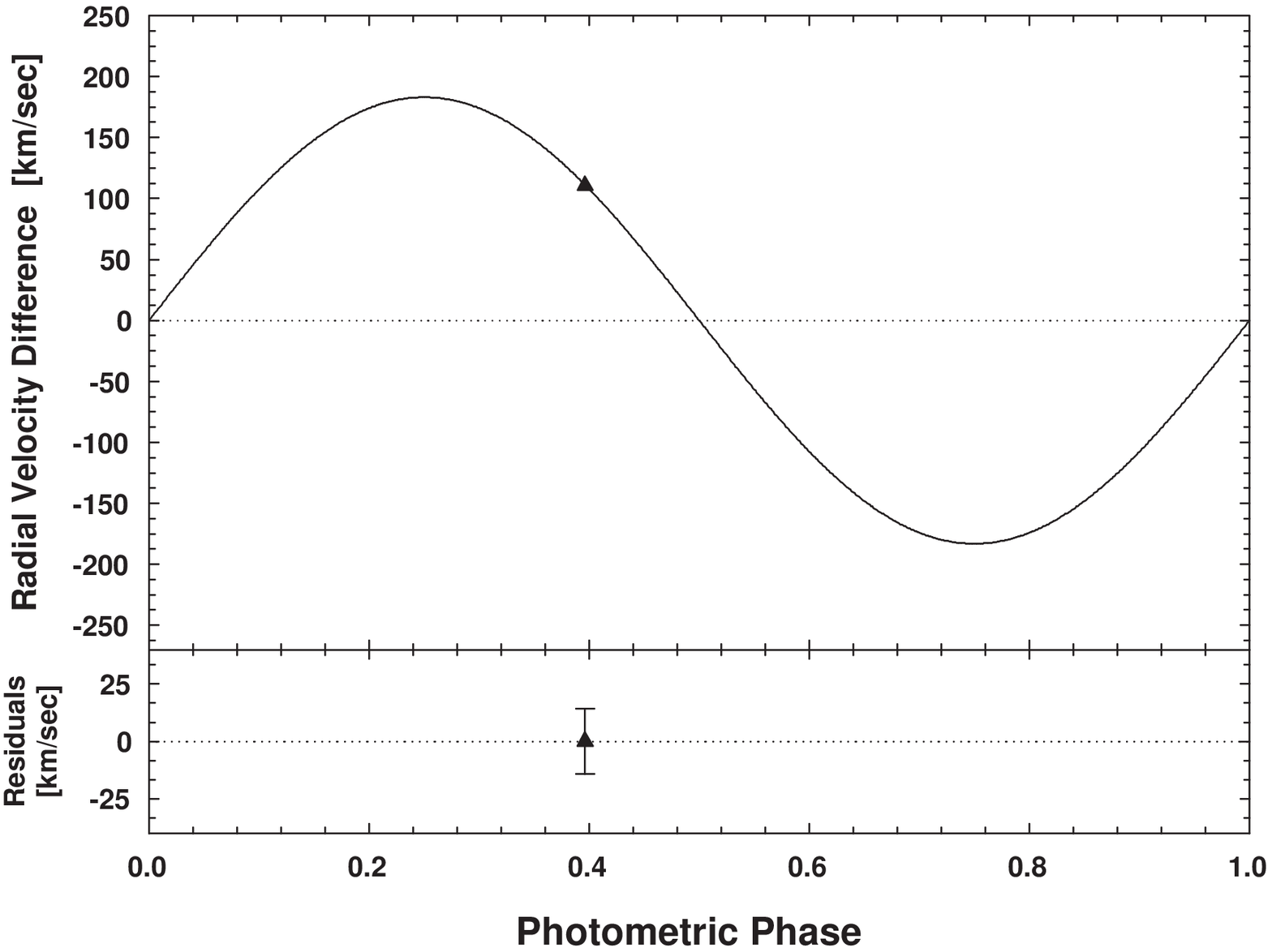} &
\includegraphics[width=3.3in]{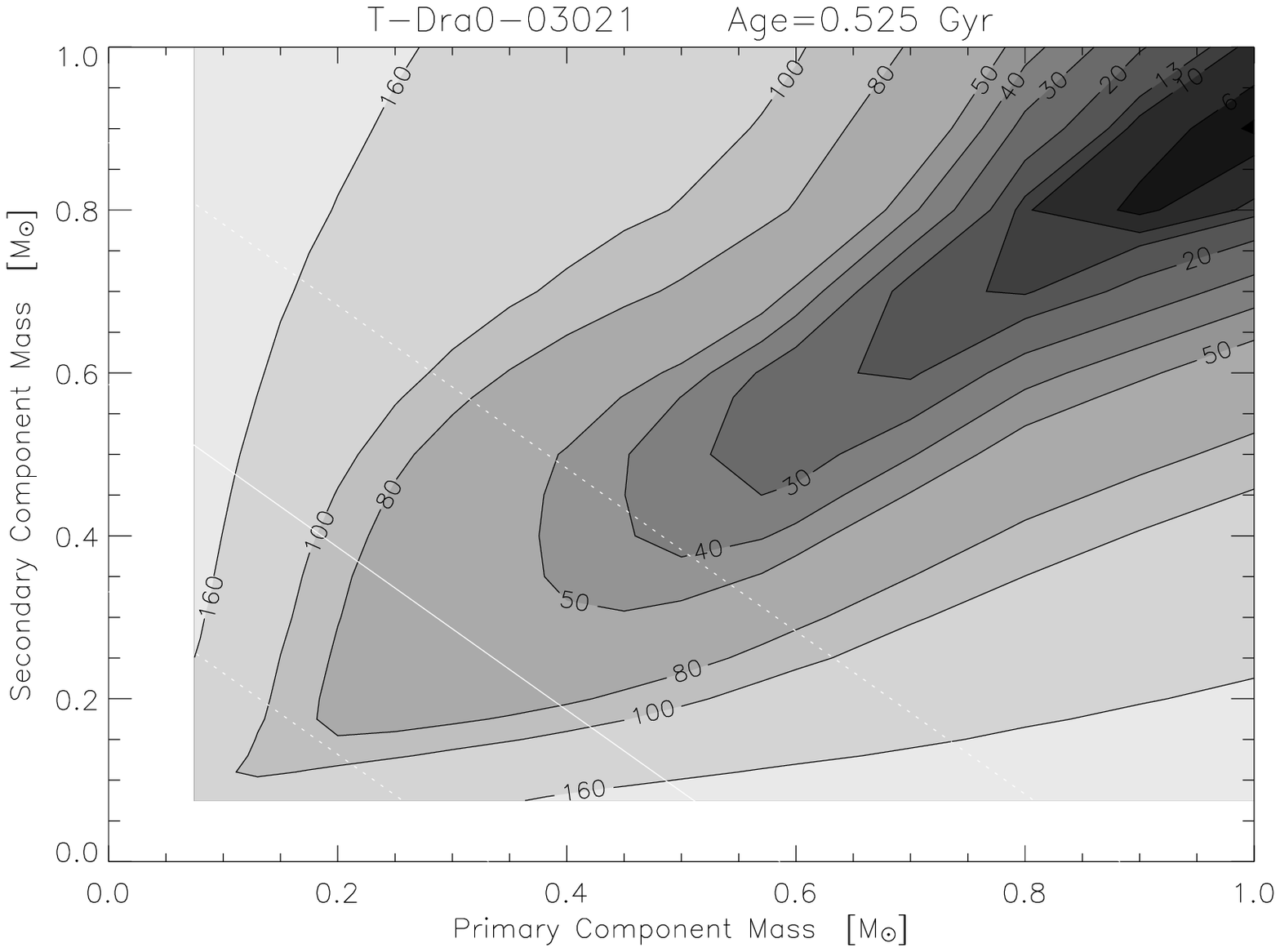}\\
a. Phased RVs with model & b. MECI likelihood contours
\end{tabular}}
\caption{The phased RVs of T-Dra0-03021 are shown with the best-fit circular orbit model to the difference between the component RVs.}
\end{figure}

\begin{deluxetable}{lc}
\tabletypesize{\tiny}
\tablecaption{Catalog information of T-Dra0-03021}
\tablewidth{0pt}
\tablehead{\colhead{Parameter} & \colhead{Value}}
\startdata
$\alpha$ (J2000)    & 17:01:03.618 \\
$\delta$ (J2000)    & 55:14:54.70\\
USNO-B \ $B$-mag    & 15.135 $\pm$ 0.2\\
GSC2.3 \ $V$-mag    & 13.68  $\pm$ 0.16 \\
USNO-B \ $R$-mag    & 13.135 $\pm$ 0.2\\
2MASS \  $J$-mag    & 11.498 $\pm$ 0.015\\
2MASS \  $H$-mag    & 10.924 $\pm$ 0.015\\
2MASS \  $K_s$-mag  & 10.771 $\pm$ 0.015\\
USNO-B $\mu_\alpha$ [${\rm mas\,yr^{-1}}$] &  20  $\pm$ 2\\
USNO-B $\mu_\delta$ [${\rm mas\,yr^{-1}}$] & -20 $\pm$ 2\\
TrES third light [$R$-mag]      &  0.045 $\pm$ 0.011
\enddata
\end{deluxetable}

\begin{deluxetable}{lcc}
\tabletypesize{\tiny}
\tablecaption{Binary parameters of T-Dra0-03021}
\tablewidth{0pt}
\tablehead{\colhead{Parameter} & \colhead{Symbol} & \colhead{Value}}
\startdata
Orbital Period [days]                                                 & $P$            &  0.9159291 $\pm$ 0.0000025 \\
Epoch of eclipse [HJD]                                                & $t_0$       & 2453128.081067 $\pm$ 0.000051\\
Number of light curve data points                                     & $N_{LC}$       & 2000 \\
Number of RV data points                                              & $N_{RV}$       &  1\\
Relative RV amplitude [${\rm km\,s^{-1}}$]                            & $K_A+K_B$ &  183.1 $\pm$ 23.3  \\
DEBiL estimate of the projection factor                               & $\sin i$     & 0.9978 $\pm$ 0.0034\\
Combined semimajor axis with projection factor [$R_{\sun}$]           & $a \sin i$   & $3.31^{+0.54}_{-0.49}$\\
Mass sum with projection factor [$M_{\sun}$]                          & $(M_A+M_B) \sin^3 i$ & $0.58^{+0.29}_{-0.25}$
\enddata
\end{deluxetable}

\newpage
\subsection{T-Dra0-07116}

As with previous EBs, this binary was found to have double-lined
spectra and nearly identical eclipses, which suggests that the
components are likely similar. Though this system has two RV
measurements, they are spaced approximately a half a phase apart,
thus preventing them from providing a strong verification of the
RV model's phase. The RV and MECI analyses are comparably well
constrained, but are marginally consistent with one another. The
RV analysis indicates that the components are either early
M-dwarfs or late K-dwarfs, while the MECI analysis suggests that
they are likely K-dwarfs. We thus conclude that this system is
probably a late K-dwarf binary.

In principle, since this system has two double-lined spectra, we
can use the \citet{Wilson41} method to estimate the binary's mass
ratio ($q$) and barycenteric RV ($V_\gamma$), independently of its
orbital phase and period. With this method, one fits a linear
regression to a plot of the binary components' RVs (see
Figure~\ref{WilsonMethodDra7116}), and determines $q=-1/b_1$,
where $b_1$ is the slope of the regression. Similarly,
$V_\gamma=b_0 / (1-b_1)$, where $b_0$ is the offset of the
regression. Applying this method to T-Dra0-07116, we find that $q
= 0.785 \pm 0.070$, and $V_\gamma = -6.4 \pm 1.9\:{\rm
km\,s^{-1}}$ relative to GJ~182. From this, we can estimate that
$M_1 = 0.50 \pm 0.15 \: M_{\sun}$ and $M_2 = 0.38 \pm 0.12 \:
M_{\sun}$. However, this small value of $q$ is inconsistent with
our previous conclusions that the binary components are similar
(i.e. $q \approx 1$). We therefore chose not to adopt the Wilson
method results, which with only two data points are not expected
to be reliable.

\begin{figure}
\includegraphics[width=5in]{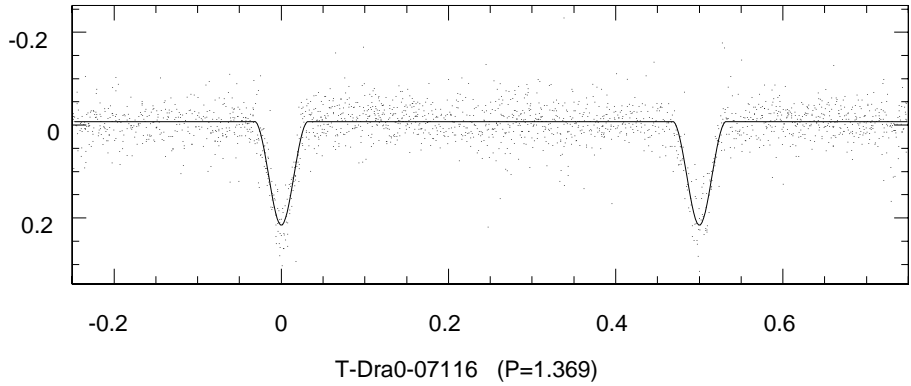}
\caption{The phased TrES light curve ($r$-band), with the best-fit MECI model (solid line).}
\end{figure}

\begin{figure}
\centerline{
\begin{tabular}{cc}
\includegraphics[width=3.1in]{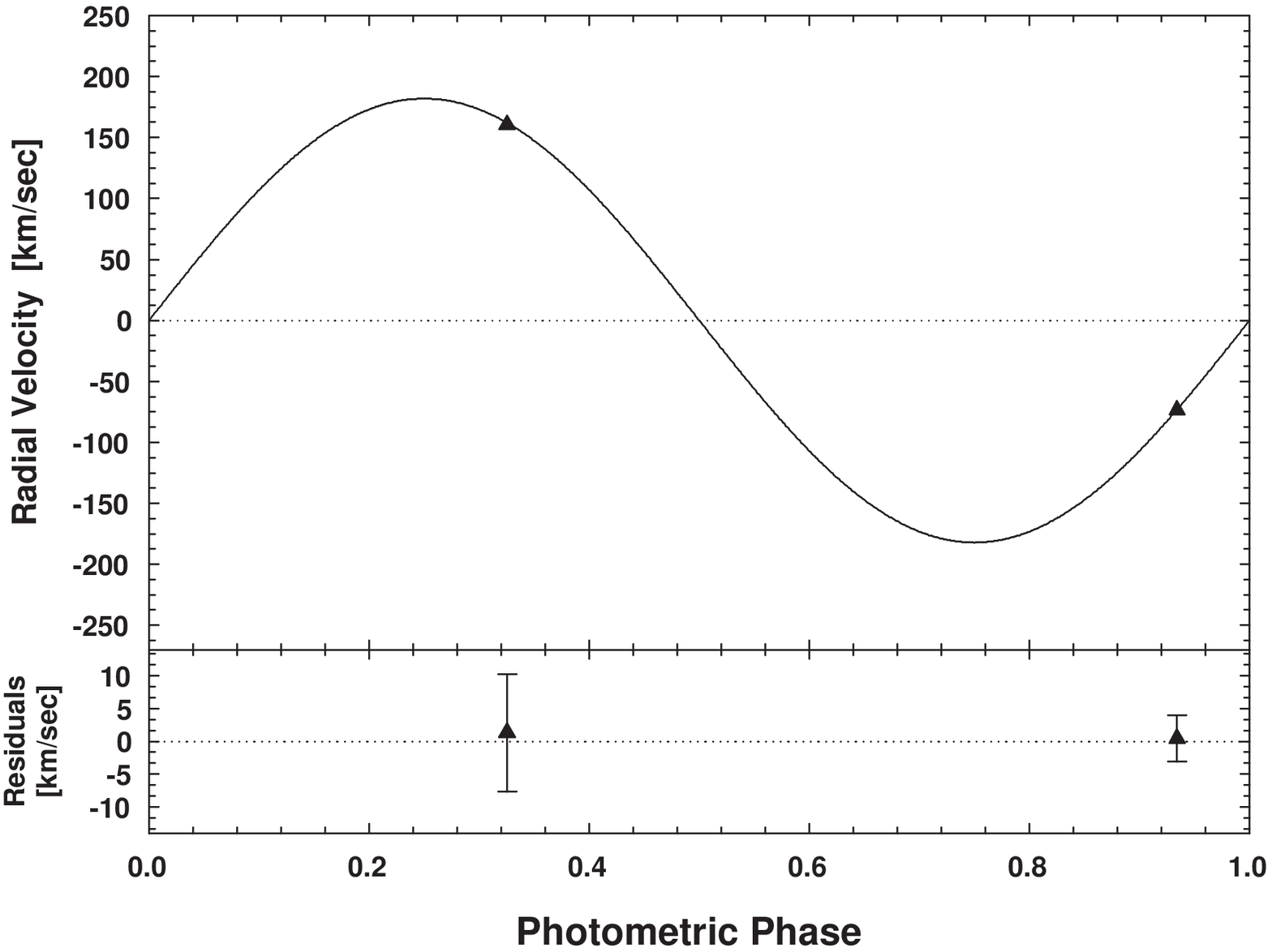} &
\includegraphics[width=3.3in]{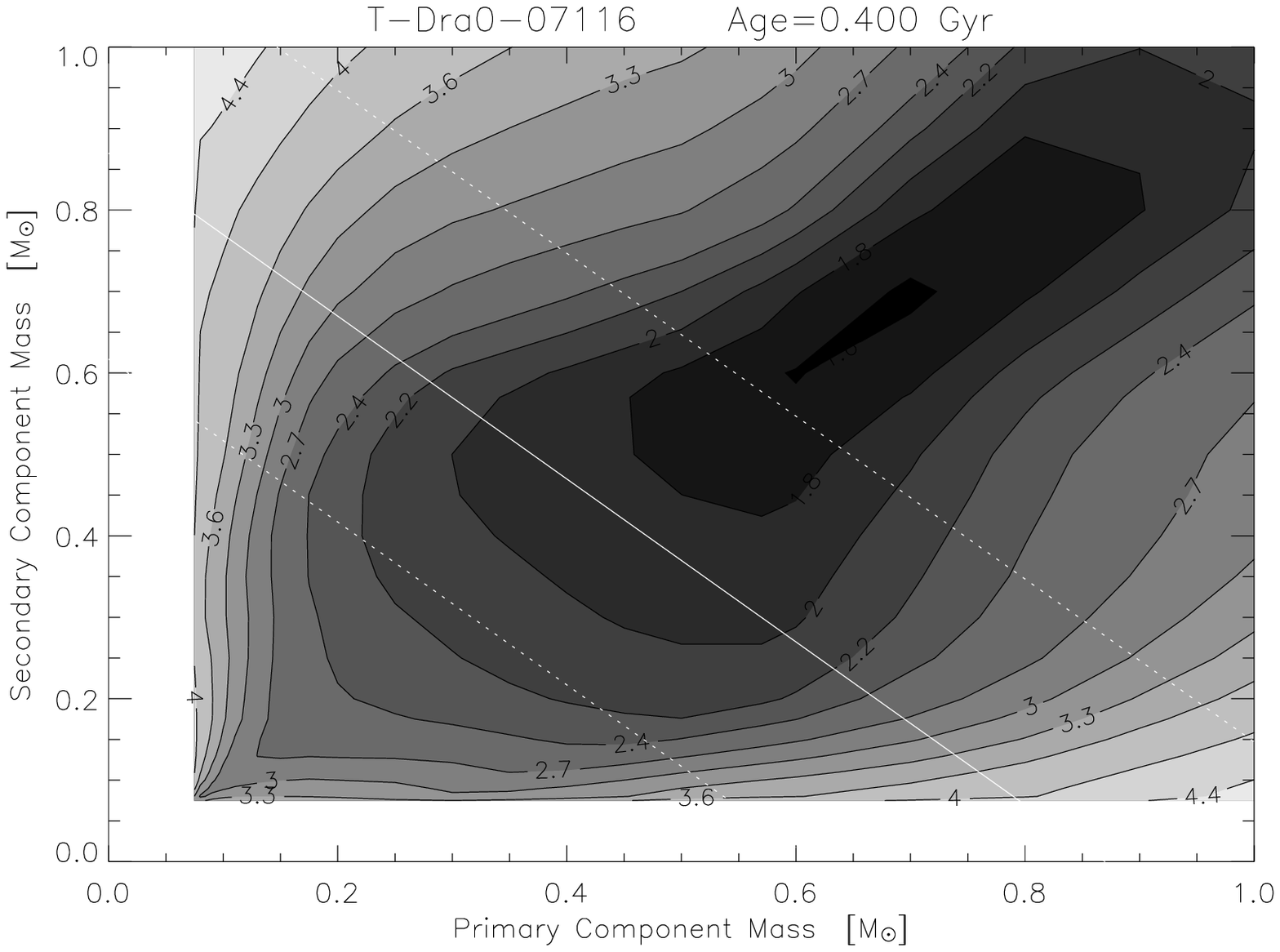}\\
a. Phased RVs with model & b. MECI likelihood contours
\end{tabular}}
\caption{The phased RVs of T-Dra0-07116 are shown with the best-fit circular orbit model to the difference between the component RVs.}
\end{figure}

\begin{deluxetable}{lc}
\tabletypesize{\tiny}
\tablecaption{Catalog information of T-Dra0-07116}
\tablewidth{0pt}
\tablehead{\colhead{Parameter} & \colhead{Value}}
\startdata
$\alpha$ (J2000)    & 17:02:53.025\\
$\delta$ (J2000)    & 55:07:47.44\\
USNO-B \ $B$-mag    & 16.805 $\pm$ 0.2\\
GSC2.3 \ $V$-mag    &  15.33  $\pm$ 0.16 \\
USNO-B \ $R$-mag    &  14.350 $\pm$ 0.2\\
2MASS \  $J$-mag    & 12.620  $\pm$ 0.015\\
2MASS \  $H$-mag    & 11.964 $\pm$ 0.015\\
2MASS \  $K_s$-mag  & 11.830 $\pm$ 0.015\\
USNO-B $\mu_\alpha$ [${\rm mas\,yr^{-1}}$] & -2  $\pm$ 1\\
USNO-B $\mu_\delta$ [${\rm mas\,yr^{-1}}$] & -16 $\pm$ 1\\
TrES third light [$R$-mag]          & 0.258 $\pm$ 0.049
\enddata
\end{deluxetable}

\begin{deluxetable}{lcc}
\tabletypesize{\tiny}
\tablecaption{Binary parameters of T-Dra0-07116}
\tablewidth{0pt}
\tablehead{\colhead{Parameter} & \colhead{Symbol} & \colhead{Value}}
\startdata
Orbital Period [days]                                                 & $P$       &  $1.368910_{-0.000019}^{+0.000026}$\\
Epoch of eclipse [HJD]                                                & $t_0$     &  $2453517.2157_{-0.0054}^{+0.0073}$ \\
Number of light curve data points                                     & $N_{LC}$       & 2000 \\
Number of RV data points                                              & $N_{RV}$       &  2\\
MECI analysis primary mass [$M_{\sun}$]                               & $M_A^{MECI}$ & 0.71 $\pm$ 0.21  \\
MECI analysis secondary mass [$M_{\sun}$]                             & $M_B^{MECI}$ & 0.69 $\pm$ 0.21  \\
MECI analysis binary age [Gyr]                                        & $T^{MECI}$ & 0.4 $\pm$ 2.0\\
Relative RV amplitude [${\rm km\,s^{-1}}$]                            & $K_A+K_B$ &  182.03 $\pm$ 6.63  \\
DEBiL estimate of the projection factor                               & $\sin i$     & 0.9945 $\pm$ 0.0072\\
Combined semimajor axis with projection factor [$R_{\sun}$]           & $a \sin i$   & $4.93^{+0.48}_{-0.52}$\\
Mass sum with projection factor [$M_{\sun}$]                          & $(M_A+M_B) \sin^3 i$ & $0.86^{+0.27}_{-0.25}$
\enddata
\end{deluxetable}

\begin{figure}
\includegraphics[width=5in]{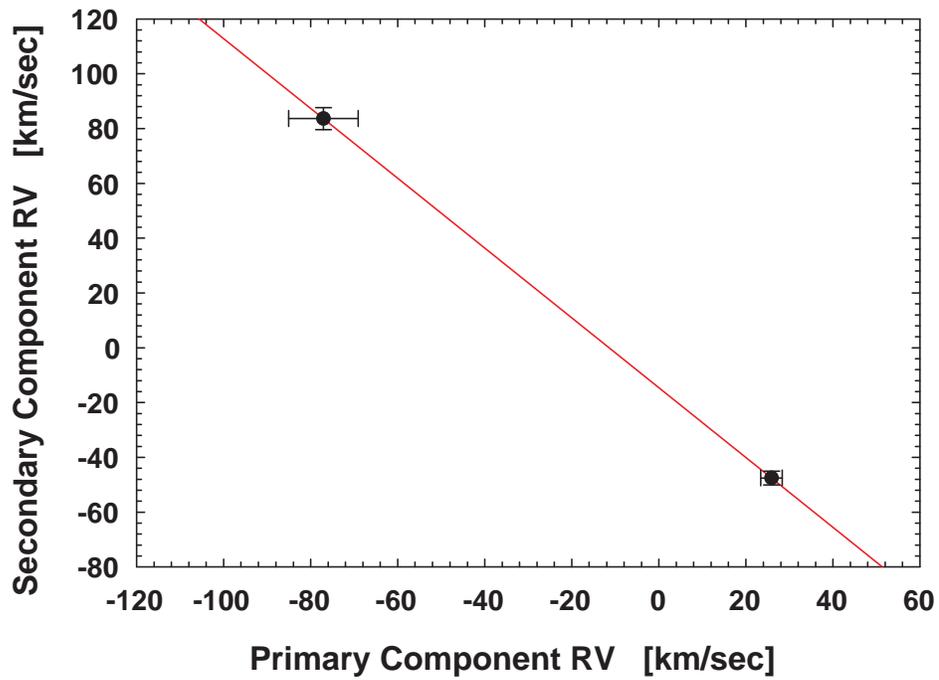}
\caption{A linear regression of a plot of T-Dra0-07116 components'
RVs. The slope and offset of this regression are used in the
\citet{Wilson41} method to determine the binary's mass ratio and
barycenteric RV.} \label{WilsonMethodDra7116}
\end{figure}

\newpage
\section{T-Cas0-07656: A Pulsating Eclipsing Binary}
\label{secCas07656pulsating}

T-Cas0-07656 is a pulsating EB that was located through a
systematic search of previously identified EBs from ten TrES
fields (see Tables \ref{tableFieldsObs} and
\ref{tableFieldsYield}). We selected LCs that exhibited strong
residual periodicities, after subtracting the best-fit DEBiL
model. To quantify these periodicities, we employed a version of
the analysis of variances period finder
[\citet{SchwarzenbergCzerny89, SchwarzenbergCzerny96}; see also
sections \S\ref{subsecCh2firstTier} and \S\ref{secMethod}], and
ignored the original EB orbital period and its harmonics. After
ranking the EBs by the periodicity strength of their residuals, we
manually inspected the most promising candidates. Of these
candidates we chose T-Cas0-07656 as the best case for an EB with a
stable pulsation. The orbital period of T-Cas0-07656 was found to
be $P = 1.560019 \pm 0.000044$ days, with an epoch of eclipse at
$t_0 = 2453252.99121 \pm 0.00068$ HJD. However, it was not
sufficiently detached to enable accurate component identification
using MECI, though its colors indicate that the system is likely a
late F-dwarf binary. Once the DEBiL model was subtracted from the
LC, the residual was found to have a 0.1-magnitude fluctuation in
the $r$-band. This fluctuation had a stable period of 1.9053 $\pm$
0.0001 hours, suggesting that one of the EB components is likely a
$\delta$-Scuti type variable. As such, by comparing and combining
the information gather through both the EB model and the pulsation
model, one could perhaps introduce new constraints and insights
into both these models.

\begin{figure}
\includegraphics[width=6in]{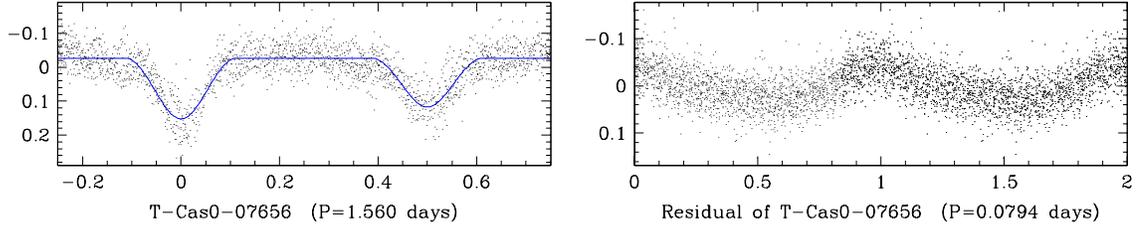}
\caption{Left: The phased LC of T-Cas0-07656. Right: The phased
residual of T-Cas0-07656, after subtracting the best-fit DEBiL
model.}
\end{figure}

\begin{deluxetable}{lc}
\tabletypesize{\tiny}
\tablecaption{Catalog information of T-Cas0-07656}
\tablewidth{0pt}
\tablehead{\colhead{Parameter} & \colhead{Value}}
\startdata
$\alpha$ (J2000)    & 00:23:55.733\\
$\delta$ (J2000)    & 50:43:47.61\\
USNO-B \ $B$-mag    & 14.125 $\pm$ 0.2\\
GSC2.3 \ $V$-mag    & 13.03  $\pm$ 0.37 \\
USNO-B \ $R$-mag    & 13.325 $\pm$ 0.2\\
2MASS \  $J$-mag    & 11.851 $\pm$ 0.015\\
2MASS \  $H$-mag    & 11.547 $\pm$ 0.015\\
2MASS \  $K_s$-mag  & 11.492 $\pm$ 0.015\\
UCAC $\mu_\alpha$ [${\rm mas\,yr^{-1}}$] &  6.0  $\pm$ 5.5\\
UCAC $\mu_\delta$ [${\rm mas\,yr^{-1}}$] & -5.4 $\pm$ 5.5 \\
TrES third light [$R$-mag]  & 0.034 $\pm$ 0.007
\enddata
\label{tableCas7656CatalogInfo}
\end{deluxetable}

\newpage
\section{T-Cyg1-03378: A Binary with Large O-C Eclipse Timing Variations}
\label{secCyg03378Timing}

\begin{figure}
\includegraphics[width=5in]{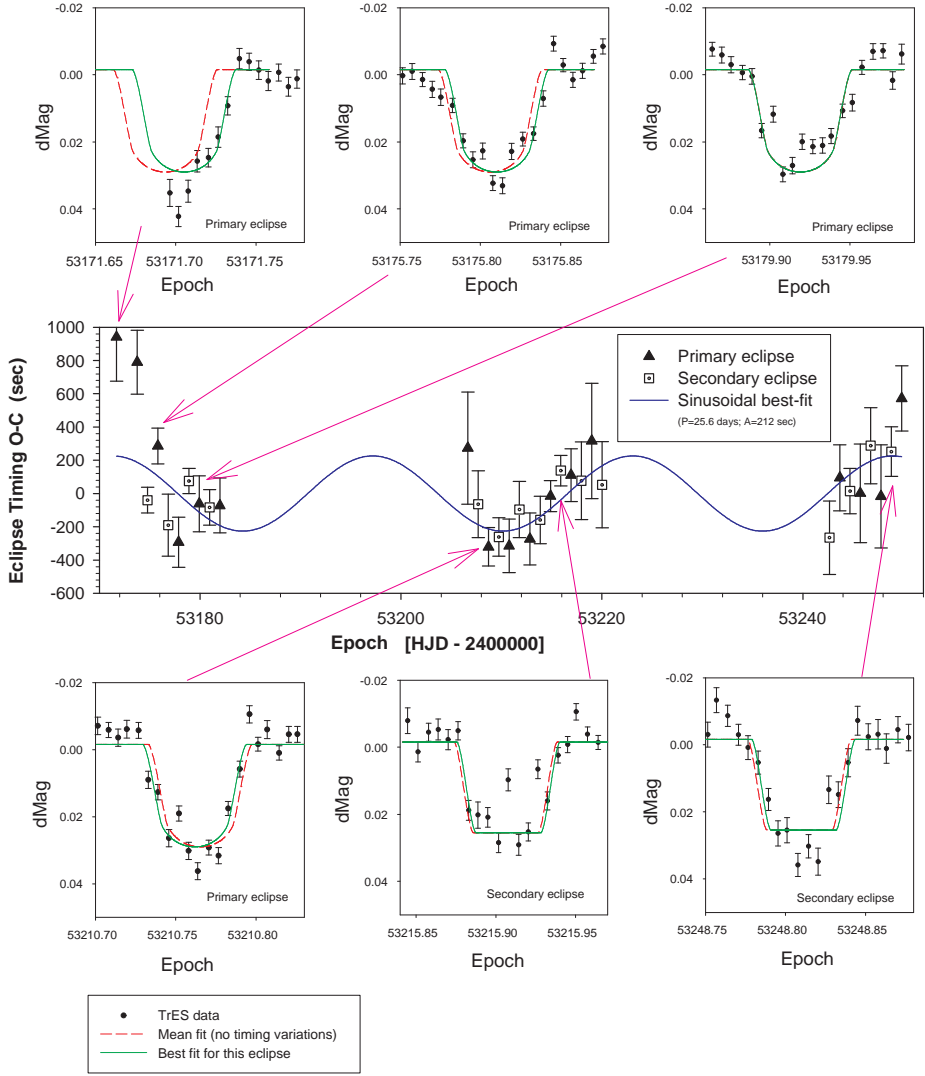}
\caption{Measured $O-C$ eclipse timing variations found in the
TrES LC of T-Cyg1-03378 (center), with details of six selected
eclipses. Each of the detailed eclipses shows both the eclipse
model with the best-fit offset to the individual eclipse data
(dotted line), and the model whose offset is determined by
extrapolating a fixed period (dashed line). The eclipse timing
plot includes $O-C$ measurements of primary eclipses (triangle
symbols) and $O-C$ measurements of secondary eclipses (square
symbols). The solid line indicates the best-fit sinusoidal model
to the TrES $O-C$ timing measurements.}
\label{fig3bodyTrES}
\end{figure}

T-Cyg1-03378 is an EB with large $O-C$ eclipse timing variation that
was found in a systematic search through previously identified EBs
from ten TrES fields (see Tables \ref{tableFieldsObs} and
\ref{tableFieldsYield}). In brief, for a given EB, we measured the
timing of each of its eclipses, provided we had at least three
observations. Fortunately, TrES operates at an effective 9-minute
cadence, thus typically capturing 10-20 observations in each
eclipse that is fully observed. For a given eclipse, we collect
all the observations made while it was in progress, as well as a
small number of out-of-eclipse observations made immediately
before and after the eclipse. We then remove all outliers, and
compare the remaining data to a generated DEBiL LC during this
eclipse, adopting the EB parameters of the best-fit model. We
compute the chi-squared value ($\chi^2$) of the DEBiL LC for these
data, and then repeat this computation in a range of time-offsets
by varying the epoch of perihelion of the DEBiL model. We find the
offset that minimizes the chi-squared value, and compute its
1$\sigma$ uncertainty. This measurement is recorded as the timing
$O-C$ during the center of the eclipse. We then repeat this
procedure for all the eclipses in the given EB's LC (see Figure
\ref{fig3bodyTrES}). Note that some LCs have long term trends that
interfere with fitting some of the eclipses. To correct this, we
normalize the magnitudes of the observations during an eclipse, so
that the median of the out-of-eclipse observations remains
constant. However, if only the ingress or egress of an eclipse are
observed, then there results a degeneracy between the magnitude
normalization and the eclipse offset, which produces large $O-C$
uncertainties.

Once all the $O-C$ of a given EB are recorded, we removed any
linear trends, which are due to having assumed an inaccurate
period\footnote{We also used this timing correction to fine-tune
the periods of EBs (see \S\ref{secMethod}).}. We then examined the
de-trended timing variation for second order effects and for
sinusoidal variations. EBs whose eclipse timing variations fit
such a pattern significantly better than a straight line (i.e. an
F-test), were selected to be examined manually. Of these selected
systems, T-Cyg1-03378 was identified as having the most convincing
sinusoidal timing variations.

T-Cyg1-03378 has an orbital period of $P = 2.05637567 \pm
0.00000104$ days, or possibly half that if the secondary eclipse
is unseen, and an epoch of eclipse at $t_0 = 2453403.037891 \pm
0.000190$. It was found to have a 3.5-minute $O-C$ amplitude,
repeating at a 25.6-day period (see Figure~\ref{fig3bodyTrES}).
This phenomenon is probably due to the orbital perturbations due
to the gravitational pull of an unusually tight tertiary
component. It is unlikely that such a short-period and
high-amplitude timing variations would be due only to a light-time
effect, as this would require a dim yet extraordinarily massive
tertiary object. This system may offer a rare test case for the
development of eclipse timing analysis methodologies, which may
ultimately be used for extrasolar planet discovery
\citep{Holman05, Agol05}. Following this discovery of T-Cyg1-03378
in the TrES dataset, we obtained additional archival observations
from PSST and HATNet (see Figure~\ref{fig3bodyTrES_HAT}). These
data have much larger $O-C$ uncertainties than TrES, however they
also seem to show the same sinusoidal timing variations, though at
certain times this pattern seems to be out of phase.

Upon further examination, this EB was found to be in an optical
binary, with the neighboring component being approximately 4"
away. Because of their proximity, these two sources are blended in
our photometric data, and we were unable to determine which of
these sources is the EB. Furthermore, this blending significantly
diluted the depth of the eclipses, thus explaining why the eclipse
is only $\sim$0.035 mag in $r$-band. We observed T-Cyg1-03378
using TRES, although we produced only single-lined spectra for
both the optical binary components. This, however, may be a result
of the fact that we were only able to observe this binary within a
few hours if its eclipse. If the relative RVs between the
components were less than a few tens of $\:{\rm km\ s^{-1}}$, we
would not be able to resolve the two template cross-correlation
peaks. However, if this binary is indeed single-lined, then we may
conclude that the secondary component is dim and low-mass, and
thus the secondary eclipse is unseen and the correct period is
half of the value listed above.

Finally, both the MECI analysis and the colors of T-Cyg1-03378
suggest that it is likely to be a late F-dwarf binary. However,
the fact that this system has a significant amount of third light
makes these determinations less certain. We thus strongly
encourage additional observations be made of this system so that
this intriguing EB may be better understood.

\begin{figure}
\includegraphics[width=5in]{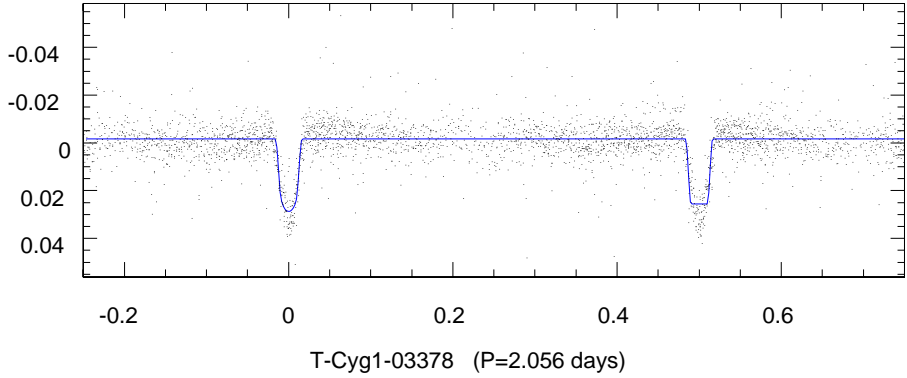}
\caption{The phased TrES light curve ($r$-band), with the best-fit DEBiL model (solid line).
The poor DEBiL model fit seems to be due to the fact that this LC contains significant third light,
which dilutes the eclipses and makes them shallower than they are supposed to be.}
\end{figure}

\begin{figure}
\includegraphics[width=5in]{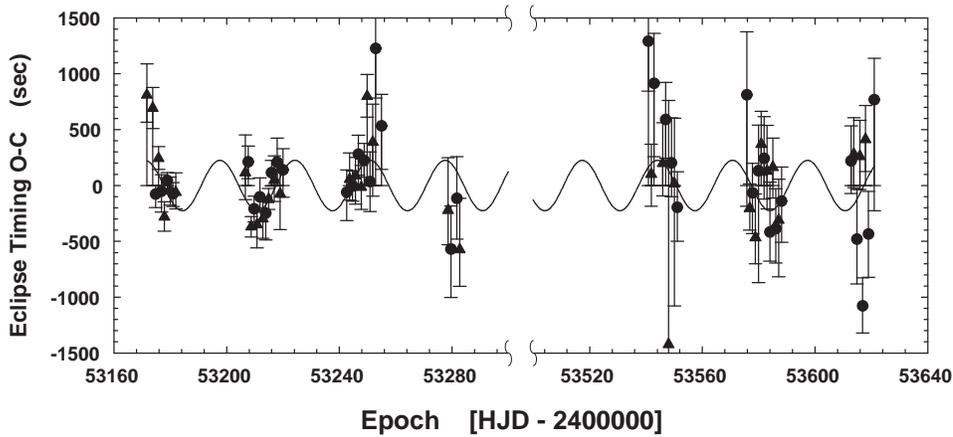}
\caption{The $O-C$ eclipse timing variations found in a light
curve constructed by combining TrES, PSST, and HATNet
observations. The triangle symbols represent timing measurements
of primary eclipses, while the circular symbols indicate timing
measurements of secondary eclipses.}
\label{fig3bodyTrES_HAT}
\end{figure}

\begin{deluxetable}{lcc}
\tabletypesize{\tiny}
\tablecaption{Catalog information of T-Cyg1-03378}
\tablewidth{0pt}
\tablehead{\colhead{Parameter} &
 \colhead{\begin{tabular}{c} Visual\\ Component A \end{tabular}} &
 \colhead{\begin{tabular}{c} Visual\\ Component B \end{tabular}}}
\startdata
$\alpha$ (J2000)    & 20:10:32.489 & 20:10:32.832\\
$\delta$ (J2000)    & 47:33:16.95  & 47:33:14.64\\
USNO-B \ $B$-mag    & 12.385 $\pm$ 0.2 & $\cdots$ \\
GSC2.3 \ $V$-mag    & 11.99 $\pm$ 0.33 & $\cdots$ \\
USNO-B \ $R$-mag    & 11.470 $\pm$ 0.2 & $\cdots$ \\
CMC14 \  $r'$-mag   & 12.755 $\pm$ 0.05  & $\cdots$ \\
2MASS \  $J$-mag    & 11.643 $\pm$ 0.015 & 12.164 $\pm$ 0.015\\
2MASS \  $H$-mag    & 11.419 $\pm$ 0.015 & 12.081 $\pm$ 0.015\\
2MASS \  $K_s$-mag  & 11.351 $\pm$ 0.015 & 12.029 $\pm$ 0.015\\
USNO-B $\mu_\alpha$ [${\rm mas\,yr^{-1}}$] &  6 $\pm$ 2 & $\cdots$ \\
USNO-B $\mu_\delta$ [${\rm mas\,yr^{-1}}$] &  4 $\pm$ 2 & $\cdots$ \\
TrES third light [$R$-mag]  & 0.189 $\pm$ 0.034 & $\cdots$
\enddata
\end{deluxetable}

\begin{deluxetable}{cccc}
\tabletypesize{\tiny} \tablecaption{T-Cyg1-03378 Eclipse Timing}
\tablewidth{0pt}
\tablehead{\colhead{Eclipse Type} & \colhead{Epoch (HJD)} & \colhead{O-C [sec]} & \colhead{Data Source}}
\startdata
Secondary  & 2453171.6943  & $811^{+279}_{-244}$ &  Sleuth \\
Secondary  & 2453173.7506  & $693^{+185}_{-184}$ &  Sleuth \\
Primary    & 2453174.7788  & $-78^{+78}_{-121}$ &  Sleuth \\
Secondary  & 2453175.8070  & $242^{+106}_{-99}$ &   Sleuth\\
Primary    & 2453176.8352  & $-52^{+150}_{-186}$ &  Sleuth \\
Secondary  & 2453177.8634  & $-281^{+135}_{-128}$ &  Sleuth \\
Primary    & 2453178.8916  & $44^{+72}_{-72}$ &  Sleuth \\
Secondary  & 2453179.9198  & $-44^{+118}_{-132}$ &  Sleuth  \\
Primary    & 2453180.9480  & $-74^{+99}_{-116}$ &  Sleuth  \\
Secondary  & 2453181.9762  & $-61^{+173}_{-150}$ & Sleuth   \\
Secondary  & 2453206.6527  & $116^{+339}_{-242}$ & Sleuth + PSST \\
Primary    & 2453207.6809  & $212^{+141}_{-151}$ & Sleuth + PSST    \\
Secondary  & 2453208.7091  & $-371^{+94}_{-90}$ &  Sleuth + PSST   \\
Primary    & 2453209.7373  & $-209^{+114}_{-117}$ &  Sleuth   \\
Secondary  & 2453210.7655  & $-350^{+150}_{-208}$ & Sleuth   \\
Primary    & 2453211.7937  & $-103^{+171}_{-134}$ &  Sleuth + PSST \\
Secondary  & 2453212.8219  & $-298^{+196}_{-177}$ &  Sleuth  \\
Primary    & 2453213.8501  & $-249^{+144}_{-239}$ & Sleuth + HATNet \\
Secondary  & 2453214.8783  & $-123^{+97}_{-93}$ &  Sleuth  + HATNet\\
Primary    & 2453215.9064  & $114^{+110}_{-120}$ & Sleuth  + HATNet \\
Secondary  & 2453216.9346  & $48^{+219}_{-190}$ &  Sleuth + PSST  \\
Primary    & 2453217.9628  & $207^{+217}_{-204}$ & Sleuth + HATNet \\
Secondary  & 2453218.9910  & $-74^{+220}_{-320}$ & Sleuth + HATNet  \\
Primary    & 2453220.0192  & $139^{+189}_{-244}$ &  Sleuth + HATNet \\
Primary    & 2453242.6394  & $-61^{+200}_{-253}$ &   Sleuth + HATNet \\
Secondary  & 2453243.6676  & $55^{+236}_{-186}$ &  Sleuth + HATNet \\
Primary    & 2453244.6958  & $-6^{+135}_{-134}$ &  Sleuth  + HATNet \\
Secondary  & 2453245.7240  & $89^{+163}_{-256}$ &  Sleuth + HATNet\\
Primary    & 2453246.7522  & $280^{+171}_{-187}$ &  Sleuth + HATNet\\
Secondary  & 2453247.7804  & $-16^{+195}_{-198}$ &  Sleuth + HATNet\\
Primary    & 2453248.8085  & $225^{+153}_{-150}$ &  Sleuth + HATNet\\
Secondary  & 2453249.8367  & $799^{+195}_{-186}$ &  Sleuth + HATNet\\
Primary    & 2453250.8649  & $34^{+265}_{-268}$ &  HATNet \\
Secondary  & 2453251.8931  & $389^{+341}_{-484}$ &  HATNet \\
Primary    & 2453252.9213  & $1227^{+296}_{-445}$ & HATNet  \\
Primary    & 2453254.9777  & $533^{+283}_{-389}$ &   HATNet \\
Secondary  & 2453278.6261  & $-223^{+472}_{-308}$ &  HATNet \\
Primary    & 2453279.6543  & $-570^{+386}_{-431}$ &  HATNet \\
Primary    & 2453281.7106  & $-115^{+374}_{-365}$ &  HATNet \\
Secondary  & 2453282.7388  & $-574^{+460}_{-329}$ &  HATNet \\
Primary    & 2453540.8147  & $1292^{+315}_{-447}$ &  HATNet   \\
Secondary  & 2453541.8428  & $101^{+269}_{-286}$ &   HATNet \\
Primary    & 2453542.8710  & $914^{+449}_{-655}$ &   HATNet  \\
Secondary  & 2453545.9556  & $197^{+366}_{-289}$ &   HATNet \\
Primary    & 2453546.9838  & $590^{+333}_{-347}$ &   HATNet  \\
Secondary  & 2453548.0120  & $-1422^{+2185}_{-414}$ &   HATNet  \\
Primary    & 2453549.0402  & $202^{+402}_{-302}$ &   HATNet  \\
Secondary  & 2453550.0684  & $16^{+585}_{-1096}$ &   HATNet  \\
Primary    & 2453551.0966  & $-196^{+320}_{-303}$ &   HATNet  \\
Primary    & 2453575.7731  & $811^{+566}_{-996}$ &    HATNet \\
Secondary  & 2453576.8013  & $-205^{+219}_{-194}$ &  HATNet   \\
Primary    & 2453577.8295  & $-67^{+266}_{-364}$ &    HATNet \\
Secondary  & 2453578.8577  & $-467^{+391}_{-235}$ &   HATNet  \\
Primary    & 2453579.8859  & $132^{+409}_{-1003}$ &   HATNet  \\
Secondary  & 2453580.9141  & $370^{+296}_{-277}$ &    HATNet \\
Primary    & 2453581.9423  & $241^{+375}_{-362}$ &   HATNet  \\
Secondary  & 2453582.9705  & $130^{+408}_{-340}$ &   HATNet  \\
Primary    & 2453583.9987  & $-417^{+447}_{-258}$ &   HATNet  \\
Secondary  & 2453585.0268  & $164^{+260}_{-275}$ &    HATNet \\
Primary    & 2453586.0550  & $-387^{+357}_{-306}$ &   HATNet  \\
Secondary  & 2453587.0832  & $-310^{+366}_{-509}$ &   HATNet  \\
Primary    & 2453588.1114  & $-140^{+306}_{-368}$ &    HATNet \\
Primary    & 2453612.7880  & $219^{+315}_{-291}$ &    HATNet \\
Secondary  & 2453613.8162  & $280^{+327}_{-299}$ &   HATNet  \\
Primary    & 2453614.8444  & $-481^{+447}_{-401}$ &   HATNet  \\
Secondary  & 2453615.8726  & $261^{+322}_{-297}$ &   HATNet  \\
Primary    & 2453616.9008  & $-1078^{+251}_{-243}$ &  HATNet   \\
Secondary  & 2453617.9289  & $413^{+303}_{-288}$ &  HATNet   \\
Primary    & 2453618.9571  & $-435^{+384}_{-387}$ &   HATNet  \\
Primary    & 2453621.0135  & $768^{+371}_{-995}$ & HATNet
\enddata
\end{deluxetable}

{}


\end{document}